\RequirePackage{ifpdf}
\documentclass[hyper,12pt,letterpaper]{JHEP3}

\usepackage{graphicx}
\usepackage{latexsym,amsmath,amsfonts,amssymb}
\usepackage{mathrsfs}
\usepackage[makeroom]{cancel}
\usepackage{bbm}
\usepackage{bm}
\usepackage{subfigure}
\usepackage{paralist}
\usepackage{url}
\usepackage[multiple]{footmisc}

\newcommand{\vol}{\mathrm{vol}}

\newcommand{\slashed}{{\bf\not}}

\newcommand{\ie}{\textit{i.e.}}

\numberwithin{equation}{section}

\newcommand{\nn}{\nonumber}
\newcommand{\mat}[1]{\begin{pmatrix} #1 \end{pmatrix}}

\newcommand{\be}{\begin{equation}} 
\newcommand{\ee}{\end{equation}}
\newcommand{\bea}{\begin{equation} \begin{aligned}} \newcommand{\eea}{\end{aligned} \end{equation}}

\newcommand{\bit}{\begin{itemize}} 
\newcommand{\eit}{\end{itemize}}

\newcommand{\cC}{\mathcal{C}}

\newcommand{\cF}{\mathcal{F}}
\newcommand{\cG}{\mathcal{G}}
\newcommand{\cH}{\mathcal{H}}
\newcommand{\cI}{\mathcal{I}}

\newcommand{\cL}{\mathcal{L}}
\newcommand{\cM}{\mathcal{M}}
\newcommand{\cN}{\mathcal{N}}

\newcommand{\cS}{\mathcal{S}}

\newcommand{\cW}{\mathcal{W}}

\newcommand{\bH}{\mathbb{H}}

\newcommand{\bQ}{\mathbb{Q}}

\newcommand{\Q}{\mathbb{Q}}
\newcommand{\Z}{\mathbb{Z}}
\newcommand{\C}{\mathbb{C}}
\newcommand{\R}{\mathbb{R}}

\renewcommand{\t}{\widetilde }
\renewcommand{\d}{\partial }

\renewcommand{\b}{\bar }

\newcommand{\half}{{1\over 2}}

\newcommand{\bz}{{\b z}}

\newcommand{\CA}{\mathcal{A}}

\newcommand{\CC}{\mathcal{C}}
\newcommand{\CD}{\mathcal{D}}

\newcommand{\CF}{\mathcal{F}}
\newcommand{\CG}{\mathcal{G}}
\newcommand{\CH}{\mathcal{H}}
\newcommand{\CI}{\mathcal{I}}
\newcommand{\CJ}{\mathcal{J}}
\newcommand{\CK}{\mathcal{K}}
\newcommand{\CL}{\mathcal{L}}
\newcommand{\CM}{\mathcal{M}}
\newcommand{\CN}{\mathcal{N}}
\newcommand{\CO}{\mathcal{O}}

\newcommand{\CQ}{\mathcal{Q}}
\newcommand{\CR}{\mathcal{R}}
\newcommand{\CS}{\mathcal{S}}
\newcommand{\CT}{\mathcal{T}}

\newcommand{\CV}{\mathcal{V}}
\newcommand{\CW}{\mathcal{W}}

\newcommand{\CZ}{\mathcal{Z}}

\newcommand{\FR}{\mathfrak{R}}
\newcommand{\Fg}{\mathfrak{g}}

\newcommand{\GG}{\mathbf{G}}

\newcommand{\GH}{\mathbf{H}}

\newcommand{\rk}{{{\rm rk}(\GG)}}

\newcommand{\m}{\mathfrak{m}}
\newcommand{\n}{\mathfrak{n}}
\newcommand{\fq}{\mathfrak{q}}
\newcommand{\q}{\mathfrak{q}}
\newcommand{\kk}{\mathrm{k}}

\newcommand{\ba}{{\b a}}
\newcommand{\bb}{{\bb b}}

\newcommand{\h}{\hat}

\newcommand{\Mgp}{\CM_{g,p}}
\newcommand{\pif}{{\Pi}}

\DeclareMathOperator{\Tr}{Tr}

\DeclareMathOperator{\sign}{sign}

\newcommand{\SL}{{\mathscr L}}

\newcommand{\ov}{\over}

\newcommand{\fM}{\mathfrak{M}}

\newcommand{\dilog}{{\text{Li}_2}}

\newcommand{\floor}[1]{ \left\lfloor{ #1} \right\rfloor}

\newcommand{\Pic}{{\rm Pic}}
\newcommand{\bd}{{\bf d}}

\title{
Seifert fibering operators in 3d $\CN=2$  theories
}

\author{Cyril~Closset,$^\flat$ Heeyeon~Kim$^\sharp$ and Brian Willett$^{\natural}$\\

{}$^{\flat}$Theory Department, CERN\\
CH-1211, Geneva 23, Switzerland\\
{}$^{\sharp}$ Mathematical Institute, University of Oxford \\
Woodstock Road, Oxford, OX2 6GG, United Kingdom   \\
{}$^{\natural}$ Kavli Institute for Theoretical Physics\\
 University of California, Santa Barbara, CA 93106
}

\preprint{CERN-TH-2018-156}

\abstract{We study 3d $\CN=2$ supersymmetric gauge theories on closed oriented Seifert manifolds---
circle bundles over an orbifold Riemann surface---, with a gauge group $\GG$ given by a product of simply-connected and/or unitary Lie groups. Our main result is an exact formula for the supersymmetric partition function on any Seifert manifold, generalizing previous results on  lens spaces. We explain how the result for an arbitrary Seifert geometry can be obtained  by combining simple building blocks,  the   ``fibering operators.''  These  operators are half-BPS line defects, whose insertion along the $S^1$ fiber has the effect of changing the topology of the Seifert fibration.
We also point out that most supersymmetric partition functions on Seifert manifolds admit a discrete refinement, corresponding to the freedom in choosing a three-dimensional spin structure.
As a strong consistency check on our result, we show that the Seifert partition functions match exactly across infrared dualities. The duality relations are given by intricate (and seemingly new) mathematical identities, which we tested numerically. 
Finally, we discuss in detail the supersymmetric partition function on the lens space $L(p,q)_b$ with rational squashing parameter $b^2 \in \mathbb{Q}$, comparing our formalism to previous results, and explaining the relationship between the fibering operators and the three-dimensional holomorphic blocks. 
}

\begin{document}

\tableofcontents

\section{Introduction}\label{sec: intro}
Any local quantum field theory (QFT) can be studied on a non-trivial space-time geometry. This is done by coupling the stress-energy tensor to a background metric.~\footnote{At first order in the background metric. The higher-order terms are (partially) constrained by diffeomorphism invariance.} In this paper, we study  three-dimensional Euclidean QFTs with $\CN=2$ supersymmetry---that is, four Poincar\'e supercharges---on Seifert three-manifolds.

In any supersymmetric QFT, the stress-energy tensor sits together with the supersymmetry current in a supercurrent multiplet,  which can be canonically coupled to a supergravity multiplet. Given a choice of supercurrent \cite{Komargodski:2010rb, Dumitrescu:2011iu} and of the corresponding off-shell supergravity, one can classify the supersymmetry-preserving geometries systematically \cite{Festuccia:2011ws}. 
In the case of 3d $\CN=2$ supersymmetric theories with an exact $U(1)_R$ symmetry \cite{Klare:2012gn, Closset:2012ru, Kuzenko:2013uya}, one finds a large class of compact half-BPS geometries, $\CM_3$, which preserve two supercharges, $\CQ$ and $\b\CQ$, satisfying the curved-space supersymmetry algebra:
\be\label{QQb intro}
\CQ^2= 0~, \qquad\quad \b\CQ^2=0~, \qquad \quad \{ \CQ, \b \CQ\} =-2i \left(\CL_K+ Z\right)~.
\ee
Here, $Z$ is the real central charge of the 3d $\CN=2$ supersymmetry algebra on $\R^3$, and $\CL_K$ generates an isometry of the Riemannian manifold $\CM_3$ along a real Killing vector~$K$. A necessary and sufficient condition for such a half-BPS background to exist is that $\CM_3$ be a Seifert manifold \cite{Closset:2012ru}.

Given any half-BPS geometric background, one can, in principle, compute the {\it supersymmetric partition function}:
\be\label{ZM3 intro 0}
Z_{\CM_3}(\nu)~,
\ee
of any UV-free $\CN=2$ supersymmetric field theory, using supersymmetric localization techniques. (See {\it e.g.} \protect\cite{Pestun:2016zxk} for a recent review.) On general grounds, the quantity \eqref{ZM3 intro 0} is renormalization-group (RG) invariant, therefore it gives access to non-perturbative information about the strongly-coupled infrared (IR) of the theory. It is also a function of background vector multiplets for the flavor symmetries of the theory, in particular through some complex parameters $\nu$, as indicated in \eqref{ZM3 intro 0}. Moreover,  $Z_{\CM_3}(\nu)$ is locally holomorphic in those parameters, $\nu$ \cite{Closset:2012vg, Closset:2013vra}.

Explicit localization formulas are known for \eqref{ZM3 intro 0} when $\CM_3$ is a lens space---see {\it e.g.} \cite{Kapustin:2009kz, Gang:2009wy, Jafferis:2010un,Hama:2010av, Hama:2011ea, Benini:2011nc, Kallen:2011ny, Beem:2012mb, Dimofte:2014zga}.
However, localization computations become increasingly complicated to carry out as the topology of $\CM_3$ becomes less trivial, in part because the sum over topological sectors becomes more involved. In this work, we will bypass  such difficulties by thinking of $Z_{\CM_3}$ as an observable in an auxilliary two-dimensional topological field theory, the so-called ``3d A-model.''~\footnote{Note that, in the terminology of this paper, the ``{\it 3d} A-model'' is really a {\it 2d} TQFT. Related works that used a 3d TQFT-like approach include {\it e.g.} \protect\cite{Kapustin:2010ag, Cecotti:2013mba, Closset:2014uda, Gukov:2016gkn, Aganagic:2017tvx}.}

The basic idea is the following. Let us first consider the product space:
\be
\CM_3 \cong \Sigma \times S^1~,
\ee
with $\Sigma$ a Riemann surface. One can preserve the supersymmetry algebra \eqref{QQb intro} on this product three-manifold by performing a topological A-twist along the Riemann surface \cite{Witten:1988xj}. The two supercharges $\CQ$ and $\b\CQ$ are the A-twisted version of the flat-space supercharges $Q_-$ and $\b Q_+$ in  the 3d $\CN=2$ algebra---equivalently,  if we consider the theory on $\R^2 \times S^1$, $Q_-$ and $\b Q_+$ are part of the 2d $\CN=(2,2)$ supersymmetry algebra along $\R^2$. We then consider the operators, $\SL$, that commute with them:
\be
[Q_-, \SL]=0~,\qquad \qquad [\b Q_+, \SL]=0~. 
\ee
These are the {\it twisted chiral operators}, in the 2d nomenclature. Note that the twisted-chiral condition is not Lorentz covariant in dimension larger than two, so that $\SL$ cannot be a local operator in 3d. In the three-dimensional $\CN=2$ theory on $\R^2 \times S^1$, the twisted chiral operators are {\it half-BPS line operators}, wrapping the $S^1$ and localized at a point on $\Sigma$. Typical examples are the half-BPS Wilson loops in 3d $\CN=2$ gauge theories. The {\it 3d A-model} is defined as the two-dimensional topological quantum field theory (TQFT) on $\Sigma$  obtained by viewing the 3d theory as a 2d theory with an infinite number of fields (the $S^1$ Fourier modes), and by performing the topological A-twist. The TQFT is obtained by passing to the (simultaneous) cohomology of the scalar supercharges $\CQ, \b\CQ$. The 3d A-model observables are of the form:
\be
\left\langle \SL_i \SL_j \SL_k \cdots \right\rangle_{\Sigma\times S^1}~,
\ee
where the insertion points of the lines can be omitted, since the theory is topological along $\Sigma$.
These observables encode the algebra of half-BPS line operators---see {\it e.g.} \cite{Kapustin:2013hpk, Closset:2016arn} for detailed discussions of the Wilson loop algebras. In this work, we will view the supersymmetric partition functions as 3d A-model observables:
\be\label{ZM3 as VEV intro}
Z_{\CM_3} = \big\langle \CG_{\CM_3}\big\rangle_{\Sigma \times S^1}~,
\ee
where the line $\SL= \CG_{\CM_3}$ is a particular line defect, or {\it geometry-changing line operator}, whose insertion is equivalent to introducing a non-trivial fibration of the $S^1$ over $\Sigma$, giving rise to the Seifert manifold $\CM_3$. In a previous work \cite{Closset:2017zgf}, we carried out this program for a restricted class of Seifert geometries. In the present paper, we define the geometry-changing line operator for any Seifert manifold, in the case of 3d $\CN=2$ gauge theories. This gives us a compact formula for the supersymmetric partition functions, \eqref{ZM3 intro 0}, for supersymmetric gauge theories on any half-BPS Seifert geometry. Even in the previously-understood cases when $\CM_3$ is a lens space, our results offers a new perspective on some well-known matrix integrals obtained by supersymmetric localization. We also clarify a number of more subtle points along the way.

\

\noindent In the remainder of this introduction, we review some necessary background material and spell out our main results in some detail.

\subsection{Seifert geometry and surgery}\label{sec:intro surgery}
By a Seifert manifold $\CM_3$, we mean a  closed,  oriented three-manifold, equipped with a Seifert fibration:
\be\label{seifert fib intro}
\pi \; : \;  \CM_3 \rightarrow \h\Sigma_{g,n}~.
\ee
For our purposes, a  Seifert fibration is simply an $S^1$ bundle over a two-dimensional orbifold  \cite{orlik1972seifert, BLMS:BLMS0401}. Here the orbifold base of the fibration, $\h\Sigma = \h\Sigma_{g,n}$, is a genus-$g$  Riemann surface with $n$ marked points, the ``orbifold points,'' $x_i$, which have  conical local neighborhoods $U_i \cong \C/\Z_{q_i}$, with $q_i>0$. 
In the absence of orbifold points, the Seifert fibration is a principal circle bundle over a smooth Riemann surface $\Sigma_g$, which is fully determined by its degree (or first Chern number) ${\bf d}\in \Z$. More generally,  we must also specify what happens at the fibers above the orbifold points, the so-called ``exceptional fibers.''  
The tubular neighborhood of each exceptional fiber in $\CM_3$ is a solid fibered torus, which is characterized by a pair of co-prime integers $(q_i,p_i)$.~\footnote{A solid fibered torus $T(q,t)$ is obtained by gluing together the two disk boundaries of a solid cylinder, with a relative rotation of ${2\pi t\ov q}$, for $q$ and $t$ some co-prime integers. One can equivalently describe $T(q,t)$ in terms of the coprime integers $(q,p)$ such that $p t= 1$ mod $q$, as we are doing here.} Note that, while a generic $S^1$ fiber has a fixed radius $\beta$, the radius jumps to $\beta/q_i$ over the orbifold point $x_i$.

In this way, the Seifert manifold $\CM_3$ is fully characterized by a finite numbers of integers, known as the ``Seifert symbols:''
\be\label{def M3 intro}
\CM_3 \cong \big[\bd~; \, g~; \, (q_1, p_1)~, \cdots~, (q_n, p_n)\big]~.
\ee
Here $\bd$ is the degree of the Seifert fibration, $g$ is the genus of the base, and $(q_i,p_i)$ are the so-called {\it Seifert invariants}  of the exceptional fibers.~\footnote{More precisely, $\bf d$ is the degree if the invariants $(q_i, p_i)$ are normalized such that $q_i>0$ and $0 \leq p_i<q_i$. We will give a detailed introduction to Seifert geometry in Section~\protect\ref{sec: seifert backgs}.}
Any such $\CM_3$ can be constructed by surgery on the product manifold:
\be
\Sigma_g \times S^1 \cong \big[0~; \, g~; \, \big]~, 
\ee
seen as a trivial Seifert fibration. Indeed, given any Seifert manifold $\CM_3^{(n)}$ with $n$ exceptional fibers, one can add an exceptional $(q,p)$ fiber by Dehn surgery along a generic Seifert fiber at $x_{n+1} \in \h\Sigma$, by removing a tubular neighborhood of the fiber, resulting in a manifold $\t\CM_3^{(n)}$ with a boundary $\d \t\CM_3^{(n)} \cong T^2$, and constructing a new compact three-manifold:
\be
\CM_3^{(n+1)} \cong \t\CM_3^{(n)} \cup_g (D^2 \times S^1)~,
\ee
by gluing $\t\CM_3$ to a solid torus  with an $SL(2, \Z)$ twist:
\be
g\; : \; \d\t\CM_3\rightarrow \d(D^2\times S^1)~, \qquad g= \mat{q& -t \\ p & s}\in SL(2, \Z)~.
\ee
Shifting the degree ${\bf d}$ to ${\bf d}+1$ in \eqref{def M3 intro} can be done similarly, with $g \in SL(2,\Z)$ corresponding to $(q,p)=(1,1)$.

In fact, any Seifert manifold can be constructed by elementary surgery operations on the genus-zero trivial fibration $S^2\times S^1$. One may consider the following operations (in whatever order;  all these operations are  reversible):
\begin{itemize}
\item Add handles to the base $\h\Sigma_{g,n}$, for instance going from $S^2\times S^1$ to $\Sigma_g \times S^1$. This operation only affects the base of the fibration.
\item Change the degree ${\bf d}$ of the fibration. This operation leaves the base invariant.
\item Add an exceptional $(q,p)$ fiber. This operation modifies both the base and the fibration.
\end{itemize}
These three topological operations can be implemented in $\CN=2$ supersymmetric field theories on $\CM_3$,  by the insertion of geometry-changing line operators along the Seifert fiber~\cite{Closset:2017zgf}, as anticipated in  \eqref{ZM3 as VEV intro}.
 In the case of a smooth base $\h\Sigma= \Sigma_g$, the line operators for the first and second operations were discussed in \cite{Nekrasov:2014xaa, Benini:2016hjo, Closset:2016arn} and \cite{Closset:2017zgf}, respectively. The main goal of this paper is to explain how to carry out the third operation---the insertion of an exceptional Seifert fiber---, thus allowing us to study 3d $\CN=2$ theories on any Seifert manifold.

\subsection{Gauge theories on Seifert manifolds}
Let us choose a Riemannian metric on $\CM_3$ compatible with the Seifert fibration \eqref{def M3 intro}. In particular, a compatible metric admits a Killing vector $K$ whose orbits are the Seifert fibers. In this paper, we will {\it assume} that $K$ is the Killing vector entering the curved-space supersymmetry algebra \eqref{QQb intro}. As we will see, this assumption is less restrictive than it might appear.

We consider any 3d $\CN=2$ gauge theory with gauge group $\GG$, and  with $\Fg= {\rm Lie}(\GG)$ its Lie algebra. Here, $\GG$  can be a compact, simply-connected, simple Lie group, $\GG_\gamma$, a  unitary group, or a product thereof:
\be\label{GG intro}
\GG = \prod_\gamma \GG_\gamma \times  \prod_I U(N_I)~.
\ee
The inclusion of non-simply-connected gauge groups (other than $U(N)$) requires additional care, and we leave it for future work.

We may decompose any 3d field in Kaluza-Klein (KK) modes along the Seifert fiber. In particular, the zero-mode of the 3d gauge field gives us a two-dimensional gauge field on $\h\Sigma$, which sits in a 2d $\CN=(2,2)$ vector multiplet $\CV_{(\rm 2d)}$. It will be natural to write down an effective field theory for the complex scalar $u$ in $\CV_{(\rm 2d)}$ on the classical Coulomb branch:
\be
u = {\rm diag}(u_a)~, \qquad a=1, \cdots, \rk~,
\ee
by integrating out all the other massive fields at generic values of $u_a$, and with various mass parameters, $\nu$, for the flavor symmetries turned on. The topological twist of this effective two-dimensional field theory will be our ``3d A-model.''

Then, any half-BPS line operators $\SL$ in the A-model will be expressed as functions of the gauge parameters, $u$, and of the flavor parameters, $\nu$:
\be\label{def SL intro}
\SL= \SL(u, \nu)~.
\ee
As an example, consider $\SL= W_\FR$, a supersymmetric Wilson loop in a representation $\FR$ of $\GG$, wrapped along the $S^1$ fiber. It takes the form:
\be
 W_{\FR} = \Tr_{\FR} {\rm Pexp}{\left( -i \int_{S^1} \left(a_\mu dx^\mu - i \beta \sigma d\psi\right)\right)}~,
\ee
with $\sigma$ the real scalar in the 3d vector multiplet, and $\psi$ the circle coordinate.
In the 3d A-model, this is simply the character of the representation:
\be
 W_{\FR}(u) =  \Tr_{\FR}(e^{2\pi i u})~.
\ee
Here, we are interested in defect line operators that change the topology of $\CM_3$.
The geometry-changing line operators: 
\be\label{hgo and f intro}
\CH=\CH(u, \nu)~, \qquad\qquad  \CF= \CF(u, \nu)~,
\ee
were discussed in previous papers \cite{Nekrasov:2014xaa, Gukov:2015sna, Benini:2016hjo, Closset:2016arn, Closset:2017zgf}.
The handle-gluing operator $\CH$ \cite{Vafa:1990mu, Nekrasov:2014xaa} has the effect of adding one handle to the base $\h\Sigma$ of the Seifert fibration $\CM_3$. The ``ordinary'' fibering operator $\CF$  \cite{Closset:2017zgf} has the effect of shifting the degree of the Seifert fibration by one, ${\bf d} \rightarrow {\bf d}+1$. In this paper, we define and compute the {\it $(q,p)$-fibering operator}:
\be\label{qp fibering op intro}
\CG_{q,p}= \CG_{q,p}(u, \nu)~,
\ee
which introduces an exceptional fiber of type $(q,p)$. 
Given these building blocks, we can write the supersymmetric partition function \eqref{ZM3 as VEV intro} as:
\be \label{INTROZM3CL}
Z_{\CM_3} = \left\langle \SL_{\CM_3}\right\rangle_{S^2 \times S^1}~, \qquad   \SL_{\CM_3} \equiv  \CH^g\, \CG_{\CM_3}= \CF^{\bf d}\, \CH^g\,\prod_{i=1}^n \CG_{q_i, p_i}~,
\ee
schematically, for any Seifert fibration. Here, the $S^2 \times S^1$ background on which we insert $\SL_{\CM_3}$ corresponds to the topologically twisted index \cite{Benini:2015noa}. The main object of this work is to understand and compute the {\it Seifert fibering operator:}
\be
 \CG_{\CM_3}\equiv \CF^{\bf d}\,\prod_{i=1}^n \CG_{q_i, p_i}~,
\ee
associated to an arbitrary Seifert manifold $\CM_3$, with Seifert symbols \eqref{def M3 intro}.

\subsection{R-charge and spin-structure dependence of $Z_{\CM_3}$}\label{intro: subsec: LR}
As part of our choice of supersymmetric background on $\cM_3$, we must specify the $U(1)_R$ line bundle, ${\bf L}_R$, associated to the background R-symmetry gauge field.  This line bundle is defined by the condition:
\be \label{LR cond intro}
{\bf L}_R^{\otimes 2} = \CK_{\CM_3}~, 
\ee
where ${\bf {\cal K}}_{\cM_3}$ is the ``canonical line bundle'' of the Seifert manifold $\CM_3$ seen as a transversely holomorphic foliation (THF) \cite{Closset:2012ru, Closset:2013vra}. For our purposes, $\CK_{\CM_3}$ can simply be defined as the pull-back of the canonical line bundle on the orbifold $\hat{\Sigma}$ along the Seifert fibration \eqref{seifert fib intro}:
\be
 \CK_{\CM_3}\cong  \pi^\ast(\CK_{\h\Sigma})\ .
\ee
  There are, in general, many solutions to \eqref{LR cond intro} for a fixed $\CM_3$. The choices are in one-to-one (but non-canonical) correspondence with the group $H^1(\cM_3,\Z_2)$, namely:
\be \label{INTROlrh1}
 \big\{ \text{valid R-symmetry line bundles ${\bf L}_R$ on $\cM_3$} \big\} \;\;\; \cong   \;\;\; H^1(\cM_3,\Z_2)~.
 \ee
Note this group also determines the allowed set of {\it spin structures} on $\cM_3$.  In fact, these two choices, of a line bundle ${\bf L}_R$ and of a spin structure on $\cM_3$, are correlated.  The supersymmetric background on $\CM_3$ includes a pair of Killing spinors, $\zeta$ and $\b\zeta$, of R-charge $\pm 1$, respectively, which solve the generalized Killing spinor equations:
\be \label{INTROKSE}
 (\h\nabla_\mu - i\CA^{(R)}_\mu) \zeta = 0~,\qquad\qquad (\h\nabla_\mu + i \CA^{(R)}_\mu) \b{\zeta} = 0~,
 \ee
where $\h\nabla_\mu$ is a particular connection adapted to the Seifert geometry, and $\CA^{(R)}_\mu$ is the connection on ${\bf L}_R$. 
Whenever we change the R-symmetry line bundle on $\CM_3$, the old and new $U(1)_R$ gauge fields are distinguished by $\Z_2$-valued holonomies along certain $1$-cycles, determined by an element of $H^1(\cM_3,\Z_2)$.  In order to retain the two solutions to the Killing spinor equations \eqref{INTROKSE}, we must simultaneously shift the spin structure in such a way as to cancel the holonomy incurred by the Killing spinors.  In this sense, different elements in \eqref{INTROlrh1} correspond to different choices of a spin structure on the Seifert manifold.

The partition function of a 3d $\cN=2$ gauge theory depends in a subtle way on the choice of spin structure, as we will see in later sections.  In particular, in the special case of ($\CN=2$ supersymmetric) Chern-Simons (CS) theories, this reproduces the  spin-structure dependence expected in certain cases, whenever the CS theory is known to be a spin-TQFT  \cite{Dijkgraaf:1989pz}. 

An additional feature of the choice of R-symmetry line bundle is that it determines the allowed R-charges for matter fields.  Namely, the R-charges, $r$, must be such that the bundle ${\bf L}_R^{\otimes r}$ is well-defined.  This typically forces the R-charges to be integer-quantized, $r\in \Z$. 
In some special cases, we may relax this condition.  In particular, in the special case of a topologically trivial R-symmetry line bundle:
\be\label{LR trivial INTRO}
{\bf L}_R \cong {\cal O}~,
\ee
we may allow arbitrary real R-charges, $r\in \R$. Such backgrounds are particularly important if we want to study the $\CN=2$ superconformal field theories (SCFT) that may appear in the IR of UV-free gauge theories. 
Indeed, the superconformal R-charges are often irrational, due to mixing of the UV R-charge with abelian flavor symmetries.~\footnote{The SCFT R-charges can often be determined by F-maximization \protect\cite{Jafferis:2010un, Closset:2012vg}.} 
A topologically trivial $U(1)_R$ line bundle is allowed only on certain choices of $\CM_3$. Important examples are the so-called squashed sphere $ S^3_b$, and more generally the squashed  lens spaces $L(p,-1)_b$.
These examples belong to more general family of Seifert manifolds with topologically trivial ${\bf L}_R$, the ``spherical manifolds,'' which are described in Section~\ref{subsec: spherical manifolds}.  
(Other interesting examples are the torus bundles described in Section~\ref{subsec: torus bundle}.)

\subsection{Parity anomaly, Chern-Simons contact terms and supersymmetry}\label{intro:subsec:parity}
In this work, we are careful to treat fermions in a manner consistent with gauge invariance and the parity anomaly \cite{Redlich:1983dv, Niemi:1983rq, AlvarezGaume:1984nf}---see \cite{Closset:2012vp, Witten:2015aba, Witten:2016cio, Seiberg:2016gmd} for detailed recent discussions.
Since our treatment differs from most of the supersymmetric localization literature, let us elaborate on this point here. For completeness, we provide more background material on 3d fermions, the parity anomaly, and supersymmetric CS terms, in Appendix~\ref{app: parity anomaly}.

\paragraph{Parity anomaly and CS contact terms.}
Consider a massless Dirac fermion $\psi$  coupled to a background  gauge field $A_\mu$. The parity anomaly is the statement that we cannot quantize $\psi$ while preserving {\it both} three-dimensional parity~\footnote{On $\R^3$ with Euclidean signature, parity acts by inverting the sign of a single coordinate.} and gauge invariance. In this work, we always wish to preserve gauge invariance, and therefore the effective action $S_{\rm eff}[A]$ obtained after integrating over the (possibly massless) fermions generally violates parity.

The relevant parity-violating terms contribute to the imaginary part of $S_{\rm eff}[A]$. They are conveniently captured by parity-odd contributions to the two-point functions of conserved currents  \cite{Closset:2012vp}, with coefficients denoted by $\kappa\in \R$. Consider various abelian symmetries $U(1)_a$ coupled to our fermions. (The generalization to non-abelian symmetries is straightforward.) In a general theory, we have the contributions:
\be\label{kappa intro}
\kappa_{ab}~, \qquad \qquad \kappa_g~,
\ee
where $\kappa_{ab}$ is the contribution from the two-point functions of $U(1)_a$ conserved currents, and $\kappa_g$ is the gravitational contributions (from the two-point function of the stress-energy tensor). The $\kappa$ coefficients in \eqref{kappa intro} are called the ``Chern-Simons contact terms,'' by a slight abuse of notation. They are physical {\it modulo integers.} This is because we always have the freedom of adding Chern-Simons terms to the effective action:
\be\label{shift Seff}
S_{\rm eff}[A] \;\rightarrow\; S_{\rm eff}[A] + k\, S_{\rm CS}[A]+ k_g\, S_{\rm grav}[g]~,
\ee 
where $S_{\rm CS}$ and $S_{\rm grav}$ are the $U(1)$ and gravitational CS actions, respectively (with $g$ a background metric). The CS levels $k$ and $k_g$ are integer-quantized, as required by gauge invariance. The shift \eqref{shift Seff} induces a shift of the CS contact terms:
\be\label{CS shifts intro}
\kappa \rightarrow \kappa+ k~, \qquad \qquad \kappa_g \rightarrow \kappa_g + k~.
\ee

In the UV, the gauge theory is free, and the only contribution to $\kappa$ is from free fermions coupled to gauge fields, whether dynamical gauge fields or background gauge fields for global symmetries, and from the CS terms themselves.~\footnote{Of course, along the RG flow,  we should distinguish between dynamical and background gauge fields, but in the far UV we are just quantizing free fermions in the background of some arbitrary gauge fields. 
The integration over the dynamical gauge fields should be done at a later stage.} Consider then a single free fermion $\psi$ coupled to $A_\mu$ with $U(1)$ charge $1$. One can consider the so-called ``$U(1)_{-\half}$ quantization'' for $\psi$, which corresponds to having the UV contact terms:~\footnote{The notation $U(1)_{-\half}$ comes from the fact that $\kappa=-\half$ is often called the ``effective CS level.'' In this paper, we distinguish carefully between the ``CS contact term'' $\kappa$, which is real (and half-integer in the free UV), and the CS level $k$, which is always integer-quantized.}
\be\label{kappa UV intro}
\kappa=-\half~, \qquad \qquad \kappa_g=-1~.
\ee
Any other choice of quantization is related to this one by a shift of the UV CS levels, as in \eqref{CS shifts intro}, and is a matter of convention. For instance, the ``$U(1)_\half$ quantization'' would correspond to $\kappa=\half$ and $\kappa_g=1$. Importantly, for a single Dirac fermion, there exists no gauge-invariant scheme in which $\kappa=0$ in the UV.~\footnote{Similarly, we have $\kappa_g=-\half$ mod $1$ for a Majorana fermion. In this work, we only consider Dirac fermions (that is, an even number of Majorana fermions), so that $\kappa_g$ will always be integer in the UV.} We refer to Appendix~\ref{app: parity anomaly} for a more detailed discussion. 

Finally, we may also consider adding a ``real mass'' $m$ for $\psi$, which breaks parity explicitly. In that case, we can integrate out the fermion in the IR, which has the effect of shifting the CS contact terms according to:
\be\label{kappa shift intro}
\delta \kappa = \half \sign(m)~, \qquad \qquad \delta \kappa_g = \sign(m)~.
\ee
In particular, for a free fermion in the ``$U(1)_{-\half}$ quantization'' and with a positive mass $m>0$, we obtain the net CS terms $\kappa=\kappa_g=0$ in the IR, since the shift \eqref{kappa shift intro} cancels the UV contribution \eqref{kappa UV intro}. If $m<0$ instead, we clearly obtain $\kappa=-1$ and $\kappa_g=-2$ in the IR.

\paragraph{Quantizing the $\CN=2$ chiral and vector multiplets.}
Given the above discussion, let us state our conventions for quantizing fermions in $\CN=2$ supersymmetric theories with a $U(1)_R$ symmetry. 
In the supersymmetric theory, we distinguish between the gauge (dynamical or flavor) Chern-Simons terms, the mixed gauge-$R$ CS terms, the $R$-$R$ CS term, and the gravitational CS term. The supersymmetrization of these terms is reviewed in Appendix~\ref{app: parity anomaly}. Given some $U(1)_a$ gauge (or flavor) symmetries, we denote the corresponding CS contact terms by: 
\be
\kappa_{ab}~,\qquad \quad \kappa_{aR}~, \qquad \quad  \kappa_{RR}~,\qquad  \quad \kappa_g~,
\ee
respectively.

Consider first a chiral multiplet $\Phi$ of $U(1)_a$ charges $Q^a$ and $R$-charge $r$.
Unless otherwise stated, we always use the ``$U(1)_{-\half}$ quantization'' described above for the Dirac fermion $\psi$ in $\Phi$, which then contributes to the CS contact terms as:
\be\label{Phi U12 quant intro}
\Phi \; : \; \begin{cases}
\quad\delta\kappa_{ab}=- \half Q^a Q^b~,\qquad\qquad & \delta\kappa_{RR}=-\half (r-1)^2~,\\
  \quad \delta\kappa_{aR}=- \half Q^a (r-1)~,\qquad &\delta\kappa_g=-1~.
\end{cases}
\ee
Consider next the vector multiplet $\CV$. It is convenient to decompose the gauge fields into abelian gauge fields in vector multiplets $\CV_a$ along a maximal torus $\GH\cong  \prod_a U(1)_a$, and into the components along the non-trivial roots. Then, we choose a so-called ``symmetric quantization'' for the gauginos, such that they contribute trivially to the contact terms involving the gauge symmetry:
\be\label{V quanti intro 1}
\CV\; : \; \qquad \delta\kappa_{ab}=0~, \qquad\qquad \delta\kappa_{a R}=0~.
\ee 
We must also specify the $U(1)_R$ and gravitational CS contact terms. In our conventions, each gaugino component contributes $\kappa_{RR}= \half$ and $\kappa_g=1$. Therefore, the adjoint gaugino in the full vector multiplet contributes:
\be\label{kappaRR intro vec}
\CV\; : \; \qquad \delta\kappa_{RR}= \half {\rm dim}(\GG)~, \qquad \qquad
 \delta\kappa_{g}={\rm dim}(\GG)~,
\ee
in the UV. We explain our motivation for this particular choice in Appendix~\ref{app: parity anomaly}.

Concretely, the quantization requirements \eqref{Phi U12 quant intro} and \eqref{V quanti intro 1}-\eqref{kappaRR intro vec}  constrain the regularization of the various one-loop determinants that appear in supersymmetric localization formulas, as we will see in later sections. Implicitly, a lot of the supersymmetric localization literature used regularizations that set every CS contact term to zero, in the UV, $\kappa^{\rm (UV)}=0$, thus preserving parity but violating gauge invariance. In this paper, as in \cite{Closset:2017zgf}, we are careful to regulate the one-loop determinants consistently with gauge-invariance.
This leads to subtle corrections with respect to many previous results in the literature. Those corrections turn out to be important when performing finer checks of the supersymmetric partition functions, for instance when testing supersymmetric dualities.

\subsection{Supersymmetric partition functions and sum over Bethe vacua}
As described above \eqref{INTROZM3CL}, we expect  that the supersymmetric partition function on a general Seifert manifold, $\CM_3$, can be computed as the expectation value of a suitable ``geometry changing line operator'' $\SL_{\cM_3}$ inserted along the circle on the A-twisted $S^2 \times S^1$ geometry.  Consequently, the explicit expression for the partition function $Z_{\CM_3}$ has a similar form to that of the expectation value of line operators in $S^2 \times S^1$.

The partition function of an $\CN=2$ supersymmetric gauge theory on $S^2 \times S^1$, with the topological A-twist on $S^2$,  was computed in \cite{Benini:2015noa} using supersymmetric localization in the UV, and in \cite{Nekrasov:2014xaa} using topological field theory methods.  The $S^2 \times S^1$ computation has been generalized to the product space $\Sigma_g\times S^1$  in \cite{Nekrasov:2014xaa, Benini:2016hjo,Closset:2016arn}. More recently, we considered the case of  the three-manifold $\cM_{g,{\bf d}}$,  a principal $S^1$ bundle of degree ${\bf d}$ over the smooth Riemann surface $\Sigma_g$  \cite{Closset:2017zgf}.~\footnote{See also \protect\cite{Ohta:2012ev} for an early computation on that same geometry.}  In all cases, the partition function can be computed using two complementary methods. Let us discuss them in turn.

\subsubsection{TQFT computation}
The first method exhibits the partition function as an observable in the 3d A-model \cite{Nekrasov:2014xaa, Gukov:2016gkn, Closset:2017zgf}.  As for any two-dimensional TQFT, A-model observables can be written as a trace over a suitable basis of field theory vacua.  In the case of the 3d A-model, these two-dimensional vacua are called the ``Bethe vacua,'' because the equations determining them coincide with the Bethe equations for a certain class of integrable spin chains \cite{Nekrasov:2009uh,Nekrasov:2014xaa}.  

%

The 3d A-model on $\Sigma_g$ is fully characterized by the two-dimensional {\it twisted superpotential,} $\cW(u,\nu)$, on the one hand, and by the {\it effective dilaton}, $\Omega(u,\nu)$, on the other hand. These two functions  are determined by the UV Lagrangian, and control the low-energy effective action on the Coulomb branch of any effective 2d $\CN=(2,2)$ gauge theory.  They depend (locally) holomorphically on the gauge parameters $u$ and on the global-symmetry parameters $\nu$.~\footnote{ 
In addition to the flavor symmetry parameters,  there is also an dependence on the $U(1)_R$ background gauge field introduced in Section~\protect\ref{intro: subsec: LR}, as we will see in detail later in the paper.}   The ``Bethe equations'' determining the supersymmetric vacua are written in terms of the twisted superpotential alone, according to:
\be
 \pif_a(u,\nu) = \exp \bigg( 2 \pi i \frac{\partial \cW(u,\nu)}{\partial u_a} \bigg) = 1~, \qquad a= 1,\cdots,\rk~.
 \ee
We denote by $\cS_{BE}$  the set of Bethe vacua:
\be\label{SBE def intro}
\cS_{BE} = \Big\{ \hat{u}_a \;\; \big| \;\; \Pi_a(\hat{u},\nu)  =1~,  \quad w \cdot \h u \neq \h u, \;\; \forall w \in W_\GG \; \Big\} / W_\GG~.
\ee
Here, we exclude solutions $\hat{u}$ which are fixed by some Weyl group elements, and count the remaining solutions up to the Weyl group action.  
Then, the partition  function on $\Sigma_g\times S^1$---also known as the genus-$g$ twisted index---can be computed as \cite{Nekrasov:2014xaa, Benini:2016hjo,Closset:2016arn}:
\be
Z_{\Sigma_g \times S^1}(\nu) =  \sum_{\hat{u} \in \cS_{BE}} \cH(\hat{u},\nu)^{g-1}~,
\ee
with $\CH$ is the handle-gluing operator introduced in \eqref{hgo and f intro}. Its explicit expression in the 3d A-model is  \cite{Nekrasov:2014xaa}:
\be\label{HGO intro}
 \cH(u,\nu) = e^{2 \pi i \Omega(u,\nu)} \det_{a,b} \frac{\partial^2 \cW(u,\nu)}{\partial u_a \partial u_b}~.
\ee
Note that, in this approach, we {\it assume} that there are a finite number of isolated Bethe vacua. This is the case in many interesting theories. In particular, in the presence of enough flavor symmetries, one can  turn on generic real mass parameters and the vacua are then isolated.

\paragraph{A-model observables and geometry-changing line defects.} A general A-model observable can be computed as:
\be
\langle \SL_i \SL_j \cdots \rangle_{\Sigma_g \times S^1} =  \sum_{\hat{u} \in \cS_{BE}} \SL_i(\h u, \nu) \SL_j(\h u, \nu)\cdots\; \cH(\hat{u},\nu)^{g-1}~,
\ee
with the insertion of any half-BPS line \eqref{def SL intro} in the 3d A-model. 
In \cite{Closset:2017zgf}, we considered the principal $S^1$ bundle $\CM_{g, {\bf d}}$, which is realized by inserting the so-called ``ordinary'' fibering operator, $\SL=\CF$. It can be written explicitly in terms of the twisted superpotential $\CW$:
\be\label{OFO intro} 
\cF(u,\nu) = \exp \bigg( 2 \pi i \bigg( \cW(u,\nu) -u_a \frac{\partial \cW}{\partial u_a} - \nu_\alpha \frac{\partial \cW}{\partial \nu_\alpha} \bigg) \bigg)~,
\ee
where the sum over repeated indices is implicit. Therefore, we have \cite{Closset:2017zgf}:
\be
Z_{\CM_{g, {\bf d}}}(\nu) =  \sum_{\hat{u} \in \cS_{BE}}  \CF(\hat{u},\nu)^{\bf d}\,  \cH(\hat{u},\nu)^{g-1}~.
\ee

In this paper, we want to generalize those results to any half-BPS background $(\CM_3, {\bf L}_R)$, with $\CM_3$ a Seifert manifold \eqref{def M3 intro}, and $ {\bf L}_R$ the R-symmetry line bundle discussed above, which determines a choice of spin structure over $\CM_3$. 
We do this by introducing the  ``$(q,p)$ fibering operator'' \eqref{qp fibering op intro}. More precisely, we should consider the object:
\be\label{Gqpmn intro}
\cG_{q,p}(u,\nu)_{\n,\m}~,
\ee 
with $q$ and $p$ some mutually prime integers. Without loss of generality, we take $q>0$.
The integers $\n$ and $\m$ in \eqref{Gqpmn intro} refer to ``fractional fluxes'' localized at the exceptional fiber, for the gauge and global symmetries, respectively, described in more detail below.  Unlike the handle-gluing operator \eqref{HGO intro} or the ``ordinary'' fibering operator \eqref{OFO intro}, the fibering operator \eqref{Gqpmn intro} cannot be expressed in a simple way in terms of the twisted superpotential and effective dilaton alone.  We can nevertheless write down explicit expressions for this operator in terms of the UV Lagrangian of the theory. For instance, the contribution of a $U(1)$ Chern-Simons term at level $k$ is given by:
\be \label{INTRO CGqp CS}
 \cG_{q,p}^{\rm CS}(u)_\n = (-1)^{\n k (1+ t+ l^R t +2 \nu_R s)}\, \exp\left({- {\pi i k \ov q}\left(p u^2- 2 \n u + t \n^2\right)}\right)~. 
\ee
Here, $t$ and $s$ are integers such that $q s+ p t=1$, while the parameters $l^R$ and $\nu_R$ are related to the choice of $U(1)_R$ bundle ${\bf L}_R$, as we will explain in detail later on. 
As another example, a chiral multiplet $\Phi$ of unit $U(1)$ charge contributes:~\footnote{Here we choose a vanishing R-charge $r=0$, and we turn off the fractional fluxes, for simplicity. The general expressions are given in Section~\protect\ref{sec: Bethe vac sum}.}
\be\label{INTRO CGqp0 Phi}
\CG_{q,p}^\Phi(u) = \exp \bigg( \sum_{l=0}^{q-1}\left\{ {p\ov 2 \pi i} \dilog(e^{2\pi i {u+ t l\ov q}}) + {pu + l \ov q} \log\left(1-e^{2\pi i {u+ t l\ov q}}\right)\right\}~ \bigg)~,
\ee
with the integer $t$ defined as before. 
In a general theory, there can be contributions from various $U(1)_R$ and gravitational Chern-Simons terms, and there is also an important contribution from the vector multiplet. These are described in full detail in Section~\ref{sec: Bethe vac sum}. 
The Seifert fibering operator for a general gauge theory is built by assembling these building blocks, for fixed gauge and flavor fluxes $\n$ and $\m$. Then, the ``physical'' $(q,p)$ fibering operator of the gauge theory is obtained by summing over the fractional gauge fluxes $\n$, according to:
\be\label{INTRO CG full sum def}
\CG_{q,p}(u, \nu)_{\m} =  \sum_{\n \in \Gamma_{\GG^\vee}(q)} \CG_{q,p}^{\rm CS}(u, \nu)_{\n, \m}\;  \CG^{\rm matter}_{q,p}(u, \nu)_{\n, \m} \; \CG_{q,p}^{\rm vector}(u)_{\n}~.
\ee
Here, $\Gamma_{\GG^\vee}(q)$ is the $\Z_q$ reduction of the lattice of magnetic fluxes, namely:
\be\label{cochar mod q def intro}
\Gamma_{\GG^\vee}(q)=\big\{\n \in {\mathfrak h}\big|~ \rho(\n)\in \mathbb{Z},\; \; \forall\rho \in\Lambda_{\rm char}~;\; \n \sim \n+ q \lambda,\;\; \forall \lambda\in \Lambda_{\rm cochar}\big\}~,
\ee
with $\Lambda_{\rm char}$ and $\Lambda_{\rm cochar}$ the character and co-character lattices of $\GG$, respectively.

\paragraph{The $\CM_3$ partition function.}
Given the above discussion, the geometry-changing line operator \eqref{INTROZM3CL} can be written as:
\be
\SL_{\CM_3}(u, \nu)_\m \equiv \CG_{\CM_3}(u, \nu)_\m \, \CH(u, \nu)^g~,
\ee
with $\CG_{\CM_3}$ the {\it Seifert-fibering operator}, which is determined by the Seifert fibration \eqref{def M3 intro}:
\be\label{def CGM3 intro}
\CG_{\CM_3}(u, \nu)_\m \equiv \CF(u, \nu)^{\bf d}\,\prod_{i=1}^n \CG_{q_i, p_i}(u, \nu)_{\m_i}~.
\ee
Here, the $(q,p)$ fibering operators is given as in \eqref{INTRO CG full sum def}.
Note that the ordinary fibering operator is a special case of the $(q,p)$ fibering operator, with $(q,p)=(1,1)$. 
We should also anticipate that the individual fibering operators appearing in \eqref{def CGM3 intro} are generally not completely well-defined for every choice of ${\bf L}_R$ in \eqref{INTROlrh1}. Nonetheless, their product in \eqref{def CGM3 intro} is always well-defined (assuming the half-BPS geometry itself is   well-defined globally). 
Thus, we have obtained an explicit expression for the supersymmetric partition on any Seifert manifold, as a sum over the Bethe vacua of the 3d $\CN=2$ gauge theory:
\be\label{ZM3 bethe sum INTRO}
Z_{\CM_3}(\nu)_\m =  \sum_{\hat{u} \in \cS_{BE}} \CG_{\CM_3}(\h u, \nu)_\m \,  \CH(\h u,\nu)^{g-1}~.
\ee
This is the main result of this paper.

\subsubsection{Supersymmetric localization computation}
An alternative approach to the TQFT computation, which can be shown to be completely equivalent, is to follow the standard localization procedure and to deform the UV action by a suitable ${\cal Q}$-exact term.  This concentrates the path-integral to the neighborhood of a finite dimensional space, $\cM_{BPS}$, which consist of field configuration that satisfies the BPS equations:
\be
\sigma = (\text{constant})~,\qquad f_{01} = f_{0\bar 1} = 0~, \qquad D=2if_{1\bar 1}+\sigma H~.
\ee
Here, $H$ is a supergravity background scalar proportional to the rational number:
\be\label{c1L0 def intro}
c_1(\CL_0)= \mathbf{d} + \sum_{i=1}^n\frac{p_i}{q_i}~.
\ee
The quantity \eqref{c1L0 def intro} is a topological invariant of the Seifert fibration. The case $c_1(\CL_0) =0$ and $c_1(\CL_0)\neq 0$ are qualitatively different. Let us first assume that $c_1(\CL_0)  \neq 0$. By a standard abelianization procedure \cite{Blau:1993tv,Blau:1994rk}, the path integral reduces to a finite dimensional integral over the complex variable $u \in {\frak h}_{\mathbb{C}}$, valued in the complexified Cartan subalgebra of $\GG$. 
The BPS equations also allow non-trivial gauge line bundles on the base $\h \Sigma$ of the Seifert fibration, and we should sum over all the lines bundles ${\bf L}$ on $\CM_3$ that can be obtained as pull-backs of the orbifold line bundles $L$ on $\h \Sigma$---that is, ${\bf L} = \pi^\ast(L)$. These line bundles form a group, which we denote by:
\be
\t\Pic(\CM_3) \cong \pi^\ast\left( \Pic(\h\Sigma)\right)~, 
\ee
with $ \Pic(\h\Sigma)$ the orbifold Picard group on $\h\Sigma$.
 After integrating out the gaugino zero modes and the auxiliary field $D$ in the vector multiplet, which can be done in the same way as in \cite{Benini:2016hjo,Closset:2016arn,Closset:2017zgf}, we can write the partition function as:
\be\label{first integral formula}
Z_{\CM_3}(\nu) = \frac{1}{|W_\GG|}\sum_{\substack{(\n_0,\frak n_1,\cdots, \n_n)\\\in \widetilde{\text{Pic}}(\CM_3)}} \int_{\CC(\eta)} d^\rk u~e^{-S_{\text{CS}} (u,\nu)}\CZ^{\text{1-loop}}_{(\n_0,\frak n_1,\cdots, \frak n_n)} (u,\nu)\, H^g(u,\nu)~.
\ee
The various factors in the integrand will be defined in Section \ref{sec:localization}. As in \cite{Closset:2017zgf}, the choice of the contour $\CC(\eta)$ in  \eqref{first integral formula}  can be rigorously derived in the rank-one case, while it remains as a conjecture  for the higher rank case, due to a number of subtleties that we will review in Section \ref{sec:localization}.  

One can check that the integrand of \eqref{first integral formula} is invariant under large gauge transformation along the Seifert fiber, which acts on the gauge parameters as:~\footnote{For non-abelian $\GG$, it is understood that the shifts are by elements of $\Lambda_{\text{cochar}}$.}
\be \label{large gauge introduction}
u\rightarrow u+1~,\qquad \n_0 \rightarrow \n_0+\mathbf{d}~, \qquad \n_i\rightarrow \n_i + p_i~.
\ee
This is a trivial operation in $\t\Pic(\CM_3)$, which ensures that the summation in \eqref{first integral formula} is well-defined. On the other hand, there exists an alternative way of fixing the gauge under this large gauge transformation \cite{Blau:2006gh, Blau:2013oha}. Namely, one can take a quotient on the ``classical Coulomb branch'' spanned by the variables $u\in  {\frak h}_{\mathbb{C}}$, by restricting them to:
\be\label{u restricted intro}
u \in {\frak h}_{\mathbb{C}}/\Lambda_{\text{cochar}}~.
\ee 
In this way, one arrives at the formula:
\be\label{second integral formula}
Z_{\CM_3}(\nu)=\frac{1}{|W_{\GG}|} \sum_{\n_0 \in \Gamma_{\GG^\vee}}\sum_{\substack{\{(\frak n_1,\cdots, \frak n_n)|\\ {\frak n_i}\in \Gamma^\vee_\GG (q_i),\forall i\}}} \int_{\CC_0(\eta)} d^{\rk}u~e^{-S_{\text{CS}} (u)}\CZ^{\text{1-loop}}_{(\n_0,\frak n_1,\cdots, \frak n_n)} (u) \,H^g( u )\ ,
\ee
where the contour $\CC_0(\eta)$ can be obtained by restricting $\CC(\eta)$ in  \eqref{first integral formula}  to the ``strip'' \eqref{u restricted intro} in the Coulomb branch variables. Here, $ \Gamma_{\GG^\vee}$ is the ordinary lattice of magnetic fluxes.

Now, the formula \eqref{second integral formula} is also valid for Seifert manifolds with $c_1(\CL_0)=0$. For instance, in the case of the twisted index on $\CM_3 \cong \Sigma_g \times S^1$, the sum over the $\n_i$ with $i>0$ trivializes, while the sum over $\n_0$ is a sum over the magnetic fluxes on $\Sigma_g$, thus reproducing the localization formula derived in \cite{Benini:2015noa, Benini:2016hjo,Closset:2016arn}.

In the case of a gauge group $\GG=U(1)$, one can explicitly perform the summation over $\n_0$ in \eqref{second integral formula}, as explained in \cite{Closset:2017zgf}. We will show that the resulting expression is equivalent to the Bethe-sum formula \eqref{ZM3 bethe sum INTRO} which we obtained from the two-dimensional TQFT point of view.

We will also show that the contour $\CC(\eta)$ used in \eqref{first integral formula}
can be continuously deformed to a non-compact integral $\CC_\sigma$, the ``$\sigma$-contour,'' which connects the region Im$(u)\rightarrow -\infty$ with the region Im$(u)\rightarrow +\infty$. This reproduces the well-known expressions for the partition functions on lens spaces in earlier literature \cite{Kapustin:2009kz,Hama:2010av,Imamura:2011wg,Dimofte:2014zga}, which were given as an integral over the constant mode of the real scalar $\sigma$ in the vector multiplet. 

When $\GG$ is non-abelian and $2g-2 +n \geq 0$, the abelianized path integral becomes singular at the loci where the non-abelian symmetry enhances. 
The simple ``$\sigma$-contour'' integral formula must be modified in generic cases due to these additional singularities in the integrand. We will briefly discuss this subtlety in Section \ref{sec:localization}.
On the other hand, the Bethe-sum formula \eqref{ZM3 bethe sum INTRO} is always valid, for any gauge group of the form \eqref{GG intro}, with the Bethe vacua defined as in \eqref{SBE def intro}.  This claim is supported by a number of highly non-trivial consistency checks.

\subsection{Testing supersymmetric dualities}
\label{sec:duality}
A common application of exact results for supersymmetric partition functions is to test field theory dualities---see {\it e.g.} \cite{Kapustin:2010xq, Willett:2011gp, Benini:2011mf}.  
In particular, since the supersymmetric partition functions are RG-invariant, we can test {\it infrared dualities}---that is, the claim that two different gauge theories flow to the same infrared fixed point. A prime example of that are the 3d infrared dualities  \cite{Aharony:1997gp, Giveon:2008zn, Benini:2011mf, Aharony:2013dha} similar to 4d Seiberg duality \cite{Seiberg:1994pq}.

Given two infrared-dual theories $\CT$ and $\CT^D$, their supersymmetric partition functions must agree:
\be\label{dual rel intro}
Z^{\CT}_{\CM_3}(\nu)_\m=Z^{\CT^D}_{\CM_3}(\nu)_\m~,
\ee
for any half-BPS geometry $\CM_3$. 
By now, many three-dimensional dualities are firmly established, and therefore we can also consider verifying \eqref{dual rel intro} as a strong consistency check on our results for $Z_{\CM_3}$. This is what we will do. 
Given the 3d A-model formula \eqref{ZM3 bethe sum INTRO} for the partition function, the duality relation \eqref{dual rel intro} is equivalent to the statement that the various $(q,p)$-fibering operators agree on dual Bethe vacua, namely:
\be\label{CG dual intro}
\CG^{\CT}_{q,p}(\h u, \nu)_\m  = \CG^{\CT^D}_{q,p}(\h u^D, \nu)_\m
\ee
Here, $\h u$ denotes a solution to the Bethe equation in theory $\CT$, and $\h u^D$ denotes a solution to the dual Bethe equation in the dual theory $\CT^D$, with $\h u$ and $\h u^D$ paired by the duality map. 
In previous work, similar duality relations were checked for the handle-gluing operator \cite{Closset:2016arn} and for the ordinary fibering operator \cite{Closset:2017zgf}.~\footnote{We will revisit those cases as well, taking into account the spin-structure dependence of the answer.} The duality relations \eqref{CG dual intro} are hard to prove in general, but we were able to checked them numerically, for a very large number of pairs of mutually-prime integers $(q,p)$, and for a large number of infrared dualities.

 For instance, consider Aharony duality  \cite{Aharony:1997gp}, which is an infared duality between a $U(N_c)$ gauge theory with $N_f$ chiral multiplets in the fundamental and anti-fundamental representations (that is, $N_f$ ``flavors''), on the one hand, and a $U(N_f-N_c)$ gauge theory on the other hand,  schematically:
 \be
 \CT\; : \; U(N_c)\, +\, N_f\; (\Phi, \t\Phi) \qquad \longleftrightarrow \qquad   \CT^D\; : \; U(N_f-N_c)\, +\, N_f\; (\Phi^D, \t\Phi^D)~.
 \ee
 The Bethe equation of both theories are determined by a certain polynomial:
 \be
 P(x, y)~, \qquad  x\equiv e^{2\pi i u}~, \qquad y=e^{2\pi i \nu}~,
 \ee
 of degree $N_f$ in a single variable $x$. Here, $\nu$ denotes collectively various flavor parameters for the $SU(N_f)^2 \times U(1)^2$ global symmetry.
 A Bethe vacuum in the $U(N_c)$ theory corresponds to a choice of $N_c$ distinct roots, $\h x_a= e^{2\pi i \h u_a}$, amongst the $N_f$ roots of $P(x, y)$, the ``Bethe roots.''  The dual vacuum in the $U(N_f-N_c)$ theory corresponds to choosing $\h x_\ba^D= e^{2\pi i \h u_{\ba}^D}$ the complement of $N_f-N_c$ roots. Then, the duality relations \eqref{CG dual intro} depends on seemingly ``miraculous'' properties of the fibering operators $\CG_{q,p}(u, \nu)_\m$ in \eqref{INTRO CG full sum def} when evaluated on the Bethe roots. 
 
 It would be interesting to prove those relations analytically, presumably using some ``number theoretic'' reasoning. In any case, the exact match that we found, numerically and in many examples of 3d dualities, already provides a very strong test of our main results.

\subsection{Lens spaces and holomorphic blocks}
Lens spaces are an important class of thee-manifolds that admit supersymmetric backgrounds. Topologically, we define the lens space $L(p,q)$, for any pair of mutually prime integers $p$ and $q$ (with $p\neq 0$), as the quotient:
\be
L(p,q) \cong S^3/\Z_p~, \qquad \qquad  \Z_p \, : \,\big(z_1~,\;  z_2\big) \sim \big(e^{2\pi i q\ov p}\, z_1~,\;e^{2\pi i \ov p}\, z_2\big)~,
\ee
where the three-sphere $S^3$ is viewed as the unit sphere, $\{|z_1|^2 + |z_2|^2=1\}$, inside $\C^2$.
Important special cases are:
\be\label{example Lp intro}
S^3 \cong L(1,1)~,\qquad 
L(p,-1)~, \qquad L(p,1)\cong \CM_{0, p}~.
\ee
namely the three-sphere itself, and certain ``simpler'' lens spaces $S^3/\Z_p$.
Every lens space supersymmetric background is part of a continuous one-parameter family, generally indexed by a complex ``squashing parameter'' $b\in \C$. The supersymmetric partition function on the (squashed) three-sphere $S^3_b$ was studied in \cite{Kapustin:2009kz,Jafferis:2010un, Hama:2010av,Hama:2011ea, Imamura:2011wg, Martelli:2011fw, Closset:2012ru, Alday:2013lba, Tanaka:2013dca, Closset:2013vra}. The generalization to the (squashed) lens space $L(p,-1)_b$ was considered in \cite{Gang:2009wy, Benini:2011nc, Alday:2012au, Imamura:2012rq, Imamura:2013qxa}. 
The three-sphere $S^3_b$ and  the lens space $L(p,-1)_b$ are examples of supersymmetric backgrounds with a trivial R-symmetry line bundle, as in \eqref{LR trivial INTRO}; they are the only lens spaces with that property. The third example in \eqref{example Lp intro} consists of the degree-$p$ principal circle bundle over $S^2$ as studied in \cite{Ohta:2012ev, Closset:2017zgf, Toldo:2017qsh}.~\footnote{To avoid any possible confusion, let us note that, in \cite{Closset:2017zgf}, we chose a different naming convention for $L(p,q)$ (which is the convention more often used in the physics literature), so that $L(p,p-1)$ here was named $L(p,1)$ there, and vice-versa.} The general $L(p,q)_b$ lens space partition functon was studied in  \cite{Dimofte:2014zga}.

Lens space are rather special amongst half-BPS geometries, due to the continuous parameter $b$---most half-BPS Seifert geometries are rigid and admit no such ``squashing'' deformation.
Moreover, for generic $b\in \C$, the half-BPS background is actually {\it not} of the form studied in the present work, because the Killing vector $K$ in the supersymmetric algebra \eqref{QQb intro} does not generate the Seifert fibers. This is because, on $L(p,q)_b$ with any $b$, the Killing vector appearing in the curved-space supersymmetry algebra takes the form:
\be
K^{(b)}= b^{-1} \big(i z_1\d_{z_1} -i \bz_1 \d_{\bz_1}\big) +b \big(iz_2\d_{z_2} - i\bz_2 \d_{\bz_2}\big)~.
\ee
For a generic $b$, this Killing vector is complex. Even for $b\in \R$, its orbits are non-compact unless we impose the rationality condition:
\be
b^2 \in \Q~.
\ee
Precisely in this case, the orbits of the $U(1)$ action generated by $K^{(b)}$ span the $S^1$ fibers of a Seifert fibration structure on $L(p,q)$. More precisely, for:
\be
b^2 = {q_1\ov q_2}~, \qquad q_1, q_2\in \Z~,
\ee
the lens space $L(p,q)_b$ admits a presentation as a Seifert fibration over $S^2(q_1, q_2)$, a genus-zero Riemann surface, with two exceptional fibers:
\be
L(p,q)_b  \cong [0~;\, 0~;\, (q_1, p_1)~,\; (q_2, p_2)]~,
\ee
and with the identifications $p= p_1 q_2 + p_2 q_1$ and $q= q_1 s_2 - p_1 t_2$ between the lens space and Seifert fibration parameters.~\footnote{As above, the integers $s_i, t_i$ are defined by the condition $q_i s_i + p_i t_i=1$.}
The lens spaces are the only Seifert manifolds that admit an infinite number of inequivalent Seifert fibrations, which are all accounted for by the rational squashing parameters, $b^2 \in \Q$.

The above discussion assumed $p\neq 0$, but it will be very natural to also define the spaces:
\be
L(0, 1) \cong S_\epsilon^2\times S^1~, \qquad \qquad L(0,-1)\cong S^2\times S^1~.
\ee
Topologically, $L(0,1)$ and $L(0,-1)$ have the same topology, $S^2 \times S^1$, but they differ very much as supersymmetric backgrounds. In our notation, $L(0,1)$ corresponds to the (refined) topologically-twisted supersymmetric index  \cite{Benini:2015noa}. For rational values of the ``refinement parameters'' $\epsilon$, $L(0,1)$ is again a Seifert fibration and it fits into our formalism (in particular, for $\epsilon=0$, this gives the twisted index discussed above). On the other hand, the supersymmetric background $L(0,-1)$ admits no Seifert description. It corresponds to the ``ordinary'' supersymmetric (or superconformal) index, without topological twist \cite{Kim:2009wb,Imamura:2011su}. 

\paragraph{Holomorphic blocks and fibering operators.}
We will demonstrate that our formalism for general Seifert manifolds reproduces, in the special case above, the known results for partition functions on rationally-squashed lens spaces.  In particular, we clarify many subtle features of these partition functions, including a detailed discussion of the possible R-symmetry backgrounds.~\footnote{For $p$ even, there are really two distinct $L(p,q)_b$ backgrounds, distinguished by two different spin structures. This was first noted in \protect\cite{Martelli:2012sz, Toldo:2017qsh}.}

We also relate our results to the ``holomorphic blocks'' of Beem, Dimofte and Pasquetti  \cite{Beem:2012mb}. The holomorphic blocks can be defined as an uplift of the two-dimensional vortex partition function, or,  equivalently, as the twisted partition function on a solid torus, $D^2 \times_\tau S^1$.  In \cite{Beem:2012mb,Dimofte:2014zga,Nieri:2015yia} it was shown that the supersymmetric partition function on lens spaces can be constructed by ``fusion'' of two holomorphic blocks:
\be\label{fusion blocks intro}
Z_{L(p,q)_b}(\nu) = \sum_{\alpha\in \cS_{BE}} B^\alpha(g\cdot \nu, - g\cdot \tau) B^\alpha(\nu, \tau)~,
\ee
schematically.
Here, $\nu$ are flavor parameters, $\tau$ is a geometric parameter related to the squashing $b$, and $g$ is an $SL(2, \Z)$ element used to glue the two solid tori into a closed three-manifold---corresponding to the genus-one Heegaard splitting of $L(p,q)$ into solid tori. The sum in \eqref{fusion blocks intro} is over the Bethe vacua, and therefore \eqref{fusion blocks intro} is very reminiscent of our general result \eqref{ZM3 bethe sum INTRO} for Seifert manifolds. Indeed, we will show that, in the limit of $b^2$ rational, \eqref{fusion blocks intro} becomes equivalent to \eqref{ZM3 bethe sum INTRO}.

The holomorphic blocks are actually singular in the limit where $b^2$ becomes rational, but that singular behavior encodes interesting physics. In particular, that limit is governed by the twisted superpotential and by the effective dilaton \cite{Beem:2012mb}.
 We will show that, in the limit of rational squashing, the holomorphic blocks essentially reduce to the $(q,p)$ fibering operators.  This serves as an independent derivation of the fibering operators, and gives a new perspective on the holomorphic blocks themselves. We will also clarify some technical features of the blocks, such as their dependence on the choice of spin structure.

\subsection{Discussion and outlook}
The present work can be connected to many other lines of inquiries. First of all, as a special case of our formalism, we can study 3d $\CN=2$ supersymmetric Chern-Simons theory on Seifert manifolds. Supersymmetric CS theory is essentially equivalent to pure (non-supersymmetric) CS theory, therefore we can directly compare our results to many exact results in Chern-Simons theory \cite{Witten:1988hf, jeffrey1992, Lawrence1999, Marino:2002fk, Beasley:2005vf, Blau:2006gh, Blau:2013oha}. 
This will be discussed in a separate work \cite{CHMW2018}. See also \cite{Fan:2018wya} for some interesting recent work in that direction.

Another interesting research direction concerns the existence of many supersymmetric backgrounds that admit a topologically-trivial canonical line bundle (in addition to $S^3_b$ and the lens space $L(p,-1)_b$), which can be used to study $\CN=2$ superconformal field theories. For instance, using our results, the partition function of $\CN=2$ SCFTs can be computed explicitly on the Poincar\'e homology sphere, and the obvious challenge is to understand exactly what kind of CFT observables that quantity may encode.  We hope to return to this investigation in future work.

Supersymmetric partition functions on Seifert manifolds were previously studied in \cite{Gukov:2015sna,Gukov:2016gkn,Gukov:2017kmk}, with a particular focus on the 3d/3d correspondence \cite{Dimofte:2011ju}; in particular, the 3d A-model played a crucial role in \cite{Gukov:2016gkn}. Let us also mention that the relation between the integral formula \eqref{first integral formula} and the Bethe-sum formula \eqref{ZM3 bethe sum INTRO} first appeared, in some special instances, in studies of state integrals in complex Chern-simons theory \cite{Dimofte:2009yn, Garoufalidis:2014ifa}. It would be interesting to understand if our evaluation formula \eqref{ZM3 bethe sum INTRO}, in the case of a lens space, can provide additional insight into multi-dimensional state-integrals. More generally, it would be very interesting to better understand our results in the context of the 3d/3d correspondence, wherein the supersymmetric partition functions on a given Seifert three-manifold should be related to observables in some (possibly new) 3d TQFT.

The results of this paper can be uplifted to four-dimensional $\CN=1$ gauge theories, by considering complex four-manifolds that are also $T^2$ fibrations over a Riemann surface, $T^2 \rightarrow \CM_4\rightarrow \h\Sigma$ \cite{Dumitrescu:2012ha}. In this way, one could study the most general half-BPS 4d $\CN=1$ geometries, generalizing the approach of \cite{Closset:2017bse}.

Finally, it would be very interesting to study boundaries and boundary conditions in the 3d A-model, making contact with the work of, {\it e.g.}, \cite{Aprile:2016gvn, Dimofte:2017tpi}. Such a study would likely lead to a deeper understanding of the fibering operators as 3d defects defined in the UV by suitable boundary conditions (instead of the simpler two-dimensional definitions we adopted here), and it would likely allow us to explore more interesting coupled bulk-boundary systems exactly, using TQFT and localization techniques.

\vspace{0.5cm}\noindent
Due to its length, this paper is divided in three parts,  plus appendices. 

Part \ref{part one} gives a detailed discussion of half-BPS supersymmetric backgrounds. In Section \ref{sec: seifert backgs}, we provide an introduction to Seifert three-manifolds. On any Seifert three-manifold, we construct a half-BPS ``A-twisted'' supergravity background. The formalism of that section will be used extensively throughout the paper. In Section \ref{subsec:expls}, we spell out a number of interesting examples of half-BPS Seifert geometries. We pay particular attention to lens spaces, for which we enumerate all the possible Seifert structures and choices of spin structure. 

In Part \ref{part two}, we compute supersymmetric partition functions of 3d $\CN=2$ gauge theories on a general Seifert manifold. In Section \ref{sec: Bethe vac sum}, we derive the result \eqref{ZM3 bethe sum INTRO} from the point of view of the three-dimensional A-model, by introducing various geometry-changing line operators. In Section \ref{sec:duality test}, as a non-trivial test of our result, we study infrared dualities of three-dimensional gauge theories on Seifert manifolds. In Section~\ref{sec:localization}, we provide an alternative derivation of the partition function formula via Coulomb branch localization in the UV, which leads to the expressions \eqref{first integral formula} and \eqref{second integral formula}. 

In Part \ref{part three}, we revisit the computation of supersymmetric partition functions on the lens spaces, $L(p,q)$. We compare our results with the previous literature and we clarify various subtleties. In Section \ref{sec: S3b} and \ref{sec:s2s1top}, we study the squashed three-sphere partition function ($\CM_3= S^3_b$) and the refined twisted index ($\CM_3= S^2_\epsilon \times S^1$), respectively.  In Section \ref{sec:HB}, we discuss the general squashed lens space $L(p,q)_b$ in terms of holomorphic blocks, and we exhibit the precise relation between the holomorphic blocks and the Seifert fibering operators.

\part{Half-BPS Seifert geometry}
\label{part one}
\section{Supersymmetric backgrounds on Seifert three-manifolds}\label{sec: seifert backgs}
In this section, we give an introduction to the topology and geometry of Seifert three-manifolds, following \cite{orlik1972seifert, BLMS:BLMS0401} and  \cite{FURUTA199238, Beasley:2005vf, Blau:2013oha}. Each Seifert fibration $\CM_3$, together with a choice of spin structure on $\CM_3$, provides us with a distinct half-BPS supersymmetric background, which we will spell out in detail.

\subsection{Two-dimensional orbifolds and holomorphic line bundles}
Since a Seifert manifold~\footnote{Here and in the rest of this paper, every two- and three-manifold is orientable. In particular, by ``Seifert manifold'' we mean ``orientable Seifert manifold.'' $\CM_3$ is also taken to be a closed manifold.}
$\CM_3$ can be viewed as circle bundle $\CM_3 \rightarrow \h \Sigma_{g,n}$ over a two-dimensional orbifold $\h \Sigma_{g,n}$, we first discuss the latter in some detail.
A two-dimensional orbifold $\h \Sigma_{g,n}$ is topologically a genus-$g$ closed orientable Riemann surface $\Sigma_g$ with $n$ marked points $x_i \in \Sigma_g$, $i=1, \cdots, n$, called the orbifold (or ramification) points. In an open neighborhood $U_i$ of an orbifold point $x_i$, the coordinate system is modeled on $\C/ \Z_{q_i}$ instead of $\C$, where $q_i \in \Z_{>0}$ a positive integer. That is, in terms of a complex coordinate $z_{(i)}$ centered at $x_i$, we have a cyclic identification:
\be
z_{(i)}  \sim e^{2\pi i \ov q_i}\, z_{(i)}~.
\ee
An orbifold point $x_i$ has an ``anisotropy parameter'' $q_i >1$.
If $q_i=1$ instead, the point $x_i$ is simply a smooth marked point. We often denote the orbifold $\h \Sigma_{g, n}$ by $\h \Sigma_g(q_1, \cdots, q_n)$.

Many of the familiar geometrical and topological tools can be extended to the orbifold case, in particular, one can define vector bundles, various cohomology theories, etc., similarly to the smooth case \cite{Satake359, satake1957, kawasaki1979}. In particular, one can define a $\Q$-valued orbifold Euler characteristic. It takes the numerical
value:
\be\label{chi def}
\chi(\h\Sigma_{g,n}) = \chi(\Sigma_g) + \sum_{i=1}^n {1-q_i\ov q_i}~, 
\ee
with $ \chi(\Sigma_g) = 2 - 2g$, the Euler characteristic of the underlying smooth surface. 
One can also choose a Riemannian metric compatible with the orbifold structure. A metric $g(\h\Sigma)$ is a Riemannian metric with conical singularities~\footnote{Near the point $x_i$, we have:
	\be\nn
	ds^2 = dr^2 + {r^2 \ov q_i^2} d\phi^2~,
	\ee
	in terms of the polar coodinates $z=r e^{i \phi}$, so that there is a deficit angle $2 \pi {q_i-1\ov q_i}$.}
at $x_i$, with deficit angles $2 \pi {q_i-1\ov q_i}$. The orbifold version of the Gauss-Bonnet theorem \cite{satake1957} reads:
\be\label{GBthm}
{1\ov 4 \pi} \int_{\h \Sigma_{g,n}} d^2 x \sqrt{g} \, R_{\h\Sigma}= \chi(\h\Sigma_{g,n})~.
\ee

Physicists may be more familar with the notion of an orbifold $\h \Sigma$ as the quotient, $\h \Sigma \cong \Sigma/\Gamma$, of a smooth surface $\Sigma$ by a discrete group $\Gamma$. An orbifold that can be written in this way is called ``good'', otherwise it is called ``bad''  \cite{BLMS:BLMS0401}.
Almost all two-dimensional compact orbifolds $\h \Sigma_{g, n}$ are good. The only bad orbifolds are $S^2(q_1, q_2)$, with $q_1 \neq q_2$, a two-sphere with two orbifold points of anisotropy parameters $q_1$ and $q_2$. (This includes the cases $q_1=1$ or $q_2=1$. If $q_1= q_2=q$, we have $S^2(q, q) \cong S^2/\Z_q$, which is a good orbifold.)

\paragraph{Example: the spindle.} The orbifold $S^2(q_1, q_2)$ is a sphere $S^2$ with two orbifold points, also called a ``spindle.''
Consider the angular coordinates $\theta \in [0, \pi]$ and $\phi\sim \phi +2 \pi$ on $S^2$, with the $\Z_{q_1}$ and $\Z_{q_2}$ orbifold points at the poles $\theta=0$ and $\theta=\pi$, respectively. The spindle metric can be chosen as:
\be\label{met spindle}
ds^2\big(S^2(q_1,q_2)\big) =  d\theta^2 + {\sin^2\theta \ov f(\theta)^2}  d\phi^2~,
\ee
with the function $f(\theta)$ any smooth positive function of $\theta$ such that
$f(\theta) = q_1 + O\big(\theta^2\big)$ as $\theta \sim 0$, and $f(\theta)= q_2 + O\big((\pi-\theta)^2\big)$  as $\theta \sim \pi$. Using this metric, one can check that:
\be
{1\ov 4 \pi} \int d^2 x\sqrt{g} \, R = {1\ov f(0)}+  {1\ov f(\pi)} = {1\ov q_1}+ {1\ov q_2}~,
\ee
in agreement with \eqref{GBthm}.

\subsubsection{Holomorphic line bundles over $\h\Sigma$}
One may define an orbifold holomorphic line bundle $L$ over $\h\Sigma_{g,n}$, similarly to the smooth case.  Topologically, the line bundle $L$ is fully determined by the data:
\be\label{def deg and b}
{\rm deg}(L) \in \Z~, \qquad \qquad b_i(L)  \in \Z_{q_i}~, \quad i=1, \cdots, n~.
\ee
The integer ${\rm deg}(L)$ is the degree of $L$. On the open set  $U_i$ centered at the orbifold point $x_i$, the local trivialization is modeled on $(\C \times \C^\ast)/\Z_{q_i}$, with the quotient:
\be\label{zs quotient}
\big(z_{(i)}~,\;  s_{(i)}\big) \sim  \big(e^{2\pi i \ov q_i}\, z_{(i)}~,\; e^{2\pi i  b_i(L)\ov q_i}\, s_{(i)}\big)~,
\ee
with $s_{(i)}$ the fiber coordinate. Note that $L$ is an ordinary line bundle over the underlying smooth Rieman surface $\Sigma_g$ if and only if $b_i(L)=0$ mod $q_i$, $\forall i$.
We may also introduce a connection $A$ on $L$. The (integrated) first Chern class of $L$ may be defined as:
\be
c_1(L) = {1\ov 2\pi} \int_{\h\Sigma_{g,n}} dA~.
\ee
In terms of the invariants \eqref{def deg and b}, it reads:
\be\label{c1L}
c_1(L) = \deg(L) + \sum_{i=1}^n {b_i(L)\ov q_i}~.
\ee
Unlike the degree, the first Chern class transforms simply under tensor product:
\be
c_1(L_1 \otimes L_2) = c_1(L_1) + c_1(L_2)~.
\ee
On the other hand, the invariants $b_i$ satisfy:
\be
b_i(L_1\otimes L_2\big) = b_i(L_1)+b_i(L_2) \mod q_i~, \qquad \qquad 0 \leq b_i(L_1\otimes  L_2\big)  <q_i~.
\ee
It follows that the degree of the tensor product line bundle is:
\be\label{rel deg tensor}
\deg(L_1\otimes  L_2\big) =\deg(L_1)+\deg(L_2)  + \sum_i \floor{ b_i(L_1)+b_i(L_2) \ov q_i}~.
\ee
Here, $\floor{x}$ is the floor function:
\be
\floor{x} = \max\{ n \in \Z \, |\,  n \leq x \}~.
\ee
Let us denote by:
\be
L \cong \big[d~; \, g~; \, (q_1, b_1), \cdots, (q_n, b_n)\big]~,\qquad \qquad d= \deg(L)~, \quad b_i = b_i(L)~,
\ee
the line bundle $L$ over $\h\Sigma_{g,n}$.
In the following, it will sometimes be useful to relax the constraints $0\leq b_i <q_i$ on the fiber invariants $b_i$. That is, we may take $b_i \in \Z$ in $[d~; \, g~; \, (q_i, b_i)\big]$, with the understanding that a shift of $b_i \rightarrow b_i +q_i$, at any given orbifold point, is equivalent to shifting the degree by one unit. We then have the equivalences:
\be\label{equiv Ls}
\big[d~; \, g~; \, (q_i, b_i)\big] \cong \big[d- \sum_i \m_i~; \, g~; \, (q_i, b_i + \m_i q_i)\big]~,
\ee 
for any $(\m_i) \in \Z^n$. The first Chern class \eqref{c1L} is invariant under such shifts.  

We should note that, as in the case of an ordinary line bundle, the holomorphic line bundle $L$ is also characterized by some continuous holomorphic data, corresponding to flat connections valued in $H^{0,1}(\h\Sigma_{g,n})$. 

\paragraph{Canonical line bundle and spin structures.}
The canonical line bundle $\CK$ over $\h\Sigma_{g,n}$ has the topological invariants:
\be\label{def K}
\deg(\CK) = 2 g-2~, \qquad\qquad b_i(\CK) = q_i-1~.
\ee
Its first Chern class is equal to minus the Euler characteristic \eqref{chi def},  $c_1(\CK) = - \chi(\h\Sigma)$.
A spin structure on $\h \Sigma_{g,n}$ is a line bundle $\sqrt{\CK}$ such that $\sqrt{\CK}\otimes \sqrt{\CK}\cong \CK$. Such a square root: 
\be\label{def Ksqrt}
\sqrt{\CK}\cong \big[g-1~;\, g~;\, (q_i, {q_i -1\ov 2})\big]~,
\ee
exists if and only $q_i \in 2\Z+1$, $\forall i$. More generally, we will need to consider a spin$^c$ structure on $\h\Sigma$, which always exists---that is, on a given $\h \Sigma$, we may always introduce another line bundle $L$ such that $\CK \otimes L$ possesses a well-defined square root. We will come back to this point later in the discussion.

\paragraph{Riemann-Roch-Kawasaki theorem.}
Let $h^0(L) = {\rm dim} \,H^0(\h\Sigma, L)$
denote the number of holomorphic sections of the line bundle $L$ over $\h\Sigma_{g,n}$. The Riemann-Roch-Kawasaki theorem \cite{kawasaki1979} states that:
\be\label{RRK thm}
h^0(L) - h^0(L^{-1}\otimes \CK) = \deg(L) +1 -g~,
\ee
generalizing the smooth case.

\paragraph{The orbifold Picard group.} Let us denote by $\Pic(\h\Sigma)$ the group of linearly inequivalent line bundles, with the group multiplication given by the tensor product. 
Let us denote by $L_0$ and $L_j$ the elementary line bundles:
\be\label{def L0 Li}
L_0\cong  \big[1~; \, g~; \,(q_i, 0)\big]~, \qquad \qquad L_j \cong\big[0~; \, g~; \,(q_i, \delta_{ij})\big]~.
\ee
That is, $L_0$ is an ordinary line bundle of degree $1$, while the line bundle $L_j$ is an elementary orbifold line bundle with $b_j=1$ at the orbifold point $x_j$, and $b_i=0$ for $i \neq j$.
The Picard group takes the form:
\be
{\rm Pic}(\h \Sigma) = \Big\{L_0, L_i \;  \Big|  \;  L_i^{\otimes q_i}= L_0\; \; \forall i \Big\}~.
\ee
Let us emphasize that, in general, ${\rm Pic}(\h \Sigma)$ is not freely generated. 
To summarize, any $L\in {\rm Pic}(\h \Sigma)$ can be written as:
\be\label{2d L gen}
L \cong  L_0^{\otimes \n_0}\otimes L_1^{\otimes \n_1} \otimes \cdots \otimes L_n^{\otimes \n_n}~, 
\ee
for some integers $\n_0, \n_i$, giving $d=\n_0$, $b_i= \n_i$ up to the equivalences \eqref{equiv Ls}. It follows from the above discussion that:
\be
\deg(L)= \n_0 + \sum_{i=1}^n \floor{\n_i\ov q_i}~, \qquad b_i(L)= \n_i \mod q_i~,\qquad
c_1(L) = \n_0 + \sum_{i=1}^n {\n_i \ov q_i}~.
\ee
In the following, when thinking of the connection $A$ on $L$ as an abelian gauge field, we will often refer to $\n_0$ and $\n_i$ as the ``ordinary flux'' and the ``fractional fluxes'' of  the $U(1)$ gauge field $A$, respectively. They can be thought of as gauge fluxes localized at a smooth point $x_0 \in \h\Sigma$ or at the orbifold points $x_i$:
\be
dA=2\pi\, \n_0 \,\delta^2(x-x_0) +2\pi  \sum_i {\n_i\ov q_i} \, \delta^2(x-x_i)~, \qquad c_1(L) = {1\ov 2 \pi}\int_{\h\Sigma} dA~.
\ee
The statement $ L_i^{q_i}= L_0$ is the statement that $q_i$ units of fractional flux at $x_i$ are equivalent to a single unit of ordinary flux, which can then be moved away from the orbifold point.

\subsection{Seifert three-manifolds: definition and properties}
Given any line bundle $L$ over the two-dimensional orbifold $\h\Sigma$, we may consider the associated circle bundle $S[L]$. The total space of $S[L]$ is a smooth three-manifold if and only if, at each each orbifold point $x_i$, the integers $q_i$ and $b_i(L)$ are mutually prime. This follows simply from \eqref{zs quotient}. 
A {\it Seifert manifold} is a closed three-manifold $\CM_3$ endowed with a Seifert fibration:
\be\label{seifert fib}
S^1\longrightarrow \CM_3 \stackrel{\pi}\longrightarrow\h\Sigma~,
\ee
as we explain momentarily.
Any Seifert manifold $\CM_3$ can be viewed as the circle bundle associated to a certain 
{\it defining line bundle} $\CL_0$ over the orbifold $\h\Sigma_{g,n}(q_1, \cdots, q_n)$, with topological invariants:
\be
\deg(\CL_0) =\bd~, \qquad \quad b_i(\CL_0)= p_i~, \qquad  \gcd(q_i, p_i)=1~, \; \; \forall i~.
\ee
We summarize this construction by the standard short-hand notation:
\be\label{M3 Seifert def}
\CM_3 \cong S[\CL_0] \cong \big[\bd~; \, g~; \, (q_1, p_1)~, \cdots~, (q_n, p_n)\big]~.
\ee
The integers $\bd$ and $(q_i,p_i)$ appearing in \eqref{M3 Seifert def} are the so-called normalized Seifert invariants if $1 \leq p_i < q_i$. We will often consider the unnormalized invariants with $p_i\in \Z$, taking into account the equivalences \eqref{equiv Ls}. 
Moreover, it is possible to relax the condition that $q_i >0$, taking into account the equivalence $(q_i, p_i)\cong (-q_i, -p_i)$ for each exceptional fiber.  In the following, we will always choose $q_i>0$ unless otherwise stated.
For future reference, let us write down the first Chern class:
\be\label{c1L0}
c_1(\CL_0) = \bd + \sum_{i=1}^n {p_i\ov q_i}~.
\ee
This quantity is independent of the specific normalization of the Seifert invariants, since it remains invariant under the shift $\bd \rightarrow \bd-1$, $p_i \rightarrow p_i + q_i$, as well as under the inversion $(q_i, p_i) \rightarrow (-q_i, -p_i)$, for any $i$.

\begin{figure}[t]
	\begin{center}
		\subfigure[\small $T(1,0)$.]{
			\includegraphics[height=3.2cm]{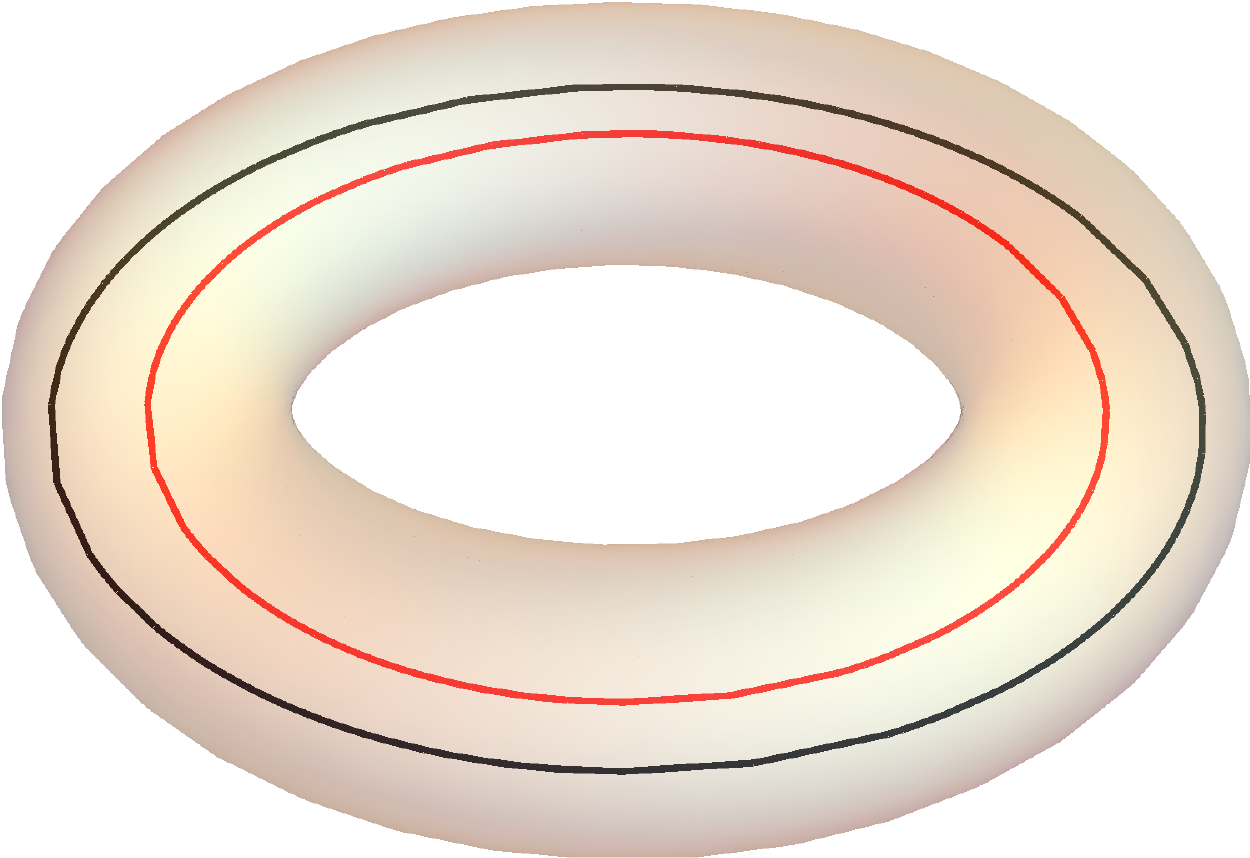}\label{fig:T10}}\quad
		\subfigure[\small $T(2,1)$.]{
			\includegraphics[height=3.2cm]{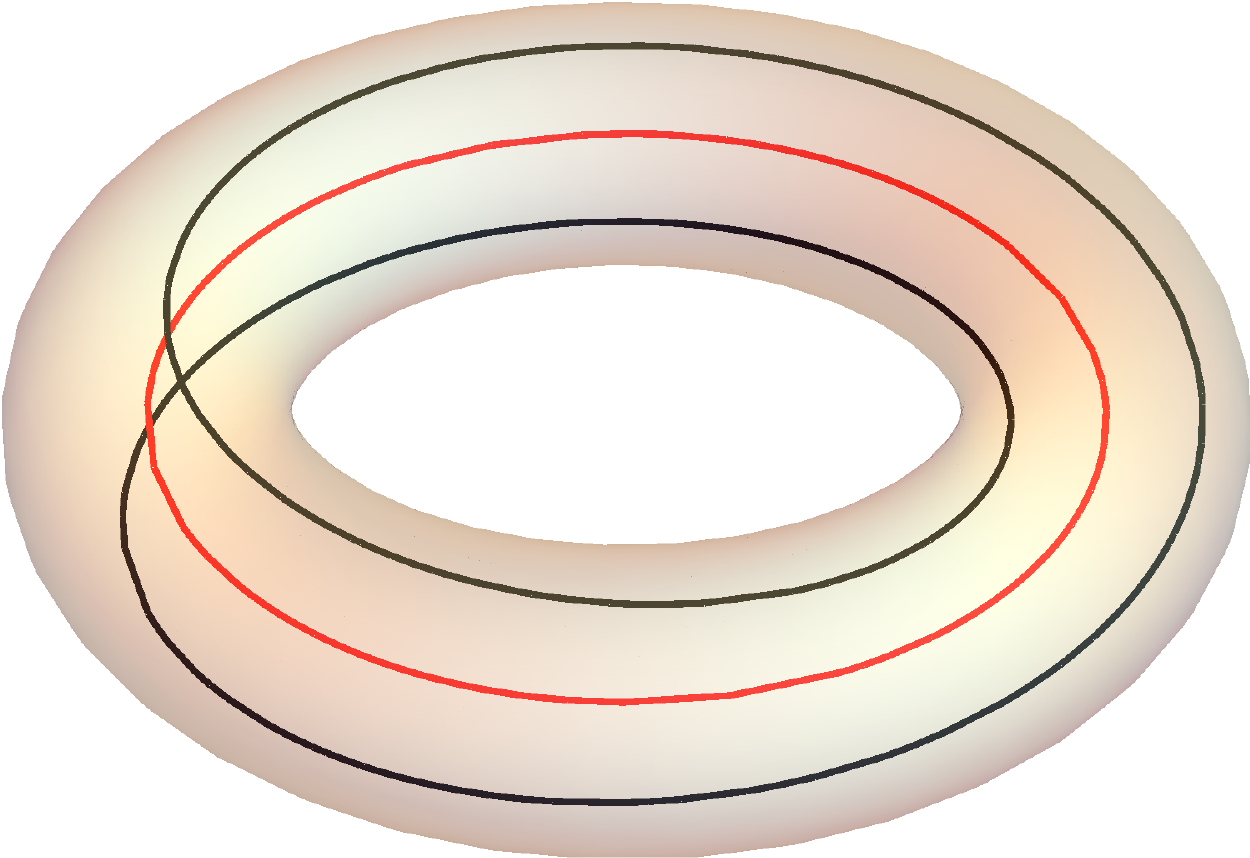}\label{fig:T21}}\quad
		\subfigure[\small $T(3,2)$.]{
			\includegraphics[height=3.2cm]{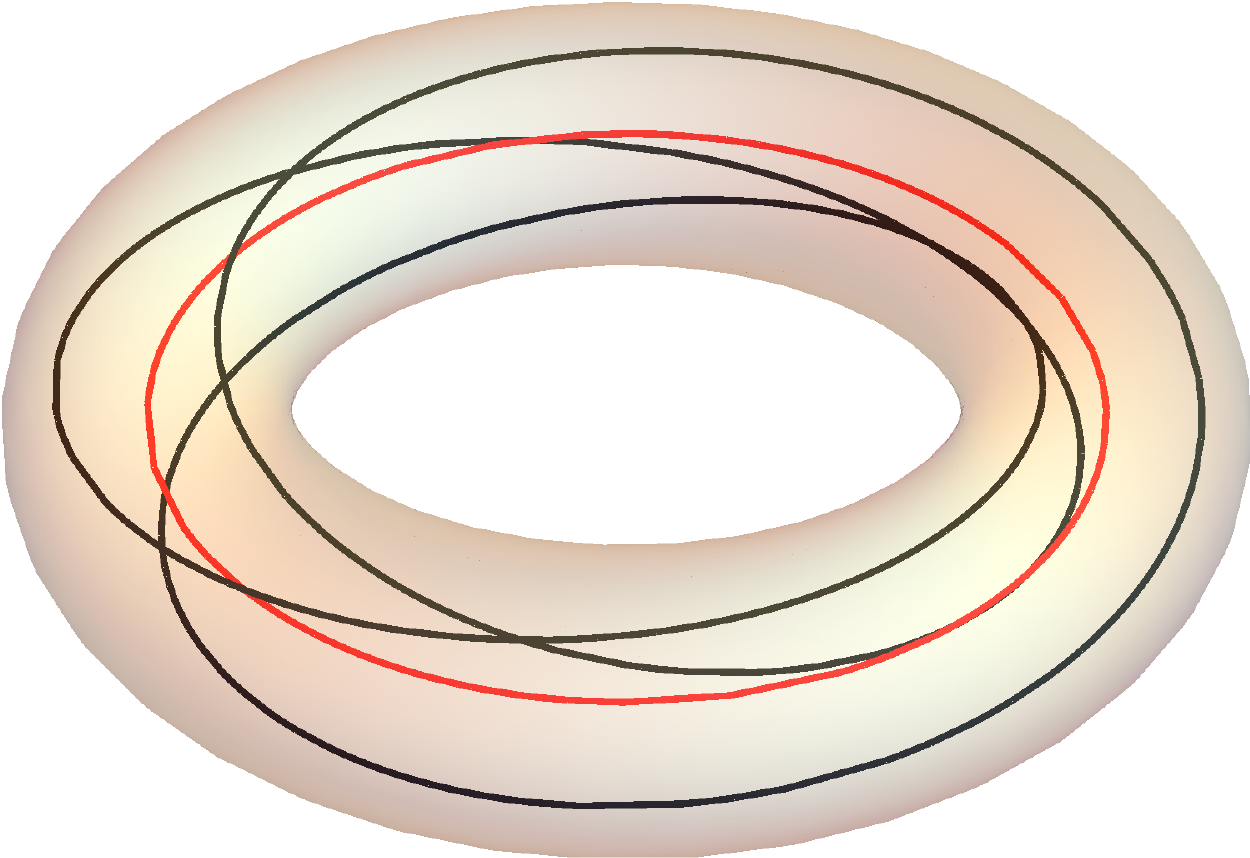}\label{fig:T32}}
		\caption{Solid fibered tori $T(q,t)$. The central fiber, shown in red, is exceptional if $q>1$. Generic fibers are shown in black.\label{fig:Tqt}}
	\end{center}
\end{figure}

A Seifert fibration \eqref{seifert fib} is a smooth map $\pi : \CM_3 \rightarrow \h \Sigma$ such that any point $x\in \h\Sigma$ has a neighborhood $D^2~\subset ~\h\Sigma$ (with $x$ at $r=0 \in D^2$, the center of the disk) whose pre-image is isomorphic to a solid fibered torus, $\pi^{-1}(D^2)~\cong~T(q,t)$. Here we define 
\be\label{Tqt def}
T(q, t) \cong D^2 \times_{t/q} S^1~,
\ee
 to be obtained by gluing the two disk  boundaries of the cylinder $D^2\times I$ with an angular twist  $2\pi t/q$, with $t\in \Z$ and $\gcd(q,t)=1$. In the local coordinates $(r, \varphi, \t\psi)$, with $r\in [0,1)$, $\varphi \in [0, 2\pi)$ the polar coordinates on the disk $D^2$, and  $\t\psi\in [0,2\pi)$ the angular coordinate on $S^1$, we identify:
\be\label{Tqk def}
\big(\t\psi~,\, \varphi\big)\sim \big(\t\psi+2 \pi~,\,  \varphi + {2\pi t\ov q}\big)~.
\ee
For $q>1$, the fiber $S^1$ at $r=0$ is an {\it exceptional Seifert fiber}. There can only be a finite number of exceptional fibers, at a finite number of points $x_i \in \h\Sigma$ ($i=1, \cdots, n$). The neighborhood of an exceptional fiber is illustrated in Figure~\ref{fig:Tqt}. At any smooth point $x_0\in \h\Sigma$, we simply have $\pi^{-1}(D^2)~\cong~T(1,0)$.

This description of $\CM_3$ is related to the $S^1$ bundle description \eqref{M3 Seifert def} as follows. For $T(1,0) \cong D^2 \times S^1$, we clearly have a (trivial) circle fibration structure over the disk, corresponding to the neighborhood of a smooth point on $\h\Sigma$. For $T(q,t)$, the neighborhood of an exceptional fiber, on the other hand, we can obtain a $\Z_q$~orbifold trivialization $D^2 \times S^1$ of the form \eqref{zs quotient}, by considering the $q$-covering $T(1, 0)\rightarrow T(t,q)$. Note that the generic fiber (at $r=r_0>0$) in $T(q,t)$ winds $q$ times around the torus. Let us define the coordinate ${\psi} =\tilde{\psi}/q$, so that the generic fiber has length $2\pi$, and the singular fiber (at $r=0$) has length $2\pi/q$. Then,  the identification \eqref{Tqk def} becomes $(\psi,\varphi) \sim  (\psi+{2\pi/q},\varphi + {2\pi t/q})$, which  is equivalent to:
\be\label{quotient 2}
\big(\varphi~,\, \psi\big)\sim \big(\varphi + {2 \pi \ov q}~,\, \psi + {2\pi p \ov q}\big)~,\qquad \text{with}\quad p t = 1 \mod q~,
\ee
reproducing the local trivialization \eqref{zs quotient} at an orbifold point.~\footnote{The integer pairs $[q,t]$ are called the {\it orbit invariants} and the integer pairs $(q, p)$ are called the {\it Seifert invariants}  \cite{orlik1972seifert}. The integer $t\in \Z_q$ is the so-called {\it modular inverse} of $p \in \Z_q$, and vice-versa.}

\paragraph{Surgery construction.} Any Seifert manifold can be constructed by some simple surgery operations, as summarized in Section~\ref{sec:intro surgery}. Consider a Seifert manifold $\CM_3$ presented as in \eqref{M3 Seifert def}. One may add a new exceptional fiber of type $(q,p)$, as follows. Consider cutting out $T(1,0)\cong D^2\times S^1$ around a generic fiber, with $D^2$ a small disk around the smooth point on the base. We thus obtain a three-manifold $\t\CM_3$ with boundary $\d \t\CM_3 \cong T^2 \cong -\d D^2 \times S^1$. One can then glue back $D^2 \times S^1$ along the boundary with an $SL(2, \Z)$ twist, to obtain a new closed three-manifold. If $(\varphi, \psi)$ and $(\varphi', \psi')$ denote the angular coordinates on the $T^2$ boundary of $\t \CM_3$ and $D^2 \times S^1$, respectively, we glue the boundaries according to:
\be\label{M mat}
\mat{\psi\\ \varphi} = M\mat{\psi'\\ \varphi'}~, \qquad M= \mat{q& -t \\ p & s}\in SL(2, \Z)~.
\ee
This introduces a new $(q,p)$ fiber, for $q$, $p$ two mutually prime integers; by convention, we choose $q>0$. In particular, this introduces a new $\Z_q$ orbifold point on the base of the Seifert fibration. The special case:
\be
M= \mat{1& 0 \\ 1 & 1}
\ee
leaves the base invariant. The introduction of such a ``$(1,1)$ fiber'' is equivalent to shifting the degree, $\bd \rightarrow \bd +1$. 
In this way, we may obtain the Seifert manifold \eqref{M3 Seifert def} starting from the trivial fibration:
\be
\Sigma_g \times S^1 \cong \big[0~; \, g~; \, ]~,
\ee
and performing surgery to introduce the exceptional fibers $(1, \bd)$ and $(q_i, p_i)$, $i=1, \cdots, n$. One may also consider the reverse process,  by performing surgery at exceptional fibers.  
Note that $s, t\in \Z$ in \eqref{M mat} are not fully determined by:
\be
q s+ p t =1~.
\ee
Given a solution $(s,t)$, we have an infinite number of solutions $(s+n p, t- n q)$, $n\in \Z$. This shift of $(s,t)$ corresponds to:
\be
M \rightarrow M T^n~, \qquad T= \mat{1 & 1\\ 0 & 1}~,
\ee
which does not affect the topology of the resulting thee-manifold.

\subsubsection{Fundamental group, (co)homology and line bundles}\label{sec:pi1 M3}
Consider the Seifert manifold $\CM_3 \cong [\bd; g; (q_i, p_i)]$. Its fundamental group $\pi_1(\CM_3)$ has the following explicit presentation:
\bea\label{pi1M3}
&\pi_1(\CM_3) =\cr
&\; \Big\langle a_l,\, b_l,\, g_i,\, h \; \Bigg|\; [a_l, h]=[b_l, h] =[g_i, h]=g_i^{q_i} h^{p_i}=1, \;\; \prod_{l=1}^g [a_l, b_l] \prod_{i=1}^n g_i = h^{\bd} \Big\rangle~,
\eea
with $l=1, \cdots g$, $i=1, \cdots, n$, and $[a,b]= aba^{-1}b^{-1}$ the group commutator.
The generators $a_l, b_l$ correspond to the $A$- and $B$-cycles of the underlying Riemann surface $\Sigma_{g}$,  $g_i$ corresponds to a loop in $\h \Sigma$ around the orbifold points $x_i$,  and $h$ corresponds to a generic Seifert fiber.

The first homology of $\CM_3$ is the abelianization of \eqref{pi1M3}.
It can be written as:
\be\label{H1 M3}
H_1(\CM_3, \Z) \cong H^2(\CM_3, \Z) \cong  \t\Pic(\CM_3) \oplus \Z^{2g}~.
\ee
Here, the free factor $\Z^{2 g}$ in $H_1(\CM_3, \Z)$ corresponds to the one-cycles $a_l, b_l$ of $\Sigma_g$.
The group $\t\Pic(\CM_3)$ which appears in \eqref{H1 M3}  can be viewed as the pull-back of the orbifold Picard group $\Pic(\h\Sigma)$ through the map: 
\be
\pi^\ast \, : \, H^2(\h\Sigma, \Z)  \rightarrow  H^2(\CM_3, \Z)~,
\ee
where we identified lines bundles over $\h\Sigma$ or $\CM_3$ with their first Chern classes in $H^2(\h\Sigma, \Z)$ or $H^2(\CM_3, \Z)$, respectively. 
One finds:%
~\footnote{Assuming that we can simply generalize the smooth case to the orbifold case, this follows from the Gysin sequence,  $\cdots \longrightarrow H^0(\h \Sigma)  \stackrel{c_1(\CL_0)}\longrightarrow H^2(\h\Sigma)  \stackrel{\pi^\ast}\rightarrow H^2(\CM_3) \stackrel{\pi_\ast}\longrightarrow \cdots$.}
\be\label{pic3 as quotient}
\t\Pic(\CM_3) \cong \Pic(\h\Sigma)/(\CL_0)~.
\ee
That is, the group $\t\Pic(\CM_3)$ is isomorphic to the two-dimensional orbifold Picard group modulo tensor products with the defining line bundle $\CL_0$. One can also show that $\t\Pic(\CM_3)$ is finite if and only if $c_1(\CL_0)\neq 0$, with:
\be
\t\Pic(\CM_3)\cong  \begin{cases} {\rm Tor} \, H_1(\CM_3, \Z) &  {\rm if}\;\; \; c_1(\CL_0)\neq 0~,\\	{\rm Tor} \, H_1(\CM_3, \Z) \oplus \Z	\qquad& {\rm if} \;\; \; c_1(\CL_0)=0~.
\end{cases}			
\ee
We also have:
\be
H_2(\CM_3, \Z) \cong H^1(\CM_3, \Z) \cong 
\begin{cases} \Z^{2 g} &  {\rm if}\;\; \; c_1(\CL_0)\neq 0~,\\	\Z \oplus \Z^{2 g}	\qquad& {\rm if} \;\; \; c_1(\CL_0)=0~.
\end{cases}	
\ee

\paragraph{Line bundles over $\CM_3$.} 
The group $\t\Pic(\CM_3)$ is the group of complex line bundles over $\CM_3$ which are the pull-backs of orbifold line bundles over $\h\Sigma$. For later purposes, it is useful to spell this out explicitly.
Let us introduce the generators:
\be
[\gamma],\; [\omega_i]\; \in\, \t\Pic(\CM_3) \subset H_1(\CM_3, \Z)~, 
\ee
such that $h= e^{\gamma}$ and $g_i = e^{\omega_i}$ in the abelianization of \eqref{pi1M3}.
Here, $\gamma$ and $\omega_i$ are viewed as representatives of first homology classes in $H_1(\CM_3, \Z)$ (which are all torsion if $c_1(\CL_0)~\neq0$). They are in one-to-one correspondence with the elementary line bundles \eqref{def L0 Li} in $\Pic(\h\Sigma)$, according to:
\be
[\gamma] \sim L_0^{-1}~, \qquad  [\omega_i] \sim  L_i^{p_i}~.
\ee
More precisely, the pull-back of $L_0$ to $\CM_3$ is a line bundle $\pi^\ast(L_0)$ represented by $-[\gamma]$, and the pull-back of $L_i^{p_i}$ is represented by $[\omega_i]$.
Therefore, the abelian group $\t\Pic(\CM_3)$ has the explicit presentation:
\be\label{Pic M3 presentation}
\t\Pic(\CM_3)  \cong \bigg\{\; [\gamma],\;  [\omega_i] \:\; \Bigg| \;\;  q_i [\omega_i]+ p_i [\gamma]=0, \; \forall i~, \;\; \sum_i [\omega_i]= \bd [\gamma]  \bigg\}~.
\ee
For each pair $(q_i, p_i)$, let us introduce the integers $s_i$ and $t_i$ such that:
\be
s_i q_i+ t_i p_i=1~.
\ee
 The elementary line bundles $L_0$,  $L_i$ pull-back to the following ordinary line bundles over $\CM_3$:~\footnote{Note that $ - s_i [\gamma]+ t_i [\omega_i]\in \t\Pic(\CM_3)$ is independent of the choise of $s_i$, $t_i$ solving $s_i q_i + t_i p_i=1$.}
\be\label{pullback L Li}
\pi^*(L_0) \cong -[\gamma]~, \qquad \qquad
\pi^*(L_i) \cong   - s_i [\gamma]+ t_i [\omega_i]~.
\ee
This directly gives us the pull-back of  an arbitrary orbifold line bundle \eqref{2d L gen} to $\CM_3$.

\subsubsection{The classification of Seifert manifolds}\label{subsec:classification seifert}
Most three-manifolds that admit a Seifert fibration $\pi : \CM_3 \rightarrow \h\Sigma$ do so in a {\it unique} way. The only---and very important---exceptions are the lens spaces (including $S^2\times S^1$), which are realized as genus-zero Seifert manifolds with $n\leq 2$ exceptional fibers. Each lens space $L(p,q)$ admits an infinite number of inequivalent Seifert fibrations. This will be discussed in greater detail below.
For the purpose of classification, it is useful to define a {\it small Seifert manifold} as one amongst the following short list \cite{orlik1972seifert}:%
~\footnote{We restrict ourselves to the case of oriented Seifert manifolds with an oriented base $\h\Sigma$. The fuller classification is discussed in~\cite{orlik1972seifert}, where it is also shown that the prism manifolds (a subset of case (ii)) and the case (iii) admit another, distinct Seifert fibration over an unorientable surface.}
\begin{enumerate}
	\item[(i)] The lens spaces $\CM_3\cong [0; 0; (q_1, p_1), (q_2, p_2)]$.
	\item[(ii)] The manifolds $\CM_3 \cong [0; 0; (q_1, p_1), (q_2, p_2), (q_3, p_3)]$ such that ${1\ov q_1}+{1\ov q_2}+{1\ov q_3}>1$.
	\item[(iii)] The manifold $\CM_3 \cong [0; 0; (2,1),(2,1),(2,-1),(2,-1)]$.
	\item[(iv)] The manifolds $\CM_3\cong [0; 1; (1,p)]$.
\end{enumerate}
Here we have set $\bd=0$ by using unnormalized Seifert symbols; these four cases will be discussed in detail  in Section~\ref{subsec:expls}. 
Any other Seifert manifold is called {\it large.} 
It turns out that any two large Seifert manifolds $\CM_3$ and $\CM_3'$ have equivalent Seifert fibrations if and only they are homeomorphic~\cite{orlik1972seifert}. Moreover, they are homeomorphic if and only if they have the same fundamental group.
Note that this topological classification of Seifert manifolds is valid up to orientation reversal, which acts on $\CM_3\cong [\bd; g; (q_i, p_i)]$ as $-\CM_3 \cong [-\bd; g; (q_i, -p_i)]$.

Interestingly, the only Seifert manifolds with a {\it finite} fundamental group \eqref{pi1M3} are the small Seifert manifolds of type (i) or (ii). Otherwise, the order of $\pi_1(\CM_3)$ is infinite.
Seifert manifolds span six out of the eight Thurston geometries \cite{BLMS:BLMS0401}.~\footnote{The hyperbolic ($\bH^3$) and Sol geometries are not Seifert.} It is convenient to distinguish three cases according to the Euler characteristic of the base. If $\chi(\h\Sigma)>0$, the Seifert manifold is modeled on $S^3$ if $\pi_1(\CM_3)$ is finite, or on $S^2 \times \R$ otherwise. If $\chi(\h \Sigma)=0$, $\CM_3$ is Euclidean if $c_1(\CL_0)=0$, and has Nil geometry otherwise. If  $\chi(\h \Sigma)<0$, the general case, then $\CM_3$ has $\t{ SL(2, \R)}$ or $\bH^2\times \R$ geometry.

\subsubsection{Adapted metric and transversely holomorphic foliation}\label{sec:THF}
Given a Seifert manifold $\CM_3$, we choose an adapted metric:
\be
ds^2(\CM_3) = \eta^2 + ds^2(\h\Sigma)~,
\ee
which admits a real Killing vector $K$ along the Seifert fiber. The one-form $\eta$ is dual to $K$---that is, $\eta_\mu =K_\mu$ in local coordinates. We introduce the local coordinates $\psi, z, \bz$, with $\psi\in [0, 2\pi)$ the angular coordinate along the fiber, and $z, \bz$ the complex coordinates along the orbifold $\h\Sigma$, so that:
\be
\eta = \beta(d\psi+ \CC(z,\bz))~, \qquad\qquad  K ={1\ov \beta} \d_\psi~, \qquad \qquad K^\mu\eta_\mu=1~.
\ee 
and:
\be\label{metric M3}
ds^2(\CM_3)= \beta^2 \big(d\psi + \CC(z,\bz)\big)^2 + 2 g_{z\bz}(z, \bz) dz d\bz~,
\ee
with $\beta>0$ the radius of a generic fiber. Let us normalize the volume of $\h\Sigma$ to $\vol(\h\Sigma) =\pi$. The connection $\CC$ is a real two-dimensional connection on the defining line bundle $\CL_0$. Its curvature reads $\d_z \CC_\bz - \d_\bz \CC_z = c_1(\CL_0) \, 2i g_{z\bz}$, with $c_1(\CL_0)$ given in \eqref{c1L0}. Therefore:
\be
{1\ov 2\pi} \int_{\h \Sigma} d\CC = c_1(\CL_0)= \bd + \sum_{i=1}^n {p_i\ov q_i}~.
\ee
Note also the useful identities:
\be\label{deta explicit}
d\eta =2 \beta c_1(\CL_0)\, d\vol(\h\Sigma)~, \qquad \qquad \eta\wedge d\eta = 2 \beta c_1(\CL_0) \, d\vol(\CM_3)~.
\ee

\paragraph{THF and the canonical line bundle of $\CM_3$.} A Seifert fibration is a special case of a {\it transversely holomorphic foliation} (THF), viewed as the foliation generated by the nowhere-vanishing vector field $K$---see \cite{Closset:2012ru, Closset:2013vra} and references therein. A metric-compatible THF can be characterized by the tensors $\eta_\mu$ and
${\Phi_\mu}^\nu= - {\epsilon_\mu}^{\nu\rho}\eta_\rho$, 
which gives a three-dimensional analogue to a complex structure. Under a change of  adapted coordinates, we have: 
\be\label{change of coord adpt}
\psi'= \psi -\lambda(z, \bz)~,\qquad z'= f(z)~,
\ee
with $\lambda$ real and $f(z)$ a holomorphic function of $z$.
Given a THF, there exists a notion of Dolbeault-like decomposition of differential forms  \cite{Closset:2013vra}. In particular, one can define the notion of a holomorphic one-form,   $\omega \in \Lambda^{1,0}\CM_3$, such that:
\be\label{def holo oneform 3d}
\omega_\mu {\Pi^\mu}_\nu= \omega_\nu~,\qquad \quad  {\Pi^\mu}_\nu =\half \left({\delta^\mu}_\nu - i {\Phi^\mu}_\nu- \eta^\mu \eta_\nu\right)~.
\ee
Thus $\omega$ has a single component, $\omega_z$, which transforms as $\omega_{z'}'=(\d_z f(z))^{-1}\omega_z$ under \eqref{change of coord adpt}. This makes  $\omega_z$  a section of a {\it holomorphic} line bundle over $\CM_3$, by definition. We call this particular holomorphic line bundle the {\it canonical line bundle} of the Seifert manifold $\CM_3$,  denoted by $\CK_{\CM_3}$.
It is also the pull-back of the canonical line bundle of $\h\Sigma$ through the Seifert fibration:
\be
\pi^\ast(\CK)= \CK_{\CM_3}~.
\ee
We see from \eqref{def K}-\eqref{def L0 Li} that $\CK \cong L_0^{2g-2}\otimes_i L_i^{q_i-1}$. Therefore, \eqref{pullback L Li} implies:
\be\label{KM3}
\CK_{\CM_3} \cong   \Big(2-2g -n+ \sum_{i=1}^n  s_i \Big)[\gamma]- \sum_{i=1}^n t_i [\omega_i] \:  \in \: \t\Pic(\CM_3)~.
\ee
In particular, the canonical line bundle of $\CM_3$ is topologically trivial if and only if $\CK_{\CM_3}\cong 0$ in  $\t\Pic(\CM_3)$.

 THFs on three-manifolds were classified in \cite{Brunella1996, Ghys1996}. It turns out that ``most'' three-manifolds that admit a THF are Seifert manifolds (in the sense that they admit at least one Seifert fibration), and can thus preserve two supercharges.~\footnote{There can also exist more general THFs on lens spaces which only preserve one supercharge, as discussed in \cite{Closset:2013vra}.}
We also expect that most half-BPS Seifert backgrounds do not admit any continuous ``squashing'' deformations, in the following sense. In order to affect supersymmetric observables, any half-BPS ``squashing'' should modify the supersymmetry algebra \eqref{QQb intro}, deforming $K$ to $K+\epsilon K'$, with $K'$ a distinct Killing vector on $\CM_3$. This $K'$ must  be the pull-back of a Killing vector along the base $\h \Sigma_{g, n}$, which can only exist at genus $0$ with $n \leq 2$ (giving us the lens space case) or at genus $1$. It would be interesting to verify this by a direct analysis of the THF moduli of half-BPS Seifert manifolds.

\subsubsection{Spin structures on $\CM_3$}
Recall that any orientable three-manifold is spin, and that the distinct spin structures are in one-to-one (non-canonical) correspondence with elements of $H^1(\CM_3, \Z_2)$. Using the universal coefficient theorem and Poincar\'e duality, one finds:
\be
H^1(\CM_3, \Z_2) \cong \left(H^1(\CM_3, \Z) \otimes \Z_2\right) \oplus {\rm Tor}\big(H_1(\CM_3, \Z), \Z_2\big)~.
\ee
Here ${\rm Tor}(A, \Z_2)\cong \{ a\in A\,|\, 2 a=0\}$ for any abelian group $A$---the group of elements that square to zero. 
From Subsection~\ref{sec:pi1 M3}, we find:
\be
\label{M3 spin structures} 
H^1(\CM_3, \Z_2)\cong  \Z^{2g}_2 \oplus \left(\t\Pic(\CM_3)\otimes \Z_2\right)~.
\ee
The factor $\Z^{2g}_2$ corresponds to the familiar choice of spin structure on the underlying genus-$g$ Riemann surface. The supersymmetric observables will be independent of that choice. The factor $\t\Pic(\CM_3)\otimes \Z_2$ reflects the further possibility of choosing the periodic or anti-periodic boundary conditions for fermions along all the other non-trivial one-cycles in $\CM_3$.
Note that:~\footnote{For any abelian group $A$ of rank ${\rm r}$, one has $A \otimes \Z_2 \cong {\rm Tor}(A, \Z_2) \oplus \Z_2^{\rm r}$.}
\be\label{PicM3 tensor Z2}
\t\Pic(\CM_3)\otimes \Z_2  \cong \begin{cases}
{\rm Tor}\big(\t\Pic(\CM_3), \Z_2\big) \qquad &  {\rm if}\;\; \; c_1(\CL_0)\neq 0~,\\
{\rm Tor}\big(\t\Pic(\CM_3), \Z_2\big) \oplus \Z_2 \qquad&  {\rm if}\;\; \; c_1(\CL_0)= 0~.
  \end{cases}
\ee
Now, let us sketch a more explicit description of the spin structure. 

\paragraph{Spin structures on $\h \Sigma$.} 
On a smooth Riemann surface $\Sigma_g$, choosing a spin structure is equivalent to choosing a square root of the canonical line bundle:
\be
\sqrt{\CK}~.
\ee
This square root always exists in that case, with $c_1(\CK)=g-1$. There is a $\Z_2^{2g}$ ambiguity corresponding to the boundary conditions for the fermions along the $A$- and $B$-cycles. Equivalently, we can introduce a $\Z_2$-valued flat connection on $\Sigma_g$, which couples non-trivially to fields of half-integer spin, as follows. The two-dimensional connection $\CA_\mu^\CK$ on the canonical line bundle over $\h\Sigma$ is given by the spin connection:
\be\label{spin conn 2d def}
\CA_\mu^{\CK} =\omega^{(2d)}_\mu~.
\ee
Then, choosing a spin structure with periodic boundary conditions for fermions around a one-cycle $C_l$ is equivalent to choosing a connection $\half \omega^{(2d)}_\mu$ on $\sqrt{\CK}$ with a non-trivial $\Z_2$ holonomy around $C_l$. 
For a two-dimensional orbifold $\h\Sigma(q_i)$,  on the other hand, we saw in \eqref{def Ksqrt} that the square root $\sqrt{\CK}$ only exists if all the $q_i$'s are odd integers. When it exists, $\sqrt{\CK}$ has a $\Z_2^{2g}$ ambiguity as in smooth case---see {\it e.g.} \cite{geiges2012}.

While $\sqrt{\CK}$ generally does not exist, we may be able to define a spin$^c$ structure for some spinors, in the sense that there might exist a line bundle $L \in \Pic(\h\Sigma)$ such that 
$\sqrt{\CK \otimes L}$ exists. As we will show, this is precisely what happens in the 3d A-model.

\paragraph{Spin structures on $\CM_3$.} 
Consider a three-dimensional Seifert manifold $\CM_3$ with an adapted metric. Similarly to the 2d case, it is natural to consider the square roots of the three-dimensional canonical line bundle \eqref{KM3},
\be\label{def sqrt K3d}
{\mathscr S} \,\in  \t\Pic(\CM_3)\qquad \text{such that} \quad  {\mathscr S}\otimes  {\mathscr S}  = \CK_{\CM_3}~.
\ee
Since $\CK_{\CM_3} = \pi^\ast(\CK)$ by definition, we may view the ``holomorphic'' cotangent bundle of $\CM_3$ as:
\be
T^{(1,0)}\CM_3 \cong  \R \oplus \CK_{\CM_3}~,
\ee
with fibers $\R \oplus \C$ adapted to the THF. The 3d spin bundle, denoted by $S$, is a $\C^2$ vector bundle  over $\CM_3$ with structure group ${\rm Spin}(3)=SU(2)$, such that:
\bea\label{S square}
&S\otimes S &\cong&\;  \CO \oplus T^\ast \CM_3\cr
&& \cong&\; \CO \oplus \CO \oplus \CK_{\CM_3} \oplus \CK_{\CM_3}^{-1}~.
\eea
In other words, for any two Dirac spinors $\psi_\alpha$ and $\t\psi_\alpha$ valued in $S$, we have the obvious decomposition $\psi_\alpha \otimes \t\psi_\alpha \cong \psi \oplus \psi_\mu$ into a scalar and a vector.%
~\footnote{Our conventions for spinors are summarized in Appendix~\protect\ref{app:geom}. We will always choose an adapted frame $(e^0, e^{1}, e^{\b 1})= (\eta, \sqrt{2 g_{z\bz}}dz, \sqrt{2 g_{z\bz}}d\bz)$.
In the second equation in \protect\eqref{S square}, we view $\CK_{\CM_3}$ as the associated $U(1)$ line bundle, so that $\CO \oplus \CK_{\CM_3} \oplus \CK_{\CM_3}^{-1}$ corresponds to the decomposition of a vector into components $v_0, v_1, v_{\b 1}$ in the adapted frame.}
The spin bundle over $\CM_3$ can then be written as:
\be\label{S as S Sin}
S\cong {\mathscr S} \oplus {\mathscr S}^{-1}~,
\ee
with ${\mathscr S}$ defined in \eqref{def sqrt K3d}. The decomposition \eqref{S as S Sin} also naturally corresponds to the decomposition of a 3d Dirac spinor $\psi_\alpha$ into the two-dimensional Weyl spinors $\psi_\pm$. (We refer to Appendix D of  \protect\cite{Martelli:2012sz} and to \cite{Toldo:2017qsh} for related discussions of $S$ some special cases.)

Therefore, choosing a spin structure on $\CM_3$ is equivalent to choosing a square root ${\mathscr S}$ of $\CK_{\CM_3}$.~\footnote{Here and in the following, we ignore the additional choice of spin structure on the underlying smooth Riemann surface $\Sigma_g$. As we mentioned, the supersymmetric observables are independent of that choice. Heuristically, this is because, after the topological A-twist, all the A-twisted fields have integer spins on $\Sigma_g$ and therefore do not depend on the spin structure on the Riemann surface.} Let us assume that we have found a solution to \eqref{def sqrt K3d}, denoted by ${\mathscr S}_0$. If $c_1(\CL_0)\neq 0$ for the defining line bundle $\CL_0$ over $\h\Sigma$,   $\t\Pic(\CM_3)$ is a finite group and any ${\mathscr S}$ can be written as a product of ${\mathscr S}_0$ with a nilpotent element:
\be
{\mathscr S}= {\mathscr S}_0 \otimes {\mathscr N}~, \qquad\quad  {\mathscr N}\in {\rm Tor}\big(\t\Pic(\CM_3), \Z_2\big)~.
\ee
If $c_1(\CL_0)=0$, there is an additional twofold freedom associated with the free generator of $\t\Pic(\CM_3)$. We will show this more explicitly in the next subsection. In this way, the set of spin structures on $\CM_3$ are indeed in one-to-one correspondence with elements of the group \eqref{PicM3 tensor Z2}.

\subsection{A-twist and supersymmetric backgrounds on Seifert manifolds}
\label{subsec: M3 backgd}
On $\CM_3$ with the adapted metric \eqref{metric M3}, we may preserve two supercharges $\CQ$ and $\b\CQ$, corresponding to the generalized Killing spinors $\zeta$ and $\b\zeta$ of $R$-charge $1$ and $-1$, respectively  \cite{Closset:2012ru, Klare:2012gn}. 
A rigid supersymmetric background on $\CM_3$ consists of supersymmetry-preserving background values for the bosonic fields in the 3d $\CN=2$ new-minimal supergravity multiplet \cite{Festuccia:2011ws, Closset:2012ru, Kuzenko:2013uya}:
\be
\CH = \left(g_{\mu\nu}~, \; A_\mu^{(R)}~, \, H~, \, V_\mu\right)~,
\ee
where $A_\mu^{(R)}$ is a background gauge field for the $U(1)_R$ symmetry. The metric $g_{\mu\nu}$ is given by~\eqref{metric M3}, and the other background fields are:
\be\label{AVH sugra}
A_\mu^{(R)}= \CA_\mu^{(R)} + \beta c_1(\CL_0) \eta_\mu~,
\qquad
H= i \beta c_1(\CL_0)~, \qquad V_\mu = -2 \beta c_1(\CL_0) \, \eta_\mu~, 
\ee 
with the $U(1)_R$ gauge field $\CA^{(R)}$ given by the expression:
\be\label{def CAR}
\CA^{(R)}_\mu= {1\ov 4} {\Phi_\mu}^\nu \d_\nu \log\sqrt{g} + \d_\mu s~,
\ee
in the adapted coordinates $\psi, z, \bz$. Using a modified (torsionfull) connection $\h\nabla$ such that $\h\nabla \eta=0$, as defined in Appendix~\ref{app:geom}, the Killing spinor equations take the simple form:
\be
(\h\nabla_\mu - i \CA_\mu^{(R)})\zeta=0~, \qquad \qquad  (\h\nabla_\mu +i \CA_\mu^{(R)})\t\zeta=0~,
\ee
The THF-adapted connection $\h\nabla$ has $U(1)$ holonomy, which can be compensated by the holonomy of the $U(1)_R$ connection \eqref{def CAR}, generalizing the familiar topological A-twist from two to three dimensions \cite{Closset:2012ru, Closset:2014uda, Closset:2017zgf}. The Killing spinors read:
\be\label{KS explicit}
(\zeta_\alpha)=e^{i s}\mat{0\cr 1}~, \qquad\qquad (\b\zeta_\alpha)=e^{-i s}\mat{1\cr 0}~,
\ee
in the adapted frame. The function $s$ appearing in \eqref{def CAR} and \eqref{KS explicit} encodes the freedom in choosing the spin structure associated to our supersymmetric background, as we will now explain.

\subsubsection{Three-dimensional A-twist and spin structure}
Let ${\bf L}_R$ be the $U(1)_R$ line bundle on $\CM_3$. Given the Killing spinor $\b\zeta$, we may define a spinor bilinear valued in ${\bf L}_R^{-2}$:
\be
P_\mu = \b\zeta \sigma_\mu \b\zeta~.
\ee
One can show that $P_\mu$ is in fact a holomorphic one-form \cite{Closset:2012ru}---it satisfies  \eqref{def holo oneform 3d}. It follows that:
\be\label{Pz explicit}
P_z = e^{-2 i s} \sqrt{2 g_{z\bz}}
\ee
is a nowhere-vanishing section of $\CK_{\CM_3}  \otimes {\bf L}_R^{-2}$. We thus have:
\be
\CK_{\CM_3} \cong {\bf L}_R^2~.
\ee
Therefore, the more precise statement of the three-dimensional A-twist \cite{Closset:2017zgf} is that we choose the $R$-symmetry line bundle ${\bf L}_R$ to be {\it  any} well-defined square root of the canonical line bundle, as in \eqref{def sqrt K3d}:
\be\label{3d A twist another def}
 {\bf L}_R \cong {\mathscr S}~.
\ee
The connection \eqref{def CAR} on ${\bf L}_R$ can be written as:
\be\label{CAR rewritten}
\CA^{(R)}_\mu= \half \pi^\ast(\omega_\mu^{(2d)}) + \CA^{(R),\rm flat}_\mu~,
\ee
in terms of the spin connection \eqref{spin conn 2d def} on $\h\Sigma$ and of a flat connection $\CA^{\rm flat}_\mu$ with $\Z_2$ holonomies:
\be
e^{-i \int_\CC \CA^{(R),\rm flat}}= \pm 1~, \qquad\quad \forall [\CC]\in H_1(\CM_3, \Z)~.
\ee
Since we have $\CA_\mu^{(R),\rm flat}= \d_\mu s$, the ``function'' $e^{is}$ in \eqref{KS explicit} should really be seen as a non-trivial map: 
\be 
e^{i s}: \CM_3 \rightarrow \C^\ast~.
\ee
Once we have fixed some $s_0$ so that \eqref{CAR rewritten} is well-defined, we can also find any other maps $e^{is}$ related to $e^{is_0}$ by a non-trivial $\Z_2$-valued homotopy \cite{Closset:2014uda}.
The A-twist \eqref{3d A twist another def} correlates this choice of $\CA_\mu^{(R),\rm flat}$ with the choice of spin structure on $\CM_3$, precisely so that the Killing spinors \eqref{KS explicit} are always invariant. In other words, for every one-cycle on $\CM_3$, we might choose either the periodic or the anti-periodic boundary condition for the fermions, as long as we also choose the $R$-symmetry background gauge field with $\Z_2$-valued holonomies in such a way that the Killing spinors $\zeta$ and $\b\zeta$, of $R$-charges $\pm1$, remain periodic and therefore globally defined. 

In this way, we obviously preserve supersymmetry. Even though a particular fermionic dynamical  field might be either periodic or anti-periodic in some spin structure, the $U(1)_R$ background ensures that all the scalars in the same supersymmetry multiplet have the same boundary conditions, thus preserving supersymmetry.  Note that, for a vector multiplet, both the gauge field and the gauginos are periodic in any spin structure, since the gauge field has zero $R$-charge while the gauginos have $R$-charge $\pm 1$ just like the Killing spinors.

We should also note that the two-dimensional connection $\half \omega_\mu^{(2d)}$ in \eqref{CAR rewritten} is not well-defined by itself if $\sqrt{\CK}$ does not exist on $\h\Sigma$. When that happens, the flat connection $\CA^{(R),\rm flat}_\mu$ will be similarly ill-defined such that the full three-dimensional $\CA^{(R)}_\mu$ is well-defined. This construction can be made very explicit from the two-dimensional point of view, as we now explain.

\subsubsection{Constructing $L_R$ on $\Sigma_g$}\label{subsec: def LR}
We can view the Seifert supersymmetric background $(\CM_3, {\bf L}_R)$ as the pull-back of the topological A-twist background on $\h\Sigma$ \cite{Dumitrescu:2012ha, Closset:2012ru, Closset:2017zgf}. In particular, there exists a two-dimensional $R$-symmetry line bundle $L_R\in \Pic(\h\Sigma)$ such that:
\be
\pi^\ast(L_R) \cong {\bf L}_R \cong {\mathscr S}~,
\ee
with ${\mathscr S}$ defined in \eqref{def sqrt K3d}.
In the rest of this paper, it will be most useful to consider any three-dimensional $\CN=2$ supersymmetric field theory as a two-dimensional theory with $\CN=(2,2)$ supersymmetry, topologically twisted along $\h\Sigma$ \cite{Closset:2017zgf}. The two-dimensional field theory has an infinite number of fields, corresponding to the Kaluza-Klein (KK) modes of the 3d fields along the Seifert fiber:
\be\label{KK expansion gen}
\varphi = \sum_{\kk \in \Z} \varphi_\kk(z, \bz) e^{i \kk \psi}~.
\ee
For any field $\varphi$ a section of a vector bundle ${\bf E}$ over $\CM_3$, the two-dimensional modes are sections:
\be
\varphi_\kk \in \Gamma[E \otimes \CL_0^\kk]~,
\ee
where $E$ is such that $\pi^\ast(E)= {\bf E}$. 

The Killing spinors \eqref{KS explicit} need {\it not} be constant along the Seifert fiber. We generally have:
\be\label{introduce nuR}
s = -\nu_R \psi + \cdots~, \qquad   \qquad \nu_R \in \half\Z~.
\ee
in local coordinates, where the ellipsis denotes $\psi$-independent term. 
The parameter $\nu_R$ encodes the $\Z_2$ holonomy of the flat connection $\CA_\mu^{(R),\rm flat}$ along the generic Seifert fiber:
\be
-{1\ov 2\pi}\int_\gamma \CA^{(R),\rm flat} = \nu_R~.
\ee
Somewhat formally, the 3d spinors $\b\zeta$ is the pull-back of the following nowhere-vanishing section on $\h\Sigma$:
\be
\b\zeta_- \in \Gamma\left[ \sqrt{\CK} \otimes \CN \otimes (\CL_0)^{\nu_R} \otimes L_R^{-1}\right]~,
\ee
and similarly for $\zeta$. Here, $\sqrt{\CK}$ is the naive square root defined in \eqref{def Ksqrt}, $\CN$ is some formally nilpotent 2d line bundle in $\Pic(\h\Sigma)$, while $\CL_0$ pulls back to the trivial line bundle in $\t\Pic(\CM_3)$. In general, both $\sqrt{\CK}$ and $\CN \otimes (\CL_0)^{\nu_R}$ are ill-defined, in such a way that $L_R$ is well-defined. 
This gives a slight generalization of the usual A-twist on a Riemann surface, with the identification:
\be\label{LR 2d general definition}
L_R \cong  \sqrt{\CK} \otimes \CN \otimes (\CL_0)^{\nu_R}~.
\ee
We can now characterize this $L_R$ explicitly, as an element of $\Pic(\h\Sigma)$. We have:
\be\label{LR full}
L_R \cong L_0^{\t\n_0^R}\, \bigotimes_{i=1}^n  L_i^{\n_i^R}~,
\ee
with $\t\n_0^R\in \Z$ the ``ordinary flux'' and $\n_i^R\in \Z$ the ``fractional fluxes'' for $U(1)_R$. 
For future reference, let us define another integer $\n_0^R$ by:
\be\label{param LR}
\t\n_0^R=g-1+ \n_0^R~.
\ee  
We then introduce the parameterization:
\be\label{param LR ii}
\n_0^R \equiv { l_0^R\ov 2}+ \nu_R \bd~, \qquad\qquad \n_i^R \equiv {q_i-1\ov 2}+ {l_i^R q_i\ov 2}+  \nu_R p_i ~,
\ee
Here, we introduced the integer-valued parameters:
 \be\label{sum lR zero}
l_0^R\in \Z~, \qquad l_i^R\in \Z\; \;\;(i=1, \cdots, n)~, \qquad \text{such that:}\qquad \quad l_0^R + \sum_{i=1}^n l_i^R=0~.
\ee
They correspond to the formal line bundle $\CN$, with:
\be
\CN\otimes \CN \cong L_0^{l_0^R} \otimes \bigotimes_{i=1}^n L_i^{q_i l_i^R} \cong \CO \in \Pic(\h\Sigma)~.
\ee
Note that the (generally ill-defined) line bundle $\CN \otimes \CL_0^{\nu_R}$ in  \eqref{LR 2d general definition} cannot be trivial unless $\sqrt{\CK}$ exist. If $q_i$ is odd for every exceptional fiber, we can choose $l_i^R=0$ and $\nu_R= 0$, as in the ordinary A-twist (and as was considered in \cite{Closset:2017zgf}, for instance). On a generic Seifert background $\CM_3$, however, the parameters $l_i^R$ and $\nu_R$ are non-trivial. There are chosen precisely such that the fluxes $\n_0^R, \n_i^R$ are integers.

\subsubsection{$L_R$ and the $\CM_3$ spin structure}
By definition, we have:
\be
c_1(L_R) = g-1 + \n_0^R + \sum_{i=1}^n {\n_i^R\ov q_i}~.
\ee
From  \eqref{param LR ii}, we immediately see that:
\be
c_1(L_R) = \half c_1(\CK) + \nu_R \,c_1(\CL_0)~.
\ee
Let us fix a solution $L_R^{(0)}$ to the constraints $\n_0^R\in \Z$, $\n_i^R \in \Z$. Any other solution $L_R$ is obtained by a shift of the parameters $\nu_R$ modulo $1$ and  $l_i^R$ modulo 2. Let us denote these shifts by:
\be\label{shift param}
 \Delta\nu_R~, \qquad \text{and} \qquad {\Delta l_i^R\ov 2}~,
\ee
respectively. We have:
\be
L_R \cong L_R^{(0)} \otimes \delta L_R~,
\ee
with the first Chern class of the line bundle $\delta L_R$ given by:
\be
c_1(\delta L_R)=c_1(L_R)-c_1(L_R^{(0)}) = \Delta \nu_R\, c_1(\CL_0)~.
\ee
The shifting parameters \eqref{shift param} are constrained by:
\be
-\sum_i {\Delta l_i^R\ov 2} + \Delta \nu_R \bd \in \Z~, \qquad\qquad    {\Delta l_i^R\ov 2}+ \Delta \nu_R p_i \in \Z~.
\ee
By comparing this constraint to the presentation \eqref{Pic M3 presentation} of  three-dimensional Picard group, we see that the shifts \eqref{shift param} are in one-to-one correspondence with elements of $\t\Pic(\CM_3)\otimes \Z_2$, by identifying the generators as:
\be
 [\gamma]\quad\rightsquigarrow\quad \Delta\nu_R~, \qquad\qquad
  [\omega_i] \quad \rightsquigarrow\quad {\Delta l_i^R\ov 2}~.
\ee
Since the choice of $L_R$ is equivalent to a choice of spin structure, this completes our discussion of the relation between the choice of spin structure on $\CM_3$ and the group \eqref{PicM3 tensor Z2}. In Section~\ref{subsec:expls}, we will illustrate the above procedure in many examples.

\subsubsection{Holomorphic line bundle moduli}
Holomorphic lines bundles on $\h\Sigma$ and $\CM_3$ are determined by additional data, beyond the topological data encoded in $\Pic(\h\Sigma)$ or $\t\Pic(\CM_3)$. In two-dimensions, it is well-known that holomorphic line bundles can come in continuous families, indexed by complex moduli, corresponding to the Dolbeault cohomology $H^{0,1}(\h\Sigma)$;   the moduli correspond to (anti-holomorphic) flat connections $A_\bz^{\rm flat}$. In three-dimensions, the THF structure allows for similar holomorphic moduli for a holomorphic line bundle $\bf L$, with complex moduli valued in $H^{0,1}(\CM_3)$ \cite{Closset:2013vra}.  Given $A_\mu^F$ a 3d background gauge field and $\sigma^F$ its scalar superpartner in a vector multiplet, it is convenient to define the complexified gauge field:
\be\label{def CAF}
\CA_\mu^F = A_\mu^F -i \eta_\mu \sigma^F~.
\ee
The complex moduli of the associated bundle ${\bf L}_F$ correspond to the holonomies of \eqref{def CAF} along one-cycles. In particular, we denote by:
\be
\nu_F \equiv -{1\ov 2 \pi} \int_\gamma \CA^F~,\qquad\quad y_F \equiv e^{2\pi i \nu_F}~,
\ee
its holonomy along the Seifert fiber $[\gamma]$. The continuous parameters $\nu_F$ play a crucial role in this paper,  because the various half-BPS observables we compute are holomorphic---more precisely, meromorphic---in them.~\footnote{The other continuous moduli of ${\bf L}_F$ correspond to the holonomies of $\CA^F$ along the 1-cycles on the Riemann surface; they couple to $Q$-exact operators in the A-twisted theory, and therefore they can safely ignored, for our purposes.}  Note that we have the identifications:
\be
\nu_F \sim \nu_F+1~,
\ee
corresponding to large $U(1)$ gauge transformations along the fiber.

In summary, at least for our present purposes, any holomorphic line bundle ${\bf L}_F$ over $\CM_3$ is fully characterized by a particular element ${\bf L}_F \in \t \Pic(\CM_3)$ together with a complex parameter $\nu_F\in \C$.
From the two-dimensional point of view, the parameters $\nu_F$ are simply twisted masses in background $\CN=(2,2)$ vector multiplets on $\h\Sigma$.

\paragraph{The $U(1)_R$ complex structure modulus.}   The $U(1)_R$ line bundle ${\bf L}_R$ is also a holomorphic line bundle. While the background gauge field $A_\mu^{(R)}$ is part of the supergravity multiplet, it is also part of a smaller vector multiplet, which includes the gauge field $A_\mu^{(R)}+V_\mu$ and the scalar $\sigma^R = H$ \cite{Kuzenko:2013uya, Closset:2014uda}.  Thus, we see that the $U(1)_R$ analogue of \eqref{def CAF} is precisely the gauge field $\CA_\mu^{(R)}$ in \eqref{def CAR}:
\be
\CA_\mu^{(R)}= A_\mu^{(R)}+V_\mu- i \eta_\mu H~.
\ee
 It follows that the $U(1)_R$ complex structure modulus is given by the parameter $\nu_R$ introduced in \eqref{introduce nuR}. In addition to $\nu_R$, we have seen above that it is important to keep track of the parameters $l_i^R$, corresponding to $\Z_2$ holonomies of $\CA_\mu^{(R), \rm flat}$ along the one-cycles $\omega_i$, or equivalently along the exceptional fibers.

\subsubsection{Quantization condition on the $R$-charge}
By definition, any 3d field $\varphi$ of $R$-charge $r\in \Z$ is valued in $({\bf L}_R)^r$, and the corresponding two-dimensional KK modes $\varphi_n$ take value in $L_R^r \otimes \CL_0^n$. For the three-dimensional A-model to be well-defined, those bundles should be well-defined. This condition can be written as:
\be\label{dirac quant}
L_R^r \in \Pic(\h\Sigma)~,
\ee
with $L_R$ the two-dimensional $U(1)_R$ line bundle defined in Subsection~\ref{subsec: def LR}.
If we assume that $L_R$ exists, we generally have the integrality condition:
\be
r\in \Z~,
\ee
for the $R$-charges of the matter fields.  We may also consider the case for which only $L_R^2$ exists---in such a case, we must impose the quantization condition $r\in 2\Z$ for all 3d chiral multiplets. In such a case, the partition function becomes independent of a choice of spin structure.
More general $R$-charges might also be allowed on a given background. For instance, if the $U(1)_R$ fluxes $\t\n_0^R$ and $\n_i^R$ have a common divisor $m$, the Dirac quantization condition will simply be $m r\in \Z$.

\subsubsection{Topologically trivial $\CK_{\CM_3}$ and real $R$-charges}
A very special but very important case occurs when $L_R$ is topologically trivial:
\be\label{LR trivial}
L_R \cong \CO~.
\ee
This is equivalent to the conditions:
\be\label{vanishing flux}
\t\nu^R_0 = g-1 + \nu_R \bd -\half \sum_{i=1}^n l_i^R=0~, \qquad\quad
\n_i^R= {q_i-1\ov 2}+ {q_i l_i^R \ov 2}+ \nu_R p_i =0~.
\ee
This system of $n+1$ equations for the $n+1$ variables $2\nu_R$, $l_i^R$, generally does not have any solution over the integers. By construction, there exists a solution if and only if the three-dimensional line bundle ${\bf \CK}_{\CM_3}$ in \eqref{KM3} is trivial in $\t\Pic(\CM_3)$.  When $c_1(\CL_0)\neq 0$, this fixes $\nu_R$ to be:
\be\label{nuR for scft cL0 nonzero}
\nu_R = - \half {c_1(\CK)\ov c_1(\CL_0)}~.
\ee
When $c_1(\CL_0)=0$, there exists a linear relation between the equations \eqref{vanishing flux}, and we then have an additional freedom in choosing a topologically trivial $L_R$.

When \eqref{vanishing flux} holds, there is no Dirac quantization condition and we may consider arbitrary real R-charges $r\in \R$ for the matter fields. Such supersymmetric backgrounds are particularly interesting in order to study 3d  $\CN=2$ superconformal field theories, which generally have irrational R-charges.  When $\CK_{\CM_3} \cong \CO$ and $c_1(\CL_0)\neq 0$, there exists a unique spin structure on $\CM_3$ compatible with real $R$-charges.

\section{Examples of supersymmetric Seifert backgrounds}\label{subsec:expls}
In this section, we discuss some important examples of Seifert manifolds and their associated supersymmetric backgrounds, to illustrate the formalism of Section \ref{sec: seifert backgs}. After reviewing the principal-bundle case of \cite{Closset:2017zgf}, we discuss in detail all the possible Seifert-fibration backgrounds on lens spaces. We then present a few other interesting examples of manifolds with more general topology.

\subsection{The principal bundle $\Mgp$}\label{subsec:Mgp}
As a first example, consider the family of manifolds $\Mgp$ studied in \cite{Closset:2017zgf}. In the Seifert notation, we have:
\be\label{def Mgp}
\Mgp \cong \big[p~;\, g~;\, \big]\cong \big[0~;\, g~;\, (1, p)\big]~.
\ee
This is a principal circle bundle of first Chern class $p\in \Z$ over the smooth Riemann surface $\Sigma_g$:
\be
S^1 \longrightarrow \Mgp \longrightarrow \Sigma_g~.
\ee
The 2d Picard group $\Pic(\Sigma_g) \cong \{ L_0\} \cong \Z$ is freely generated. The defining line bundle is:
\be
\CL_0 \cong L_0^p~, 
\ee
and therefore:
\be
\t\Pic(\CM_3) \cong \{L_0 \}/(\CL_0) \cong \Z_p~.
\ee
To fully specify the supersymmetric background, we need to choose a $U(1)_R$ line bundle over $\Sigma_g$:
\be
L_R \cong L_0^{g-1 + \nu_R p}~.
\ee
For $p$ odd, we must take $\nu_R$ integer. The case $\nu_R=0$ was called the ``A-twist gauge'' in \cite{Closset:2017zgf}. When $p$ is even, we can choose $\nu_R$ to be either integer or half-integer. In particular, taking $2\nu_R$ even or odd corresponds to the choice between the two distinct spin structures on $\Mgp$ \cite{Martelli:2012sz, Toldo:2017qsh}. Indeed, we have:
\be
H^1(\Mgp, \Z_2) \cong  \Z_2^{2g}\oplus \Z_{\gcd(p, 2)}~,
\ee
when $p$ is even. Note that, if $p=0$, we have $\CM_{g,0} = \Sigma_g \times S^1$ and $H^1(\CM_{g,0}, \Z_2) \cong \Z_2^{2g}\times \Z_2$ as well. In that case, $\CL_0 \cong \CO$, and we can still turn on the $U(1)_R$ fugacity $\nu_R \in \half \Z$. Thus, even the ``ordinary'' twisted index $Z_{\Sigma_g\times S^1}$ admits a $\Z_2$ refinement keeping track of the choice of spin structure.~\footnote{In earlier literature on the twisted index  \cite{Nekrasov:2014xaa, Benini:2015noa, Benini:2016hjo, Closset:2016arn}, the role of the spin structure was not discussed. In in \cite{Closset:2017zgf}, we restricted ourselves to the periodic spin structure.}

\subsection{The twisted $S^2_\epsilon \times S^1$}\label{subsec: S2S1 back}
Consider the three-manifold $S^2 \times S^1$. There exists a one-parameter family of half-BPS backgrounds, denoted by $S^2_\epsilon \times S^1$, with non-trivial $U(1)_R$ flux through the $S^2$. 
The corresponding supersymmetric partition function is the refined twisted index of \cite{Benini:2016hjo}.
The case $\epsilon=0$ corresponds to the standard A-twist on $S^2$.

The THF associated to this family of supersymmetric backgrounds is best understood as the quotient of $S^2 \times \R$ by:
\be\label{THF S2S1 def}
(\tau~,\;   z)\sim (\tau+2 \pi~, \; e^{2 \pi i \epsilon} z)~, \qquad \epsilon \in \R~,
\ee
with $\tau$, $z, \b z$ the adapted coordinates, and $\epsilon$ the THF modulus  \cite{Closset:2013vra}. Let us introduce the real coordinates $\theta, \varphi, \tau$, with $\theta \in [0, \pi]$, $\varphi \in [0, 2\pi)$ the standard angular coordinates on $S^2$. We have:
\be
z= \tan{\theta\ov 2} e^{i \left(\varphi + \epsilon \tau\right)}~.
\ee
In these angular coordinates,  the quotient \eqref{THF S2S1 def} simply corresponds to the identification $\tau \sim \tau + 2\pi$. For definiteness, we consider the same adapted metric as in \cite{Benini:2016hjo}:
\be\label{met S2S1}
ds^2(S^2_\epsilon \times S^1) = \t\beta^2 d\tau^2 + {R_0^2 dz d\bz\ov (1+|z|^2)} =\t\beta d\tau^2 + {R_0^2\ov 4} \left(d\theta^2 + \sin^2 \theta (d\varphi + \epsilon d\tau)^2\right)~.
\ee
The Killing vector entering in the curved-space supersymmetry algebra reads:
\be\label{K S2S1}
K = {1\ov \t\beta}\left(\d_\tau - \epsilon \d_\varphi\right)~.
\ee
We have $\eta = \t\beta d\tau$ with $\eta_\mu = K_\mu$. 
Note that the orbits of $K$ only close if $\epsilon$ is a rational number:
\be\label{eps tq}
\epsilon = {t\ov q}~, \qquad\qquad t, q\in \Z~, \qquad\qquad \gcd(q, t)=1~.
\ee
Let us restrict ourselves to this case. We claim that, with the identification \eqref{eps tq},  $K$ generates a Seifert fibration:
\be
S^1  \longrightarrow S^2_\epsilon \times S^1 \longrightarrow S^2(q, q)~,
\ee
with $S^2(q, q)\cong S^2/\Z_q$. In the standard notation, we have:
\be\label{S2S1M3 standard}
\CM_3\cong S^2_\epsilon \times S^1 \cong \big[0~; \, 0~; \, (q, p)~, (q, -p) \big]~, \qquad \quad q s+ p t=1~.
\ee
In particular, we have the defining line bundle:
\be
\CL_0 \cong L_1^p L_2^{-p}~
\ee
over the spindle $S^2(q,q)$, with $L_1$ and $L_2$ the elementary line bundles associated to the two orbifold points at the north and south poles, respectively. Note that $c_1(\CL_0) = 0$. One can show that \eqref{S2S1M3 standard} gives the complete list of Seifert fibrations of $S^2 \times S^1$ \cite{2016arXiv160806844G}.

The fibration \eqref{S2S1M3 standard} is given explicitly by:
\be
\pi\; : \; S^2\times S^1 \rightarrow S^2\; : \; (\tau, z) \mapsto z^q~.
\ee
This map is obviously invariant under \eqref{THF S2S1 def}  if $\epsilon = {t/q}$.
We can also understand the fibration structure directly from the metric  \eqref{met S2S1} and Killing vector \eqref{K S2S1}. Consider the change of coordinates:
\be
\psi = s \tau - p \varphi~, \qquad \phi = t \tau + q\varphi~.
\ee
For $q, s, p, t \in \Z$ such that $q s+ p t=1$, this is an $SL(2, \Z)$ transformation and the new angles $\psi, \phi$ have periodicities $2\pi$. Let us also define $\beta= q \t\beta$. In the new coordinates, we have:
\be
K= {1\ov \beta}\d_\psi~, \qquad \eta= \beta\left(d\psi + {p\ov q}d \phi\right)~,
\ee
and
\be\label{met S2S1 seifert}
ds^2 = \eta^2 + ds^2\left(S^2(q,q)\right)~, \qquad  ds^2\left(S^2(q,q)\right)= {R_0^2\ov 4} \left(d\theta^2 + {\sin^2 \theta\ov q^2} d\phi^2 \right)~.
\ee
We may choose the radius $R_0 = \sqrt{q}$ so that $\vol(S^2(q,q))=\pi$. This brings this geometric background to the general form \eqref{metric M3}, with the flat $\CL_0$ connection $\CC= {p\ov q}d\varphi$. 
The spindle metric in \eqref{met S2S1 seifert} is a special case of \eqref{met spindle}. 
We have the holonomies:
\be
{1\ov 2\pi}\int_{\omega_1} \CC ={p\ov q}~, \qquad \qquad {1\ov 2\pi}\int_{\omega_2} \CC =-{p\ov q}~, 
\ee
with $\omega_{1}$, $\omega_2$ the generators of the orbifold fundamental groups---the loops around the poles. This reproduces \eqref{S2S1M3 standard}.

\paragraph{Line bundles and choice of $L_R$.}
Finally, let us discuss the orbifold bundles on $\h\Sigma \cong S^2(q,q)$. We have: 
\be
\Pic(\h\Sigma) = \big\{L_0, L_1, L_2\big| \; L_1^q = L_2^q=L_0 \big\}~,
\ee
which is not freely generated. These orbifold line bundles pull-back to the three-dimensional line bundles in:
\bea
&\t\Pic(S^2_\epsilon\times S^1) \cong \Pic(\h\Sigma)/(\CL_0) \cr
&\qquad\cong \Big\{[\gamma],[\omega_1], [\omega_2]\; \Big| \; q [\omega_1] + p [\gamma]=0~, q [\omega_2] - p [\gamma]=0~, [\omega_1]+[\omega_2]=0  \Big\} \cr
&\qquad \cong \Big\{[\gamma],[\omega]\; \Big|\;  q [\omega] + p [\gamma]=0~  \Big\}
\cong \Big\{ [\Omega]\Big\}  \cong \Z~.
\eea
In the third line, we used $[\omega]\equiv [\omega_1]= - [\omega_2]$ and defined the generator:
\be
[\Omega]\equiv  t [\omega] - s [\gamma]~.
\ee
Of course,   $[\Omega]$ is the single generator of $H^2(S^2\times S^1, \Z) \cong \Z$. Note that we have:
\be
[\Omega]\cong \pi^\ast(L_1) \cong  \pi^\ast(L_2)~, 
\ee
and $L_1 \cong L_2\otimes \CL_0^t$.
We have the 2d and 3d canonical line bundles:
\be
\CK \cong L_0^{-2} L_1^{q-1}L_2^{q-1} \cong L_1^{-1}L_2^{-1}~, \qquad \qquad
\CK_{\CM_3} \cong -2 [\Omega]~.
\ee
We thus obviously have ${\bf L_R} \cong -[\Omega]$ as an element of $\t\Pic(\CM_3)$. However, there  remains a twofold choice of square root of the canonical line bundle over $\CM_3$, corresponding a choice of flat connection along the $S^1$ in $S^2\times S^1$. Using the parameterization of Section~\ref{subsec: def LR}, let us consider the two-dimensional line bundle:
\be
L_R \cong \sqrt\CK \otimes \CN\otimes \CL_0^{\nu_R}~.
\ee
with integer $U(1)_R$ fluxes:
\be\label{nR S2 S1}
\n_0^R = -{l_1^R+ l_2^R\ov 2}~, \qquad \n_1^R= {q-1\ov 2} + {l^R_1 q\ov 2}+ \nu_R p~,
\qquad \n_2^R= {q-1\ov 2} + {l^R_2 q\ov 2}-\nu_R p~.
\ee
Note that $l_1^R$ and $l_2^R$ must have the same parity. There are three different cases, depending on the parity of  $p$ and $q$. One easily finds:
\be\label{LR S2S1 cases}
\begin{cases}
 (-1)^{2\nu_R}=(-1)^{l_i^R}= \pm 1 \quad & \text{for}\; \; q\; \text{and}\; p \,\, \text{odd}\,,\cr
 (-1)^{2\nu_R}=\pm 1 \, , \quad (-1)^{l_i^R}=1 \quad & \text{for}\; \; q \,\, \text{odd},\,  p\, \,\text{even}\,,\cr
 (-1)^{2\nu_R}=-1 \, , \quad (-1)^{l_i^R}=\pm 1 \; & \text{for}\; \; q \,\, \text{even},\, p \,\, \text{odd}\,.
\end{cases}
\ee
These two distinct solutions, for any given $(q, p)$, are in one-to-one correspondence with the two spin structures on $S^2\times S^1$.

\subsection{The three-sphere $S^3_b$}
\label{subsec: S3b background}
Consider the three-sphere $S^3$. There exists a one-parameter family of supersymmetric backgrounds preserving two supercharges, indexed by the so-called squashing parameter $b\in \C$ \cite{Hama:2011ea, Imamura:2011wg, Alday:2013lba, Closset:2013vra}. Viewing $S^3$ as a $T^2$ fibered over an interval, with $\varphi, \chi$ the angular coordinates on the $T^2$, let us consider the background of  \cite{Hama:2011ea} with $b^2\in \R$, for definiteness. The supersymmetry algebra involves the Killing vector:
\be
K = b \d_\varphi + b^{-1} \d_\chi~.
\ee
There is a natural symmetry under $b \rightarrow b^{-1}$, which corresponds to exchanging the two cycles, $\varphi$ and $\chi$, of the torus.  For generic $b^2 \in \R$, the orbits of $K$ do not close and this $K$ does not define a Seifert structure (it only defines a THF). However, in the special case:
\be\label{b2q1q2}
b^2 = {q_1\ov q_2} \in \Q~,
\ee
the orbits close and the $S^3_b$ geometry can be viewed as a Seifert fibration over a sphere with two orbifold points:
\be
S^1 \longrightarrow S^3_b  \stackrel{\pi}\longrightarrow S^2(q_1, q_2)~.
\ee
The integers $q_1$ and $q_2$ should be mutually prime, and exchanging their roles corresponds to the equivalence  $b\rightarrow b^{-1}$. We then have:
\be\label{S3b as M3}
S^3_b \cong [0~;\, 0~;\, (q_1, p_1)~, (q_2, p_2)]~,\quad \qquad q_1 p_2+ q_2 p_1=1~,
\ee
for  $b^2= {q_1\ov q_2}$ a rational squashing parameter.~\footnote{According to \protect\eqref{b2q1q2}, $q_1$ and $q_2$ (and therefore $b^2$) could be of either sign. However, using the equivalence $(q_i, p_i) \cong (-q_i, -p_i)$ at each exceptional fiber, we can always write \protect\eqref{S3b as M3} in the convention in which the anisotropy parameters are all positive.}
We will demonstrate \eqref{S3b as M3} momentarily, in the more general case of the lens space.
This can also be understood by bringing any metric on the three-sphere, with the Killing vector $K$, to the form \eqref{metric M3}---we give an example of that approach in Appendix~\ref{app: S3b metric}.  Moreover, the Seifert fibrations \eqref{S3b as M3}, modulo trivial equivalences, coincide with the set of all Seifert fibrations of the three-sphere \cite{orlik1972seifert}.

Consider the spindle $S^2(q_1, q_2)$ with $\gcd(q_1, q_2)=1$. 
We have the defining line bundle:
\be
\CL_0 \cong L_1^{p_1} L_2^{p_2}~.
\ee
The orbifold Picard group of $S^2(q_1, q_2)$  is freely generated, with $\CL_0$ the single generator:
\be
\Pic(\h\Sigma) = \{ L_0, L_1, L_2 \; |\; L_1^{q_1}= L_2^{q_2} = L_0 \} \cong \{ \CL_0 \}~,
\ee
with $L_1 \cong \CL_0^{q_2}$ and $L_2 \cong \CL_0^{q_1}$. It directly follows that the three-dimensional Picard group is trivial:
\be
\t\Pic(S^3_b) \cong 0~,
\ee
in agreement with the fact that $H^2(S^3, \Z)=0$. Therefore, we have $\CK_{S^3_b}\cong 0$, and the $U(1)_R$ line bundle over $S^3_b$ is topologically trivial. On the other hand, we have the two-dimensional canonical line bundle:
\be
\CK\cong L_1^{-1} L_2^{-1} \cong \CL_0^{-q_1-q_2}~.
\ee
We may choose:
\be
L_R \cong \sqrt\CK \otimes \CL_0^{\nu_R} \cong \CL_0^{\nu_R-{q_1+q_2\ov 2}}~.
\ee
We can choose any $\nu_R$ such that $2\nu_R-q_1-q_2$ is even. Note that the parity $2\nu_R$ is fixed by the parity of $q_1+q_2$, which here corresponds to the fact that there is a unique spin strucure on $S^3$.
 A particularly interesting choice is:
\be\label{nuR for S3b}
\nu_R= {q_1+ q_2\ov 2}~.
\ee
such that $L_R$ is trivial in $\Pic(\h\Sigma)$ as well.
This is the background that can be directly compared to the squashed-sphere backgrounds considered in the literature \cite{Hama:2011ea, Imamura:2011wg}, as we will further discuss in Section~\ref{sec: S3b}. For this choice of $L_R \cong \CO$, we have $\n_0^R= \n^R_{1}= \n_2^R=0$ and the Dirac quantization on the $R$-charge is trivial. This allows us to consider real $R$-charges for the matter fields.

\subsection{The lens space $L(p,q)_b$}\label{sec:lpqgeo}
We define the  lens space $L(p,q)$ as the quotient of $S^3$:
\be\label{Lpq quotient}
\big(z_1~,\;  z_2\big) \sim \big(e^{2\pi i  q\ov p} \, z_1~,\;e^{2\pi i \ov p}\, z_2\big)~, \qquad  \quad \gcd(p,q)=1~,
\ee
with $S^3$ a three-sphere inside $\C^2 \cong \{(z_1, z_2)\}$.~\footnote{\label{footnoteLpq}In this paper, we are using the ``mathematicians' definition'' for $L(p,q)$. The lens space $L(p,q)$ as defined in many physics papers, for instance in \cite{Benini:2011nc, Alday:2012au, Closset:2017zgf}, corresponds to $L(p,-q)$ here.} 
The associated supersymmetric backgrounds come in one-parameter families $L(p,q)_b$, indexed by the so-called ``squashing parameter'' $b$ \cite{Hama:2011ea, Benini:2011nc, Closset:2013vra}. We already discussed the special case $L(1,1)_b \cong S^3_b$ in the previous subsection. Another very special case is $L(0,1) \cong S^2 \times S^1$, which we analysed separately; here we consider $p \neq 0$.

As mentioned above, lens spaces are the only compact three-manifolds which admit more than one Seifert fibration (with the base $\h \Sigma_g$ orientable); in fact, they admit an infinite number of distinct Seifert fibrations, which are fully classified \cite{orlik1972seifert, 2016arXiv160806844G}. 
Any Seifert fibration of genus $0$ with $n \leq 2$ exceptional fibers is a lens space, and every Seifert fibration over a lens space can be constructed in that way. We have:
\be\label{M3lens}
L(p,q)_b  \cong [0~;\, 0~;\, (q_1, p_1)~,\; (q_2, p_2)]~,
\ee
with the identifications:
\be\label{p and q for lens}
p= p_1 q_2 + p_2 q_1~, \qquad q= q_1 s_2 - p_1 t_2~, \qquad \quad b^2 = {q_1 \ov q_2}~.
\ee
Here, as usual, the integers $s_i, t_i$ are such that $q_i s_i + p_i t_i=1$. The pair of integers $(s_2, t_2)$ entering in \eqref{p and q for lens} is only defined up to shifts by $(p_2, -q_2)$. This has the effect of shifting $q$ to $q+p$, which gives the same lens space. More generally, two lens spaces $L(p,q)$ and $L(p,q')$ are related by an orientation-preserving diffeomorphism if and only if $q= q'$ mod $p$, or $qq'= 1$ mod $p$. This second equivalence is realized in \eqref{M3lens}-\eqref{p and q for lens} by exchanging the roles of the two exceptional fibers, which also takes $b \rightarrow b^{-1}$.~\footnote{Taking $q'= q_2 s_1 -p_2 t_1$ and $q$ as in \protect\eqref{p and q for lens}, one can check that $q q'= 1 - p(t_1 s_2 +s _1 t_2)$.}

Note also the following special cases: For $q_1=q_2=1$, there are no orbifold points on $S^2$ and we have the principal bundle $L(p,1)_1 \cong \CM_{0, p}$ as defined in \eqref{def Mgp}, where $p=p_1+p_2$. The other special case is when  there is only one orbifold point, with $q_1=q$ and $q_2=1$ or $q_1=1$ and $q_2=q$, then:
\be
L(p,q)_b  \cong [0~;\, 0~;\, (q, p)]~, \qquad \text{with} \quad b^2= q \;\; \text{or} \;\; b^2= {1\ov q}~,
\ee
respectively.

\paragraph{Seifert fibration from Hopf surface.} To understand the identification of the Seifert structures \eqref{M3lens} with the known supersymmetric backgrounds, it is useful to consider the four-dimensional uplift to the secondary Hopf surface, which is diffeomorphic to $L(p,q) \times S^1$. The primary Hopf surface $\CM_4^{{\bf p}, {\bf q}} \cong S^3 \times S^1$ is defined as the quotient of $\C^2-\{(0,0)\}$ by:
\be\label{hopf quot}
\big(z_1~,\;  z_2\big) \sim \big({\bf p}\, z_1~,\; {\bf q}\,  z_2\big)~.
\ee
The parameters ${\bf p}$, ${\bf q}$ are complex structure moduli of the Hopf surface---they give rise to the one-parameter family of THFs in 3d, which we can parameterize by ${\bf p}= e^{- 2 \pi b \beta }$,  ${\bf q}= e^{-2 \pi b^{-1} \beta}$, with $\beta$ the radius of the $S^1$   \cite{Aharony:2013dha, Closset:2013vra}.
The secondary Hopf surface $\CM_4 \cong L(q,p) \times S^1$ is obtained by a further  $\Z_p$ quotient of $\CM_4^{{\bf p}, {\bf q}}$ as in \eqref{Lpq quotient}.  
As in the three-sphere case, the THF is {\it also} a Seifert fibration if and only if $b^2 \in \Q$:
\be\label{b2 cond lens hopf}
b^2 = {q_1\ov q_2}\qquad \Leftrightarrow \qquad {\bf q}^{q_1} = {\bf p}^{q_2}~.
\ee  
The lens space Seifert fibration is given explicitly by:
\be\label{def z lens}
\pi  \; : \;  L(p,q) \rightarrow S^2(q_1, q_2)  \; : \; (z_1, z_2) \mapsto z= {z_1^{q_2} z_2^{-q_1}}~,
\ee
with $z$ the local coordinate on the $S^2$ base. The map  \eqref{def z lens} is obviously compatible with the quotient \eqref{hopf quot} when \eqref{b2 cond lens hopf} holds. It is also compatible with  \eqref{Lpq quotient} provided that $q q_2- q_1=0$ mod $p$; one can check that this automatically follows from \eqref{p and q for lens}. The map \eqref{def z lens} also makes manifest the presence of ramification points at the poles, so that the base of the fibration is a spindle $S^2(q_1, q_2)$. 
The defining line bundle of the Seifert fibration \eqref{M3lens} has a first Chern class:
\be
c_1(\CL_0) ={q_1\ov p_1}+{q_2\ov p_2}= {p\ov q_1 q_2}~.
\ee
Unlike the case of $S^3_b$, the integers $q_1$ and $q_2$ may not be coprime; in Appendix \ref{app:lens space algo}, we review an algorithm to constructs all the Seifert fibrations of any given $L(p,q)$ \cite{2016arXiv160806844G}.

\paragraph{Line bundles and $R$-symmetry backgrounds.}
The base space for the Seifert fibration \eqref{M3lens} of $L(p,q)$ is the spindle $S^2(q_1, q_2)$.  Since $q_1$ and $q_2$ need not be relatively prime, the two-dimensional orbifold Picard group $\Pic(S^2(q_1,q_2))$ is not necessarily freely generated.  In the three-dimensional notation, it has the presentation:
\be
\Pic\big(S^2(q_1,q_2)\big)\cong \Big\{ [\gamma],[\omega_1],[\omega_2]\;\; \Big| \;\;  p_i [\gamma] + q_i [\omega_i] = 0~, \; \; i=1,2\Big\}~. \ee
Following \eqref{pullback L Li}, we denote the pull-backs of the elementary orbifold line bundles $L_i$ by:
\be
 [\delta_i]  \equiv \pi^*(L_i)  \cong - s_i [\gamma] + t_i [\omega_i]~, \qquad i=1,2~.
  \ee
One may check that:
\be
 [\gamma] = -q_1 [\delta_1]= -q_2 [\delta_2]~,\qquad\qquad
   [\omega_1] =  p_1 [\delta_1]~,\qquad \qquad    [\omega_2] =  p_2 [\delta_2]~,
 \ee
as relations inside the 2d Picard group,  thus $\Pic\big(S^2(q_1,q_2)\big)$ is also generated by $[\delta_1]$ and $[\delta_2]$.
The three-dimensional Picard group $\t\Pic\big(L(p,q)\big)$ is the quotient of $\Pic\big(S^2(q_1,q_2)\big)$ by the relation:
\be
 [\cL_0]= [\omega_1] + [\omega_2]= p_1 [\delta_1]+ p_2 [\delta_2]\cong 0~.
\ee
One can check that:
\be\label{relations in Pic Lens}
q [\delta_1] = [\delta_2]  -t_2 [\cL_0]~, \qquad\qquad
 p [\delta_1] = q_2 [\cL_0]~,
 \ee
as relations in $\Pic\big(S^2(q_1,q_2)\big)$, with $p$ and $q$ given by \eqref{p and q for lens}.
The first relation implies that $[\delta_2]$ lies in the group generated by $[\delta_1]$ in $\t\Pic\big(L(p,q)\big)$, and therefore $\t\Pic(L(p,q))$ can be generated by the single generator $[\delta_1]$.  The second relation in \eqref{relations in Pic Lens} then implies that:
\be\label{PicM3 Lens}
 \t\Pic\big(L(p,q)\big) \cong \Big\{ [\delta_1]\; \Big| \; p[\delta_1]=0 \Big\} \cong \Z_p~, 
\ee
in agreement with  $H^1(L(p,q), \Z)\cong \Z_p$.

Now, consider the $R$-symmetry line bundle. The canonical line bundle of the spindle is given by $\CK\cong L_1^{-1} L_2^{-1}$, which can  be written as:
\be
 [{\cal K}] \cong   -[\delta_1]  - [\delta_2]
 \ee
in $\Pic\big(S^2(q_1,q_2)\big)$. Using the relations \eqref{relations in Pic Lens}, we may rewrite this as:
\be 
[{\cal K}] = -(q + 1) [\delta_1] - t_2 [\cL_0]~.
\ee
In particular, this is trivial in \eqref{PicM3 Lens} if and only if:
\be \label{lensrsymtriv} 
q + 1 =  0 \mod p~.
\ee
In the general case, the 2d $R$-symmetry line bundle takes the form:
\be
L_R \cong \sqrt{\CK} \otimes \CL_0^{\nu_R} \cong L_1^{-{q+1\ov 2}} \, \CL_0^{-{t_2\ov 2}+\nu_R}~.
\ee
In the special case \eqref{lensrsymtriv}, it is natural to choose $L_R\cong \CO$ topologically trivial, so that we may consider any real $R$-charges. 
Writing $q+1=n p$ for some integer $n$,  we see that:
\be 
[{\cal K}] = -(n q_2+ t_2) [\cL_0]~.
\ee
It follows that $L_R$ is topologically trivial if we choose:
\be\label{nuR LR trivial Lpq} 
\nu_R ={n q_2 + t_2\ov 2} = {(q+1)q_2 + t_2 p\ov 2 p} = {q_1+ q_2\ov 2 p}~,
\ee
where we used \eqref{p and q for lens}. Equivalently, this follows from \eqref{nuR for scft cL0 nonzero}. From \eqref{lensrsymtriv}, we also see that a lens space admits a trivial $R$-symmetry line bundle if and only if has the form $L(p,p-1)$.
This obviously generalizes the case of the $S^3_b$ background with $\nu_R$ given by \eqref{nuR for S3b}.
 This $L(p,p-1)$ supersymmetric background was the one studied by supersymmetric localization in \cite{Benini:2011nc, Alday:2012au, Closset:2017zgf}.~\footnote{It was called $L(p,1)$ in those papers, due to our different conventions. See footnote~\protect\ref{footnoteLpq}.}
 
\paragraph{Choice of $L_R$ and spin structure.} 
On a  general lens space $L(p,q)$, we have:
\be
c_1\left(\CK_{L(p,q)}\right) \cong -q-1  \in \Z_p~,
\ee
 and the R-symmetry line bundle cannot be topologically trivial unless \eqref{lensrsymtriv} holds  \cite{Closset:2013vra}.  Nonetheless, we can always choose $\nu_R$ and  $l_i^R$ such that arbitrary integer $R$-charges satisfy the Dirac quantization condition \eqref{dirac quant}. Consider the general parameterization of $L_R$ as in \eqref{param LR}-\eqref{param LR ii}, with the 2d integer fluxes:
\be\label{constraints nuR lR Lpq}
\t\n_0^R = -1 - {l_1^R+ l_2^R\ov 2}~, \qquad 
\n^R_i= {q_i-1\ov 2} + \nu_R p_i + {l_i^R q_i\ov 2}~, \quad i=1, 2~.
\ee
That implies:
\be\label{constraint on nuR Lpq}
q_1 q_2\, c_1(L_R) =   p \nu_R - {q_1+ q_2\ov 2} \in \Z~.
\ee
The special case \eqref{nuR LR trivial Lpq} with $L_R \cong \CO$ corresponds to this integer being zero.

We should carefully distinguish between the cases with $p$ odd or even. If $p$ is odd, \eqref{constraint on nuR Lpq} implies that $\nu_R$ mod $1$ is fixed by the parity of $q_1+q_2$---that is, $\nu_R\in \half + \Z$ if $q_1+q_2$ is odd, and $\nu_R \in \Z$ if $q_1+q_2$ is even. If $p$ is even, on the other hand, we see that $q_1+q_2$ must be even~\footnote{This is easily seen by contradiction. Assume that $p$ is even and $q_1+q_2$ odd. Then we must have that $q_1$ is odd and $q_2$ even (or vice-versa). That would imply that $p_2$ is odd (since $\gcd(q_2, p_2)=1$), and then $p=q_1 p_2+ q_2 p_1$ is odd, a contradiction.} and the constraint \eqref{constraint on nuR Lpq} can be solved with $\nu_R$ either integer or half-integer. Therefore:
\be
(-1)^{2\nu_R}=\begin{cases} (-1)^{q_1+ q_2} \qquad & \text{if}\; \quad p \in 2 \Z+1\,,\cr
1 \; \text{or} \; -1  & \text{if}\; \quad p \in 2 \Z\,.
\end{cases}
\ee
While any such choice of $\nu_R$ mod $1$ satisfy \eqref{constraint on nuR Lpq}, we must also solve the stronger constraints \eqref{constraints nuR lR Lpq}.  From the constraint $\t\n_0^R \in \Z$,  we see that $l_1^R$ and $l_2^R$ have the same parity. Then, one can check that, if $p$ is odd, that parity is determined by the fibration:
\be
p \in 2\Z+1\quad \Rightarrow \quad(-1)^{l_i^R}= (-1)^{p_1+p_2+1}~,
\ee
while if $p$ is even, we have the following three cases:
\be\label{choices p even Lpq}
p \in 2\Z \; \Rightarrow\; \begin{cases}
 (-1)^{2\nu_R}=(-1)^{l_i^R}= \pm 1 \quad & \text{for}\; \; q_1, q_2, p_1, p_2 \,\, \text{odd}\,,\cr
 (-1)^{2\nu_R}=\pm 1 \, , \quad (-1)^{l_i^R}=1 \quad & \text{for}\; \; q_1, q_2 \,\, \text{odd},\,  p_1, p_2\, \,\text{even}\,,\cr
 (-1)^{2\nu_R}=-1 \, , \quad (-1)^{l_i^R}=\pm 1 \; & \text{for}\; \; q_1, q_2 \,\, \text{even},\,  p_1, p_2 \,\, \text{odd}\,.
\end{cases}
\ee
The important point is that there are two and only two distinct possibilities, for any Seifert fibration of $L(p,q)$.
This nicely agrees with the geometric result:
\be
H^1(L(p,q), \Z_2) =\begin{cases} 0 \qquad & \text{if}\; \quad p \in 2 \Z+1\,,\cr
\Z_2  & \text{if}\; \quad p \in 2 \Z\,.
\end{cases}
\ee
For $p$ even, the two distinct spin structures of $L(p,q)$ correspond to the two choices in \eqref{choices p even Lpq}, generalizing the $\CM_{0,p} \cong L(p,1)$ example discussed in Section~\ref{subsec:Mgp}.

\subsection{Torus bundles over the circle}\label{subsec: torus bundle}
Another interesting class of examples consists of Seifert manifolds which are also surface bundles over the circle. It turns out that a Seifert manifold, $\CM_3$, fibers over the circle if and only if $c_1(\CL_0)=0$~\cite{orlik1972seifert}.~\footnote{We already studied a trivial example of this, the manifold $S^2 \times S^1$.} Here we consider two simple examples of {\it torus bundles.} (Interestingly, there are only five torus bundles which are also oriented Seifert manifolds, including the trivial case $T^3\cong [0; 0;]$. In addition to the $S$- and $C$-twisted torus bundles to be discussed below, the two last examples can be found in \cite{Hatcher}.)

\subsubsection{$S$-twisted torus bundle}
A torus bundle $\CM_3^A$ is a three manifold which is obtained from the product $T^2 \times I$, with $I\in [0,1]$ an interval, by gluing the end-points of the interval into a circle while identifying the tori with a non-trivial automorphism $A$:
\be\label{def CMA torus bundle}
\CM_3^A \cong (T^2 \times I)/\sim_A~, \qquad A\in GL(2, \Z)  
\ee
Let us choose the automorphism to be:
\be
S= \mat{0& 1\\ -1&0} \in SL(2, \Z)~, \qquad S^4={\bf 1}~.
\ee
Consider $T^2$ to be the square torus $T^2\cong \C/\sim$, with identifications $z \sim z+1\sim z+i$. The order-$4$ element $S$ acts by rotation by $90$ degrees. The points $z=0$ and $z=\half + {i\ov 2}$ have a $\Z_4$ stabilizer (i.e. they are left invariant), while the points $z=\half$ and $z={i\ov 2}$ have a $\Z_2$ stabilizer (generated by $C\equiv S^2$) and are identified by $S$. The fundamental domain for $\Z_4$ is the square with corners $z=0, \half, \half + {i\ov 2}, {i\ov 2}$. The corresponding Seifert three-manifold takes the form:
\be
S^1 \longrightarrow \CM_3^S \longrightarrow S^2(2,4,4)~,
\ee
where the three orbifold points on the genus-zero base correspond to $z=\half \sim {i\ov 2}$, $z=0$ and $z=\half + {i\ov 2}$, respectively. More precisely, it is possible to show that:
\be
 \CM_3^S  \cong  [0~;\, 0~;\, (2,1)~,\; (4,-1)~, \; (4,-1)]~.
\ee
Note that the base has vanishing orbifold Euler character, and the first Chern class of the defining line bundle vanishes as well:
\be
\chi\big(\h S^2(2,4,4)\big)= -c_1(\CK) = 0~,\qquad c_1(\CL_0)=0~,
\ee
as anticipated. The 2d Picard group takes the form
\be
\Pic\big(S^2(2,4,4)\big) \cong \left\{L_0, L_1, L_2, L_3\; \big| \; L_1^2 = L_2^4= L_3^4 =L_0 \right\}~,
\ee
and we have:
\be
\CK \cong \CL_0 \cong L_1 L_2^{-1} L_3^{-1}~.
\ee
The 2d $U(1)_R$ line bundle therefore takes the form:
\be
L_R \cong \left(\CL_0\right)^{\half + \nu_R}~,
\ee
so that $\nu_R$ must be half-integer. We will consider this square-root a little bit more carefully below.

\paragraph{The 3d Picard group.}  The three-dimensional Picard group takes the form:
\be
\t\Pic\big(\CM_3^S\big) \cong \left\{[\omega_1]~, \; [\omega_2]\; \big|\;  2 [\omega_1] + 4 [\omega_2]=0 \right\}~.
\ee
By changing the basis of generators to:
\be
[\lambda_1] = [\omega_1]+ [\omega_2]~, \qquad
[\lambda_2] = [\omega_1]+ 2[\omega_2]~,
\ee
we find that:
\be\label{M3S Pic3}
\t\Pic\big(\CM_3^S\big) \cong  \left\{[\lambda_1]~, \; [\lambda_2]\; \big|\;  2[\lambda_2]=0 \right\}  \cong \Z \times \Z_2~.
\ee
Note that the free generator $[\lambda_1]$ corresponds to $L_1^2 L_2^4$ in $\Pic\big(S^2(2,4,4)\big)$. It follows that:
\be\label{spin structures M3S}
H^1(\CM_3^S, \Z_2) \cong \Z_2 \times \Z_2~,
\ee
which comes entirely from \eqref{M3S Pic3}.
In other words, there are four distinct spin structures compatible with supersymmetry on $\CM_3^S$.

\paragraph{Choice of $L_R$ and spin structures.}  Consider a general $L_R$. We have the $U(1)_R$ fluxes:
\bea
&\t \n_0^R = - 1- {l_1^R+l_2^R +l_3^R\ov 2}~, \quad && \n_1^R =\half + \nu_R +l_1^R~, \cr
&\n_2^R = {3\ov 2}-\nu_R + 2 l_2^R~, \quad
&&\n_3^R = {3\ov 2}-\nu_R + 2 l_3^R~, 
\eea
which must be integers. It follows that $\nu_R \in \half + \Z$, as already mentioned, together with:
\be
l_1^R + l_2^R + l_3^R \in 2\Z~.
\ee
There are four possibilities for $l_i^R$ mod $2$, namely $(l_i^R)= (0,0,0)$, $(0,1,1)$, $(1,1,0)$ or $(1,0,1)$. This matches exactly the expectation from \eqref{spin structures M3S}.

It is also very natural to choose a line bundle $L_R \cong \CO$,  with vanishing fluxes. This is possible for:
\be
\nu_R = -\half + 2k~, \qquad (l_1^R, l_2^R, l_3^R)= (-2 k, -1+k, -1+k)~, \qquad \forall k\in \Z~.
\ee
Interestingly, we can have a topologically-trivial $L_R$ (and therefore choose any real $R$-charge for the matter fields) for two out of the four choices of spin structures, namely for the two spin structures corresponding to $(l_i)$ mod $2$ equal $(0,0,0)$ and $(0,1,1)$.

\subsubsection{$C$-twisted torus bundle}
Another interesting torus bundle is obtained by taking the order-two automorphism $C=-{\bf 1}$ in \eqref{def CMA torus bundle}. It corresponds to a genus-zero Seifert fibrations with four exceptional fibers:
\be
 \CM_3^C  \cong  [0~;\, 0~;\, (2,1)~,\; (2,1)~, \; (2,-1)~, \; (2,-1)]~.
\ee
This is also the ``small'' Seifert manifold (iii) in Section~\ref{subsec:classification seifert}.
It has $c_1(\CK)=0$ and $c_1(\CL_0)=0$ as in the last example. We easily find the three-dimensional Picard group:
\be
\t\Pic\big(\CM_3^C\big) \cong  \left\{[\lambda_1]~, \; [\lambda_2]~, \; [\omega_3]\; \big|\;  2[\lambda_1]=0~, \, 2[\lambda_2]=0 \right\}  \cong \Z_2 \times \Z_2\times \Z~,
\ee
with $[\lambda_i]=[\omega_i]+ [\omega_3]$ for $i=1,2$. It follows that:
\be
H^1(\CM_3^S, \Z_2) \cong \Z_2 \times \Z_2\times \Z_2~,
\ee
so that there are eight distinct spin structures on $\CM_3^C$. They can all be probed by our supersymmetric backgrounds. Indeed, the conditions on the $U(1)_R$ flux are equivalent to:
\be
\nu_R= \half \mod 1~, \qquad l_1^R+ l_2^R+ l_3^R+ l_4^R \in 2 \Z~,
\ee
and there are eight distinct possibilities for the parameters $l_i^R$ mod $2$. There are also ``conformal'' backgrounds for which $L_R$ is topologically trivial, with $\nu_R=- \half+k$ and $(l_i^R)=(-k,-k,-1+k,-1+k)$, spanning two out of the eight spin structures.

\subsection{Spherical manifolds $S^3/\Gamma_{\rm ADE}$}\label{subsec: spherical manifolds}
Another interesting family of genus-zero Seifert manifolds are the so-called spherical three-manifolds, which are quotients of the three-sphere, $\CM_3 \cong S^3/ \Gamma$,
with $\Gamma$ a freely-acting finite subgroup of $SO(4) \cong SU(2)_L \times SU(2)_R$.  Here we consider the simple case in which $\Gamma$ is a subgroup of $SU(2)_L$.  These background are particularly interesting from the point of view of 3d $\CN=2$ superconformal theories. Recall that one can preserve the superalgebra:
\be
 OSp(2|2) \times SU(2)_L
\ee
on the round $S^3$, where the $SU(2)_L$ factor can be taken to act by left multiplication.  
This is a subgroup of the full flat-space superconformal algebra $OSp(2|2,2)$ which contains four supercharges and is compatible with $\CN=2$ massive deformations of the theory on $S^3$ \cite{Jafferis:2010un}.
It is immediately clear that, if we quotient $S^3$ by $\Gamma \subset SU(2)_L$, we do not break any further supersymmetry. Therefore, such spherical three-manifolds can also be used to construct 3d $\CN=2$ backgrounds preserving  four superchages \cite{Closset:2012ru}.

\begin{table}[t]
\centering
\be\nn
\begin{array}{c||ccccc}
ADE  & \;\;A_{p-1}\;\;\  &\; \;D_{n+2} \;\;&\;\; E_6\;\;&\;\; E_7\;\; &\;\; E_8 \\
\hline
\Gamma & \Z_p & \h D_{n} & \h T & \h O& \h I\\
|\Gamma|  & p & 4 n & 24 & 48 & 120
\end{array}
\ee
\caption{ADE classification of the finite subgroups of $SU(2)$.}
\label{tab:ADE groups}
\end{table}
It is well-known that the finite subgroups $\Gamma \subset SU(2)$ are in one-to-one correspondence with the finite Dynkin diagrams of type $ADE$, according to:
\be
\Gamma_{ADE}= \big(\Z_p~, \; \h D_{n}~, \;  \h T~, \; \h O~, \; \h I\big) \quad \leftrightarrow\quad \big(A_{p-1}~, \; D_{n+2}~, \; E_6~, \; E_7~, \; E_8\big)~.
\ee
Here, $\h D_{4n}$ is the binary dihedral group, and $\h T$, $\h O$ and $\h I$ are the binary tetrahedral, binary octahedral and binary icosahedral groups, respectively. Their orders are given in Table~\ref{tab:ADE groups}.
Let us also introduce the notation:
\be
\CS^3[ADE] \cong S^3/\Gamma_{ADE}~.
\ee
By construction, spherical three-manifolds have a finite fundamental group: 
\be
\pi_1(\CS^3[ADE])= \Gamma_{ADE}~.
\ee
 The corresponding Seifert manifolds can be written as \cite{orlik1972seifert, nishiharuko:1998}:
\bea\label{M3 for S3Gamma}
&\CS^3[A_{p-1}]  &\cong &\; [0~;\, 0~;\, (q_1,1)~,\; (q_2,1)]~, \qquad& & p=q_1+q_2\geq 2~, \cr
&\CS^3[D_{n+2}]  &\cong &\; [0~;\, 0~;\, (2,-1)~,\; (2,1)~, \; (n,1)]~, \qquad && n \geq 1  \cr
&\CS^3[E_{m+3}]  &\cong &\; [0~;\, 0~;\, (2,-1)~,\; (3,1)~, \; (m,1)]~,  \qquad && m=3, 4, 5~.
\eea
Extrapolating the $E$-series to $m=1, 2$, we have the equivalences $E_4 = A_4$ and $E_5= D_5$, which are clearly realized by the Seifert geometries \eqref{M3 for S3Gamma}. (We also have $D_3=A_3$.) In all cases, we find the relation:
\be
\CK \cong \CL_0^{-1} \in \Pic(\h\Sigma)~,
\ee
with $c_1(\CL_0) \neq 0$,  and therefore there exists a ``superconformal'' background with the $U(1)_R$ fugacity:
 \be
 \nu_R =-\half {c_1(\CK) \ov  c_1(\CL_0)}= \half~,
 \ee so that $L_R \cong \CO$. The $A$-series consists of the lens spaces $L(p, p-1)$ with squashing $b^2= q_1/q_2$, which we discussed in Section~\ref{sec:lpqgeo}. Let us briefly discuss the other cases.

\subsubsection{The $D$-series}
Consider the Seifert fibrations $\CS^3[D_{n+2}]$ as defined in \eqref{M3 for S3Gamma}. One easily finds that:
\be
\t\Pic(\CS^3[D_{n+2}]) \cong \begin{cases}
 \Z_2 \times \Z_2 \qquad & \text{if}\; \; n \; \; \text{is even,}\cr
  \Z_4 \qquad & \text{if}\; \; n \; \; \text{is odd.}
\end{cases}
\ee
When $n$ is even, the two generators of $\Z_2\times \Z_2$ can be taken to be $[\omega_1]$ and $[\omega_1]+[\omega_2]$. When $n$ is odd, the generator of $\Z_4$ can be taken to be $[\omega_1]$.
Correspondingly, there are four distinct spin structures if $n$ is even, and two distinct spin structures if $n$ is odd. The $U(1)_R$ fluxes are:
\bea
&\n_0^R = -\frac{l_1^R + l_2^R + l_3^R}{2}~, \qquad
&&  \n_1^R=\half + l_1^R - \nu_R~,\cr
&  \n_2^R=\half + l_2^R + \nu_R~, \qquad
&&   \n_3^R=\frac{n-1}{2} + {l_3^R n\ov 2} + \nu_R~.
\eea
If $n$ is even, we have $\nu_R =\half$ mod $1$ and there are four distinct choices of $l_i^R$ mod $2$. If $n$ is odd, we also have $\nu_R =\half$ mod $1$ but we also need $l_3^R=1$ mod $2$ and there then only two distinct choices of the $U(1)_R$ fractional fluxes, matching the number of spin structures.

\subsubsection{The $E$-series}
For the Seifert fibrations $\CS^3[E_{m+3}]$ defined in \eqref{M3 for S3Gamma}, we find:~\footnote{Note that, for $m=1$, this agrees with the Picard group $\Z_5$ of the $A_4$ case, and for $m=2$ this agrees with the $\Z_4$ of the $D_5$ case, as expected.}
\be
\t\Pic(\CS^3[E_{m+3}]) \cong \Z_{6-m}~,
\ee
where the single generator can be take to be $[\omega_1]+[\omega_2]$. In other words, we have the homologies:
\bea
&H_1(\CS^3[E_6], \Z) = \Z_3~, \qquad \quad && H_2(\CS^3[E_6], \Z) =0~,\cr
&H_1(\CS^3[E_7], \Z) = \Z_2~, \qquad \quad && H_2(\CS^3[E_7], \Z) =0~,\cr
&H_1(\CS^3[E_8], \Z) = 0~, \qquad \quad && H_2(\CS^3[E_8], \Z) =0~.
\eea
It also follows that the $E_6$ and $E_8$ spherical three-manifolds admit a unique spin structure, while there are two distinct spin structures in the $E_7$  case. The Seifert manifold:
\be
\CS^3[E_8] \cong  [0~;\, 0~;\, (2,-1)~,\; (3,1)~, \; (5,1)]
\ee
is the celebrated Poincar\'e homology sphere---it has the same homology as the three-sphere, but its fundamental group (the binary icosahedral $\h I$) is non-trivial.

Beyond this complete list of spherical three-manifolds preserving four supercharges, it is natural to extend the $E$-series in \eqref{M3 for S3Gamma} to $m>5$. The bound $m\leq 5$ is equivalent to:
\be
\chi(\h\Sigma) = \half + {1\ov 3}+ {1\ov m}-1 >0~.
\ee
The ``$E_{m+3}$ manifolds'' with $m>5$ are examples of rational homology spheres (discussed in the next subsection), which generally do not have spherical geometry.  For instance,  the Seifert manifold $\CS^3[E_9]$ has $\chi(\h\Sigma)=0$ and $c_1(\CL_0)=0$ and Euclidean geometry (according to the classification reviewed in Section~\ref{subsec:classification seifert}).
The manifold $\CS^3[E_{10}]$ provides an example of  an integral homology sphere with $SL(2,\R)$ Thurston geometry.

\subsection{Homology spheres}
A three-dimensional ``integral homology sphere''  ($\Z$HS) is a three-manifold $\CM_3$ with the same integral homology as the three-sphere:
\be
H_0(\CM_3, \Z)= H_3(\CM_3, \Z)= \Z~, \qquad H_1(\CM_3, \Z)= H_2(\CM_3, \Z)=0~.
\ee
More generally, a ``rational homology sphere''  ($\Q$HS)  has $H_1(\CM_3, \Z_d)=H_2(\CM_3,\Z_d)=0$ for some integer $d$, with  $d=\pm 1$ in the integral case.

Seifert rational homology spheres can be constructed as the genus-zero fibration:
\be\label{ZHS seifert}
\CM_3\cong [0~;\; 0~;\, (q_1, p_1), \cdots, (q_n, p_n)]~,
\ee
such that:
\be
c_1(\CL_0) = \sum_{i=1}^n {p_i \ov q_i}=  \pm {1\ov \prod_{i=1}^n q_i}~.
\ee
This implies that the anisotropies $q_i$ should be mutually prime:
\be
\gcd(q_i, q_j)=1~, \quad \forall i, j~.
\ee
For instance, that is the case for both the examples $\CS^3[E_{8}]$ (the Poincar\'e homology sphere) and $\CS^3[E_{10}]$ mentioned above, with $(q_i)=(2,3,5)$ and $(2,3,7)$, respectively.
Then, $\Pic(\h\Sigma)\cong \Z$ is freely generated by:
\be
\CL_0 \cong \prod_i L_i^{p_i}~,
\ee 
and $\t\Pic(\CM_3)$ is trivial. In particular, there is a unique choice of $L_R$ and of spin structure. 
In their studies of CS theories,  \cite{Beasley:2005vf, Blau:2013oha} also considered certain simple $\Q$HS's obtained by a $\Z_d$ quotient of a $\Z$HS. They can be obtained by replacing $p_i $ by $d p_i$, $\forall i$, in   \eqref{ZHS seifert}. 

\

\noindent To conclude this survey of Seifert backgrounds, let us note  that, while all the examples of Seifert fibration in this section had an a underlying genus-zero base $\h\Sigma_{0,n}$, the generalization to $\h\Sigma_{g, n}$ is automatic. Indeed, changing the genus of the base does not affect the discussion of $\t\Pic(\CM_3)$ nor the determination of the $R$-symmetry line bundle. In the field theory discussion of the next sections, this will be reflected in the fact that the handle-gluing operator and the Seifert fibering operator are independent from each other.

\part{Seifert fibering operators}
\label{part two}
\section{Fibering operators and partition functions}\label{sec: Bethe vac sum}
Supersymmetric partition functions of 3d $\CN=2$ gauge theories on Seifert manifolds $\CM_3$ can be computed in the language of the ``3d A-model,'' a topologically twisted 2d $\CN=(2,2)$ effective field theory  for abelianized vector multiplets on the base $\h \Sigma$ of the Seifert fibration \cite{Closset:2017zgf}.

In this section, we first review this construction in the case of a principal bundle $\CM_3 = \CM_{g,{\bf d}}$, while also discussing the spin-structure dependence of 3d A-model. We then introduce the $(q,p)$-fibering operator, whose insertion at a smooth point on $\h \Sigma$ corresponds to adding a $(q,p)$ exceptional fiber to $\CM_3$. This allows us to elegantly write down the exact supersymmetric partition on any Seifert manifold.

\subsection{Twisted superpotential and Bethe vacua}
Consider a 3d $\CN=2$ gauge theory with gauge group $\GG$, a product of semi-simple simply-connected  and unitary groups,
 coupled to chiral multiplets $\Phi$ in representations $\CR$ of $\GG$. We consider the theory on $\R^2 \times S^1$, with a finite circle $S^1$ of radius $\beta$, and we focus on the effective field theory on the two-dimensional Coulomb branch. The low-energy modes consist of 2d $\CN=(2,2)$ twisted chiral multiplets $U_a$, $a= 1, \cdots, \rk$---which are the field strength multiplets of 2d gauge fields---with complex scalar $u_a$. We have $u=i \beta \sigma - a_0$ in terms of the 3d real scalar $\sigma$ and the Polyakov loop $a_0 = {1\ov 2 \pi} \int_{S^1} A$, which gives the equivalences:
\be
u_a \sim u_a +1~,
\ee
under large gauge transformations along the maximal torus of $\GG$.

The theory might also have a flavor group $\GG_F$, with maximal torus $\prod_\alpha U(1)_\alpha \subset \GG_F$, and we may turn on chemical potentials $\nu_\alpha$  ($\alpha= 1, \cdots, {\rm rk}(\GG_F)$)---{\it i.e.} background gauge fields---for these flavor symmetries. We have the periodicities $\nu_\alpha \sim \nu_\alpha +1$ under large gauge transformations of the background gauge fields. 
Let us  introduce the convenient notation:
\be
{\bf u}_{\bf a}= (u_a, \nu_\alpha)~, 
\ee
with the index ${\bf a}= (a, \alpha)$ running over both the gauge and flavor groups. 
We also consider the $U(1)_R$ fugacity:
\be
\nu_R \in \half \Z~,
\ee
as discussed in previous sections. Choosing either $\nu_R=0$ or $\nu_R= \half$ (mod $1$), is correlated with choosing the periodic or anti-periodic spin structure, respectively, for 3d fermions along the $S^1$ fiber.

\subsubsection{Twisted superpotential of the 3d gauge theory on $\R^2\times S^1$}
Up to $Q$-exact terms, the 2d low energy action on $\R^2$ is fully determined by the twisted superpotential \cite{Nekrasov:2014xaa, Closset:2017zgf}:
\be\label{fullW3d}
\CW({\bf u}; \nu_R) = \CW_{\rm CS}({\bf u}; \nu_R) + \CW_{\rm matter}({\bf u}; \nu_R) + \CW_{\rm vector}({\bf u}; \nu_R)~.
\ee
On general grounds, the twisted superpotential is only defined up to linear shifts \cite{Closset:2017zgf}:
\be\label{linear shifts}
\CW \rightarrow \CW + n^a u_a+ n^\alpha u_\alpha+ n_0~, \qquad n^a, n^\alpha, n_0 \in \Z~.
\ee

\paragraph{Classical contribution.} The first term in \eqref{fullW3d} is the contribution from the Chern-Simons terms (for the gauge and flavor groups) in the UV action.  It is given by:
\bea\label{WCS full}
&\CW_{\rm CS}({\bf u}; \nu_R) & =&\; \half \sum_{\substack{{\bf a, b} \\{\bf a} \neq {\bf b}}}k^{\bf ab} {\bf u}_{\bf a} {\bf u}_{\bf b} +\half (1+ 2 \nu_R) \sum_{{\bf a}} k^{\bf aa} {\bf u_a} \cr
&&&\; +\sum_{\bf a} k^{{\bf a} R}   {\bf u_a} \nu_R + \cdots~.
\eea
Here, $k^{\bf ab}$ are the CS levels for the gauge and flavor groups, including possible,  mixed gauge-flavor CS levels in the abelian sector; $k^{{\bf a}R}$ are the mixed $U(1)_{\bf a}$-$U(1)_R$ CS levels.
All the Chern-Simons levels are integer-quantized. The ellipsis in \eqref{WCS full} corresponds to ${\bf u}$-independent terms, which will not concern us in the following. 
Note that $\nu_R$ enters in \eqref{WCS full} in two conceptually different ways: In the second line, it enters as a chemical potential for $\nu_R$ in the mixed gauge-R CS term. In the first line, it enters as a shift of the linear term in ${\bf u}$, corresponding to a choice of spin structure on $\R^2 \times S^1$. In fact, the classical action for a $U(1)_k$  $\CN=2$ Chern-Simons theory on $\Sigma_g\times S^1$, with $\n$ units of flux through $\Sigma_g$, gives:
\be
e^{-S_{U(1)_k}}= (-1)^{k \n (1+ 2 \nu_R)} e^{2\pi i k \n u} = e^{2 \pi i \n {\d \CW_{\rm CS}\ov \d u}}~, \qquad \CW_{\rm CS}(u) = {k\ov 2}\left(u^2+ (1+ 2 \nu)u\right)~,
\ee 
for $u$ constant,  where the sign depends on whether we choose the periodic ($\nu_R=0$) or anti-periodic ($\nu_R = \half$) spin structure.~\footnote{This is explained in Appendix C of \cite{Closset:2017zgf}.}  Note that the sign dependence disappears if $k$ is even, in agreement with the fact that $U(1)_k$ CS theory only depends on the spin structure if $k$ is odd, while it is a fully topological theory for $k$ even \cite{Dijkgraaf:1989pz}.

\paragraph{Matter contribution.} The second term in the twisted superpotential \eqref{fullW3d} is a one-loop-exact contribution from the chiral multiplets. For a single chiral multiplet of $U(1)$ gauge charge $1$ and $R$-charge $r\in \Z$, we have the formal contribution:
\be
 -{1\ov 2 \pi i} \sum_{\kk \in \Z}(u + \nu_R r + \kk)\left(\log(u + \nu_R r + \kk)-1\right)~,
\ee
where the infinite sum is over the KK modes on the circle. 
This gives:
\be
 \CW^\Phi(u+ \nu_R r)~, \qquad \text{with} \qquad \CW^\Phi(u) \equiv {1\ov (2\pi i)^2} \dilog\left(e^{2 \pi i u}\right)~.
\ee
This is the result in the ``$U(1)_{-\half}$ quantization,'' consistent with the parity anomaly~\cite{AlvarezGaume:1984nf}. In the large-$\sigma$ limit, we have:
\bea
&\lim_{\sigma \rightarrow +\infty}\CW^\Phi(u+\nu_R r) =0~, \cr
&\lim_{\sigma \rightarrow -\infty}\CW^\Phi(u+\nu_R r)= - \half \left(u^2+(1+2 \nu_R) u\right)- (r-1)u\nu_R+ \cdots~.
\eea
The limit  $\sigma \rightarrow +\infty$ corresponds to an empty theory at vanishing CS levels. The limit 
$\sigma \rightarrow -\infty$ reproduces the integer CS levels:
\be\label{k shifts sigma}
k_{GG} = -1~, \qquad  k_{GR}= -(r-1)~, \qquad k_{RR}= -(r-1)^2~, \qquad k_{g}=-2~,
\ee
as expected. We refer to  \cite{Closset:2017zgf} for further discussion of the parity anomaly in this context.~\footnote{The $U(1)_R$ CS level $k_{RR}$ and the gravitational CS level $k_g$ are omitted in \protect\eqref{WCS full}, since they are ${\bf u}$-independent. Below we will discuss how they contribute to the partition function on general Seifert three-manifolds.}  The matter contribution $\CW_{\rm matter}$ in \eqref{fullW3d}  is the sum over all the chiral multiplets, weighted by their charges:
\be
\CW_{\rm matter}({\bf u}; \nu_R) = \sum_{\omega \in \FR_F}\sum_{\rho \in \FR} \CW^\Phi(\rho(u) + \omega(\nu)+ \nu_R r_\omega)~,
\ee
with $r_\omega$ the $R$-charge of the fields $\Phi_\omega$ in some representation $\FR$ of the gauge group.
Here and henceforth, $\rho=(\rho^a)$ and $\omega= (\omega^\alpha)$ denote the gauge and flavor weights, respectively. We also use the short-hand notation $\rho({\bf u})= \rho(u) + \omega(\nu)$.

\paragraph{Vector multiplet contribution.}  The last term in \eqref{fullW3d} is the contribution from the vector multiplet. Consider first the W-bosons $\CW_\alpha$, with $\alpha\in \Delta$ the roots of $\Fg= {\rm Lie}(\GG)$.
In the abelianized theory on the Coulomb branch, $\CW_\alpha$ contributes exactly like a chiral multiplet of gauge charges $\alpha^a$ and R-charge $r=2$ \cite{Closset:2015rna}. For each pair of roots $\alpha$ and $-\alpha$, we choose the ``symmetric quantization,'' such that there is a vanishing net contribution to the contact terms $\kappa_{GG}$, $\kappa_{GR}$, $\kappa_{RR}$ and $\kappa_g$. Each positive root then contributes:
\be\label{Walpha contrib}
 {1\ov (2 \pi i)^2}\Big(\dilog\big(e^{2\pi i( \alpha(u) +2 \nu_R)}\big)+
\dilog\big(e^{2\pi i( -\alpha(u) +2 \nu_R)}\big)\Big) + \half\big(\alpha(u)^2+ (1+2\nu_R)\alpha(u)\big)
\ee
to the twisted superpotential.  Up to $u$-independent terms, the expression \eqref{Walpha contrib} is equivalent to $\nu_R \alpha(u)$. Therefore, we find the very simple result:
\be\label{Wvec}
\CW_{\rm vector}({\bf u}; \nu_R)=\nu_R \sum_{\alpha\in \Delta^+} \alpha(u) = 2 \nu_R\, \rho_W(u)~,
\ee
with $\Delta^+$ the set of positive roots, and $\rho_W=\half \sum_{\alpha\in \Delta^+}\alpha$ the Weyl vector. 
For $\GG$ a simply-connected simple gauge groups, the Weyl vector is a weight, so that $\rho_W^a \in \Z$ and $\CW_{\rm vector}$  can be set to zero by a linear shift \eqref{linear shifts}. For $\GG= U(N)$, on the other hand, $\CW_{\rm vector}$ will contribute some non-trivial signs to physical observables if and only if $N$ is an even integer (and $\nu_R=\half$ mod $1$). 

As already anticipated in the introduction, with our choice of quantization the vector multiplet also contributes non-trivially to the $\kappa_{RR}$ and $\kappa_g$ CS contact terms, as in \eqref{kappaRR intro vec}. These contribute to the constant piece in the twisted superpotential, which we ignored in the above.~\footnote{This is because, for our purposes in this paper, we will only need the derivatives of $\CW$ with respect to $u$ or $\nu$, not the ``constant'' pieces that are either numerical constants or depend on the parameter $\nu_R$.} They do affect the more general geometry-changing operators to be introduced below, as we will see.

\subsubsection{Bethe equations and Bethe vacua}
Consider the so-called ``gauge flux operators:''
\be\label{def gfo}
\pif_a(u, \nu; \nu_R) \equiv \exp\left(2 \pi i {\d \CW\ov \d u_a}\right)~.
\ee
The Bethe vacua are the two-dimensional vacua of the effective field theory for the abelianized 2d vector multiplets. These Bethe vacua are obtained from the Bethe equations:
\be\label{S BE 3d}
\cS_{BE} =\left\{ \;\h u_a \; \bigg| \; \Pi_a(\h u, \nu; \nu_R) = 1~, \;\; \forall  a~, \quad w \cdot \h u \neq \h u, \;\; \forall w \in W_\GG \; \right\} /W_\GG~,
\ee
with the identification $u_a \sim u_a +1$. Here $W_\GG$ denotes the Weyl group of $\GG$, and $w\cdot u$ the Weyl group action on $\{u_a\}$---that is, we should discard any solution to the equations $\{\pif_a=1\}$ that is not acted on freely by the Weyl group \cite{Hori:2006dk,Aharony:2016jki}. 

Note that the Bethe equations generally depend on $\nu_R$. In the special case when all $U(1)$ CS levels and all chiral-multiplet R-charges are {\it even}, the dependence on $\nu_R$ drops out and the Bethe equations are the same as in the ``A-twist background'' studied in \cite{Closset:2017zgf}. When the $R$-charges are all integer-quantized (as is the case, for instance, on $\Sigma_g\times S^1$), the dependence on $\nu_R$ only appears through subtle signs.

\subsection{Handle-gluing operator and twisted index}~\label{subsec: HGO}
Consider the supersymmetric partition function of the 3d $\CN=2$ theory on $\Sigma_g \times S^1$, also known as the twisted index \cite{Nekrasov:2014xaa, Benini:2015noa, Benini:2016hjo, Closset:2016arn}. It is given by a sum over Bethe vacua, as:
\be\label{twisted index expl}
Z_{\Sigma_g \times S^1}(\nu; \nu_R) = \sum_{\h u \in \CS_{\rm BE}}  \CH(\h u, \nu; \nu_R)^{g-1}~.
\ee
This is a sum over the solutions $u= \h u$ of the Bethe equations \eqref{S BE 3d}, with the {\it handle-gluing operator} $\CH$ evaluated on the vacua. The handle-gluing operator of a 3d gauge theory reads:
\be\label{hgoFull}
\CH({\bf u}; \nu_R)  = e^{2 \pi i \Omega({\bf u}; \nu_R)} \; \det_{ab}\left(\d_{u_a}\d_{u_b} \CW\right)~.
\ee
Here, the ``effective dilaton'' $\Omega$ is given by:
\be
\Omega({\bf u}; \nu_R) = \Omega_{\rm CS} + \Omega_{\rm matter}+ \Omega_{\rm vector}~,
\ee
with:
\be
\Omega_{\rm CS}({\bf u}; \nu_R) = \sum_{\bf a} k^{{\bf a} R} {\bf u_a} + \half k_{RR}
\ee
the contribution from the gauge or flavor-$R$ and $RR$ CS levels.
The matter contribution reads:
\be\label{Omega Phi}
 \Omega_{\rm matter}({\bf u}; \nu_R) = -\sum_{\omega, \rho}{r_\omega-1\ov 2\pi i}\log\left(1-e^{2\pi i(\rho({\bf u})+\nu_R r_\omega)}\right)~,
\ee
and the vector multiplet contribution reads:
\be\label{Omega vec}
 \Omega_{\rm vector}({\bf u}; \nu_R) = -{1\ov 2\pi i} \sum_{\alpha \in \Fg}\log\left(1-e^{2\pi i \alpha(u)}\right)~.
\ee
This last term is $\nu$- and $\nu_R$-independent, and gives rise to a factor:
\be
e^{2\pi i (g-1)  \Omega_{\rm vector}} =  \prod_{\alpha\in \Fg} (1-e^{2\pi i \alpha(u)})^{1-g}~.
\ee
At genus $g=0$, this is an ordinary Weyl determinant. The last factor in \eqref{hgoFull} is the Hessian determinant of the twisted superpotential \eqref{fullW3d} as a function of the gauge parameters $u_a$. It can  be written as:
\be\label{detW and Pif}
 \det_{ab}\left(\d_{u_a}\d_{u_b} \CW\right)=  \det_{ab}\left({1\ov 2\pi i }\d_{u_a}\log\pif_b\right)~,
\ee
in terms of the gauge flux operators \eqref{def gfo}.

One can also insert background flux $\n_\alpha\in \Z$ for some $U(1)_\alpha$ flavor symmetry through $\Sigma_g$. That background flux can be localized at a point, in which case it can be viewed as another local operator in 2d, the flavor flux operator \cite{Closset:2017zgf}, given by:
\be\label{def ffo}
\pif_\alpha(u, \nu; \nu_R)^{\n_\alpha} \equiv \exp\left(2 \pi i \n_\alpha {\d \CW\ov \d \nu_\alpha}\right)~.
\ee
In three dimensions, this is a line operator wrapping the $S^1$, at a point on $\Sigma_g$ \cite{Kapustin:2012iw,Drukker:2012sr}.
One can also insert a supersymmetric Wilson loops $W_\FR$ in representations $\FR$ of the gauge group, wrapping the $S^1$, which corresponds to the characters:
\be
W_\FR(u) = \Tr_\FR(e^{2\pi i u})= \sum_{\rho\in \FR} e^{2\pi i \rho(u)}~.
\ee
The operators $\pif_\alpha$ and $W_\FR$ are the simplest examples of half-BPS lines operators, also known as twisted-chiral operators, which can be inserted at any point on $\Sigma_g$ while preserving the same supercharges $\CQ_-$ and $\b\CQ_+$ as the geometric background. The correlations functions of twisted-chiral  operators are topological---{\it i.e.} independent of the insertion points on $\Sigma_g$. Therefore, the most general correlation function in the 3d A-model is the expectation value of a general line operator $\SL$ wrapped over $S^1$, which is given explicitly by the formula:
\be
\left\langle  \SL \right\rangle_{\Sigma_g \times S^1} = \sum_{\h u \in \CS_{\rm BE}}  \SL(\h u, \nu; \nu_R)\,  \CH(\h u, \nu; \nu_R)^{g-1}~.
\ee
The twisted index \eqref{twisted index expl} is the very special case $\SL= \mathbf{1}$.

\subsection{Fibering operators for Seifert manifolds}
In this work, we study a more general class of half-BPS line operators in the 3d A-model, the {\it geometry-changing line operators}, whose insertion on $\Sigma_g$ is equivalent (in $\CQ$-cohomology) to considering the field theory on a different three-manifold with $S^1$ non-trivially fibered over $\Sigma_g$ \cite{Closset:2017zgf}. 
For any supersymmetric background $(\CM_3, L_R)$ on a Seifert manifold $\CM_3$, we expect that the supersymmetric partition function can be written (at least formally) as an expectation value:
\be\label{ZM3 gen introduce LM3}
Z_{\CM_3} = \left\langle  \SL_{\CM_3} \right\rangle_{S^2 \times S^1}=  \sum_{\h u \in \CS_{\rm BE}}   \CH(\h u, \nu; \nu_R)^{-1}\, \SL_{\CM_{3}}(\h u, \nu; \nu_R)~,
\ee
with $\SL_{\CM_3}$ the line operator associated to the Seifert geometry $\CM_3$, from the point of view of $S^2\times S^1$.  In \cite{Closset:2017zgf}, we studied the case of $\CM_3 \cong \CM_{g, {\bf d}}$ a principal circle bundle, in which case:
\be
\SL_{\CM_{g, {\bf d}}}(u) = \CF(u)^{\bf d}\, \CH(u)^{g}~,
\ee
with $\CF$ the so-called fibering operator. In the present work, we will write down the geometry-changing line operator associated to any Seifert geometry \eqref{M3 Seifert def} as:
\be
\SL_{\CM_3}({\bf u}) =\CF({\bf u})^{\bf d}\,  \CH({\bf u})^{g} \,  \prod_{i=1}^n \CG_{q_i, p_i}({\bf u})~, 
\ee
schematically. Each {\it $(q,p)$-fibering operator} $\CG_{g,p}$ corresponds to adding an exceptional fiber with the Seifert invariants $(q,p)$. Using unnormalized Seifert invariants and the fact that the degree ${\bf d}$ can be viewed as a $(1,{\bd})$ exceptional fiber, we write a general Seifert fibration as:
\be
\CM_3 \cong \left[0~;\, g~;\, (1, {\bf d})~, \, (q_1, p_1)~, \, \cdots~, \, (q_n,p_n)\right]~,
\ee 
and we will use the notation:  
\be
(q_0, p_0)\equiv (1, {\bf d})~, \qquad \quad  \CG_{1, {\bf d}}({\bf u})=\CF({\bf u})^{\bf d}
\ee
for the ``ordinary'' fibering operator, $\CF({\bf u})$. We then write:
\be\label{def GM3}
\SL_{\CM_3}({\bf u}) =  \CH({\bf u})^{g} \, \CG_{\CM_3}({\bf u})~, \qquad   \CG_{\CM_3}({\bf u})\equiv  \prod_{i=0}^n \CG_{q_i, p_i}({\bf u})~.
\ee
 Note that the $(q_i,p_i)$-fibering operators (with $q_i>1$) should be inserted at  {\it distinct} points $x_i \in \Sigma_g$, corresponding to the $n$ ramification points of the two-dimensional orbifold $\h\Sigma_g(q_1, \cdots, q_n)$ at the base of the Seifert fibration $\CM_3$. The position of the ramification points is otherwise arbitrary.  Since the correlation function does not depend on the position of these points, we may take them arbitrarily close to one another, and so may effectively treat the operator ${\SL}_{\cM_3}$ as a single line operator.

\paragraph{Fractional fluxes and large gauge transformations.}  From the point of view of the 2d orbifold $\h\Sigma$, we may consider any holomorphic line bundle as in \eqref{2d L gen}. We refer to the corresponding integers $\n_i$ as the ``fractional fluxes'' at the orbifold points $x_i$, while a flux $\n_0$ localized at a smooth point is an ``ordinary flux.'' Then, considering a single $(q,p)$ exceptional fiber, we denote by:
\be\label{introduce frac flux in G}
\CG_{q,p}({\bf u})_{\n}~,
\ee
with $\n\in \Z$ the fractional flux. (From here onwards, the dependence on $\nu_R$ is left implicit.) 
Strictly speaking, the fractional flux is valued in $\Z_q$, and we must  have:
\be\label{pif from G}
\CG_{q,p}({\bf u})_{\n+ q\n_0} =\pif({\bf u})^{\n_0}\, \CG_{q,p}({\bf u})_{\n}~,
\ee
with $\pif({\bf u})$ the ordinary flux operator defined above. (Treating gauge and flavor fluxes democratically.) In particular, we have:
\be\label{CG as Fpi}
\CG_{1,{\bf d}}({\bf u})_{\n}=\pif({\bf u})^{\n} \CF({\bf u})^{\bf d}~,
\ee
for the $(1,{\bf d})$ ``exceptional fiber'' encoding the degree of $\CL_0$. We will use the notation $\CG_{q,p}({\bf u})$ for \eqref{introduce frac flux in G} with $\n=0$.

The fractional fluxes encode all the relevant (supersymmetry-preserving) non-trivial lines bundles over $\CM_3$ in a redundant manner, according to \eqref{pic3 as quotient}. Tensoring by $\CL_0$ corresponds to a {\it large gauge transformation} in the A-model. Considering a configuration with fractional fluxes $(\n_0, \n_i)$ and holomorphic modulus ${\bf u}\in \C$, we have:
\be\label{lgt action}
L \rightarrow L \otimes \CL_0 \qquad \Leftrightarrow \qquad u \rightarrow u+ 1~, \quad \n_0 \rightarrow \n_0 + {\bf d}~, \quad \n_i \rightarrow \n_i + p_i~.
\ee
By gauge invariance of the quantum field theory, the shift \eqref{lgt action} should be an invariance of the various fibering operators, viewed as functions of ${\bf u}\in \C$ and $\n\in \Z$:
\be\label{diff eq CG}
\CG_{q,p}({\bf u}+1)_{\n + p}= \CG_{q,p}({\bf u})_{\n}~.
\ee
We will verify this in the explicit results below.~\footnote{There is an important subtlety for general $L_R$, which we will address below.} In the special case \eqref{CG as Fpi}, the difference equation \eqref{diff eq CG} is equivalent to:
\be\label{diff FPif}
\CF({\bf u}-1)=\pif({\bf u})\, \CF({\bf u})~,
\ee
which was a crucial ingredient in the analysis of \cite{Closset:2017zgf}.  In the following subsections, we present explicit expressions for the building blocks of the fibering operators for general 3d $\CN=2$ gauge theories.

\subsubsection{Contribution from Chern-Simons terms}\label{subsec: CSterms}
Let us first consider the contribution from the classical action to the fibering operators. It comes entirely from Chern-Simons terms. (The relevant supersymmetric Lagrangians can be found in \cite{Closset:2017zgf}, following \cite{Closset:2012ru}.)

\paragraph{$U(1)_k$ CS term.} Consider first the supersymmetric Chern-Simons action for a single $U(1)$ vector multiplet, at level $k$. We claim that each $(q,p)$-fibering operator contributes:
\be
\CG^{U(1)_k}_{q,p}(u)_\n= \Big(\CG^{\rm GG}_{q,p}(u)_\n\Big)^k
\ee
with the function:
\be\label{CS GG}
\CG^{\rm GG}_{q,p}(u)_\n \equiv (-1)^{\n (1+ t+ l^R t +2 \nu_R s)}\, \exp\left({- {\pi i \ov q}\left(p u^2- 2 \n u + t \n^2\right)}\right)~.
\ee
Here, $\n\in \Z$ denotes the fractional flux, $l^R\in \Z$ corresponds to the parameterization \eqref{param LR} for the R-symmetry line bundle $L_R$, and the integers  $s$ and  $t$ are such that $q s+ p t=1$, as usual.
It follows from \eqref{param LR}  that, for a given $(q,p)$,  \eqref{CS GG}  is independent of the choice of $(s,t)$.~\footnote{Under a shift $(s,t)\rightarrow (s-p, t+q)$, the expression \protect\eqref{CS GG} gets multiplied by a trivial sign $(-1)^{\n(q-1 +2 p \nu_R+ l^R q)}=1$, where we have used the fact that $q-1 +2 p \nu_R+ l^R q= 2 \n^R$ is even.}  

The function \eqref{CS GG} is invariant under the large gauge transformation $(u, \n)\rightarrow (u+1,  \n+p)$ {\it up to a sign}:
\be\label{GG lgt}
\CG^{\rm GG}_{q,p}(u+1)_{\n+p} =(-1)^{l^R}\,  \CG^{\rm GG}_{q,p}(u)_\n~,
\ee
as one can easily check. When taking the product of all the fibering operators to obtain a geometry-changing operator, as in \eqref{def GM3}, the signs $(-1)^{l^R}$ from \eqref{GG lgt} cancel out due to \eqref{sum lR zero}. Therefore the operator $\CG_{\CM_3}({\bf u})$ for the $U(1)_k$ theory is gauge invariant, as expected.  We can also check that:
\be
\CG^{\rm GG}_{q,p}(u)_{\n+ q \n_0}= \pif^{\rm GG}(u)^{\n_0} \, \CG^{\rm GG}_{q,p}(u)_{\n}~,
\ee
with:
\be
 \pif^{\rm GG}(u)\equiv (-1)^{1+ 2\nu_R} \, e^{2\pi i u}
\ee
the ordinary flux operator for the $U(1)_{k=1}$ term.

\paragraph{Mixed abelian CS term.} Consider a mixed abelian CS term at level $k_{12}\in \Z$ for some $U(1)_1\times U(1)_2$ vector multiplets. At level $k_{12}=1$, we have:
\be\label{CS G1G2}
\CG^{\rm G_1G_2}_{q,p}(u_1, u_2)_{\n^{(1)},\, \n^{(2)}} \equiv \exp\left({-{2 \pi i \ov q}\left(p u_1 u_2 - \n^{(1)} u_2 - \n^{(2)} u_1+ t \n^{(1)} \n^{(2)} \right)}\right)~,
\ee
with $\n^{(1)}, \n^{(2)}$ the fractional fluxes for the two $U(1)$ factors.
This expression is fully gauge-invariant and satisfies:
\be
\CG^{\rm G_1G_2}_{q,p}(u_1,  u_2)_{\n^{(1)}+q \n_0^{(1)}, \, \n^{(2)}+ q \n_0^{(2)}}= e^{2 \pi i \left(\n_0^{(1)} u_2 +\n_0^{(2)}u_1\right)}\, \CG^{\rm G_1G_2}_{q,p}(u_1, u_2)_{\n^{(1)},\, \n^{(2)}}~,
\ee
in agreement with the form of the corresponding flux operators \cite{Closset:2017zgf}. Note that \eqref{CS G1G2} is independent of $\nu_R$ and $l^R$, in agreement with the fact that the mixed CS action is spin-structure independent.

\paragraph{Non-abelian CS terms.} The Chern-Simons contribution for a non-abelian gauge group $\GG$  can be obtained similarly. For $\GG= U(N)$ at level $k$, we simply have:
\be
\CG^{U(N)_k}_{q,p}(u)_\n= \left( \prod_{a=1}^N \CG^{\rm GG}_{q,p}(u_a)_\n\right)^k~.
\ee
For $\GG$ a simply-connected simple Lie group, the signs in front of \eqref{CS GG} cancel out, in agreement with the expected spin-structure-independence of the answer, and one obtains:
\be\label{GG nonabelian}
\CG^{\GG_k}_{q,p}(u)_\n=  \exp\left({- {\pi i \, k\, h^{ab} \ov q}\left(p u_a u_b - \n_a u_b- \n_b u_a + t \n_a \n_b\right)}\right)~,
\ee
with $h^{ab}$ the Killing form of $\Fg= {\rm Lie}(\GG)$.

\paragraph{Mixed gauge-R CS term.}  We may also have a mixed CS term between an abelian gauge field and the $R$-symmetry background gauge field on $\CM_3$, at level $k_{GR}\in \Z$. This is given in terms of:
\be\label{CS GR}
\CG^{\rm GR}_{q,p}(u)_\n \equiv \exp\left({-{2 \pi i \ov q}\left(p u\nu_R -\n \nu_R- \n^R  u + t \n^R  \n \right)}\right)~,
\ee
to the power $k_{GR}$. Here the $R$-symmetry ``fractional flux'' $\n^R$ at the exceptional fiber $(q,p)$ is given as in \eqref{param LR}, namely:
\be\label{nR qp def}
\n^R = {q-1\ov 2}+ \nu_R p+ {l^R q\ov 2}~.
\ee
 The expression \eqref{CS GR} can be understood as  a special case of \eqref{CS G1G2}, essentially because the $R$-symmetry gauge field also sits in a vector multiplet, which is a submultiplet of the 3d $\CN=2$ ``new minimal'' supergravity multiplet \cite{Closset:2014uda}.
Using \eqref{nR qp def}, the expression \eqref{CS GR} can also be written as:
\be\label{CS GR bis}
\CG^{\rm GR}_{q,p}(u)_\n= (-1)^{\n(l^R t + 2 \nu_R s)}\, e^{\pi i u l^R} \, e^{\pi i {q-1\ov q} (u- t\n)}~.
\ee

\paragraph{The $RR$ CS term.}
We can also have a $U(1)_R$ CS term at level $k_{RR}\in \Z$. It contributes to the fibering operator through:
\be\label{CS RR}
\CG^{\rm RR}_{q,p} \equiv (-1)^{\n^R (1+ t+ l^R t +2 \nu_R s)}\, \exp\left({- {\pi i \ov q}\left(p \nu_R^2- 2 \n^R \nu_R + t (\n^R)^2\right)}\right)~,
\ee
to the power $k_{RR}$, with $\n^R$ given by \eqref{nR qp def}. Since $\nu_R$ is real, $\CG^{\rm RR}_{q,p}$ is a pure phase. Obviously, this can also be understood as a special case of \eqref{CS GG}.

\paragraph{The gravitational CS term.}  The last supersymmetric Chern-Simons term is the supersymmetrization of the gravitational CS term, with level $k_g\in \Z$ \cite{Closset:2012vp}. For future reference, let us first introduce the phase:
\be\label{def CG0}
\CG^{(0)}_{q,p}\equiv \exp\left(\pi i \left({p\ov 12 q}- s(p,q)\right)\right)~,
\ee
with $s(p,q)$ the Dedekind sum:
\be
s(p,q)= {1\ov 4 q}\sum_{l=1}^{q-1}\cot{\left({\pi l\ov q}\right)}\cot{\left({\pi l p\ov q}\right)}~.
\ee
We will denote by:
\be\label{CGkg}
\left(\CG^{\rm grav}\right)^{k_g}
\ee
the contribution of the gravitational CS term at level $k_g$. We find:
\be\label{Ggrav2 full}
\left(\CG^{\rm grav}\right)^2= \left(\CG^{\rm RR}_{q,p}\right)^{-1} \, \left(\CG^{(0)}_{q,p}\right)^2~,
\ee
with the phase $\CG^{\rm RR}_{q,p}$ given by \eqref{CS RR}.  Note that we could only determine the precise phase \eqref{CGkg} for $k_g$ even. (In other words, we determined $\CG^{\rm grav}$ up to a sign. In any case, in this paper at least, we will only ever need to consider $k_g$ even.)

\subsubsection{Contribution from chiral multiplets}\label{subsec: GPhiqp}
Next, we consider the contribution of chiral multiplets to the fibering operator. This can be extracted from the one-loop determinant of a free chiral multiplet on  the supersymmetric background $(\CM_3, L_R)$. Let $\Phi$ be a chiral multiplet coupled to a $U(1)$ background vector multiplet with charge $1$, of $R$-charge $r\in \Z$. We can compute the one-loop determinant by considering the KK expansion \eqref{KK expansion gen} of the fields along the Seifert fiber. By supersymmetry, the only modes that give a net contribution to the one-loop determinant are holomorphic sections on $\h\Sigma$ \cite{Closset:2017zgf}. Let $L= L_0^{\n_0}\otimes_i L_i^{\n_i}$ denote the gauge bundle. The KK modes $\phi_\kk$ for the scalar field $\phi$ in $\Phi$ are then valued in the holomorphic line bundles:
\be\label{Lrkk}
L_{(r,\kk)} = L\otimes (L_R)^r \otimes (\CL_0)^\kk = L_0^{\n_0 + \t\n^R_0 r + {\bf d} \kk}\, \bigotimes_{i=1}^n L_i^{\n_i + \n_i^R r + p_i \kk} 
\ee
Using the Riemann-Roch-Kawasaki theorem \eqref{RRK thm}, we obtain the simple result:
\be\label{ZM3 Phi product}
Z_{\CM_3}^\Phi = \prod_{\kk\in \Z}\left(1\ov u + \nu_R r + \kk\right)^{{\rm deg}(L_{(r,\kk)}) + 1-g}~,
\ee
with:
\be
{\rm deg}(L_{(r,\kk)})= \n_0 + \t\n^R_0 r + {\bf d} \kk+ \sum_{i=1}^n \floor{\n_i + \n_i^R r + p_i \kk\ov q}~.
\ee
The formal infinite product \eqref{ZM3 Phi product} needs to be regularized.

\paragraph{Vanishing R-charge.} Consider first the case of a chiral multiplet of $R$-charge $r=0$.
As in \cite{Closset:2017zgf}, we define the ordinary flux and fibering operators for a free chiral by:
\be\label{flux fiber Phi}
\pif^\Phi(u) \equiv  \prod_{\kk\in \Z} {1\ov u+ \kk}~, \qquad\qquad 
\CF^\Phi(u)  \equiv  \prod_{\kk \in \Z}\left(1\ov u+ \kk\right)^\kk~.
\ee
Similarly, we find the $(q,p)$-fibering operator for a free chiral multiplet:
\be\label{Gqp product}
\CG_{q,p}^\Phi(u)_\n \equiv \prod_{\kk\in \Z}\left(1\ov u+ \kk\right)^{\floor{\n+ p \kk\ov q}}~.
\ee
With these definitions, we find:
\be
Z_{\CM_3}^\Phi \Big|_{r=0}=\pif^\Phi(u)^{\n_0 +1 -g}\, \CF^\Phi(u)^{\bf d}\,  \prod_{i=1}^n  \CG^\Phi_{q_i, p_i}(u)_{\n_i} = \pif^\Phi(u)^{1 -g}\, \prod_{i=0}^n  \CG^\Phi_{q_i, p_i}(u)_{\n_i}~.
\ee
In the $U(1)_{-\half}$ quantization for the Dirac fermions, we obtain: 
\be
\pif^\Phi(u) \equiv {1\ov 1-e^{2\pi i u}}~, \qquad
\CF^\Phi(u)  \equiv  \exp\left({1\ov 2 \pi i} \dilog(e^{2\pi i u}) + u \log\left(1- e^{2\pi i u}\right)\right)~,
\ee
upon regularizing \eqref{flux fiber Phi}.
Note that $\CF^\Phi(u)$ is a meromorphic function of $u$ which satisfies $\CF^\Phi(u-1)=\pif^\Phi(u)  
\CF^\Phi(u) $, as in \eqref{diff FPif}  \cite{Closset:2017zgf}.
We can similarly regularize the infinite product \eqref{Gqp product}---this is elaborated upon in Appendix~\ref{app:oneloop det}.
In the absence of fractional flux ($\n=0$),  we find:
\be\label{CGqp0 Phi}
\CG_{q,p}^\Phi(u) \equiv \exp \sum_{l=0}^{q-1}\left\{ {p\ov 2 \pi i} \dilog(e^{2\pi i {u+ t l\ov q}}) + {pu + l \ov q} \log\left(1-e^{2\pi i {u+ t l\ov q}}\right)\right\}~,
\ee
with $t$ the modular inverse of $p$ mod $q$.
Note that $\CG_{q,p}^\Phi(u)$ is a meromorphic function of $u$ with poles or zeros at $u =- \kk \in \Z$, as expected from \eqref{Gqp product}. (One can show that the various branch cut ambiguities in the definition \eqref{CGqp0 Phi} cancel out entirely, so that $\CG_{q,p}^\Phi(u)$ is single-valued.) This generalizes the ordinary flux operator, which corresponds to:
\be
\CG_{1, {\bf d}}^\Phi(u)= \CF^\Phi(u)^{\bf d}~.
\ee
The function \eqref{CGqp0 Phi} can be also written as:
\be\label{CGqp0 Phi Bis}
\CG_{q,p}^\Phi(u) = e^{{p\ov q}\left({1\ov 2 \pi i} \dilog(e^{2\pi i u})+ u \log\left(1-e^{2\pi i u}\right)\right) } \, \prod_{l=1}^{q-1} \left(1-e^{2\pi i {u+ t l\ov q}}\right)^{l \ov q}~.
\ee
The contribution from the fractional fluxes can be conveniently written in term of  Pochhammer symbols:
\be \label{qpochn}
 \pif_{q,p}^\Phi(u)_\n \equiv \left(e^{2\pi i u \ov q}; e^{2\pi i t\ov q}\right)_{-\n}= 
  \begin{cases} \prod_{l=0}^{\n-1} \left({1\ov 1-e^{2 \pi i {u+t(l-\n)\ov q}}}\right) &  {\rm if}\;\; \; \n>0~,\\ \\	\prod_{l=1}^{|\n|} \left(1-e^{2 \pi i {u-t(l+\n)\ov q}}\right)	\qquad& {\rm if} \;\; \; \n<0~.								\end{cases}
\ee
Let us note the identity:
\be
 \pif_{q,p}^\Phi(u)_{\n+ \m} =  \pif_{q,p}^\Phi(u)_\n \,  \pif_{q,p}^\Phi(u- t \n)_\m~.
\ee
The full $(q,p)$ fibering operator for $\Phi$ is then given by:
\be
\CG_{q,p}^\Phi(u)_{\n} \equiv \pif_{q,p}^\Phi(u)_\n \, \CG_{q,p}^\Phi(u)
\ee
One can check that:
\be
\CG_{q,p}^\Phi(u)_{\n+ q}= \pif^\Phi(u) \,  \CG_{q,p}^\Phi(u)_{\n}~, \qquad 
\CG_{q,p}^\Phi(u+1)_{\n+ p}=\CG_{q,p}^\Phi(u)_{\n}~,
\ee
as expected on general grounds.

\paragraph{General R-charge.} For a general $R$-charge $r\in \Z$, we simply have:
\be\label{chiral multiplet contribution}
Z_{\CM_3}^\Phi = \pif^\Phi(u+ \nu_R r)^{(g-1)(r-1)}\, \prod_{i=0}^n \CG_{q_i, p_i}(u+ \nu_R r)_{\n_i + \n^R_i r}~,
\ee
with the $R$-symmetry line bundle over $\h\Sigma$ parameterized as in \eqref{param LR} and \eqref{param LR ii}.
In particular, the chiral multiplet simply contributes:
\be
\CG_{q,p}^\Phi(u+ \nu_R r)_{\n+  \n^R r}
\ee
to the $(q,p)$-fibering operator.

\paragraph{Large $\sigma$ limits:}
In the limit $\sigma \rightarrow \infty$, we find:
\be
\lim_{u \rightarrow i \infty} \CG_{q,p}^\Phi(u+ \nu_R r)_{\n+  \n^R r} =1~,
\ee
in agreement with having an empty theory in the IR. In the opposite limit $\sigma \rightarrow- \infty$, one can show that:
\bea\label{CS limit of GPhi}
&\lim_{u \rightarrow -i \infty} \CG_{q,p}^\Phi(u+ \nu_R r)_{\n+  \n^R r} 
=\; e^{-\pi i (u+r \nu_R)l^R}  \cr
&\qquad\qquad\qquad\qquad\times \left(\CG_{q,p}^{GG}\right)^{-1}\,  \left(\CG_{q,p}^{GR}\right)^{-(r-1)}\,  \left(\CG_{q,p}^{RR}\right)^{-(r-1)^2}\,  \left(\CG_{q,p}^{\rm grav}\right)^{-2}~,
\eea
with the right-hand-side given in terms of the CS fibering operators defined in Subsection \ref{subsec: CSterms}.~\footnote{See Appendix~\protect\ref{app:oneloop det} for details on the derivation of \protect\eqref{CS limit of GPhi}.} This is exactly as expected from \eqref{k shifts sigma}. The $l^R$-dependent term in front of the CS terms in \eqref{CS limit of GPhi} cancels out in the full fibering operator $\CG_{\CM_3}$ due to the constraint $\sum_{i=0}^n l_i^R=0$.

\paragraph{Complex mass term.}  Similarly, we can consider two chiral multiplets $\Phi_1, \Phi_2$ with gauge charges $\pm 1$ and $R$-charges $r$ and $2-r$, such that we can write down a superpotential mass term $W= \Phi_1 \Phi_2$. One finds: 
\bea\label{phi12}
& \CG_{q,p}^\Phi(u+ \nu_R r)_{\n+  \n^R r} \; \CG_{q,p}^\Phi(-u+ \nu_R (2-r))_{-\n+  \n^R (2-r)} 
=\; \t g(u+\nu_R r)^{ l^R}  \cr
&\qquad\qquad\qquad\qquad\times \left(\CG_{q,p}^{GG}(u)_\n\right)^{-1}\,  \left(\CG_{q,p}^{GR}(u)_\n\right)^{-(r-1)}\,  \left(\CG_{q,p}^{RR}\right)^{-(r-1)^2}\,  \left(\CG_{q,p}^{\rm grav}\right)^{-2}~,
\eea
with:
\be\label{def tgu}
\t g(u)\equiv -{e^{\pi i u}\ov 1- e^{2\pi i u}}~,
\ee
 a $(q,p)$-independent function, which therefore cancels out from  $\CG_{\CM_3}$.
In the limit $u \rightarrow -i \infty$, we have $\t g(u) \sim e^{-\pi i u}$ and \eqref{phi12} obviously reduces to \eqref{CS limit of GPhi}.
The relation \eqref{phi12} nicely matches our expectations---the two chiral multiplets are given in the ``$U(1)_{-\half}$ quantization'' and, upon integrating them out with a superpotential mass term,  we remain with the integer-quantized CS contact terms \eqref{k shifts sigma}, including the $U(1)_{-1}$ CS term.

\subsubsection{Contribution from the vector multiplet}\label{subsec: vector contrib to G}
Let us now consider the vector multiplet $\CV$ for the gauge group $\GG$ with Lie algebra $\Fg$. The $(q,p)$ fibering operator has contributions both from the 3d $\GH=U(1)^\rk$ vector multiplets $\CV_a$ in the maximal torus of $\GG$, and from the massive ``W-bosons''  on the Coulomb branch.

\paragraph{W-boson contribution.} Consider first the W-bosons, which contribute like chiral multiplets of gauge charges $\alpha^a$ and $R$-charge $2$. For each pair of roots $\alpha$, $-\alpha$, we choose the symmetric quantization as discussed around equation \eqref{Walpha contrib}.  This gives:
\be\label{GW def 0}
\t\CG^\CW_{q,p}(u)_\n \equiv  \prod_{\alpha\in \Delta^+}  {\CG^\Phi(\alpha(u)+ 2 \nu_R)_{\alpha(\n) + 2 \n^R}\ov \CG^\Phi(\alpha(u))_{\alpha(\n)}}\, \CG^{\rm GR}_{q,p}(\alpha(u))_{\alpha(\n)}~.
\ee
Here we have used the fact that, in the symmetric quantization, we have to turn on some bare CS levels $k= \alpha^2$---as in \eqref{Walpha contrib}---as well as $k_{RR}= 1$ and $k_g=2$. Then, using the relation \eqref{phi12} leads us to \eqref{GW def 0}.~\footnote{Up to some $l^R$-dependent factor which cancels from $\CG_{\CM_3}$.}  This expression can be massaged to:
\be\label{GW def new}
\t\CG^{\CW}_{q,p}(u)_\n \equiv \prod_{\alpha\in \Delta^+}  (-1)^{l^R} \t g(\alpha(u))^{l^R}   \; 
\CG^{\CW_0}_{q,p}\big(\alpha(u)\big)_{\alpha(\n)}~,
\ee
with the function $\t g(u)$ defined in \eqref{def tgu}. Here we defined:
\bea\label{def GW0}
& \CG^{\CW_0}_{q,p}(u)_\n& \equiv& \; (-1)^{\n(t+ l^R t+ 2\nu_R s)} {e^{-\pi i {u-t\n\ov q}}\ov e^{-\pi i u}} \, {1- e^{2\pi i  {u-t\n\ov q}}\ov 1- e^{2\pi i u}} \cr
&&=&\; (-1)^{\n(t+ l^R t+ 2\nu_R s)} {\sin\Big({\pi (u-t\n)\ov q}\Big)\ov \sin(\pi u)}~,
\eea
a function of a single set of parameters $u\in \C$, $\n\in \Z$.
The factors $(-\t g)^{l^R}$ in \eqref{GW def new} can be dropped, since they cancels out in $\CG_{\CM_3}$. We are then left with a rather simple expression for the W-boson contribution to the fibering operator:
\bea\label{GW full}
&\CG^\CW_{q,p}(u)_\n &=&\; \prod_{\alpha\in \Delta^+} \CG^{\CW_0}_{q,p}\big(\alpha(u)\big)_{\alpha(\n)} \cr
&&=&\; (-1)^{2 \rho_W(\n)  (t+l^R t+ 2\nu_R s)} \,{e^{-{2 \pi i \ov q} \rho_W(u-t \n)}\ov e^{-2 \pi i \rho_W(u)}} \;
\prod_{\alpha\in \Delta^+} 
{1- e^{{2\pi i\ov q} \alpha(u -t  \n)}\ov  1- e^{2 \pi i \alpha(u)}}~.
\eea
Note that, while $\t\CG^\CW_{q,p}$ is gauge-invariant by construction,  $\CG^\CW_{q,p}$ shifts by a sign under large gauge transformations, due to the relation:
\be
\CG^{\CW_0}_{q,p}(u+1)_{\n+p} = (-1)^{l^R} \; \CG^{\CW_0}_{q,p}(u)_{\n}~.
\ee
This follows either from the definition \eqref{def GW0}, or from the factorization \eqref{GW def new} together with the fact that $\t g(u+1)= - \t g(u)$. As a small consistency check on \eqref{GW full}, one can also check that:
\be\label{GW shift n by q}
\CG^\CW_{q,p}(u)_{\n+ \delta_a q} = e^{4\pi i \nu_R \rho_W^a}\; \CG^\CW_{q,p}(u)_\n~,
\ee
where $\n+\delta_a q$ corresponds to the shift $\n_a \rightarrow \n_a + q$. This shift is due to the twisted superpotential term \eqref{Wvec}, which contributes to the gauge-flux operator a sign:
\be
\pif_a^{\CW}=e^{4\pi i \nu_R \rho_W^a}~.
\ee
Therefore, \eqref{GW shift n by q} is in perfect agreement with \eqref{pif from G}.

\paragraph{$U(1)$ contribution.}
Consider next a $U(1)$ vector multiplet, and more generally the maximal torus $\GH$ of $\GG$. On general grounds, the one-loop fluctuations of $\CV_a$ could contribute to the fibering operator. We claim that each $U(1)$ vector multiplet does contribute a pure number:
\be\label{CG U1 contrib}
\CG^{\CV}_{q,p} ={1\ov \sqrt{q}}\, \CG^{(0)}_{q,p}
\ee
to the fibering operator, with $\CG^{(0)}_{q,p}$ defined by \eqref{def CG0}. We interpret the phase $ \CG^{(0)}_{q,p}$ as the result of quantizing the gaugino $\lambda$ such that it shifts the RR and gravitational contact terms by $\delta\kappa_{RR}= \half$ and  $\delta\kappa_{g}=1$, as we explained in Section~\ref{intro:subsec:parity}.
Indeed, we see from \eqref{Ggrav2 full} that:
\be
\left( \CG^{(0)}_{q,p}\right)^2=\CG^{\rm RR}_{q,p}\, \left(\CG^{\rm grav}\right)^2~,
\ee
therefore it is natural to assign the contact terms:
\be\label{contact terms gaugini}
\kappa_{RR}= \half~, \qquad \kappa_g= 1
\ee
 to \eqref{CG U1 contrib}. The factor of ${1/\sqrt{q}}$ in \eqref{CG U1 contrib} is more {\it ad-hoc}, but it can be argued for by requiring consistency with many well-established results. In particular, we can argue for \eqref{CG U1 contrib} by comparing our results to well-known results for pure CS theory \cite{Marino:2002fk}. This will be discussed elsewhere \cite{CHMW2018}.

Following our discussion in Section~\ref{intro:subsec:parity} (and in Appendix~\ref{app: parity anomaly}), it is also convenient to introduce a factor of $(\CG^{(0)}_{q,p})^2$ for each pair of W-bosons  $\CW_\alpha$, $\CW_{-\alpha}$---in that case, this is simply a different choice of quantization of the vector multiplet, since it corresponds to adding the integer-quantized RR and gravitational CS terms (with level $k_{RR}=1$ and $k_g=2$) to the UV action, for each $\alpha \in \Delta_+$. We then have an overall contribution:
\be
\left({1\ov \sqrt{q}}\right)^{\rk}\, \left( \CG^{(0)}_{q,p}\right)^{{\rm dim}(\GG)}
\ee
from the vector multiplet, in addition to the W-boson contribution \eqref{GW full}.

\subsection{Supersymmetric partition on $\CM_3$ as a sum over Bethe vacua}
Combining all the ingredients above, we can now write down the full formula for the geometry-changing operator, for any Seifert manifolds, in the general class of gauge theories considered here. 

\subsubsection{The full $(q,p)$-fibering operator: summing over fractional fluxes}
Consider a 3d $\CN=2$ supersymmetric gauge theory with gauge group $\GG$ and matter fields $\Phi$ in chiral multiplets, with generic background gauge fields turned on for the global symmetry group $\GG_F$. 
Let $\n_a$ and $\m_\alpha$ denote the gauge and flavor fluxes, respectively. 
The gauge and flavor chemical potentials are denoted by $u_a$ and $\nu_\alpha$, respectively.
At fixed gauge flux $\n_a= (\n_{a,0}, \n_{a,i})$ on $\CM_3$, the fibering operator reads:
\be\label{Gqp nm gen}
\CG_{q,p}(u, \nu)_{\n, \m} = \CG_{q,p}^{\rm CS}(u, \nu)_{\n, \m}\;  \CG^{\rm matter}_{q,p}(u, \nu)_{\n, \m} \; \CG_{q,p}^{\rm vector}(u)_{\n}~.
\ee
The first factor $\CG_{q,p}^{\rm CS}$ is the classical contribution from Chern-Simons terms, including CS contact terms for global symmetries, which can be constructed from the results in Section \ref{subsec: CSterms}. The second factor is the contribution from matter fields, which reads:
\be
\CG^{\rm matter}_{q,p}(u, \nu)_{\n, \m} =\prod_\omega \prod_{\rho \in \FR} \CG_{q,p}^\Phi(\rho(u)+ \omega(\nu) + r_\omega \nu_R)_{\rho(\n) + \omega(\m)+ \n^R r_\omega}
\ee
with the gauge and flavor weights $\rho^a$ and $\omega^\alpha$, respectively, $r_\omega$ the $R$-charges, and the function $\CG_{q,p}^\Phi$ defined in \eqref{CGqp0 Phi}.  
The third factor in \eqref{Gqp nm gen} is the contribution from the vector multiplet, which reads:
\be\label{CG vec full}
\CG_{q,p}^{\rm vector}(u)_{\n} =\left({1\ov \sqrt{q}}\right)^{\rk}\, \left( \CG^{(0)}_{q,p}\right)^{{\rm dim}(\GG)}\; \CG^\CW_{q,p}(u)_\n~,
\ee
with $ \CG^\CW_{q,p}$ given by \eqref{GW full}. Note that \eqref{CG vec full} is independent of the flavor parameters.

The expression \eqref{Gqp nm gen} depends both on the gauge theory chemical potentials $u_a$ and on the gauge fluxes $\n_a$. In the full theory, we need to ``integrate over'' the vector multiplet. In the present formalism, ``integration over $u$'' is realized by the sum over Bethe vacua in \eqref{ZM3 gen introduce LM3}. However, we also need to sum over the fractional fluxes, for each exceptional fiber $(q,p)$. For a gauge group $\GG$, the fractional fluxes are valued in a $\Z_q$ reduction of the magnetic flux lattice:
\be\label{cochar mod q def}
\Gamma_{\GG^\vee}(q)=\big\{\n \in {\mathfrak h}\big|~ \rho(\n)\in \mathbb{Z},\; \; \forall\rho \in\Lambda_{\rm char}~;\; \n \sim \n+ q \lambda,\;\; \forall \lambda\in \Lambda_{\rm cochar}\big\}~,
\ee
with $\Lambda_{\rm char}$ and $\Lambda_{\rm cochar}$ the character and co-character lattices of $\GG$, respectively---in other words, $\Gamma_{\GG^\vee}(q) \cong  \Lambda_{\rm cochar} \otimes \Z_q$.
The full fibering operator is obtained by summing over all the fractional fluxes:
\be\label{CG full sum def}
\CG_{q,p}(u, \nu)_{\m} =  \sum_{\n \in \Gamma_{\GG^\vee}(q)} \CG_{q,p}(u, \nu)_{\n, \m}~.
\ee
We will argue below that this is the correct sum over topological sectors, when we consider supersymmetric localization in Section~\ref{sec:localization}. Heuristically, the idea is that, from the two-dimensional point of view, we should sum over {\it all} orbifold $\GH$-bundles on the orbifold base $\h\Sigma$, in order to have a consistent sum over topological sectors.
The sum \eqref{CG full sum def} is well-defined when evalued on a Bethe vacuum. Indeed, if we shift any $\n_a$ by $q$ in the summand, we obtain:
\be
\CG_{q,p}(u, \nu)_{\n+\delta_a q, \m} = \pif_a(u, \nu)\,  \CG_{q,p}(u, \nu)_{\n, \m}~,
\ee
where the new prefactor trivializes on any Bethe vacuum,  $\pif_a(\h u, \nu)=1$. 

Moreover, while we saw before that the individual fibering operators $\CG_{q_i,p_i}$ are not well-defined in the presence of a general R-symmetry line bundle $L_R$ (when $l^R_i \neq 0$), these ambiguities factor out of the sum \eqref{CG full sum def}, and safely cancel in the full Seifert fibering operator:
\be\label{def GM3 finally}
\CG_{\CM_3}(u, \nu)_{\m} = \prod_{i=0}^n \CG_{q_i,p_i}(u, \nu)_{\m_i}~. 
\ee
Recall that, in this notation, $(q_0, p_0)=(1, {\bf d})$, and so $\m=(\m_i)$ includes both the ``ordinary fluxes'' $\m_0$ and the ``fractional fluxes'' $\m_i$, $i>0$.

\subsubsection{The $\CM_3$ partition function as a sum over Bethe vacua}\label{subsec: M3 part bethe vac form}
This completes the definition of the general fibering operator of a 3d $\CN=2$ gauge theory. We have found the geometry-changing line operator, given by the product of $g$ handle-gluing operators \eqref{hgoFull} with the Seifert fibering operator \eqref{def GM3 finally}, namely:
\be
\SL_{\CM_3}(u, \nu)_{\m}= \CH(u, \nu)^g \; \CG_{\CM_3}(u, \nu)_\m~.
\ee
This operator can be inserted along $S^1$ in the topologically-twisted $S^2 \times S^1$ geometry, as in \eqref{ZM3 gen introduce LM3}. We therefore found a closed exact formula for the supersymmetric partition function on any Seifert manifold:
\be\label{ZM3 Bethe formula}
Z_{\CM_3}(\nu)_\m =  \sum_{\h u \in \CS_{\rm BE}}   \CH(\h u, \nu)^{g-1}\, \CG_{\CM_3}(\h u, \nu)_{\m}~.
\ee
Note that this expression is properly gauge invariant, including under large gauge transformations for the flavor symmetry.~\footnote{Under a large gauge transformation for some flavor $U(1)_\alpha$, the parameters $(\nu_\alpha, \m_{i, \alpha})$ are shifted to $(\nu_\alpha+1, \m_{i, \alpha}+ p_i)$, which leaves the fibering operator invariant.} 
In Section \ref{sec:localization}, we will provide further evidence for \eqref{ZM3 Bethe formula}  from a localization argument in the UV.

\section{Infrared dualities on Seifert manifolds}
\label{sec:duality test}

The above results can be used to check infrared dualities between 3d $\CN=2$ gauge theories, by matching their supersymmetric partitions on any Seifert manifold. Given two dual theories $\CT$ and $\CT^D$, we must have:
\be
Z_{\CM_3}^{\CT}(\nu)_\m=Z_{\CM_3}^{\CT^D}(\nu)_\m~,
\ee
for any supersymmetric background $\CM_3$. More generally, any infrared duality implies the existence of a ``duality map'' mapping the Bethe vacua in the dual theories:
\be\label{CDuality map}
\CD \; : \; \CS_{\rm BE}[\CT] \rightarrow  \CS_{\rm BE}[\CT^D]\; : \; \h u \mapsto \h u^D~,
\ee
where the Bethe vacua are defined by \eqref{S BE 3d}.
Given the formula \eqref{ZM3 Bethe formula} for $Z_{\CM_3}$ as a sum over Bethe vacua, the equality of partition functions is equivalent to the relations:
\be\label{dual rel GM3}
\CH(\h u, \nu) = \CH^D(\h u^D, \nu)~, \qquad \qquad  \CG_{\CM_3}(\h u, \nu)_\m  = \CG^D_{\CM_3}(\h u^D, \nu)_\m  
\ee
between the handle-gluing and fibering operators of the dual theories, for every pair of dual Bethe vacua $\h u$ and $\h u^D$. These duality relations were previously proven for $\CH$~\cite{Closset:2016arn}  and  for the ``ordinary'' fibering operator $\CF= \CG_{1, 1}$~\cite{Closset:2017zgf}, for many three-dimensional Seiberg-like dualities \cite{Aharony:1997gp,deBoer:1997kr, Dorey:1999rb, Giveon:2008zn, Benini:2011mf}.~\footnote{In our previous work \protect\cite{Closset:2017zgf}, we only discussed the case $\nu_R \in \Z$ for the $U(1)_R$ background, corresponding to the ``periodic'' spin structure. Here we consider the general case $\nu_R = \half \Z$ as well.}

In the following, we verify the duality relations \eqref{dual rel GM3} in a few simple examples. This also serves to illustrate the formalism of the previous section, by spelling out the fibering operators of some well-known theories. In each case, we performed detailed numerical checks of the duality relations \eqref{dual rel GM3}. Even in the case of the simplest abelian dualities, these are very non-trivial checks, which work for any $(q,p)$-fibering operator and depend crucially on the fine structure of the theories and of the duality relations, including the relative background Chern-Simons contact terms between dual theories \cite{Closset:2012vp, Benini:2011mf}. We thus view the successful matching across dualities as very stringent consistency checks of our results, including more subtle features such as the spin-structure dependence of the supersymmetric background $(\CM_3, {\bf L}_R)$.

\subsection{Duality between the $U(1)_1$ CS theory and an almost trivial theory}
As our first example, consider a 3d $\CN=2$ supersymmetric $U(1)$ gauge theory at Chern-Simons level $k=1$, without matter. It is dual to an empty theory. More precisely, the dual theory consist only of a CS contact term for the topological symmetry $U(1)_T$, at level $k_{TT}=-1$, and of a pure gravitational CS term at level $k_g=-2$:
\be\label{U11 duality}
\CT\; : \; U(1)_1 \; \text{gauge theory}  \qquad \longleftrightarrow \qquad \CT^D\; : \; \;\;  k_{TT}= -1~, \;\; k_g= -2~.
\ee
As we discussed in Section \ref{subsec: vector contrib to G},  we quantize the gaugino in the vector multiplet in such a way that it contributes the contact terms:
\be
\kappa_{RR}= \half~, \qquad\qquad \kappa_g= 1~,
\ee
for the $U(1)_R$ and gravitational background fields, respectively.
Recall that most $\CN=2$ CS theories are equivalent to pure CS theories in the infrared---a CS term at level $k$ gives a real mass $m= - {k g^2\ov 2\pi}$ to the gaugino,~\footnote{With $g^2$ the YM coupling, or any other UV regulator.} which can be integrated out, and we are left with the gauge field only. In the present case, integrating out the single gaugino in the $U(1)$ vector multiplet shifts the CS contact terms by $-\half$ and $-1$, respectively, so that we have:
\be
\kappa_{RR}^{(IR)}=0~, \qquad\qquad \kappa_g^{(IR)}=0~.
\ee
in the deep infrared. We then have a $U(1)_1$ pure CS theory, which was recently studied in detail in \cite{Seiberg:2016gmd}. As argued there, the theory is ``almost trivial,'' being equivalent to a purely gravitational CS term at level $k_g=-2$. If we couple the topological symmetry of the $U(1)_1$ theory to a background $U(1)_T$ gauge field, we also have the $U(1)_T$ CS term at level $k_{TT}=-1$ in the dual description \cite{Hsin:2016blu}, as indicated in \eqref{U11 duality}. 

\paragraph{The Bethe vacuum.}
The $U(1)_1$ supersymmetric theory has the gauge flux operator:
\be
\pif(u, \zeta) = (-1)^{1+ 2\nu_R} e^{2\pi i u} e^{2\pi i \zeta}~.
\ee
Here $\zeta= \nu_T$ is the fugacity for $U(1)_T$, which is equivalent to a complexified FI parameter. 
We then have a single Bethe vacuum (up to large gauge transformations, $u \rightarrow u+1$):
\be\label{bethe vac U11}
\pif(\h u, \zeta)=1 \qquad \Rightarrow \qquad \h u = -\zeta + \half + \nu_R~.
\ee
A very simple observable in this theory is the $U(1)_T$ flavor flux operator, which reads:
\be\label{UTflux op U11}
\pif_T(u) = e^{2 \pi i u}~.
\ee
Plugging in the Bethe solution, we obtain:
\be\label{UTflux op U11 dual}
\pif_T(\h u) = (-1)^{1+2\nu_R} e^{-2\pi i \zeta}~,
\ee
with is precisely the $U(1)_T$ flux operator generated by a $U(1)_T$ CS level $k_{TT}=-1$, in agreement with the right-hand-side of \eqref{U11 duality}. It is interesting to note that the $U(1)_T$ flux operator \eqref{UTflux op U11} is equivalent to a Wilson line of charge $1$ wrapping a generic Seifert fiber. Even as we turn off the $U(1)_T$ fugacity, setting $\zeta=0$, we remain with a non-trivial sign in \eqref{UTflux op U11 dual}, which depends on the spin structure of $\CM_3$ restricted to the Seifert fiber.~\footnote{That is, assuming the Seifert fiber corresponds to a non-trivial element of $H^1(\CM_3, \Z_2)$. Otherwise there is a unique choice of $\nu_R$ mod 1 consistent with supersymmetry.} This is in agreement with the known properties of this line operator \cite{Seiberg:2016gmd}.

\paragraph{Matching the fibering operators.}  In the present theory, we have $\CH=1$ for the handle-gluing operator, so it trivially matches across the duality. To match the supersymmetric partition functions, we then only need to match the $(q,p)$-fibering operators. In the $U(1)_1$ theory, we have:
\be\label{Gqp U11}
\CG_{q,p}^{\CT}(u, \zeta)_{\n, \m_T}={1\ov \sqrt{q} } \CG^{(0)}_{q,p}\; \CG^{\rm GG}_{q,p}(u)_\n \; \CG^{\rm G_1 G_2}_{q,p}(u, \zeta)_{\n, \m_T}~.
\ee
Here $\n$ is the gauge flux and $\m_T$ is the $U(1)_T$ background flux. The fibering operator is obtained by summing over the gauge fluxes:
\be
\CG_{q,p}^{\CT}(u, \zeta)_{\m_T}=\sum_{\n=0}^{q-1}\,\CG_{q,p}^{\CT}(u, \zeta)_{\n, \m_T}~.
\ee
In the ``dual theory,'' we have:
\be
\CG_{q,p}^{\CT^D}(\zeta)_{\m_T} = \left( \CG^{\rm GG}_{q,p}(\zeta)_{\m_T}\right)^{-1}\; \left(\CG^{\rm grav}_{q,p}\right)^{-2}
\ee
Plugging in the Bethe vacuum \eqref{bethe vac U11}, we  find:
\be\label{Gqp match U11}
\CG_{q,p}^{\CT}(\h u, \zeta)_{\m_T} = e^{{\pi i \ov 2} l^R}\; \CG_{q,p}^{\CT^D}(\zeta)_{\m_T}~.
\ee
The prefactor on the right-hand-side cancels out when we consider the full fibering operator $\CG_{\CM_3}$. The relation \eqref{Gqp match U11} therefore implies \eqref{dual rel GM3}.

We should emphasize that the factor $\CG_{q,p}^{(0)}$ in \eqref{Gqp U11}, which originates from the gaugino, is crucial in obtaining a precise match across the duality, even in this overly simple example. This and other related consistency checks give us confidence that we have identified the correct fibering operators. The above analysis can be generalized to the $\CN=2$ version of level/rank duality, $U(N)_k \leftrightarrow U(|k|-N)_{-k}$, and one again finds a perfect agreement \cite{CHMW2018}. (Supersymmetric level/rank duality can be also obtained as a limit of Aharony duality, which we study below.)

\subsection{Elementary mirror symmetry}\label{subsec: ems}
Consider the elementary $\CN=2$ mirror symmetry \cite{Dorey:1999rb} between a $U(1)$ gauge theory with one chiral multiplet and a free chiral multiplet:
\be\label{EMS duality}
\CT\; : \; U(1)_\half + \Phi\;\; \text{(gauge theory)}  \qquad \longleftrightarrow \qquad \CT^D\; : \; \;\;  T^+ \;\; \text{(free chiral)}~.
\ee
The gauge theory consists of a $U(1)$ vector multiplet coupled to a single chiral multiplet of unit charge, with effective CS level $\kappa= \half$. The dual theory is a single free chiral, denoted by $T^+$, which is identified with the gauge-invariant monopole operator of the $U(1)$ gauge theory.
If the chiral multiplet of the original theory has an $R$-charge $r\in \Z$, the dual chiral multiplet $T^+$ has $R$-charge $R[T^+]=1-r$. Moreover, we also have the relative Chern-Simons contact terms:
\be\label{kTR kRR rel mirrsym 1}
k_{TR}= -r~, \qquad\qquad k_{RR}=r^2~, 
\ee
in the dual description. These relative CS levels can be derived, for instance, by integrating out $\Phi$ with a large positive real mass, thus flowing to the duality \eqref{U11 duality}.~\footnote{In this limit, the gauge theory $\CT$ flows exactly to the $U(1)_1$ theory in \protect\eqref{U11 duality}. In the dual description, integrating out the chiral multiplet $T$ shifts the CS levels \eqref{kTR kRR rel mirrsym 1} back to zero, while it also generates the levels $k_{TT}=-1$ and $k_g=-2$.  See {\it e.g.} \cite{Closset:2012vp, Closset:2016arn, Closset:2017zgf} for detailed discussions of such decoupling limits.}

\paragraph{Bethe equation, handle-gluing operator and fibering operators.}  Consider first the gauge theory $\CT$. We have the twisted superpotential:
\be\label{WT mirsym1}
\CW^{\CT}= {1\ov (2\pi i)^2} \dilog(e^{2\pi i (u+r \nu_R)})+ \half(u^2+ (1+2\nu_R)u)+ \zeta u+ \cdots~.
\ee
Note the presence of a $U(1)$ CS level $k=1$, corresponding to the choice of $U(1)_{-\half}$ quantization for the chiral multiplet.~\footnote{That is, we must add a ``bare'' CS term at level $1$ to go from our default ``$U(1)_{-\half}$ quantization'' to the ``$U(1)_\half$ quantization'' that appears in \protect\eqref{EMS duality}, in the theory $\CT$.}
The  Bethe equation reads:
\be\label{BE mirr sym 1}
\pif(u, \zeta) = {(-1)^{1+2\nu_R} e^{2\pi i u} e^{2\pi i \zeta}\ov 1-e^{2\pi i (u+ \nu_R r)}}=1~.
\ee
Let $\h u$ denote the unique solution . We also have the non-trivial effective dilaton:
\be\label{OmegaT mirrsym1}
\Omega^{\CT}=-{r-1\ov 2\pi i}\log(1-e^{2\pi i (u+ \nu_R r)})~,
\ee
and the handle-gluing operator:
\be\label{H mirr sym 1}
\CH^{\CT}(u, \zeta)= \left({1\ov 1-e^{2\pi i (u+ \nu_R r)}}\right)^r~,
\ee
which is easily derived from \eqref{WT mirsym1} and \eqref{OmegaT mirrsym1}.
Last but not least, the $(q,p)$-fibering operator reads:
\be
\CG_{q,p}^{\CT}(u, \zeta)_{\m_T}= {1\ov \sqrt{q}}\sum_{\n=0}^{q-1} \CG^{(0)}_{q,p}\, \CG^\Phi_{q,p}(u+\nu_R r)_{\n+ \n^R r}\, \CG^{\rm GG}_{q,p}(u)_{\n}\, \CG^{\rm G_1G_2}_{q,p}(u, \zeta)_{\n, \m_T}~.
\ee
The second and third factors in the summand corresponds to the chiral multiplet in the $U(1)_\half$ quantization, and the last factor is the mixed $U(1)$-$U(1)_T$ CS term, which includes the contribution from the FI coupling.

\paragraph{Matching across the duality.} It is straightforward to check the matching of the handle-gluing operators across the duality. Plugging the solution to \eqref{BE mirr sym 1} into \eqref{H mirr sym 1}, we obtain:
\be
\CH^{\CT}(\h u, \zeta)= (-1)^r e^{-2 \pi i r \zeta}\left({1\ov 1-e^{2\pi i (\zeta+\nu_R(1-r))}}\right)^{-r}= \CH^{\CT^D}(\zeta)~.
\ee
The dual handle-gluing operator is derived from the effective dilaton:
\be
\Omega^{\CT^D} = {r\ov 2\pi i}\log(1-e^{2\pi i( \zeta+\nu_R (1-r))})- r\zeta+ {r^2\ov 2}~.
\ee
Note the contribution from the contact terms \eqref{kTR kRR rel mirrsym 1}.
Similarly, the dual $(q,p)$-fibering operator takes the form:
\be
\CG_{q,p}^{\CT^D}(\zeta)_{\m_T}= \CG^\Phi_{q,p}(\zeta+\nu_R (1-r))_{\m_T+ \n^R (1-r)} \, \left(\CG^{\rm GR}_{q,p}(\zeta)_{\m_T}\right)^{-r} \, \left(\CG^{\rm RR}\right)^{r^2}~.
\ee
We verified numerically the duality relations:
\be
\CG_{q,p}^{\CT}(\h u, \zeta)_{\m_T}= f_D(\zeta, r)^{l^R}\;  \CG_{q,p}^{\CT^D}(\zeta)_{\m_T}~,
\ee
with $f_D(\zeta, r)$ a function independent of $(q,p)$, which therefore cancels out from the physical fibering operator $\CG_{\CM_3}$.~\footnote{By checking the duality for $(q,p)=(1,0)$, one can derive $f_D(\zeta, r)^2= 1-e^{2\pi i (\zeta+ \nu_R(1-r))}$, which determines $f_D$ up to a sign. We can also easily perform numerical checks of the matching of $\CG_{\CM_3}$ across the duality for various numbers of exceptional fibers. We find perfect agreement.}  This establishes the equality of the dual supersymmetric partition functions on any $\CM_3$.

\subsection{General abelian mirror symmetry and gauging flavor symmetries}

The elementary mirror symmetry described above is the basic building block in a more general class of dualities due to \cite{Dorey:1999rb}.  These can be obtained by starting with several decoupled copies of the above duality and gauging certain combinations of the flavor symmetries on each side of the duality \cite{Kapustin:1999ha}.  Then, since the original theories were equivalent, the theories we obtain after this gauging procedure must also be equivalent.

At the level of the partition function on a Seifert manifold, $\cM_3$, suppose we have shown that two theories, $A$ and $B$, have equal $\cM_3$ partition functions.  Equivalently, this implies that all the basic building blocks match between the two theories, \ie:

\be \cH^{A}(v) = \cH^{B}(v)~, \qquad \text{and} \qquad \cG_{q,p}^A(v)_\m = \cG_{q,p}^B(v)_\m, \;\;\; \forall \; q,p~. \ee
Here, we denote by $v$ the flavor symmetry parameters, and by $\m$ the corresponding (fractional) fluxes.  We have implicitly included any relative background CS terms necessary for the duality.  Then, we expect that any pair of theories we obtain by gauging some flavor symmetry, on both sides of the duality, will also have matching partition functions.

It is clear that this last statment will follow immediately if we can construct the building blocks for any daughter theory, ${\cal T}'$, obtained by gauging some flavor symmetry of a parent theory, ${\cal T}$, entirely in terms of the building blocks of the parent theory.  In the case of the handle-gluing and ordinary fibering operators (\ie, $\cG_{1,p}$), this was shown in \cite{Closset:2017zgf}.  In fact, these ingredients were shown to be obtained from two more basic objects, the ``on-shell twisted superpotential'' and the ``on-shell effective dilaton'' of the parent theory.  For the more general $(q,p)$ fibering operators, the latter statement no longer holds, however it is still straightforward to obtain the building blocks of the daughter theory from those of the parent.  Consider gauging a $U(1)_F$ symmetry for concreteness, as the general case is a straightforward extension, and let $v$ and $\m$ be the corresponding $U(1)_F$ flavor parameter and fractional flux.  Then we first obtain the Bethe vacua of the daughter theory as the set:
\be 
\cS_{BE}^{{\cal T}'} = \{ \hat{v} \; | \; \Pi^\alpha_v(\hat{v}) = 1, \;\;\ \text{some vacuum}\; \alpha \in \cS_{BE}^{\cal T} \}~,
 \ee
where $\Pi_v$ is the ordinary $U(1)_F$ flux operator in the parent theory, and $\pif_v^\alpha$ denotes that flux operator evaluated on the Bethe vacuum $\alpha$ in the parent theory.  Then, we simply have:
\be
 {\cG_{(q,p)}^{{\cal T}'}}^\beta = \frac{1}{\sqrt{q}} \sum_{\m=0}^{q-1}  \cG_{(q,p)}^{{\cal T}}(\hat{v}_\beta)_{\m}~,\qquad\quad \beta \in \cS_{BE}^{{\cal T}'}~,
 \ee
for the ``on-shell'' fibering operators of the daughter theory.

\paragraph{Example of $\CN=2$ abelian mirror symmetry.} To illustrate the above discussion, let us consider a simple example. According to \cite{Dorey:1999rb}, we have the duality:
\begin{itemize}
	\item[\bf A:] A single $U(1)$ gauge group $U(1)_{-N_f/2}$ coupled to $N_f$ chiral multiplets of charge $1$.
	\item[\bf  B:]  A circular quiver with gauge group $U(1)_{1/2}^{N_f}/U(1)$,  with bifundamental chiral multiplets connecting adjacent nodes.
\end{itemize}
``Theory A'' can clearly be obtained by starting with $N_f$ copies of theory $\CT^D$ in \eqref{EMS duality}, and gauging the diagonal $U(1)$ flavor symmetry.  Similarly, we claim that ``Theory~B'' can be obtained by starting with $N_f$ copies of the theory $\CT$ in \eqref{EMS duality},  and then gauging a diagonal $U(1)_J$ flavor symmetry.  Implementing this gauging at the level of the fibering operator, starting from $N_f$ copies of \eqref{Gqp U11}, we obtain:
\be \label{GMS fib} \CG_{q,p}^{\CT'}(u_i,\hat{\zeta}, \zeta_i,\zeta')_{\n_i, \hat{\m}_T,\m_{i,T}, \m_T'}=\CG^{\rm G_1 G_2}_{q,p}(\hat{\zeta}, \zeta')_{\hat{\m}_T, \m_T'} \prod_{i=1}^{N_f} \CG_{q,p}^{\CT}(u_i,\hat{\zeta}+ \zeta_i )_{\n_i,\hat{\m}_T+ \m_{i,T}}~, 
\ee
where we have written the FI terms for the $U(1)$ gauge groups as $\hat{\zeta}+\zeta_i$, with the constraint $\sum_i \zeta_i=0$, and similarly for the corresponding fluxes.   We have also introduced a new FI parameter, $\zeta'$, for new gauge symmetry corresponding to $\hat{\zeta}$.  It is useful to solve the Bethe equation corresponding to the dynamical FI parameter, $\hat{\zeta}$, which is simply:

\be \label{GMSBE} e^{2 \pi i (\zeta' + \sum_i u_i)} =  1\qquad\Rightarrow \qquad\sum_i u_i= - \zeta' + n~,\qquad  n\in \Z~.
\ee
We may absorb the dependence on $n$ into a redefinition of $\zeta'$.
If we isolate the $\hat{\zeta}$ and $\hat{\m}_T$ dependence in \eqref{GMS fib}, we find:
\be \label{GMSFIdep} 
e^{-\frac{2 \pi i}{q}  (\hat{\zeta}-\hat{\m}_T)(\zeta'+\sum_i u_i  - \m_T'-\sum_i \n_i ) } =e^{-\frac{2 \pi i}{q}  (\hat{\zeta}-\hat{\m}_T)(  - \m_T'-\sum_i \n_i ) }~.
 \ee
Recall we must also sum over the fractional fluxes, $\hat{\m}_T \in \Z_q$.  This imposes:
\be \label{GMSflux}  - \m_T'-\sum_i \n_i =0 \; (\text{mod} \; q)~.
\ee
Then the $\hat{\zeta}$ dependence in \eqref{GMSFIdep} also drops out.  We may solve \eqref{GMSBE} and \eqref{GMSflux} by writing (setting $n=0$):
\be 
u_i = - \frac{\zeta'}{N_f} + u_i'-u_{i+1}'~, \qquad  \n_i = - \frac{\m_T'}{N_f} + \n_i'-\n_{i+1}'~.  
\ee
Note the second relation implies a non-trivial quantization of the fluxes $\n_i'$ when $\m_T'$ is not a multiple of $N_f$, so that $\n_i \in \Z$. 
Plugging this in to \eqref{GMS fib}, we find:
\be \label{GMS fib2}
 \CG_{q,p}^{\CT'}(u_i, \zeta_i,\zeta')_{\n_i,\m_{i,T}, \m_T'}= \prod_{i=1}^{N_f} \CG_{q,p}^{\CT}(u_i'-u_j'-\frac{\zeta'}{N_f}, \zeta_i )_{\n_i'-\n_j'-\frac{\m_T'}{N_f},\m_{i,T}}~, 
 \ee
which is the expected fibering operator for Theory B.  Since we demonstrated in Section~\ref{subsec: ems} that the original two theories have equal fibering operators, this proves the equality of the fibering operators across this more general abelian mirror symmetry.

\subsection{Aharony duality}\label{subsec: aha duality}
Next, consider $\CN=2$ SQCD with gauge group $U(N_c)$  and $N_f$ flavors---$N_f$ chiral multiplets $Q_i$ in  the fundamental representation of $U(N_c)$, and $N_f$ chiral multiplets $\t Q^j$ in the anti-fundamental, of $R$-charges $r\in \Z$, with vanishing effective CS level for the gauge group. The gauge and flavor charges of the fields are summarized in Table~\ref{tab:SQCD charges}.
\begin{table}[t]
\centering
\be\nn
\begin{array}{c|c|c|ccccc}
    &  U(N_c)&U(N_f-N_c)& SU(N_f) & SU(N_f)  & U(1)_A &  U(1)_T & U(1)_R  \\
\hline
Q_i        & \bm{N_c}&\bm{1}& \bm{\overline{N_f}} & \bm{1}& 1   & 0   &r \\
\tilde{Q}^j   & \bm{\overline{N_c}}&\bm{1}  &  \bm{1}& \bm{N_f}  & 1   & 0   &r \\
\hline
q_j   &\bm{1}&\bm{N_f-N}&    \bm{1} & \bm{\overline{N_f}} & -1   & 0   &1-r \\
\t q^i     &\bm{1} & \bm{\overline{N_f-N_c}}& \bm{N_f}    & \bm{1}    & -1   & 0   &1-r \\
{M ^j}_i    &\bm{1}  & \bm{1} &\bm{\overline{N_f}}& \bm{N_f} &  2   & 0   &2 r \\
T_+   &\bm{1}    & \bm{1} & \bm{1} & \bm{1} & -N_f   & 1   & -N_f(r-1) -N_c +1\\
T_-   &\bm{1}  & \bm{1} & \bm{1} & \bm{1} & -N_f   & -1   & -N_f(r-1) -N_c +1
\end{array}
\ee
\caption{Gauge and flavor charges of the chiral multiplets $Q, \t Q$ in  3d $\CN=2$ SQCD with $U(N_c)$ gauge group, and of the dual  flavors $q, \t q$ and gauge singlet  chiral multiplets $M, T_+, T_-$ in the Aharony-dual $U(N_f-N_c)$ theory. 
}
\label{tab:SQCD charges}
\end{table}
When $N_f \geq N_c$, the theory has a infrared-dual description discovered by Aharony \cite{Aharony:1997gp}, similar to Seiberg duality:
\be\label{Aharony duality}
\CT\; : \; U(N_c) \; +\, Q, \t Q \qquad \longleftrightarrow \qquad \CT^D\; : \; \;\;  U(N_f-N_c) \; +\,  q, \t q \;+ \;  M, T_+, T_-~.
\ee
The dual theory is a $U(N_f-N_c)$ gauge theory with $N_f$ dual flavors $q_j, \t q^i$ together with  $N_f^2+2$ gauge singlets $M^j_i$ and $T_\pm$, which are dual to the gauge-invariants mesons $M= \t Q Q$ and to the monopole operators $T_\pm$ (of topological charge $\pm1$) in the $U(N_c)$ theory, respectively. The gauge-singlet fields couple to the dual gauge sector through the superpotential:
\be
W= M^j_i \t q^i q_j + T_+ t_- + T_- t_+~,
\ee
with $t_\pm$ the elementary monopole operators of the $U(N_f-N_c)$ theory. The charges of the dual matter fields are sumarized in Table~\ref{tab:SQCD charges} as well. 

To fully define the duality, we need to specify the relative Chern-Simons contact terms for the global symmetries \cite{Benini:2011mf,Closset:2012vp}. In our present conventions, the dual theory contains the   CS levels:
\be\label{AD CS ct1}
k_{SU(N_f)}= \t k_{SU(N_f)}= N_f-N_c
\ee
for the $SU(N_f) \times SU(N_f)$ flavor symmetry,  the integer-quantized CS levels:
\bea\label{AD CS ct2}
& k_{TT}=1~,\cr
& k_{AA}=  4 N_f^2 -2 N_c N_f~,\cr
& k_{AR}=2N_f^2 +(4N_f^2 - 2 N_c N_f)(r-1)~,\cr
& k_{RR}= N_c ^2 + N_f^2 + 4 N_f^2 (r-1) + (4N_f^2 - 2 N_c N_f)(r-1)^2~,
\eea
for the abelian global symmetries, and:
\be\label{AD CS ct3}
k_g = 2 N_f (N_f-N_c)+2
\ee
for the gravitational CS level. These levels can be derived in a variety of ways. One interesting derivation consist in integrating out the matter fields so that Aharony duality reduces to 3d $\CN=2$ supersymmetric level/rank duality $U(N)_K \leftrightarrow U(K-N)_{-K}$ in the IR, with $K=N_f$---after integrating out the gaugini, one can compare the IR relative CS levels to known results for level/rank duality in pure CS theory \cite{Hsin:2016blu}, which then implies the above contact terms in the Aharony-dual theory.

In previous literature, it was found that the relative CS levels vanish for the Aharony duality,~\footnote{For instance, this is what was found in \cite{Willett:2011gp, Benini:2011mf} by looking at the $S^3_b$ partition function.} contrary to what we are claiming here. The more precise claim is that, in a scheme in which all the fermions (including the gaugini) would be quantized in such a way that gives a trivial contribution to the  flavor contact terms, $\kappa_F=0$, then we would indeed have no relative CS levels. However, that (implicit) quantization scheme is not consistent with background gauge invariance for $U(1)_R$~\cite{Closset:2017zgf}, as we already explained.
Here instead, we are following the conventions in which each chiral multiplet is taken in the ``$U(1)_{-\half}$ quantization,'' while the gaugini are quantized with the contact terms \eqref{contact terms gaugini}. 
For instance, in the $U(N_c)$ gauge theory $\CT$, the fermions in $Q, \t Q$ contribute:
\be
\delta \kappa_{U(N_c)}= -N_f~, \qquad \delta \kappa_{AA} = -N_f N_c
\ee 
to the gauge and $U(1)_A$ UV contact terms. Since those shifts are by integers, we can always cancel them by adding the bare CS levels $k_{U(N_c)}=N_f$ and $k^\CT_{AA}= N_f N_c$. For the gauge group, this is part of the definition of the theory, and we need to introduce the bare level so that we have the effective CS level $k_{U(N_c)}=0$. For the flavor symmetry $U(1)_A$, however, this is a non-physical contact term, which we can choose at will---only the relative level $k_{AA}\equiv k^{\CT^D}_{AA}-k_{AA}^{\CT}$ across the duality is physically meaningful. Here we chose $k_{AA}^{\CT}=0$. For the $U(1)_R$ symmetry, on the other hand, the gauge-invariant quantization (in the conventions of this paper) gives a contribution:
\be
\delta \kappa_{RR} = -(r-1)^2 N_f N_C + \half N_c^2
\ee
to the $U(1)_R$ CS contact term in the UV, coming from the chiral multiplets and the gaugino, respectively. We see that, for $N_c$ odd, $\delta \kappa_{RR}$ is half-integer and cannot be fully cancelled by a gauge-invariant counter-term. Nonetheless, consider for a moment setting $k^{\CT}_F=-\delta \kappa_F^{\CT}$ formally, for the full flavor symmetry $\GG_F$, and  similarly in the dual theory $\CT^D$. Then, the assumption that there are no additional relative contact terms between the dual theories, in that particular ``scheme,''  allows us to derive: 
\be\label{kF derivation 2}
k_{F}\equiv k_F^{\CT^D}- k_F^{\CT} = - \delta \kappa_F^{\CT^D}+\delta \kappa_F^{\CT}~,
\ee
which must be integer quantized. We choose $ k_F^{\CT}=0$ throughout, so that the flavor CS levels in the dual description are given by the right-hand-side of \eqref{kF derivation 2}. This reproduces \eqref{AD CS ct1}-\eqref{AD CS ct3}.

\paragraph{Bethe equations and dual Bethe vacua.}
Consider turning on generic chemical potentials for the flavor group $\GG_F= SU(N_f)\times SU(N_f)\times U(1)_A\times U(1)_T$, denoted by:
\be\label{nus SQCD}
\nu_i~,\quad \t \nu_j \quad \Big(\text{with} \; \sum_{i=1}^{N_f} \nu_i=  \sum_{j=1}^{N_f} \t\nu_j=0\Big)~, \qquad
\nu_A~, \qquad \zeta~,
\ee
respectively, and similarly for the background fluxes $\m_i, \t\m_j, \m_A, \m_T \in \Z$.  The gauge-flux operators of the $U(N_c)$ theory are given by:
\be
\pif_a(u, \nu)= \pif_0(u_a,  \nu)~, \qquad a=1, \cdots, N_c~,
\ee
with $\nu$ denoting collectively all the flavor parameters \eqref{nus SQCD}, and:
\be
 \pif_0(u,  \nu)\equiv (-1)^{2\nu_R(N_f+N_c-1)}\left(-e^{2\pi i u}\right)^{N_f} e^{2\pi i \zeta}\,{\prod_{j=1}^{N_f} \left(1- e^{2\pi i (-u+ \t \nu_j+ \nu_A + \nu_R r)}\right)\ov \prod_{i=1}^{N_f} \left(1- e^{2\pi i (u-  \nu_i+ \nu_A + \nu_R r)}\right)}~,
\ee
a function of a single variable $u$. Let us call ``Bethe roots'' the $N_f$ solutions $\{\h u_\alpha\}_{\alpha=1}^{N_f}$ to $\pif_0(\h u, \nu)=1$. They correspond to the $N_f$ roots $\h x_\alpha$ of the polynomial:
\be\label{Bethe Pol Aharony duality}
 (-1)^{N_f + 2\nu_R(N_f+N_c-1)} e^{2\pi i \zeta} \,\prod_{j=1}^{N_f} \left(x- e^{2\pi i (\t \nu_j+ \nu_A + \nu_R r)}\right)- \prod_{i=1}^{N_f} \left(1- x e^{2\pi i (-\nu_i+ \nu_A + \nu_R r)}\right)~,
\ee
of degree $N_f$ in $x$, with $\h x= e^{2\pi i \h u}$. A  Bethe vacuum:
\be
\h u \equiv \{\h u_a\}_{a=1}^{N_c} \subset \{\h u_\alpha\}_{\alpha=1}^{N_f}~,
\ee
consists of a choice of $N_c$ distinct Bethe roots. In particular, there  are:
\be\label{witten index SQCD}
|\CS_{\rm BE}|= \mat{N_f\cr N_c} 
\ee
distinct Bethe vacua. This number is the flavored Witten of $U(N_c)$ SQCD \cite{Intriligator:2013lca, Closset:2016arn}, and it is obviously invariant under $N_c \rightarrow N_f-N_c$.

Consider now the dual $U(N_f-N_c)$ gauge  theory. Using the identification $\zeta^D= - \zeta$ for the dual FI term, and the charge assignments in Table~\ref{tab:SQCD charges}, it is easy to show that the dual Bethe equations are isomorphic to the ones in the original theory. More precisely, we find:
\be
\pif_{\ba}^D(u^D, \nu)=1\qquad \Leftrightarrow \qquad \pif_0(u_\ba-\nu_R, \nu)=1~.
\ee
for the dual gauge-flux operators. Here the index $\ba=1, \cdots, N_f-N_c$ runs over the Cartan of the dual gauge group. It follows that the duality map \eqref{CDuality map}  is given by:
\be
\CD\; : \; \h u= \{\h u_a\} \mapsto  \h u^D = \{\h u_\ba\}= \{\t u_\ba+\nu_R\}~, \quad \text{with}\quad \{\t u_\ba\}=\{\h u_a\}^c \subset \{\h u_\alpha\}_{\alpha=1}^{N_f}~.
\ee
Namely, the Bethe vacuum $\h u^D$ dual to $\h u$ is obtained by taking the complement $\t u= \h u^c$ of $\h u$ in the set of Bethe roots, and shifting each $\t u_\ba$ by $\nu_R$.

\paragraph{Matching the handle-gluing operators.}  The handle-gluing operator of the $U(N_c)$ gauge theory is given by:
\bea
&\CH^{\CT}(u, \nu)= \prod_{\substack{a, b=1\\ a\neq b}}^{N_c} \left(1- e^{2 \pi i (u_a-u_b)}\right)^{-1}\;  \prod_{a=1}^{N_c} \Bigg[H(u_a, \nu)\;\cr
& \qquad \times \prod_{i=1}^{N_f} \left(1-e^{2\pi i(u_a- \nu_i + \nu_A+ \nu_R r)}\right)^{1-r} \prod_{j=1}^{N_f} \left(1-e^{2\pi i(-u_a+\t\nu_j + \nu_A+ \nu_R r)}\right)^{1-r} \Bigg]~,
\eea
with the function:
\be\label{def Hu in CH for AD}
H(u, \nu) \equiv  N_f + \sum_{i=1}^{N_f} {e^{2 \pi i(u- \nu_i + \nu_A + \nu_R r)} \ov 1- e^{2 \pi i(u- \nu_i + \nu_A + \nu_R r)}}+ \sum_{j=1}^{N_f} {e^{2 \pi i(-u+ \t\nu_j + \nu_A + \nu_R r)} \ov 1- e^{2 \pi i(-u+\t\nu_j + \nu_A + \nu_R r)}}~,
\ee
of a single variable $u$, corresponding to the Hessian $\det_{ab} \d_a \d_b \CW= \prod_a H(u_a, \nu)$.
Similarly, the handle-gluing operator of the dual gauge theory takes the form:
\be\label{CH for AD theory}
\CH^{\CT^D}(u^D, \nu)= \CH^{\rm gauge}(u^D, \nu)\,  \CH^{\rm singlets}(u^D, \nu)\,  \CH^{\rm ct}(u^D, \nu)~,
\ee
a product of three contributions.
The gauge theory contribution reads:
\bea
&\CH^{\rm gauge}=\prod_{\substack{\ba, \b b=1\\ \ba\neq \b b}}^{N_f-N_c} \left(1- e^{2 \pi i (u_\ba-u_{\b b})}\right)^{-1}\;  \prod_{\ba=1}^{N_f-N_c} \Bigg[H^D(u_\ba, \nu)\;\cr
& \qquad \times \prod_{i=1}^{N_f} \left(1-e^{2\pi i(-u_\ba+ \nu_i - \nu_A+ \nu_R(1- r))}\right)^{r} \prod_{j=1}^{N_f} \left(1-e^{2\pi i(u_\ba-\t\nu_j - \nu_A+ \nu_R (1-r))}\right)^{r} \Bigg]~,
\eea
with the function:
\be
H^D(u, \nu) \equiv  N_f + \sum_{i=1}^{N_f} {e^{2 \pi i(-u+\nu_i - \nu_A + \nu_R (1-r))} \ov 1- e^{2 \pi i(-u+ \nu_i - \nu_A + \nu_R(1- r))}}+ \sum_{j=1}^{N_f} {e^{2 \pi i(u-\t\nu_j - \nu_A + \nu_R(1- r))} \ov 1- e^{2 \pi i(u-\t\nu_j - \nu_A + \nu_R (1-r))}}~.
\ee
Note that $H^D(u, \nu)= -H(u-\nu_R, \nu)$, with $H(u, \nu)$ defined in \eqref{def Hu in CH for AD}. The contribution from the gauge-singlet fields $M$ and $T_\pm$ reads:
\be
\CH^{\rm singlets}= \prod_{i, j=1}^{N_f} \left(1- e^{2\pi i (\t\nu_j- \nu_i + 2 \nu_A + 2 r \nu_R)}\right)^{1- 2 r}\, \prod_\pm \left(1- e^{2\pi i (\pm \zeta- N_f \nu_A + \nu_R r_T)}\right)^{1-r_T}~,
\ee
with $r_T\equiv - N_f(r-1) -N_c+1$ the R-charge of $T_\pm$.
We also have a contribution from the Chern-Simons contact terms:
\be
\CH^{\rm ct}= (-1)^{k_{RR}} \; e^{2\pi i k_{AR} \nu_A}~,
\ee
with the levels $k_{RR}$ and $k_{AR}$ given in \eqref{AD CS ct2}. Note that the handle-gluing operator $\CH^{\CT}$ and $\CH^{\CT^D}$ are rational functions of $x_a= e^{2\pi i u_a}$ and $x_\ba= e^{2 \pi i u_\ba}$, respectively. By using elementary identities involving the  roots $\h x_\alpha$ of the  polynomial \eqref{Bethe Pol Aharony duality}, one can prove the matching of the handle-gluing operators across the duality \cite{Closset:2016arn}. That is, given any Bethe vacuum $\h u$ in $\CT$ and its dual Bethe vacuum $\h u^D$ in $\CT^D$, we find:
\be\label{match H AD}
\CH^{\CT}(\h u, \nu) = \CH^{\CT^D}(\h u^D, \nu)~.
\ee
This was first proven in \cite{Closset:2016arn} in the special case $\nu_R=0$, and the generalization to $\nu_R \in \half \Z$ is straightforward.~\footnote{In \protect\cite{Closset:2016arn}, the relation \protect\eqref{match H AD} was proven up to a complicated sign. Following \protect\cite{Closset:2017zgf}, we see that, after properly treating the parity anomaly, the match is exact.}  The identity \eqref{match H AD}  proves the matching of the twisted indices \eqref{twisted index expl} across Aharony duality, for both the periodic and anti-periodic spin structures.

\paragraph{Matching the fibering operators.}  We can similarly compare the $(q,p)$-fibering operators of the dual theories. Let us set $r=0$ to avoid clutter. The $R$-charge dependence can easily be restored by shifting $\nu_A \rightarrow \nu_A + \nu_R r$ and $\m_A \rightarrow \m_A + \n^R r$. For a $U(N_c)$ gauge theory with $N_f$ flavors, we have:
\be\label{UNNf gauge theory Gqp}
\CG^{N_c, N_f}_{q,p}(u, \nu_i, \t\nu_j, \nu_A, \zeta)_{\m_i, \t\m_j, \m_A, \m_T} = \sum_{\n_1=0}^{q-1}  \cdots \sum_{\n_{N_c}=0}^{q-1} \CG^{N_c, N_f}_{q,p}(u, \nu)_{\n, \m}~,
\ee
with:
\bea\nn
&\CG^{N_c, N_f}_{q,p}(u, \nu)_{\n, \m}\equiv  q^{-{N_c\ov 2}}  \left(\CG^{(0)}_{q,p}\right)^{N_c^2} \, \prod_{\substack{a, b=1\\ a>b}}^{N_c} \CG^{\CW_0}(u_a- u_b)_{\n_a- \n_b}\; \prod_{a=1}^{N_c}\Bigg[ \CG^{\rm G_1G_2}_{q,p}(u_a, \zeta)_{\n_a, \m_T} \cr
&\times \left(\CG^{\rm GG}_{q,p}(u_a)_{\n_a}\right)^{N_f}\; \prod_{i=1}^{N_f} \CG^\Phi_{q,p}(u_a -\nu_i+\nu_A)_{\n-\m_i + \m_A}\; \prod_{j=1}^{N_f} \CG^\Phi_{q,p}(-u_a +\t\nu_j+\nu_A)_{-\n_a+\t\m_j + \n_A}\Bigg]~,
\eea
and with the various building blocks introduced in Section~\ref{sec: Bethe vac sum},  including the W-boson contribution \eqref{def GW0}.
Note the presence of the bare $U(N_c)$ CS term at level $N_f$ on the second line, as discussed above. In our conventions, we then have:
\be
\CG^{\CT}_{q, p}(u, \nu)_\m = \CG^{N_c, N_f}_{q,p}(u, \nu)_{\m}
\ee
for 3d $\CN=2$ SQCD. For the dual theory, on the other hand, we have: 
\be
\CG^{\CT^D}_{q, p}(u^D, \nu)_\m= \CG^{\rm gauge}_{q, p}(u^D, \nu)_\m\, \CG^{\rm singlets}_{q, p}(u^D, \nu)_\m\, \CG^{\rm ct}_{q, p}(u^D, \nu)_\m~,
\ee
a product of three contributions, similarly to \eqref{CH for AD theory}. The gauge-theory contribution is given by:
\be
 \CG^{\rm gauge}_{q, p}(u^D, \nu)_\m = \CG^{N_f-N_c, N_f}_{q,p}(u^D, \t\nu_j, \nu_i, -\nu_A+ \nu_R, -\zeta)_{\t\m_j, \m_i, -\m_A+ \n^R, -\m_T}~,
\ee
in terms of the function \eqref{UNNf gauge theory Gqp}. The gauge-singlet contribution reads:
\bea
&\CG^{\rm singlets}_{q, p}(u^D, \nu)_\m = \prod_{i, j=1}^{N_f}  \CG^\Phi_{q,p}(\t\nu_j-\nu_i+2\nu_A)_{\t\m_j-\m_i + 2\m_A}\;\cr
&\qquad\qquad\qquad\qquad \times  \prod_\pm  \CG^\Phi_{q,p}(\pm \zeta -N_f\nu_A+\nu_R r_T)_{\pm \m_T -N_f \m_A+ \n^R r_T}~,
\eea
with $r_T$ the monopole R-charge (with $r=0$, here). The CS contact term contribution reads:
\bea
&\CG^{\rm ct}_{q, p}(u^D, \nu)_\m=
\left(\CG_{q,p}^{SU(N_f) \, {\rm CS}}(\nu)_\m\right)^{k_{SU(N_f)}}\,  
\left(\CG_{q,p}^{SU(N_f) \, {\rm CS}}(\t\nu)_{\t\m}\right)^{\t k_{SU(N_f)}}\,
\cr
& \times   \left(\CG^{\rm GG}_{q,p}(\zeta)_{\m_T}\right)^{k_{TT}}\,
  \left(\CG^{\rm GG}_{q,p}(\nu_A)_{\m_A}\right)^{k_{AA}} \, \left(\CG^{\rm GR}_{q,p}(\nu_A)_{\m_A}\right)^{k_{AR}} \, 
\left(\CG^{\rm RR}_{q,p}\right)^{k_{RR}} \, \left(\CG^{\rm grav}_{q,p}\right)^{k_{g}}~,
\eea
with the $SU(N_f)$ CS contribution:
\be
\CG_{q,p}^{SU(N_f) \, {\rm CS}}(\nu)_\m\equiv \prod_{i=1}^{N_f} \CG^{\rm GG}_{q,p}(\nu_i)_{\m_i}~, \qquad \text{such that}\quad \sum_{i=1}^{N_f}\nu_i =  \sum_{i=1}^{N_f}\m_i = 0~,  
\ee
and the flavor CS levels  \eqref{AD CS ct1}-\eqref{AD CS ct3}.  The infrared duality between the two theories implies the equality:
\be\label{CG duality relation AD}
\CG^{\CT}_{q, p}(\h u, \nu)_\m=f_D(\nu)^{l^R}\; \CG^{\CT^D}_{q, p}(\h u^D, \nu)_\m
\ee
for any pair of dual Bethe vacua $\h u$ and $\h u^D$, with $f_D(\nu)$ some function independent of $(q,p)$, which cancels from the physical fibering operator $\CG_{\CM_3}$. We checked \eqref{CG duality relation AD} numerically, for a large number of examples.  Together with \eqref{match H AD}, this identity implies the equality of the supersymmetric partition function across Aharony duality on any Seifert manifold. Note that \eqref{CG duality relation AD} was previously  proven in the special case $q=1$ \cite{Closset:2017zgf}. For $q>1$, given the sum over fractional fluxes in \eqref{UNNf gauge theory Gqp}, the duality relation \eqref{CG duality relation AD} is rather more complicated. It would be  worthwhile to find an analytic proof.  

Finally, let us note that, through various decoupling limits, we can go from 3d $\CN=2$ SQCD to more general $U(N_c)_k$ theories with $N_f$ fundamental chiral multiplets, $N_a\leq N_f$ anti-fundamental chiral multiplets, and an effective CS level $k \in \half\Z$ such that $k+\half(N_f-N_a)\in \Z$. These theories also enjoy interesting dualities~\cite{Giveon:2008zn, Benini:2011mf}. (The elementary mirror symmetry \eqref{EMS duality} is a special case.) The corresponding duality relations for the geometry-changing operators directly follow from \eqref{match H AD} and \eqref{CG duality relation AD} by taking appropriate limits on the chemical potentials \cite{Closset:2017zgf}. 

It would be interesting to extend this discussion to various other three-dimensional IR dualities, such as dualities involving adjoint fields \cite{Niarchos:2008jb, Kim:2013cma} or monopole-operator superpotentials \cite{Benvenuti:2016wet,Benini:2017dud, Giacomelli:2017vgk}.

\subsection{The ``duality appetizer''}

As our final example, we consider the duality of \cite{Jafferis:2011ns}, relating the following theories:

\begin{itemize}
	\item[\bf A:] An $SU(2)_1$ CS theory coupled to an adjoint chiral multiplet, $\Phi$.
	\item[\bf B:] A free chiral multiplet, $Z$,  tensored with a decoupled topological sector $U(1)_2$.
\end{itemize}
The chiral operators in the two theories are identified by: 
\be
 \text{Tr} \; \Phi^2 \;\;\; \leftrightarrow \;\;\; Z~.
 \ee
In particular, theory A has a $U(1)_F$ global symmetry acting on $\Phi$ with charge $1$, which maps to a symmetry acting on $Z$ with charge $2$.  In addition, theory $A$ has a $\Z_2$ $1$-form electric symmetry \cite{Gaiotto:2014kfa}, which maps to that of the decoupled $U(1)_2$ CS theory.

\paragraph{Bethe equations} 

The twisted superpotential and the Bethe equation for the $SU(2)$ theory were described in \cite{Closset:2017zgf}. The Bethe equation reads:~\footnote{In \protect\cite{Closset:2017zgf}, we worked with the A-twist background, namely $\nu_R=0$. More generally, the dependence on $\nu_R$ here is trivial, and we have absorbed it into the definition of $\mu$ below.}
\be\label{DABE}
 \Pi_0 = x^6 \bigg( \frac{1-\mu x^{-2}}{1- \mu x^2} \bigg)^2 = 1~,
 \ee
where we have defined:
\be x = e^{2 \pi i u}~, \qquad  \mu = e^{2 \pi i ( \nu + r \nu_R)}~,\ee
with $\nu$ the $U(1)_F$ parameter.
Here we have used the $U(1)_{-1/2}$ quantization for the chiral multiplets, so that the adjoint chiral  multiplet itself contributes a gauge CS contact term $\kappa=-2$ in the UV, therefore we have included a bare CS term at level $k=3$,  to recover the appropriate effective level $k=1$ of the $SU(2)_1$ theory.  Then, factoring out trivial solutions, \eqref{DABE} can be rearranged to:
\be \label{DAsol} 
 (x+x^{-1})^2 = (1+\mu)^2~.
 \ee
Up to Weyl equivalence, taking $x \rightarrow x^{-1}$, there are two solutions, $x_\pm = e^{2 \pi i\hat{u}_\pm}$, given by the two choices of sign in taking the square root of \eqref{DAsol}.

On the dual side, we must consider the Bethe equation for the $U(1)_2$ supersymmetric CS theory, which is:~\footnote{In principle one can include a background gauge field coupled to the $U(1)_J$ topological symmetry, as was suggested in \cite{Jafferis:2011ns}.  However, this explicitly breaks the $\Z_2$ $1$-form symmetry, and so we do not include this here.}
\be
 \tilde{x}^2 =1 \qquad \Rightarrow \qquad \tilde{x} = \pm 1~.
 \ee
and so there are two vacua, matching the counting in Theory A.  Note that, since a free chiral by itself has only one vacuum, this extra decoupled CS sector in  Theory B is crucial for the duality to make sense.

\paragraph{Handle-gluing and flux operators.}

Let us first consider the flux operator for the $U(1)_F$ flavor symmetry.  The adjoint chiral multiplet contributes:

\be
 \Pi_F= \frac{1}{(1-\mu)(1-\mu x^2)(1-\mu x^{-2})}~.
\ee
Plugging in the solutions \eqref{DAsol}, we find:
\be \Pi_F \bigg|_{\hat{u}_\pm} = \frac{1}{(1- \mu^2)^2}~.
 \ee
which precisely agrees with the contribution of the charge $2$ chiral $Z$ in the dual theory.

Next consider the handle-gluing operator.  Without loss of generality, we may assign the chiral multiplet $\Phi$ an R-charge of $r=1$, and then $Z$ has R-charge $2$.  Then in the $SU(2)$ theory, the handle-gluing operator is given by:
\be 
\CH_{A}(u, \nu) = {1\ov (1- x^2)(1-x^{-2})}\; H(u, \nu)~,
\ee
where the first factor is the W-boson conrtribution, and $H$ is the Hessian of the twisted superpotential:
\be 
H(u, \nu) = \frac{1}{2 \pi i} \frac{\partial}{\partial u} \log \Pi_0 = 6 + 4 \frac{\mu x^2}{1-\mu x^2} +4  \frac{\mu x^{-2}}{1- \mu x^{-2}}~.
\ee
Plugging in the solutions \eqref{DAsol}, one finds that, in both vacua, we have:
\be \label{DAHa} 
\CH_{A}(\h u_\pm, \nu) = \frac{2}{1-\mu^2}~.
\ee
On the dual side, the $U(1)_2$ CS theory contributes a factor $\CH_{U(1)_2}= 2$ to the handle-gluing operator, while the free chiral $Z$ contributes a factor $(1-\mu^2)^{-1}$, therefore we recover \eqref{DAHa}.

\paragraph{Fibering operators.}
Let us now consider the matching of the fibering operators, $\cG_{q,p}$.  For the $SU(2)$ theory, we have:
\bea
& \cG_{q,p}^A(u,\nu)_{\n,\m} &=& \frac{1}{\sqrt{q}} \left(\cG^{(0)}_{q,p}\right)^3\, \cG_{q,p}^{\cW_0}(2u)_{2\n} \,
\cG_{q,p}^{\cW_0}(-2u)_{-2\n} \,\left(\cG_{q,p}^{GG}(u)_\n\right)^6  \cr
&\qquad &&\times  \cG_{q,p}^\Phi(2u+\nu + r \nu_R)_{2\n+\m+r \n_R}\cG_{q,p}^\Phi(-2u+\nu+r n_R)_{-2\n+\m+r \n_R} \cr
&\qquad &&\times \cG_{q,p}^\Phi(\nu+r n_R)_{\m+r \n_R}~,
 \eea
 with $\n$ and $\m$ the gauge and flavor fractional fluxes, as usual. Then the full fibering operator, evaluated at the Bethe vacua \eqref{DAsol}, reads:
\be
 \cG_{q,p}^A(\nu)_{\m}^{(\pm)} = \sum_{\n=0}^{q-1}
\cG_{q,p}^A(\hat{u}_\pm,\nu)_{\n,\m}~.
\ee
For the dual theory, we find:
\be
 \cG_{q,p}^B(\tilde{u},\nu)_{\tilde{\n},\m} = \frac{1}{\sqrt{q}} \, \cG^{(0)}_{q,p}\, \left(\cG_{q,p}^{GG}(\tilde{u})_{\t\n}\right)^2\, \cG_{q,p}^\Phi(2\nu + 2r \nu_R)_{2\m+2r \n_R}~,
 \ee
The two Bethe vacua (from the $U(1)_2$ sector) correspond to $\hat{\tilde{u}}_+=0$ and $\hat{\tilde{u}}_-=\frac{1}{2}$. Thus,  we find the full ``on-shell'' fibering operators:
\be 
\cG_{q,p}^B(\nu)_{\m}^{(\pm)}  =  \sum_{\tilde{\n}=0}^{q-1}
\cG_{q,p}^B(\hat{\tilde{u}}_\pm,\nu)_{\n,\m}~.
\ee
We note that the contribution of the decoupled CS theory is quite non-trivial; in particular, unlike the flux and handle-gluing operators, the two vacua give different contributions, which must be separately matched.  We have checked numerically in several examples that the following relation holds:~\footnote{More precisely, we have shown that the two values on the LHS agree with those on the RHS, but the precise identification between the two vacua can vary, {\it e.g.}, as we cross branch cuts of the Bethe solutions, \eqref{DAsol}, in the complex $\nu$ plane. 
}
\be
 \cG_{q,p}^A(\nu)_{\m}^{(\pm)}  = 
\cG_{q,p}^B(\nu)_{\m}^{(\pm)} ~,
 \ee
giving a new strong test of this peculiar duality.  In particular, this is a strong test that the conjectured decoupled topological sector described above indeed appears. It also gives a precise handle on the relative CS contact terms between the two dual theories.  Interestingly, with the choice of quantization for the fermions used in this paper,  there are no additional relative global Chern-Simons term necessary for the duality to hold.

%


\section{Localization on Seifert manifolds}
\label{sec:localization}

The exact result for the Seifert manifold partition function can also be arrived at by a direct three-dimensional path-integral computation, as in \cite{Benini:2015noa, Benini:2016hjo, Closset:2016arn}. In this section, we derive the result \eqref{ZM3 Bethe formula} via supersymmetric localization onto the classical Coulomb branch, starting from the three-dimensional UV action. The relevant curved-space supersymmetric actions and supersymmetry multiplets \cite{Closset:2012ru} can be found in Section~3 of \cite{Closset:2017zgf}. The supersymmetric localization argument is a straightforward generalization of Section~4 of \cite{Closset:2017zgf}, which we we will briefly summarize  for completeness. The gauge group $\GG$ is chosen as in \eqref{GG intro}.

\subsection{Supersymmetric localization and integral formula}
Consider $\CM_3$ with the supergravity background as in Section~\ref{subsec: M3 backgd}. In particular, we have first Chern class:
\be
c_1(\CL_0) = {\bf d} + \sum_{i=1}^n {p_i\ov q_i}~,
\ee
for the defining orbifold line bundle. 

\paragraph{BPS configurations.} Let us first consider the BPS equation for the vector multiplet. The gaugino variations with respect to the two supersymmetries \eqref{KS explicit} vanish if and only if:~\footnote{Here we impose a reality condition on $\sigma$ and $a_\mu$, not on the auxiliary field $D$.}
\be
D_\mu\sigma = 0~, \qquad f_{01} = f_{0\bar 1} = 0~, \qquad D=2if_{1\bar 1} +\sigma H~,
\ee
with $H= i \beta c_1(\CL_0)$ the supegravity background field given in \eqref{AVH sugra}.
 In addition to the BPS equations, we impose the gauge fixing condition along the Seifert fiber:
\be
\eta^\mu (\CL_K a_\mu) = 0~,
\ee
which implies $\partial_0 a_0 = 0$, with $a_0\in \mathbb{R}$, in the adapted frame. We can diagonalize $a_0$ using the residual gauge symmetry. The equation $D_\mu\sigma=0$, together with the reality condition on $\sigma$, implies $[a_0, \sigma]=0$, and therefore we can diagonalize $a_0$ and $\sigma$ simultaneously:
\be\label{a0 sigma diag}
a_0 = \text{diag} (a_{0,a})~, \qquad \sigma = \text{diag}(\sigma_a)~,
\ee
with constant $\sigma_a$'s. This breaks the gauge group down to its maximal torus times the residual Weyl symmetry:
\be
\GG \rightarrow  \GH  \rtimes W_\GG~, \qquad \qquad \GH \equiv \prod_{a=1}^{\rk} U(1)_a~.
\ee
The fields $a_0$ and $\sigma$ are combined into the $\eta$-component of the complexified 3d gauge field:
\be
\CA_\mu = a_\mu -i \eta_\mu \sigma~,
\ee
similarly to \eqref{def CAF}. Then, the two-dimensional scalar fields $u_a$ are defined as:
\be
u_a = -{1\ov 2 \pi} \int_\gamma \CA_a~.
\ee
We localize onto the constant modes:
\be\label{u def 1 secloc}
u_a = i \beta(\sigma + i a_0)~,
\ee
which naturally span the ``classical Coulomb branch'' for the 3d theory on $\R^2 \times S^1$.

It is important to note that the BPS equations allow for non-trivial $\GH$-bundles on the two-dimensional orbifold. We must sum over all such independent line bundles, which are indexed by the 3d Picard group:
\be\label{Pic M3 secLoc}
\widetilde{\text{Pic}}(\CM_3) = \text{Pic}(\hat\Sigma)/\langle[\CL_0]\rangle\ ,
\ee
as discussed in Section \ref{sec:pi1 M3}. We will denote the gauge line bundles in $\text{Pic}(\hat\Sigma)$ by:   
\be\label{gauge line bundle}
\CF = L_0^{\otimes \n_0} \otimes L_1^{\otimes\n_1} \otimes \cdots \otimes L_n^{\otimes\n_n}\ ,
\ee
where $\n_i=(\n_{i, a})$ label the GNO-quantized magnetic fluxes:
\be
\Gamma_{\mathbf{G}^\vee} = \{\n_i \in \mathfrak h|~ \rho(\n_i) \in \mathbb{Z}~, \quad\forall\rho \in \Lambda_{\text{char}} \}\ .
\ee
Any line bundle \eqref{gauge line bundle} has the orbifold first Chern class:
\be
c_1(\CF) = \n_0 + \sum_{i=1}^n {\n_i \ov q_i}~.
\ee
Recall that these orbifold line bundles are not all independent. They are subject to the equivalence relations in Pic$(\hat \Sigma)$:
\be \label{pic2 relation}
L_0 = L_i^{\otimes q_i}~, \quad \text{for }i=1,\cdots, n~.
\ee
From now on, let us assume that $c_1(\CL_0)\neq 0$. Then the line bundles in \eqref{Pic M3 secLoc} are all torsion line bundles, and they can be characterized by their flat connections.
In each topological sector $\{\n_0,\n_1,\cdots, \n_n\}$, the localized gauge field can be written as:
\be
a = \h a_0\eta + a^{(\text{flat})}\ ,
\ee 
where $\hat a_0\in \mathbb{R}$ is a constant. The flat connection takes the form:
\be
a^{(\text{flat})} =  a^{(\text{flat})}_\psi d\psi +\alpha~,
\ee 
in local coordinates, where $\alpha$ is the flat connection on the base $\h\Sigma$.
 To find the correct expression for $ a_\psi^{(\text{flat})}$, consider the coordinate transformation:
\be\label{coordinate transformation 1}
\delta_x \,:\, \psi' = \psi - \lambda(z,\bar z)~, \quad\qquad \CC' = \CC+ d\lambda (z,\bar z)~,
\ee
so that the one-form $\eta$ is well-defined. We can also write:
\be
d\CC = c_1(\CL_0)\, \pi^*(w_{\h\Sigma})~,\qquad\quad  d a = c_1(\CF)\,\pi^*(w_{\h\Sigma})~,
\ee
where $w_{\h\Sigma}$ is the volume form on the two-dimensional base. It is convenient to define the forms $w_{\h\Sigma}^{(0)}$ and $w_{\h\Sigma}^{(1)}$  by descent with respect to the coordinate transformation $\delta_x$, as:
\be
w_{\h\Sigma} = dw_{\h\Sigma}^{(0)}~,\qquad\quad \delta_x w_{\h\Sigma}^{(0)} = dw_{\h\Sigma}^{(1)}~.
\ee
Then, we can write $\CC$ and the gauge field $a$ as:
\be
\CC = c_1(\CL_0)\, \pi^*w_{\h\Sigma}^{(0)} + db~,\qquad \quad  a = c_1(\CF)\, \pi^* w_{\h\Sigma}^{(0)} + dc~,
\ee
where $b$ and $c$ are globally-defined functions on the base. It follows that the change of coordinates acts as:
\be
\delta_x\CC = c_1(\CL_0)\, \pi^*dw_{\h\Sigma}^{(1)}~,\qquad \quad\delta_x a = c_1(\CF) \,\pi^*dw_{\h\Sigma}^{(1)}~.
\ee
Comparing this expression with \eqref{coordinate transformation 1}, we find: 
\be
a_{\psi}^{(\text{flat})} d\psi=  -\frac{c_1(\CF)}{c_1(\CL_0)} d\psi~.
\ee
This gives the holonomy:
\be
\exp\left({-i\int_\gamma a^{(\text{flat})}}\right) = \exp\left({2\pi i \frac{c_1(\CF)}{c_1(\CL_0)}}\right)
\ee
along the Seifert fiber $\gamma$ (at generic point on the base). 
Therefore, the complex variable $u$ in \eqref{u def 1 secloc} can be written as:
\be
u = i\beta(\sigma + i\hat a_0) + \frac{c_1(\CF)}{c_1(\CL_0)}~,
\ee
which is valued in $u \in {\frak h}_\mathbb{C}$. One can check that this expression is compatible with the large gauge transformation:
\be\label{lgt secLoc}
u\rightarrow u+1~, \qquad\n_0 \rightarrow \n_0+\mathbf{d}~, \qquad\n_i\rightarrow \n_i + p_i\ ,\forall i~,
\ee
which we discussed in \eqref{lgt action}. (Here, $\h a_0$ is invariant under large gauge transformations.)

If  $c_1(\CL_0)=0$ instead, the flat connection $a_\psi^{(\rm flat)} \in \R$ is a free parameter which can be reabsorbed into $\h a_0$, corresponding to the free generator in $\t \Pic(\CM_3)$. In that case,  $\h a_0$ is also subject to the large gauge transformation $\h a_0 \sim \h a_0+1$.
It is then more natural to gauge-fix that invariance by imposing that $\h a_0$ is valued in the interval $\hat a_0 \in [0,1)$. Then, the Coulomb branch variable $u$ can be defined as:
\be\label{cL0 0 case for u}
u = i\beta(\sigma + i\hat a_0)\ ,~ \text{ with }~ u \in  {\frak h}_\mathbb{C}/\Lambda_{\text{cochar}}~.
\ee
Finally, the flat connection $\alpha$ on the Riemann surface corresponds to the factor $\mathbb{Z}^{2g} \subset H^1(\CM_3)$, which we can expand as:
\be
\alpha = \sum_{I=1}^g\left(\alpha_I w_1^I dz + \t \alpha_I  w_{\bar 1}^I d\bar z\right)~,\qquad \quad  [w^I] \in H^1(\CM_3)~. 
\ee
In addition to these bosonic zero modes, this background admits $2g +2$ gaugino zero-modes, which can be written as:
\be
\Lambda = \sum_{I=1}^g\left(\Lambda_I w_1^I dz + \t \Lambda_I  w_{\bar 1}^I d\bar z\right)~, \qquad \Lambda_0~,\, \t \Lambda_0 = (\text{constant}).
\ee

\paragraph{Localization formula for the path integral.}
The bosonic and fermionic zero-modes can be organized into short supermultiplets:
\be\label{all zeromodes}
\CV_0 = (a_0,\sigma, \Lambda_0, \widetilde\Lambda_0, \hat D)~, \qquad\qquad \CV_I = (\alpha_I, \widetilde\alpha_I. \Lambda_I, \widetilde\Lambda_I)~, \quad I=1, \cdots, g~.
\ee
Here, we also introduced a constant mode $\hat D$ for the auxiliary field $D$, as a regulator for the localization computation, as studied {\it e.g.} in \cite{Benini:2013xpa,Hori:2014tda}. Then, using a standard localization argument, we can reduce the supersymmetric path integral to supersymmetric ordinary integral over the zero-modes \eqref{all zeromodes}, namely \cite{Closset:2017zgf}:~\footnote{Here and in the following, in addition to the gauge  magnetic fluxes $\n$ (that must be summed over), we can turn on background fluxes $\m$ for the flavor symmetries. We leave them implicit to avoid clutter. We similarly leave the dependence on the $L_R$ parameters implicit.}
\be\label{almost final}
Z_{\CM_3}(\nu) =\lim_{e^2,g^2\rightarrow 0}\frac{1}{|W_{\GG}|} \sum_{\substack{(\n_0,\n_i)\\ \in \widetilde{\text{Pic}}(\CM_3)}}\int d\CV_0 \prod_{I=1}^g d\CV_I ~e^{-S_{\text{CS}}(\CV_0,\CV_I)}\,\CZ^{\text{1-loop}}_{(\n_0,\n_i)} (\CV_0,\CV_I)\ ,
\ee
where $\CZ^{\text{1-loop}} (\CV_0,\CV_I)$ is the contribution from the one-loop fluctuation around the solution to the BPS equations. Here we again assumed that $c_1(\CL_0)\neq 0$.
The integration over the fermionic zero modes and the $\hat D$ integral can be done as in \cite{Closset:2017zgf}. 
We then obtain:
\be\label{JK integral formula}
Z_{\CM_3}(\nu) = \frac{1}{|W_\GG|}\sum_{\substack{(\n_0,\n_i)\\\in \widetilde{\text{Pic}}(\CM_3)}} \int_{\CC(\eta)} d^\rk u \; e^{-S_{\text{CS}}(u,\nu)}\,\CZ^{\text{1-loop}}_{(\n_0,\n_i)}(u,\nu)\, H(u,\nu)^g~,
\ee
as a holomorphic contour integral in the variables $u_a$. The factor $e^{-S_{\text{CS}}}$ in the integrand is the classical contribution from the Chern-Simons terms in the UV Lagrangian. The factor $\CZ^{\text{1-loop}} (u, \nu)$ is the one-loop contribution,~\footnote{$\CZ^{\text{1-loop}} (u, \nu)$ is specialization to $\h D=0$ of the one-loop determinant $\CZ^{\text{1-loop}}(\CV_0,\CV_I)$ appearing in \protect\eqref{almost final}. The latter is not holomorphic in $u$, but a careful integration over the gaugino zero-modes $\Lambda_0, \b\Lambda_0$, and over the bosonic mode $\h D$, leads to the holomorphic contour integral \protect\eqref{JK integral formula}---see the discussion in Appendix D of \protect\cite{Closset:2017zgf}, and references therein.} given in terms of the one-loop determinants for the chiral and vector multiplets, which we discussed in Section~\ref{sec: Bethe vac sum} (see also Appendix~\ref{app:oneloop det}). The function $H(u, \nu)$ in the integrand is the Hessian of the twisted superpotential:
\be
H(u, \nu) = \det_{a,b} \frac{\partial \CW(u, \nu)}{\partial u_a \partial u_b}~,
\ee
as in \eqref{detW and Pif}. The factor $H(u, \nu)^g$ in \eqref{JK integral formula} arises from integrating out the zero-mode multiplets $\CV_I$.

Both the classical and the one-loop factors in \eqref{JK integral formula} can be written in terms of the geometry-changing line operators of Section~\ref{sec: Bethe vac sum}. We simply have:~\footnote{The sign $(-1)^\rk$ is introduced for future convenience. It could be absorbed in the orientation of the contour integral.}
\be
e^{-S_{\text{CS}}(u,\nu)}\,\CZ^{\text{1-loop}}_{(\n_0,\n_i)}(u,\nu) =(-1)^\rk\, e^{2\pi i (g-1) \Omega(u, \nu)} \, \CG_{\CM_3}(u, \nu)_{\n}~,
\ee
with $\n=(\n_0, \n_i)$. Here, $\Omega(u, \nu)$ is the effective dilaton introduced in Section \ref{subsec: HGO}, and $\CG_{\CM_3}(u, \nu)_\n$ is the full Seifert fibering operator:
\be
\CG_{\CM_3}(u, \nu)_\n \equiv \CF(u, \nu)^{\bf d}\, \pif(u,\nu)^{\n_0} \, \prod_{i=1}^n \CG_{q_i,p_i}(u, \nu)_{\n_i}~,
\ee
at fixed gauge flux $\n$, where $\CG_{q,p}(u, \nu)_\n$ is the $(q,p)$ fibering operator given by \eqref{Gqp nm gen}, and where $\pif(u,\nu)$ and $\CF(u, \nu)= \CG_{1,1}(u, \nu)_{0}$ are the ordinary gauge-flux and fibering operators, respectively. We also used the obvious short-hand notation $\pif(u, \nu)^{\n_0} \equiv \prod_{a=1}^\rk \pif_a(u, \nu)^{\n_{a, 0}}$ for the gauge-flux operator, and suppress the flavor symmetry flux from the notation.

The contour $\CC(\eta)$ in  \eqref{JK integral formula} is a  ``JK-like'' middle-dimensional contour on the $u$-domain, closely related to the Jeffrey-Kirwan (JK) residue \cite{JK1995, 1999math......3178B}, which depends on an auxiliary real vector $\eta \in \frak h^*$. The contour can be derived in the rank-one case, while in the higher-rank case we can propose some natural conjecture~\cite{Closset:2017zgf}. We will review these contour prescriptions momentarily. We will also discuss how they can be related to a non-compact ``$\sigma$-contour'' which appeared in well-known localization formulas for the supersymmetric partition function on the three-sphere and other lens spaces \cite{Kapustin:2009kz,Hama:2010av,Imamura:2011wg, Benini:2011nc, Dimofte:2014zga}.

\paragraph{Another integral formula.}
There exists an alternative way to present the summation over the line bundles and the $u$-integral in \eqref{JK integral formula}.
The integrand in \eqref{JK integral formula} is invariant under the large gauge transformation \eqref{lgt secLoc}, as required by gauge invariance. 
Recall from Section~\ref{sec: Bethe vac sum} that this large gauge transformations corresponds to tensoring the line bundle \eqref{gauge line bundle} by the defining line bundle of the Seifert manifold:
\be\label{large gauge 2}
\CF \rightarrow \CF\otimes \CL_0
\ee
This ensures that the summation over $\widetilde{\text{Pic}}(\CM_3)$ in \eqref{JK integral formula} is well-defined. On the other hand, we could also fix the gauge freedom \eqref{large gauge 2} in a different way, so that we sum over all independent line bundles in $\text{Pic}(\hat\Sigma)$. In this scheme, we should quotient the space $u \in \frak{h}_{\mathbb{C}}$ by the action of the large gauge transformations \eqref{lgt secLoc}. This can be done by restricting the range of the Coulomb branch variables, according to:
\be\label{gauge fixing}
u \in ~\frak{h}_{\mathbb{C}}/\Lambda_{\text{cochar}}~.
\ee
Given this gauge-fixing of the large gauge transformations \eqref{large gauge 2}, we still have to sum over all the fluxes $(\n_0, \n_i)$ in $\Pic(\h\Sigma)$, instead of the smaller group $\t\Pic(\CM_3)$. Given the 2d Picard group relations \eqref{pic2 relation}, this is equivalent to:
\be\label{gauge fixing 2 flux}
\n_0 \in \Gamma_{\GG^\vee}~, \qquad 
\n_i \in \Gamma^\vee_\GG (q_i)\ ,~~ \text{ for } i=1,\cdots n~,
\ee
with $\n_0$ an ordinary magnetic flux over $\h\Sigma$, and $\n_i$ the fractional fluxes at the exceptional fibers,  as explained in detail around \eqref{cochar mod q def}.
For instance, for $\GG=U(1)$,  we should restrict the integration variables to the strip Re$(u) \in [0,1)$, while we sum over the ordinary fluxes $\n_0 \in \mathbb{Z}$ and over the fractional fluxes $\n_i \in \Z_{q_i}$.

We then arrive at the localization formula:
\be\label{almost final 2}
Z_{\CM_3}(\nu)=\frac{1}{|W_{\GG}|} \sum_{\n_0 \in \Gamma_{\GG^\vee}}\sum_{\n_i \in \Gamma^\vee_\GG (q_i),\forall i} \int_{\CC_0(\eta)} d^{\rk}u~e^{-S_{\text{CS}}(u,\nu)}\CZ^{\text{1-loop}}_{(\n_0,\n_i)}(u,\nu) H(u, \nu)^g~.
\ee
Here, $\CC_0(\eta)$ is defined as the contour $\CC(\eta)$ restricted to the quotient space \eqref{gauge fixing}. 
Thus, for $c_1(\CL_0)\neq 0$, we have replaced the finite sum over the torsion $\GH$-bundles in \eqref{JK integral formula} by an infinite sum over $\n_0$, together with a sum over the fractional fluxes. The infinite sum is essentially a sum over the images of the restricted domain \eqref{gauge fixing} under the large gauge transformations \eqref{large gauge 2}.

Interestingly, in the gauge \eqref{gauge fixing}-\eqref{gauge fixing 2 flux}, we can naturally include the case with $c_1(\CL_0)=0$, as should be clear from \eqref{cL0 0 case for u}. Thus, the integral formula \eqref{almost final 2} should be valid for any $\CM_3$.
 In this formulation, we have an infinite sum over the magnetic fluxes $\n_0$, and one should worry about whether that sum converges. The convergence ultimately follows from the properties of the contour $\CC_0(\eta)$. In the following, we will study this contour more in detail, focussing on the rank-one case where everything can be done very explicitly.

For future reference, let us note that the sum over the fractional fluxes in \eqref{almost final 2} is exactly the same sum that we considered in \eqref{CG full sum def} when computing the full fibering operators. Therefore,  we may also define:
\be
\CG_{\CM_3}(u, \nu) \equiv  \CF(u, \nu)^{\bf d}\,   \prod_{i=1}^n\left[ \sum_{\n_i \in \Gamma^\vee_\GG (q_i)} \CG_{q_i,p_i}(u, \nu)_{\n_i}\right]=  \CF(u, \nu)^{\bf d}\,   \prod_{i=1}^n \CG_{q_i,p_i}(u, \nu)~,
\ee
the product of the ``full'' $(q,p)$-fibering operators, so that the integral formula \eqref{almost final 2} reads:
\be\label{almost final 3}
Z_{\CM_3} =\frac{{(-1)^\rk}}{|W_{\GG}|} \sum_{\n_0 \in \Gamma_{\GG^\vee}} \int_{\CC_0(\eta)} d^{\rk}u\; 
\pif(u, \nu)^{\n_0}\, \CG_{\CM_3}(u, \nu)\, e^{-2\pi i \Omega(u, \nu)} \, \CH(u, \nu)^g~.
\ee
Here, $\CH(u, \nu) = e^{2\pi i \Omega(u, \nu)}H(u, \nu)$ is the handle-gluing operator \eqref{hgoFull}.  In this form the partition function is exhibited as the expectation value of the geometry changing line operator, $\SL_{\cM_3}$, in the $S^2 \times S^1$ partition function \cite{Benini:2015noa}, as anticipated in \eqref{INTROZM3CL}.

\subsection{Contour prescriptions and the Bethe-sum formula}
\label{Bethe sum}

In the rest of this section, we explain the relation between the above contour-integral formulas and the Bethe-sum formula \eqref{ZM3 Bethe formula}. For a rank-one gauge theory, we will explicitly show that the two types of expressions agree, when certain additional conditions are satisfied. We also study the relation to the ``$\sigma$-contour" formula, which has been considered in previous works on lens spaces \cite{Kapustin:2009kz,Hama:2010av,Imamura:2011wg,Dimofte:2014zga}.

\subsubsection{Singularities of the integrand}
\label{singularities}

Let us denote the integrand in \eqref{JK integral formula} by:
\be
\CJ_{(\n_0,\n_i)}(u, \nu) =(-1)^\rk \, \CG_{\CM_3}(u, \nu)_\n\,  e^{-2\pi i \Omega(u, \nu)} \, \CH(u, \nu)^g~.
\ee
The function $\CJ_{(\n_0,\n_i)}(u, \nu)$ has various singularities on the domain $u \in \frak h_{\mathbb{C}}$, which define four different types of hyperplanes \cite{Closset:2017zgf}, including ``hyperplanes at infinity.''

\paragraph{Chiral multiplet singularities.}
For each chiral multiplet with gauge charge $\rho$ and R-charge $r$, we have singular hyperplanes where the chiral multiplet becomes massless, defined by
\be
H_{\rho,r,n} = \{u \in \frak h_{\mathbb{C}}~|~ \rho(u)+ \nu_R r + \kk=0\ ,~\kk\in \mathbb{Z}\}
\ee
The order of pole at $H_{\rho,r,\kk}$ is
\be
N_{\rho, r, \kk} = \text{deg}(L_{r,\kk})+1-g = \rho(\n_0) + \widetilde{\n}_0^R r + \mathbf{d}n + \sum_{i=1}^n \left\lfloor\frac{\rho(\n_i) + \n_i^R r + p_i n}{q_i}\right\rfloor +1-g~,
\ee
with the orbifold line bundle $L_{r, \kk}$ as in \eqref{Lrkk}.
 
\paragraph{Large Im($u$) regions (monopole) singularities.}
The integrand may diverge in the large  Im($u$) region. The hyperplanes defined by these singularities are
\be
H_{a\pm}= \{ u\in {\frak h}_{\mathbb{C}}~|~ \text{Im}(u_a) \rightarrow \pm\infty \}\ .
\ee
The behaviour of the integrand in this limit depends on the charges of the monopole operators \cite{Closset:2016arn, Closset:2017zgf}. Equivalently, the behaviour of the integrand in this limit depends on the value of the effective Chern-Simons levels as $\sigma \rightarrow \pm \infty$---the limit is given explicitly in  \eqref{CS limit of GPhi} for the chiral-multiplet contribution to the fibering operator, and similarly for the effective dilaton.

\paragraph{Large Re($u$) regions.}
When $c_1(\CL_0)\neq 0$, the integrand is not periodic in the Re$(u_a)$ directions, and can be divergent as we take the limit Re$(u_a)\rightarrow \pm\infty$. Let us define the integer $Q=\prod_{i=1}^n q_i>0$. We then have:~\footnote{Here we consider shifting a single $u_a$ at the time.}
\be
 \CG_{\CM_3}(u_a+QN, \nu)_\n=  \pif_a(u, \nu)^{c_1(\CL_0) QN}  \CG_{\CM_3}(u, \nu)_\n~,
\ee
for any $N\in \Z$. (The effective dilaton and the handle-gluing operators are periodic.)
When $c_1(\CL_0)>0$, we can see that the integrand diverges in the Re$(u)\rightarrow \infty$ region, on the segments of the contour $\CC(\eta)$ where $|\pif_a(u)|>1$. Similarly, it diverges in the Re$(u_a)\rightarrow -\infty$ direction on the segments of the contour where $|\pif_a(u)|<1$. The opposite holds for $c_1(\CL_0)<0$.

\paragraph{W-boson singularities.}
For a non-abelian gauge group, when $2-2g-n>0$, we also have potential singularities at each root $\alpha \in \frak g$:
\be
H_\alpha = \{ u \in \frak h_{\mathbb{C}}~|~\alpha(u)=n\ ,n\in \mathbb{Z}\}\ .
\ee
This is the locus where the non-abelian gauge symmetry enhances. As in \cite{Closset:2017zgf}, our prescription for the contours $\CC$ and $\CC_0$ is to define them such that we do not pick any contributions from poles which intersects with $H_\alpha$.

\subsubsection{Contour integral for $U(1)$ theories.}
Let us make this contour more precise in the case $\GG= U(1)$.
For a $U(1)$ gauge theory, we can explicitly derive the contour $\CC$ and prove the equivalences among different expressions for the partition functions.

\paragraph{JK contour and the Bethe-sum formula.}
In the $U(1)$ case, the formula \eqref{almost final 3} simplifies to:
\be\label{JK contour formula U(1)}
Z_{\CM_3} =  \sum_{\n_0\in \Z} \int_{\CC_0(\eta)} du\; 
\pif(u)^{\n_0} \, \CI(u)~, 
\ee
with:
\be\label{definition of I flux summed}
\CI(u)\equiv  (-1)\,\CG_{\CM_3}(u)\, e^{-2\pi i \Omega(u)} \, \CH(u)^g~.
\ee
Here and in the following, we suppress the dependence on the flavor parameters $\nu$, for simplicity of notation. The contour $\CC_0(\eta)$ is given by:~\footnote{The derivation of the contour is essentially the same as that for the $\CM_{g,p}$ partition function. This is discussed in Appendix D of \protect\cite{Closset:2017zgf}. }
\be\label{define contour 1}
\CC_0(\eta) =\big\{ u \in \partial\hat{\fM_0} |~ \text{sign}(\text{Im}(\partial_u\CW)) = -\text{sign}(\eta)\big\}~.
\ee 
Here, $\hat{\fM_0}$ is defined to be the complex $u$-plane restricted to the strip Re($u)\in [0,1)$, with the $\epsilon$-neighbourhoods of the singularities listed in Section \ref{singularities} removed. 
We also remove the region outside of the large box with the boundaries $\text{Im}(\sigma)=\pm R$, Re$(u)=0,1$, with $R$ arbitrarily large.
Then, the boundary $\partial\hat{\fM_0}$ consists of small circles along the ``bulk'' singularities, plus the boundary of the strip.  The prescription \eqref{define contour 1} is to take the subset of $\partial\hat{\fM_0}$ such that $|\pif(u)|>1$ if $\eta>0$, or such that $|\pif(u)|<1$ if $\eta<0$.
In this way, $\CC_0(\eta)$ defines a contour integral which is equivalent to the JK-residue prescription for singularities at finite $u$, with additional contributions from the boundary integral.
Here, the parameter $\eta \in \R$ can be chosen arbitrarily, but the final answer is independent of that choice.

To relate the expression \eqref{JK contour formula U(1)} to the Bethe-sum formula \eqref{ZM3 Bethe formula}, we should simply sum over $\n_0 \in \Z$ explictly. By choosing $\eta<0$ for $\n_0 \geq 0$ sector and $\eta>0$ for $\n_0<0$, we have:
\bea\label{summation over n0}
Z_{\CM_3} &= \sum_{\n_0=-\infty}^{-1}  \int_{\CC_0(\eta>0)} du~\Pi(u)^{\n_0} \CI(u) + \sum_{\n_0=0}^\infty \int_{\CC_0(\eta<0)}du~\Pi(u)^{\n_0}\CI(u) \\
&= \left(-\int_{\CC_0(\eta>0)} du + \int_{\CC_0(\eta<0)} du\right)  \frac{\CI(u)}{1-\Pi(u)} = \oint_{\CC_{\text{BE}}} du~  \frac{\CI(u)}{1-\Pi(u)}~.
\eea
Note that the two geometric series in \eqref{summation over n0} converge, since $|\Pi(u)|< 1$ for $\eta<0$ and $|\Pi(u)|>1$ for $\eta>0$. In the last expression, $\CC_{\text{BE}}$ denotes the contour enclosing the poles of the new integrand, which are located at $\Pi(u)=1$ \cite{Closset:2017zgf}. Thus, we pick up the residues:
\be
\oint_{u= \h u} du {\CI(u)\ov 1 - \pif(u)}= -2\pi i \oint_{u= \h u} {du\ov 2\pi i} {\CI(u)\ov (u- \h u) \d_u \pif(u)}~,
\ee
at the Bethe roots $\h u$ such that $\pif(\h u)=1$. Using the fact that $H(u)={1\ov 2\pi i} \d_u \log \pif(u)$, we directly obtain:
\be\label{Final Bethe 1}
Z_{\CM_3}=  - \sum_{\h u \in \CS_{\rm BE}}   \CI(\h u)  H(\h u)^{-1}
 =\sum_{\h u \in \CS_{\rm BE}}   \CH(\h u)^{g-1}\, \CG_{\CM_3}(\h u)~,
 \ee
thus reproducing the Bethe-sum formula \eqref{ZM3 Bethe formula}. This can be considered a derivation of the TQFT formula from a UV localization computation, in this special case.

\paragraph{The $\sigma$-contour integral formula.}

For $c_1(\CL_0) \neq 0$, we have seen that there exists an alternative way of fixing the gauge under the large gauge transformation \eqref{lgt secLoc}, which leads to the expression \eqref{JK integral formula}. Let us define:
\be\label{def CIni integrand}
\CI_{(\n_i)}(u) \equiv (-1)^\rk \, e^{-2\pi i \Omega(u)} \, \CH(u)^g\, \CF(u)^{\bf d}\, \prod_{i=1}^n \CG_{q_i,p_i}(u)_{\n_i}~,
\ee
which depends on the fractional fluxes $\n_i$ but not on the ``ordinary flux'' $\n_0$.
For a $U(1)$ gauge group, the formula  \eqref{JK integral formula} can then be written as:
\be
Z_{\CM_3}  = \sum_{\substack{(\n_0, \n_i)\\\in \widetilde{\text{Pic}}(\CM_3)}} \int_{\CC(\eta)} d u\,  \pif(u)^{\n_0}\, \CI_{(\n_i)}(u)~,
\ee
with the contour:
\be
\CC(\eta) =\{ u \in \partial\hat{\frak M}\, |\; \text{sign}(\text{Im}(\partial_u\CW)) = -\text{sign}(\eta)\}~.
\ee 
Here, $\hat{\frak M}$ is defined in the same way as in $\hat{\frak M}_0$ of \eqref{define contour 1}, but now $u$ is valued in the entire complex plane, $u\in \mathbb{C}$.
Let us choose:
\be
c_1(\CL_0) >0~,  
\ee
for definiteness, and without loss of generality.~\footnote{Changing the sign of $c_1(\CL_0)$ corresponds to flipping the orientation of $\CM_3$, which can be achieved by the replacement $({\bf d}, p_i)\rightarrow (-{\bf d}, -p_i)$.}
One can then show that the contour $\CC(\eta)$ can be deformed to a non-compact $\sigma$-contour integral connecting Im($u)\rightarrow -\infty$ to  Im($u)\rightarrow +\infty$ \cite{Closset:2017zgf}.
To see this, let us first write:
\be
\int_{\CC(\eta)} du = \sum_{\m\in \mathbb{Z}} \int_{\CC_\m(\eta)} du\ ,
\ee 
where we decompose the contour $\CC(\eta)$ into components in vertical strips $\hat{\mathfrak M}_\m$ with Re$(u)\in [\m, \m+1]$, as:
\be
\CC_{\m}(\eta) = \{u\in \partial\hat{\mathfrak M}_\m~|~\text{sign}(\text{Im}(\partial_u\CW)) = -\text{sign}(\eta)\}~,\; \text{Re}(u) \in [\m,\m+1]\}~,
\ee
by closing the contour on the boundary of each strip $\hat{\mathfrak M}_\m$ in the obvious way.
We then have:
\bea\label{ZM3 sigma interm}
Z_{\CM_3}  &= \sum_{\substack{(\n_0,\n_i)\\\in \widetilde{\text{Pic}}(\CM_3)}} \sum_{\m\in\mathbb{Z}}\int_{\CC_\m(\eta)} d u\, \pif(u)^{\n_0}\, \CI_{(\n_i)}(u) \cr
&= \sum_{\substack{(\n_0,\n_i)\\\in \widetilde{\text{Pic}}(\CM_3)}} \sum_{\m\in\mathbb{Z}}\int_{\CC_0(\eta)} d u\, \pif(u)^{\n_0}\, \CI_{(\n_i)}(u-\m)~,
\eea
with $\CC_0(\eta)$ as above. Now, let us choose $\eta$ as:
\be\label{sum of contours}
\sum_{\m\in\mathbb{Z}}\int_{\CC_0(\eta)} du =\sum_{\m=-\infty}^0\int_{\CC_0(\eta>0)} du +  \sum_{\m=1}^\infty \int_{\CC_0(\eta<0)} du~.
\ee
Then, under some convenient assumptions about the flavor parameters, one can show~\footnote{We refer to Section 4.6.1 in \protect\cite{Closset:2017zgf}, whose argument can be repeated verbatim.} that most of the contributions along the contour cancel out between adjacent terms, and that the only non-vanishing contribution comes from the contour along the imaginary axis at Re$(u)=0$:
\bea\label{sigma contour}
Z_{\CM_3}  &= \sum_{\substack{(\n_0,\frak n_1,\cdots, \n_n)\\\in \widetilde{\text{Pic}}(\CM_3)}} \left(-\int_{\substack{\CC_0(\eta<0),\\\text{Re}(u)=0}} d u +\int_{\substack{\CC_0(\eta>0),\\\text{Re}(u)=0}} d u\right)~\pif(u)^{\n_0}\, \CI_{(\n_i)}(u) \\
&= \sum_{\substack{(\n_0,\frak n_1,\cdots, \n_n)\\\in \widetilde{\text{Pic}}(\CM_3)}}\int_{\text{Re}(u)=0} du~\pif(u)^{\n_0}\, \CI_{(\n_i)}(u)\ .
\eea 
Note that this simple contour along the imaginary axis is valid under certain assumptions on the charges of the chiral multiplets and monopole operators, as studied in \cite{Closset:2017zgf}. In general, the resulting contour $\CC_\sigma$ is deformed with respect to the Re$(u)=0$ contour, in such a way that the non-compact integral \eqref{sigma contour} converges. We will see explicit examples of this $\sigma$-contour in later sections.

\paragraph{From the $\sigma$-contour to the Bethe-sum formula.}
We can also derive the Bethe-sum formula \eqref{ZM3 Bethe formula} directly from the $\sigma$-contour formula \eqref{sigma contour}. More precisely, let us start from the equivalent expression \eqref{ZM3 sigma interm}, namely:
\be
Z_{\CM_3}  =\sum_{\substack{(\n_0,\n_i)\\\in \widetilde{\text{Pic}}(\CM_3)}} \sum_{\m\in\mathbb{Z}}\int_{\CC_0(\eta)} d u\, \pif(u)^{\n_0}\, \CI_{(\n_i)}(u-\m)~.
\ee
Using the gauge-invariance of the fibering operator, as in \eqref{diff eq CG} and \eqref{diff FPif}, this equals:
\be
Z_{\CM_3}  =\sum_{\substack{(\n_0,\n_i)\\\in \widetilde{\text{Pic}}(\CM_3)}} \sum_{\m\in\mathbb{Z}}\int_{\CC_0(\eta)} d u\, \pif(u)^{\n_0+ \m {\bf d}}\, \CI_{(\n_i+\m p_i)}(u)~.
\ee
Now, reparameterizing of the fluxes:
\be
\n_0' \equiv \n_0 + \m\mathbf{d}~, \qquad \n_i' = \n_i + \m p_i~,
\ee
 the sum over $(\n_0,\n_i)\in\widetilde\Pic(\CM_3)$ and the sum over $\m\in\Z$ can be replaced by a summation over all the element $(\n_0',\n_i')$ in $\Pic(\h\Sigma)$. Taking into account the relations \eqref{pic2 relation}, we find:
\be\label{formula CC0 2}
Z_{\CM_3}  =\sum_{\n_0' \in \Z} \sum_{\substack{\{(\n_1',\cdots, \n_n')|\\ \n_i' \in \Z_{q_i}\}}}
\int_{\CC_0(\eta)} du~\pif(u)^{\n_0'}\, \CI_{(\n_i')}(u) = 
  \sum_{\n_0\in \Z} \int_{\CC_0(\eta)} du\; 
\pif(u)^{\n_0} \, \CI(u)~, 
\ee
The last expression is obtained by performing the sum over the fractional fluxes to obtain the ``full'' fibering operator. Thus we obtain the same expression as in \eqref{JK contour formula U(1)}, which was shown to be equivalent to \eqref{Final Bethe 1}.

\paragraph{Non-abelian generalization}

For a non-abelian gauge theory, we lack of a complete derivation of the contour $\CC(\eta)$.  We expect that the Bethe-sum formula:
\be
Z_{\CM_3}(\nu) = \sum_{\hat u \in \CS_{\text{BE}}} \sum_{\substack{\{(\frak n_1,\cdots, \frak n_n)|\\ {\frak n_i}\in \Gamma^\vee_\GG (q_i),\forall i\}}}\,\CH(\hat u, \nu)^{g-1}\,\CF(\hat u, \nu)^{\bf d}\,\prod_{i=1}^n\CG_{q_i,p_i}(\hat u,\nu)_{\n_i}~,
\ee
provides the correct answer for any gauge group $\GG$ of the type considered in this paper. This claim, while not rigorously proven, has been corroborated by numerous highly non-trivial consistency checks. 

Recall that, when looking for the solutions $u= \h u$ to the Bethe equations $\pif_a(u, \nu)=1$, we need to exclude the would-be solutions that are located on a Weyl chamber boundary. Correspondingly, in the integral formula \eqref{almost final 3}, the contour $\CC_0(\eta)$ must be such that we do not pick any higher-dimensional residues from the W-boson singularities. 

\subsection{Higher-dimensional $\CC_0$-contour}
Somewhat formally, the contour $\CC_0$ can be defined in terms of a $\rk$-dimensional residue at the ``poles'' defined by an intersection of $r$ independent singular hyperplanes (including ``poles'' at infinity), with $r \geq \rk$. This can always be decomposed into bulk and boundary contributions:
\be
\CC_0 = \CC_0^{\text{bulk}}+ \CC_{0}^{\text{boundary}}~,
\ee
where the contour $\CC_0^{\text{bulk}}$ captures the contribution from the residues at finite $u$, while the $\CC_{0}^{\text{boundary}}$ captures the singularities that intersect the hyperplanes ``at  infinity.''
We will not derive the precise contribution from the boundary contour $\CC_0^{\text{boundary}}$ for a higher-rank gauge group in this paper.

The bulk contribution is given by the JK residue, which can be derived as in \cite{Benini:2015noa, Closset:2016arn,Closset:2017zgf}. However, when the singularity has an intersection with the hyperplane $H_\alpha$, the JK residue is not well-defined.
At these loci, where $\alpha(u) = 0$ for some root $\alpha$, the non-abelian gauge symmetry enhances and the path integral becomes potentially singular. This happens when $2g-2+n \geq 0$, in the $\CC_0$-contour integral formula. 
We claim that we should always exclude such poles, which are fixed by the Weyl group $W_{\GG}$. This is true already in the case of the trivial fiber bundle, $\cM_3=\Sigma_g \times S^1$, and the principal $S^1$ bundle, $\cM_3=\cM_{g,p}$, as discussed in \cite{Benini:2015noa,Benini:2016hjo,Closset:2017zgf}. 

\paragraph{Non-abelian $\sigma$-contour.}
For non-abelian gauge groups, the existence of additional singularities due to the W-bosons also modifies the $\sigma$-contour formula.  Namely,  the integrand in \eqref{formula CC0 2} may contain a potential singularity when $\alpha(u)=\ell\in \Z$, in addition to the acceptable solutions to the Bethe equations. In the rank-one case, the general relation between the $\sigma$-contour and the Bethe-sum formula is: 
\bea\label{ZM3 loc alpha contrib}
Z_{\CM_3} &= \sum_{\hat u \in \CS_{\text{BE}}} \CH(\hat u)^{g-1} \GG_{\CM_3}(\hat u)\\
&= \frac{1}{|W_{\GG}|}\sum_{\substack{(\n_0,\frak n_1,\cdots, \n_n)\\\in \widetilde{\text{Pic}}(\CM_3)}}
\int_{\CC_{\sigma}} du~\pif(u)^{\n_0}\, \CI_{(\n_i)}(u) 
-\frac{1}{|W_{\GG}|}\oint_{\substack{\alpha(u)=\ell ,\\ \text{Re}(u) \in [0,1)}} \frac{\CI(u)}{1-\Pi(u)}\ ,
\eea
with $\CI_{(\n_i)}(u)$ and $\CI(u)$ defined in \eqref{def CIni integrand} and \eqref{definition of I flux summed} respectively, including the W-boson contribution, which we may write as (extracting the contribution from the handle-gluing and fibering operators):
\bea \label{sigmacontourW0}
& \cI^{\cW_0} =  \prod_{\alpha \in \Delta^+} (2\sin(\pi \alpha(u)))^{2(1-g)} \prod_{i=1}^n (-1)^{\alpha(\n_i)(t_i+ l_i^R t_i+ 2\nu_R s_i)} {\sin\Big({\pi \alpha(u-t_i\n_i)\ov q_i}\Big)\ov \sin(\pi \alpha(u))}\\
&\;\;= \prod_{\alpha \in \Delta^+} 2^n (2\sin(\pi \alpha(u)))^{2-2g-n} \prod_{i=1}^n (-1)^{\alpha(\n_i)(t_i+ l_i^R t_i+ 2\nu_R s_i)} \sin\Big({\pi \alpha(u-t_i\n_i)\ov q_i}\Big)~,
\eea
For $2g-2+n < 0$, the exponent of the first factor is positive, and vanishes at points with $\alpha(u) \in \Z$, and so the contribution to the second term in \eqref{ZM3 loc alpha contrib} vanishes.  In the marginal case, $2g-2+n=0$, we note that if $\alpha(u)=n \in \Z$, then the set $\{\alpha(\frac{\hat{u}-t_i \n_i}{q_i}) \}$, as we vary $\n_i $, is invariant under reflections, $x \rightarrow - x$.  Thus when we sum over fractional fluxes, $\n_i$, the terms come in pairs related by a sign, due to the second factor in \eqref{sigmacontourW0}, and so their contribution cancels out.  Then we again find that the second term in  \eqref{ZM3 loc alpha contrib} vanishes.

The two cases above correspond precisely to the Seifert fibrations over $S^2$ with at most two exceptional fibers, which are precisely the lens spaces.  We see that, in these cases, the Bethe sum agrees with the standard $\sigma$-contour formula found in the supersymmetric localization literature  \cite{Kapustin:2009kz, Gang:2009wy, Jafferis:2010un,Hama:2010av, Hama:2011ea, Benini:2011nc, Kallen:2011ny, Beem:2012mb, Dimofte:2014zga}. We will discuss the lens-space partition functions in more detail in the next sections.

Finally, in the general case, $2g-2+n>0$, the second term in \eqref{ZM3 loc alpha contrib} is generically non-trivial, and so the Bethe-sum formula is different from the naive $\sigma$-contour formula. The discrepancy is given by the residue at $\alpha(u) =n\in\mathbb{Z}$ in the strip. Interestingly, the formula \eqref{ZM3 loc alpha contrib} agrees with a formula derived by Lawrence and Rozansky \cite{Lawrence1999} for a pure CS theory with gauge group $\GG=SU(2)$ on a Seifert homology sphere. It will be interesting to explore this point further \cite{CHMW2018}.

\part{Lens space partition functions}
\label{part three}

\section{The $S^3_b$ partition function}\label{sec: S3b}

In this final part of the paper, we compare our results above to the well-known localization results for supersymmetric partition functions on lens spaces.  We prove that our results agree with the known results, in the cases where they overlap, and clarify various subtleties related to the choice of spin structure and to the contributions from CS contact terms.

We start in this section with the squashed three-sphere, $S^3_b$.   For a 3d $\CN=2$ gauge theory, the $S^3_b$ partition function can be written as the integral formula \cite{Hama:2011ea, Imamura:2011wg}:
\be\label{ZSb loc formula}
Z_{S^3_b}(\h m)= {1\ov |W_\GG|} \int_{\CC_{\h \sigma}} \prod_{a=1}^\rk d\h\sigma_a \; Z_{S^3_b}^{\rm CS}(\h\sigma, \h m)\, Z_{S^3_b}^{\rm vector}(\h\sigma)\, Z_{S^3_b}^{\rm matter}(\h\sigma, \h m)~,
\ee
where $|W_\GG|$ is the order of the Weyl group of $\GG$, and $\h m$ stands for the real masses associated to the flavor symmetry.
The classical piece $Z^{\rm CS}(\h\sigma,\h m)$ comes from CS terms, and takes the general form:
\be\label{ZS3b cl full}
Z_{S^3_b}^{\rm CS} = Z_{S^3_b}^{\rm GG}(\h\sigma)^{k_{GG}}\; Z_{S^3_b}^{\rm G_1G_2}(\h\sigma_1,\h\sigma_2)^{k_{G_1 G_2}}\; Z_{S^3_b}^{\rm GR}(\h\sigma)^{k_{GR}}\; (Z_{S^3_b}^{\rm RR})^{k_{RR}}\; (Z_{S^3_b}^{\rm grav})^{k_g}~.
\ee
The various supersymmetric CS actions evaluated on the $S^3_b$ background give the contributions \cite{Hama:2011ea, Imamura:2011wg, Closset:2012vp}:
\bea\label{CS terms S3b}
&  Z_{S^3_b}^{\rm GG}(\h\sigma) = e^{\pi i \h\sigma^2}~, \qquad && Z_{S^3_b}^{\rm G_1G_2}(\h\sigma_1,\h\sigma_2)= e^{2\pi i \h\sigma_1 \h\sigma_2}~, \cr
& Z_{S^3_b}^{\rm GR}(\h\sigma)= e^{2\pi i \h\sigma_R \h\sigma}= e^{\pi \left(b+ b^{-1}\right) \h\sigma}~, \qquad 
&& Z_{S^3_b}^{\rm RR}= e^{\pi i \h\sigma_R^2}= e^{-{\pi i \ov 4} \left(b^2 + b^{-2} +2\right)}~, \cr
& Z_{S^3_b}^{\rm grav}= e^{{\pi i \ov 24}\left(b^2+ b^{-2}\right)}~.
\eea
Here we defined the parameter:
\be\label{def sigmaR}
\h\sigma_R \equiv -{i\ov 2}(b+ b^{-1})~,
\ee
which is the effective ``real mass'' for the R-symmetry \cite{Closset:2014uda}. The generalization to any non-abelian CS term is straightforward.
One can similarly write down CS contact terms for the flavor symmetries, by replacing the gauge parameters $\h\sigma_a$ by the flavor parameters $\h m_\alpha$ appropriately.

The remaining contributions to the integrand in  \eqref{ZSb loc formula} are one-loop determinants around the  supersymmetric background with constant $\h\sigma$. The vector multiplet contributes a term:
\be\label{ZS3b vec}
Z_{S^3_b}^{\rm vec}(\h\sigma)= \left(Z^{(0)}_{S^3_b} \right)^{{\rm dim}(\GG)} \; \prod_{\alpha \in \Delta_+}  4 \sinh\left({\pi b \alpha(\h\sigma)}\right)\, \sinh\left({\pi b^{-1} \alpha(\h\sigma)}\right)~.
\ee
Here the product is over the positive roots of $\Fg$.
Note that, in addition to the standard result from \cite{Hama:2011ea, Imamura:2011wg}, we introduced a phase:
\be
Z_{S^3_b}^{(0)} \equiv  \big(Z_{S^3_b}^{\rm RR}\big)^{\half}\,  Z_{S^3_b}^{\rm grav}= e^{-{\pi i \ov 12} \left(b^2+ b^{-2} +3\right)}~,
\ee
which is the contribution $2\delta \kappa_{RR}= \delta \kappa_g= 1$ to the contact terms from each gaugino, consistently with our conventions.

The matter contribution to  \eqref{ZSb loc formula} is given by a product over all the chiral multiplets of gauge and flavor charges $\rho, \omega$, respectively, and $R$-charges $r_\omega$:
\be
Z_{S^3_b}^{\rm matter}(\h\sigma, \h m)=\prod_\omega \prod_{\rho \in \FR} Z_{S^3_b}^\Phi(\rho(\h\sigma)+ \omega(\h m)+ \h\sigma_R r_\omega)~,
\ee
with $Z_{S^3_b}^\Phi$ the partition function for a single chiral multiplet of unit gauge charge, and with $\h\sigma_R$ defined in \eqref{def sigmaR}. The partition for a free chiral multiplet of $R$-charge $r$  can be written as:
\be\label{ZS3b chiral}
Z_{S^3_b}^\Phi(\h\sigma+\h\sigma_R r) = \t\Phi_b\big(\h\sigma+ \h\sigma_R(r-1)\big)~,
\ee
in terms of the quantum dilogarithm  \cite{Faddeev:1995nb}:~\footnote{To be exact, we defined the new function:
	\be
	\t\Phi_b(x)\equiv  \Phi_b(-x)^{-1}~,
	\ee
	with $\Phi_b(x)$ is the standard quantum dilogarithm, as discussed for instance in~\cite{Garoufalidis:2014ifa}. This is partly a matter of convention: $\Phi_b(\h\sigma)$ would be precisely the contribution from a chiral multiplet in the ``$U(1)_\half$ quantization,'' while we are considering the ``$U(1)_{-\half}$ quantization.'' The definition \protect\eqref{def Phib quantum dilog} (with the $(a, q)_\infty \equiv \prod_{k=0}^\infty (1-a q^k)$ the q-Pochhammer symbol) only holds for ${\rm Im}(b^2)>0$, but admits an analytic continuation to more general $b$, and in particular to $b\in \R_{>0}$. The one-loop determinant \protect\eqref{ZS3b chiral} has appeared under various names in the physics literature---it can be conveniently written in terms of the quantum dilogarithm $\Phi_b$~\cite{Faddeev:1993rs, Faddeev:1995nb},  the double-sine function $s_b$~\cite{kurokawa1991}, or the hyperbolic gamma function $\Gamma_h$~\cite{Ruijsenaars, vdbult_thesis}, among other names. }
\be\label{def Phib quantum dilog}
\t\Phi_b(\h\sigma) \equiv \left(e^{-{2\pi\ov b} \h\sigma}\, e^{-\pi i \left({1\ov b^2}+1\right)}; e^{-2\pi i b^{-2}}\right)_\infty  \; \left(e^{-2\pi b\h\sigma}\, e^{\pi i \left(b^2+1\right)}; e^{2\pi i b^{2}}\right)^{-1}_\infty~.
\ee
The field-theory computation actually gives us the formal infinite product \cite{Hama:2011ea}:
\be
Z_{S^3_b}^\Phi(\h\sigma+ \h\sigma_R r) =  \prod_{n_1=0}^\infty  \prod_{n_2=0}^\infty {n_2  b + n_1b ^{-1}- i\h\sigma +i \h\sigma_R(2-r) \ov n_2 b + n_1 b^{-1} +i \h\sigma + i \h\sigma_R r }~,
\ee
which must be regularized. We claim that \eqref{ZS3b chiral} is the correct gauge-invariant regularization, corresponding to the ``$U(1)_{-\half}$ quantization'' scheme and consistent with the parity anomaly \cite{AlvarezGaume:1984nf, Closset:2017zgf}.
To confirm this, we consider the limits:
\be
\t\Phi_b(\h\sigma)\sim e^{-\pi i \h\sigma^2} e^{-{\pi i\ov 12} \left(b^2+ b^{-2}\right)} \quad \text{as}\;\; \h\sigma \rightarrow -\infty~, \qquad \qquad
\t\Phi_b(\h\sigma)\sim 1 \quad \text{as} \;\;\h\sigma \rightarrow \infty~,
\ee
which correspond to integrating out the chiral multiplet of $R$-charge $r=1$ with a large real mass $\h\sigma$. By comparing with \eqref{CS terms S3b}, we see that we generate the gauge and gravitational CS terms at levels $k= -1$  and $k_g=-2$ in the limit $\h\sigma \rightarrow -\infty$, while the theory is trivial in the limit $\h\sigma \rightarrow \infty$. This is exactly as expected. Finally, the contour $\CC_{\h \sigma}$ is defined by a non-compact real $\rk$-dimensional contour that connects $\h \sigma\rightarrow \infty$ and $\h \sigma\rightarrow -\infty$ region, which is properly deformed in such a way that the integral converges.

\subsection{Comparison with the Seifert-manifold formalism}
Setting $b^2= q_1/q_2$, with $q_1, q_2 \in \Z$ some positive integers (for definiteness), we can compare the standard result \eqref{ZSb loc formula} to our new formalism. Let us consider $S^3_b$ as discussed in Section~\ref{subsec: S3b background}, a Seifert fibration of genus zero with two exceptional fibers $(q_1, p_1)$ and $(q_2, p_2)$, with ${\bf d}=0$ (for convenience). We must have:
\be
q_1 p_2 + q_2 p_1=1~.
\ee
We choose the two-dimensional $R$-symmetry line bundle to be trivial, $L_R \cong \CO$. Setting $\t \n_0^R = -1 + \n_0^R=0$ and $\n_1^R= \n_2^R=0$, this gives:
\be
\nu_R = {q_1 + q_2\ov 2}~, \qquad l_0^R= 2~, \quad l_1^R= p_2-p_1-1~, \quad l_2^R= p_1-p_2-1~.
\ee
This allows us to consider any real $R$-charge for the chiral multiplets, $r\in \R$. In our formalism, we have the gauge and flavor parameters $u= i \beta \sigma$ and $\nu=i \beta m$, and the R-symmetry chemical potential $\nu_R$, which are identified with the parameters $\h\sigma$, $\h m$ and $\h\sigma_R$ appearing in \eqref{ZSb loc formula} according to~\cite{Hama:2010av}:
\be\label{sigma to u S3b}
u = i \sqrt{q_1 q_2}\,\h\sigma~, \qquad  \nu = i \sqrt{q_1 q_2}\, \h m~, \qquad 
\nu_R = i \sqrt{q_1 q_2} \h\sigma_R
\ee
One can then write the integrand of \eqref{ZSb loc formula} as:
\be\label{Z to H G S3b}
Z_{S^3_b}(\h \sigma, \h m) =\left(-i  \sqrt{q_1 q_2}\right)^\rk \, e^{-2\pi i \Omega(u, \nu)}\, \t\CG_{S^3_b}(u, \nu)~,
\ee
with the identification \eqref{sigma to u S3b}. The $S^3_b$ fibering operator takes the simple form:
\be\label{G S3b def}
\t\CG_{S^3_b}(u, \nu)= \CG_{1,0}(u, \nu)_{r}\, \t\CG_{q_1, p_1}(u, \nu)\, \t\CG_{q_2, p_2}(u, \nu)~, \qquad q_1 p_2+ q_2 p_1=1~.
\ee
Note that all these fibering operators are evaluated at zero flux, $\n=\m=0$, except for $\CG_{1,0}$ which has a contribution from the effective $R$-symmetry flux $\n^R_0= {l_0^R\ov 2}=1$ (so that $\t \n_0^R=g-1+ \n_0^R=0$), as indicated schematically in \eqref{G S3b def}.
For instance, for the $U(1)$ CS term we have $\CG^{\rm GG}_{1,0}(u)= 1$ and $\CG^{\rm GG}_{q,p}(u)= e^{-\pi i {p\ov q} u^2}$, so that:
\be
\CG^{\rm GG}_{S^3_b}(u)= e^{-\pi i {p_1\ov q_1} u^2}\,  e^{-\pi i {p_2\ov q_2} u^2} = e^{-\pi i {u^2\ov q_1 q_2}}= e^{\pi i \h\sigma^2}~.
\ee
This CS term does not contribute to the effective dilaton.
For the $U(1)_R$ CS term, we find:
\be
e^{-2\pi i \Omega^{\rm RR}}= -1~, \qquad \CG^{\rm RR}_{1,0}= -1~, \qquad \CG^{\rm RR}_{q_1,p_1}\CG^{\rm RR}_{q_2,p_2}= e^{-{\pi i \ov 4}\left({q_1\ov q_2}+{q_2\ov q_1}+2\right)}~.
\ee
Note here that $\CG^{\rm RR}_{1,0}=-1$ is non-trivial, due to the non-zero parameter $l^R_0=2$. That sign cancels the sign from $e^{-2\pi i \Omega^{\rm RR}}$. The other CS terms can be checked similarly. Therefore we find:
\be\label{Z to H G S3b CS}
Z_{S^3_b}^{\rm CS}(\h \sigma, \h m) =  e^{-2\pi i \Omega^{\rm CS}(u, \nu)}\, \CG^{\rm CS}_{S^3_b}(u, \nu)~,
\ee
for the classical contribution \eqref{ZS3b cl full}.
For the vector-multiplet contribution, we have:
\bea
&e^{-2\pi i \Omega_{\rm vector}}=  \prod_{\alpha\in \Delta_+} 4\left(\sin{\pi \alpha(u)}\right)^2~, \cr
&\CG^{\rm vector}_{S^3_b}(u)=\left({1\ov  \sqrt{q_1 q_2}}\right)^\rk\, \left(\CG^{(0)}_{S^3_b}\right)^{{\rm dim}(\GG)}\,  \prod_{\alpha\in \Delta_+}  {\sin\left({\pi \alpha(u) \ov q_1}\right)  \sin\left({\pi \alpha(u) \ov q_2}\right)  \ov \left(\sin{\pi \alpha(u)}\right)^2  }~,
\eea
with:
\be
\CG^{(0)}_{S^3_b}= \CG^{(0)}_{q_1, p_1} \CG^{(0)}_{q_2, p_2}= e^{-{\pi i \ov 12} \left({q_1\ov q_2}+{q_2\ov q_1}-3\right)}~,
\ee
so that:
\be
Z^{\rm vector}_{S^3_b}(\h \sigma, \h m) =\left(-i  \sqrt{q_1 q_2}\right)^\rk\,  e^{-2\pi i \Omega_{\rm vector}(u, \nu)}\, \CG_{S^3_b}^{\rm vector}(u, \nu)~,
\ee
Note the factor of $-i  \sqrt{q_1 q_2}$ for each element of the Cartan, which will be important below. 
Finally, for the chiral-multiplet contribution, we find the nice factorization formula:
\be\label{ZPhiS3b factorization}
Z_{S^3_b}^\Phi(\h\sigma+ \h\sigma_R r)= \pif^\Phi(u + \nu_R r) \,  \t\CG^\Phi_{q_1, p_1}(u+\nu_R r)\,  \t\CG^\Phi_{q_2, p_2}(u+\nu_R r)~, 
\ee
if $b^2={q_1/q_2}$ and $q_1 p_2+ q_2 p_1=1$.
Here the first factor is the product of a contribution from the effective dilaton and from $\CG_{1,0}$:
\be
e^{-2\pi i \Omega^\Phi}= \pif^\Phi(u+ \nu_R r)^{1-r}~, \qquad \CG_{1,0}^\Phi(u+\nu_R r)_r = \pif^\Phi(u+\nu_R r)^r~.
\ee
We give an explicit proof of the factorization formula \eqref{ZPhiS3b factorization} in Appendix~\ref{app:subsec:factorization of S3b}.  The identity  \eqref{ZPhiS3b factorization} is equivalent to some previously-known factorization formula for the quantum dilogarithm at rational values of $b^2$ \cite{Garoufalidis:2014ifa}.

\subsection{Integral formula and its evaluation}
Using the above relations in the case $b^2={q_1\ov q_2}$, we can write the integral formula \eqref{ZSb loc formula} for the $S^3_b$ partition function in the canonical form:
\be\label{ZS3b rat integral}
Z_{S^3_b}(\nu) = {(-2\pi i)^\rk\ov |W_\GG|}\, \int_{\CC_\sigma} \prod_a {du_a\ov 2\pi i} \, e^{- 2\pi i \Omega(u, \nu)}\, \t\CG_{S^3_b}(u, \nu)~,
\ee
with the $\CG_{S^3_b}$ the zero-flux three-sphere fibering operator defined in \eqref{G S3b def}, and $\CC_\sigma$ the $\sigma$-contour defined in Section~\ref{sec:localization}.  
On the other hand, our Bethe-sum formula for the partition function reads:
\be\label{ZS3b evaluation formula}
Z_{S^3_b}(\nu) = \sum_{\h u \in \CS_{\rm BE}} \CH(\h u, \nu)^{-1} \, \CG_{S^3_b}(\h u, \nu)~,
\ee
with $\CG_{S^3_b}$ the {\it full} fibering operator for the squashed three-sphere, including the sum over fractional fluxes:
\be
\CG_{S^3_b}(u, \nu)= \CG_{1,0}(u, \nu)_{r}\,\sum_{\n_1=0}^{q_1-1}  \CG_{q_1, p_1}(u, \nu)_{\n_1}\, \sum_{\n_2=0}^{q_2-1} \CG_{q_2, p_2}(u, \nu)_{\n_2}~.
\ee
This gives an explicit evaluation formula for the squashed-sphere partition function \eqref{ZS3b rat integral} of any $\CN=2$ supersymmetric gauge theory.~\footnote{Note that, while we had to restrict the choice of gauge group on a general $\CM_3$ (to be simply connected or unitary), the three-sphere partition function is insensitive to the global structure of $\GG$.}

\paragraph{Direct computation.} One can check that \eqref{ZS3b rat integral} equals \eqref{ZS3b evaluation formula}  by an explicit computation. Here we consider the case $\GG= U(1)$, for simplicity. Let us write \eqref{ZS3b rat integral} as:
\be
Z_{S^3_b} = -\int_{i \R} d u  \, e^{-2\pi i \Omega(u)} \t\CG_{S^3_b}(u) {1- \pif(u)\ov 1-\pif(u)}~,
\ee
where we suppressed the dependence on $\nu$ to avoid clutter, and we introduce a trivial factor $1={x\ov x}$ in the integrand. The contour is taken along the imaginary axis (for appropriate choices of the flavor parameters). Now, using the difference equation:
\be
\t\CG_{S^3_b}(u- q_1 q_2)=  \pif(u) \t \CG_{S^3_b}(u)~,
\ee
we obtain: 
\be
Z_{S^3_b} =- \left(\int_{i \R}  - \int_{(i \R -q_1 q_2)}\right)du \, {e^{-2\pi i \Omega(u)} \t\CG_{S^3_b}(u)\ov 1-\pif(u)} =  \int_{\t\CC_{\rm BE}}  du \, {e^{-2\pi i \Omega(u)}\t \CG_{S^3_b}(u)\ov \pif(u)-1}~.
\ee
One can argue that the contour $\t\CC_{\rm BE}$ encloses all the poles at $\pif(u)=1$ located in the strip Re$(u)\in [0, q_1 q_2)$. We then find:
\be\label{ZS3b to BS intermed}
Z_{S^3_b} = \sum_{\h u \in \CS_{\rm BE}} \sum_{l=0}^{q_1 q_2} \CH(\h u - l)^{-1}  \t\CG_{S^3_b}(\h u-l)=  \sum_{\h u \in \CS_{\rm BE}}\CH(\h u)^{-1}  \sum_{l=0}^{q_1 q_2}  \t \CG_{S^3_b}(\h u-l)~.
\ee
In the last equation, we used the fact that the handle-gluing operator is periodic under $u\sim u+1$. Finally, one can check that the contribution from the fibering operator factorizes as:
\bea
\sum_{l=0}^{q_1 q_2-1}   \t\CG_{S^3_b}(\h u-l)&=  \sum_{l_1=0}^{q_1-1}  \t\CG_{q_1, p_1}(\h u-l_1)\; \sum_{l_2=0}^{q_2-1}  \t\CG_{q_2, p_2}(\h u-l_2)\\
&= \sum_{l_1=0}^{q_1-1} \CG_{q_1,p_1}(\h u)_{p_1l_1}\sum_{l_2=0}^{q_2-1} \CG_{q_2,p_2}(\h u)_{p_2l_2}\\
&=\sum_{l_1=0}^{q_1-1} \CG_{q_1,p_1}(\h u)_{l_1}\sum_{l_2=0}^{q_2-1} \CG_{q_2,p_2}(\h u)_{l_2}\ .   
\eea
Here we used the transformation property of $\CG_{q,p}(u)$ under the large gauge transformation and the Bethe equation $\Pi( \h u)=1$. Written in this way,  \eqref{ZS3b to BS intermed} becomes equivalent to the formula \eqref{ZS3b evaluation formula}.

\paragraph{Further comments.}
It is interesting to note that, in some very special cases, this evaluation formula \eqref{ZS3b evaluation formula} has appeared before in a different context.  Namely, in complex Chern-Simons theory, the integral  \eqref{ZSb loc formula} for an abelian gauge group (in the supersymmetric language)  is known as a ``state integral'' \cite{2007JGP....57.1895H, Dimofte:2009yn}, and it has been studied in the literature in parallel to the development of 3d $\CN=2$ localization methods. (This apparent coincidence between the two subjects is explained by the 3d/3d correspondence \cite{Dimofte:2011ju} in string theory.)   The state integral corresponding to an $U(1)_k$ supersymmetric CS theory coupled to $N_f$ chiral multiplets of unit charge was given an evaluation formula for $b^2$ rational equivalent to \eqref{ZS3b evaluation formula}  in \cite{Garoufalidis:2014ifa}.~\footnote{More precisely, Theorem 1.1 of \protect\cite{Garoufalidis:2014ifa} is equivalent to \protect\eqref{ZS3b evaluation formula} in that case. Their state integral $\CI_{A,B}$ corresponds to a $U(1)_k$ theories with $N_f=B$ flavors and effective CS level $k= \half B-A$.}

The evaluation formula \eqref{ZS3b evaluation formula} renders manifest a number of properties of $Z_{S^3_b}$ which are less than obvious from the integral expression. One property is that, for any theory without any supersymmetric vacuum,  $Z_{S^3_b}=0$. This follows from \eqref{ZS3b evaluation formula} and the fact that the number of Bethe vacua---that is, the mass-regulated Witten index \cite{Intriligator:2013lca}---is zero in that case. For instance, for  $U(N_c)$ SQCD with $N_f< N_c$ flavors, the Witten index \eqref{witten index SQCD} vanishes and so does $Z_{S^3_b}$. Another property which is obvious from \eqref{ZS3b evaluation formula} is that supersymmetric Wilson loops wrapping generic Seifert fibers~\footnote{In  this case, these correspond to $(q_1,q_2)$-torus knots on $S^3_b$, for $b^2=\frac{q_1}{q_2}$ as above.} satisfy the correct twisted chiral ring relations \cite{Kapustin:2013hpk, Closset:2016arn}.

\section{The refined twisted index}\label{sec:s2s1top}
Next we consider the refined topologically twisted index of \cite{Benini:2015noa}, computed as a supersymmetric partition function on the supersymmetric background $S_\epsilon^2 \times S^1$ discussed in Section~\ref{subsec: S2S1 back}.  In particular we have the metric \eqref{met S2S1}, where the ``refinement'' parameter:
\be
\epsilon \in \C
\ee
plays the role of a chemical potential for the azimuthal momentum on $S^2$. 
Importantly, there is a non-trivial $U(1)_R$ flux across the $S^2$:
\be
\m_R \equiv {1\ov 2\pi}\int_{S^2} dA^{(R)}= -1~.
\ee
This implies that the $R$-charges must be integer-quantized, $r\in \Z$. We also introduce a $U(1)_R$ flat connection along the $S^1$:
\be
v_R \equiv -{1\ov 2 \pi} \int_{S^1} A^{(R)}~, \qquad  {\rm with} \quad v_R \in \half \Z~,
\ee
which is correlated with a choice of spin structure on $S^2 \times S^1$.
If $v_R=0$ mod $1$, we choose the periodic boundary condition for fermions along $S^1$, while if $v_R=\half$ mod $1$, we choose the anti-periodic boundary condition. Thus, the fugacity $v_R$ introduces a further $\Z_2$ refinement of the twisted index by the choice of spin structure.
Following \cite{Benini:2015noa, Closset:2015rna}, we can compute the supersymmetric partition function on this background by supersymmetric localization:~\footnote{This is the result derived in \protect\cite{Benini:2015noa}, after taking into account our conventions for quantizing fermions, and  keeping track of the spin structure dependence. The parameters $v$ and $\epsilon$ here correspond to ${u\ov 2 \pi}$ and ${\varsigma\ov 2\pi}$, respectively, in Section 4 of \protect\cite{Benini:2015noa}.}\footnote{In this section, the parameters $\m$ and $\m^F$ denote gauge and flavor fluxes, respectively, unlike in previous sections. We hope this will cause no confusion.}
\be\label{Zepsilon full}
Z_{\epsilon}(v_F)_{\m_F} ={1\ov |W_\GG|} \sum_{\m\in \Gamma_{\GG^\vee}} \oint_{\rm JK} \prod_{a=1}^\rk {dv_a \ov 2 \pi i} \; \CZ_\epsilon(v, v_F)_{\m, \m_F}~,
\ee
with the integrand:
\be\label{CZeps gen}
\CZ_\epsilon(v, v_F)_{\m, \m_F}= (-2 \pi i)^\rk \,Z^{\rm CS}_\epsilon(v, v_F)_{\m, \m_F} \, Z^{\rm vector}_\epsilon(v)_{\m} \, Z^{\rm matter}_\epsilon(v, v_F)_{\m, \m_F}
\ee
Here and in the following, $v$ denotes the gauge parameters to be integrated over, $v_F$ denotes the flavor chemical potentials, $\m_F$ denotes the  flavor background fluxes, and we leave the dependence on the $U(1)_R$ parameter $v_R$ implicit to avoid clutter. In \eqref{Zepsilon full}, the sum is over the gauge fluxes $\m_a$, and the $v$-integral is a particular middle-dimensional contour integral in the strip-like region:
\be 
{\rm Re}(v_a) \in [0,1)~,
\ee
that implements a modified JK residue prescription---see \cite{Benini:2015noa, Benini:2016hjo, Closset:2016arn}. The integrand is periodic under $v_a \sim v_a+1$,  corresponding to large gauge transformations along the $S^1$ in $S^2_\epsilon \times S^1$.

The classical contribution to the integrand is given by the CS contributions, of the schematic form:
\be
Z_{\epsilon}^{\rm CS}(v)_\m = Z_{\epsilon}^{\rm GG}(v)_\m^{k_{GG}}\; 
Z_\epsilon^{\rm G_1G_2}(v_1, v_2)_{\m_1, \m_2}^{k_{G_1 G_2}}\; 
Z_\epsilon^{\rm GR}(v)_\m^{k_{GR}}\; (Z_\epsilon^{\rm RR})^{k_{RR}}~,
\ee
with:
\bea
& Z_\epsilon^{\rm GG}(v)_\m= (-1)^{\m(1+ 2v_R)} e^{2\pi i v \m}~, \qquad && Z_{\epsilon}^{\rm G_1G_2}(v_1,v_2)_{\m_1, \m_2} = e^{2\pi i (v_1 \m_2 + v_2 \m_1)}~, \cr
& Z_\epsilon^{\rm GR}(v)_\m=(-1)^{2 v_R \m}\, e^{-2\pi i v}~, \qquad && Z_\epsilon^{\rm RR}= -1~.
\eea
The gravitational CS term is trivial on this background.
Note the spin-structure dependence of the gauge CS term, corresponding to $v_R=0$ or $v_R=\half$.~\footnote{The RR CS term is similar, but in that case $v_R$ enters both as the choice of spin structure and as the $U(1)_R$ fugacity, and the effect cancels out---we have $Z_\epsilon^{RR}= (-1)^{\m_R (1+ 2v_R)}e^{2\pi i v_R \m_R}=(-1)^{\m_R}=-1$.}
The vector multiplet contribution to the integrand \eqref{CZeps gen} is given by:
\be
Z^{\rm vector}_\epsilon(v)_{\m}= \prod_{\alpha\in \Delta^+} (-1)^{2 v_R \alpha(\m)} \sin\left(\pi \big(\alpha(v)- {\epsilon\ov 2}\alpha(\m)\big)\right)\,\sin\left(\pi \big(\alpha(v)+ {\epsilon\ov 2}\alpha(\m)\big)\right)~.
\ee
The matter contribution to  \eqref{CZeps gen}, as usual, is a product over the chiral multiplets of the theory:
\be
Z^{\rm matter}_\epsilon(v, v_F)_{\m, \m_F}= \prod_\omega \prod_{\rho \in \FR} Z_\epsilon^\Phi(\rho(v) + \omega(v_F) + v_R r_\omega)_{\rho(\m) + \omega(\m_F) -r_\omega +1}~.
\ee
Here we defined the function:
\be\label{Zeps v def}
Z_\epsilon^\Phi(v)_\m \equiv \left(e^{2\pi i \left(v- {\epsilon\ov 2} \m\right)}; e^{2\pi i \epsilon} \right)_{\m+1}^{-1}~,
\ee
in terms of the $q$-Pochhammer symbol $(x; q)_n$, defined in \eqref{qpochn}.
which is the contribution of a chiral multiplet of $U(1)$ gauge charge $1$ and $R$-charge $r=0$. For a single chiral multiplet of $R$-charge $r\in \Z$, in particular, we have the contribution:
\be\label{Zeps Phi gen r}
Z_\epsilon^\Phi(v+ v_R r)_{\m-r} = \left( (-1)^{2 v_R r} e^{2\pi i \left(v- {\epsilon\ov 2} (\m-r)\right)}; e^{2\pi i \epsilon} \right)_{\m-r+1}^{-1}~,
\ee
including a subtle spin-structure dependence through $v_R$ when $r$ is odd. As a consistency check on this result, we should consider the decoupling limits $v \rightarrow \pm i \infty$. For a chiral multiplet of $R$-charge $r=1$, we find:
\bea
&Z_\epsilon^\Phi(v+ v_R)_{\m-1} \sim Z_\epsilon^{\rm GG}(v)_\m^{-1}  \qquad&& \text{as}\;\; v \rightarrow -i\infty~, \cr
&Z_\epsilon^\Phi(v+ v_R)_{\m-1} \sim 1 \quad &&\text{as} \;\;v \rightarrow i\infty~.
\eea
In the limit $v\rightarrow -i \infty$, we thus generate a gauge CS term $k=-1$, as expected in the $U(1)_{-\half}$ quantization. Note that, for $\epsilon=0$, the chiral multiplet contribution simplifies to:
\be
Z_{\epsilon=0}^\Phi(v+ v_R r)_{\m-r} =\left({1\ov 1-e^{2\pi i (v+ v_R r)}}\right)^{\m-r+1}~,
\ee
and the supersymmetric partition function \eqref{Zepsilon full} becomes the ordinary genus-zero twisted index $Z_{S^2\times S^1}$, given by \eqref{twisted index expl} with $g=0$. (In that $\epsilon=0$ limit, we can identify the parameters $v, v_F, v_R$ here with $u, \nu_F, \nu_R$.)

\subsection{Comparison to the Seifert fibration result for $\epsilon$ rational}
As we explained in Section~\ref{subsec: S2S1 back},  the $S^2_\epsilon \times S^1$ geometry is a Seifert fibration over the orbifold $S^2(q,q)\cong S^2/\Z_q$ if and only if the deformation parameter $\epsilon$ is rational, with:
\be\label{eps tq comp}
\epsilon = {t\ov q}~, \qquad\qquad \gcd(q,t)=1~.
\ee
The Seifert fibration is then:
\be\label{S2S1 M3 again}
S^2_\epsilon \times S^1 \cong \big[0~; \, 0~; \, (q, p)~, (q, -p) \big]~, \qquad \quad q s+ p t=1~.
\ee
In the Seifert description, we have the JK-residue formula \eqref{almost final 2}, which gives:
\be\label{Zepsilon Seifert full}
Z_{\epsilon}(\nu)_{\n^F} ={(-1)^\rk\ov  |W_\GG|}  \sum_{\n_0 \in \Gamma_{\GG^\vee}}\sum_{\n_1, \n_2 \in \Gamma_{\GG^\vee(q)}} \oint_{\CC_0(\eta)} d^\rk u\;   e^{-2 \pi i \Omega(u, \nu)} \, \CG_{\epsilon}(u, \nu)_{\n_0, \n_1, \n_2, \n^F}~.
\ee
Here, $\n_0\equiv \n_{0,a}$ and $\n_1\equiv \n_{1, a}, \n_2\equiv \n_{2,a}$ denote the ordinary fluxes and the fractional fluxes at the two $\Z_q$ orbifold points, respectively, and similarly for the flavor background fluxes $\n^F$. The contribution $\CG_\epsilon$ to the integrand of \eqref{Zepsilon Seifert full} is the fibering operator for the Seifert fibration \eqref{S2S1 M3 again} before performing the sum over fractional gauge fluxes:
\be
\CG_{\epsilon}(u, \nu)_{\n_0, \n_1, \n_2, \n_F} \equiv  \CG_{1,0}(u, \nu)_{\n_0, \n_{0}^F} \;  \CG_{q,p}(u, \nu)_{\n_1, \n_{1}^F}\;  \CG_{q,-p}(u, \nu)_{\n_2, \n_{2}^F}~.
\ee
The complete fibering operator is:
\be\label{CGeps full}
\CG_{\epsilon}(u, \nu)_{\n_0, \n_F} = \sum_{\n_1, \n_2 \in \Gamma_{\GG^\vee(q)}}  \CG_{\epsilon}(u, \nu)_{\n_0, \n_1, \n_2, \n_F}~,
\ee
and \eqref{Zepsilon Seifert full} can then be written as:
\be\label{Zeps n0sum only}
Z_{\epsilon}(\nu)_{\n^F} ={(-1)^\rk\ov  |W_\GG|}  \sum_{\n_0 \in \Gamma_{\GG^\vee}}\oint_{\CC_0(\eta)} d^\rk u\,  \; e^{-2 \pi i \Omega(u, \nu)} \, \CG_{\epsilon}(u, \nu)_{\n_0, \n_F}~.
\ee
We note the important property that the integrand in \eqref{Zepsilon Seifert full} is periodic under $u \sim u+q$. More precisely, the effective dilaton contribution is invariant under $u\sim u+1$, while we have:
\be
\CG_{\epsilon}(u+q, \nu)_{\n_0, \n_1, \n_2, \n_F}=  \CG_{\epsilon}(u, \nu)_{\n_0, \n_1, \n_2, \n_F}~,
\ee
and similarly for $\nu$, due to the fact that $c_1(\CL_0)=0$.

\paragraph{Back to the Bethe sum.}  By summing over $\n_0$ in \eqref{Zeps n0sum only} like  in  \eqref{summation over n0}, and picking the poles at the solutions to the Bethe equations, we recover the Bethe-sum formula:
\be
Z_{\epsilon}(\nu)_{\n^F} = \sum_{\h u \in \CS_{\rm BE}}  \CG_\epsilon(u, \nu)_{\n^F} \, \CH(\h u, \nu)^{-1}~,
\ee
 where $\CG_\epsilon(u, \nu)_{\n^F}$ is the full Seifert fibration \eqref{CGeps full} at $\n_0=0$.

\paragraph{Comparison to the integral formula \eqref{Zepsilon full} for $\epsilon$ rational.}
One can compare \eqref{Zepsilon Seifert full} to \eqref{Zepsilon full} and find perfect agreement at $\epsilon= {t\ov q}$, given the following identification of the parameters:
\bea\label{v to u S2S1}
&v={u\ov q}- {t\ov 2 q} (\n_1 - \n_2) + {t \n_0 \ov 2}~, \qquad && \m = q\n_0 + \n_1 + \n_2~,\cr
&v_F={\nu\ov q}- {t\ov 2 q} (\n^F_1 - \n^F_2) + {t \n^F_0 \ov 2}~, \qquad  &&\m^F = q\n^F_0 + \n^F_1 + \n^F_2~,\cr
&v_R={\nu_R\ov q}- {t\ov 2 q} (\n_1^R - \n_2^R) + {t \t\n_0^R \ov 2}~, \qquad&& \m^R = q\t \n_0^R + \n_1^R + \n_2^R~.
\eea
The matching of the fluxes between the two descriptions is clear from:
\be
\pi^\ast(L_0^{\n_0} L_1^{\n_1} L_2^{\n_2}) = (q \n_0 + \n_1 + \n_2) [\Omega]~.
\ee
The matching between the continuous parameters $v$ and $u$ is obtained by comparing the supersymmetric Wilson loops in the two descriptions \cite{Benini:2015noa}.
Note that the parameters $v, v_F, v_R$ are defined modulo $1$. Using the parameterization \eqref{nR S2 S1} for the $U(1)_R$ fluxes, with $\t\n^R_0= -1 + \n_0^R$, we find:
\be
v_R= \nu_R s + {t\ov 2}(l^R+1)  \qquad \qquad \m_R= -1~.
\ee
Here $l_R\equiv l_1^R= l_2^R$ mod $2$, where we used the fact that the parities of $l_1^R$ and $l_2^R$ must be equal for the line bundle $L_R$ to be well-defined. As a consistency check, it is interesting to note that:
\be
(-1)^{2 v_R}= (-1)^{2\nu_R s + t (l^R+1)}= \pm 1~,
\ee
where, for any fixed $(q,p)$, the two signs correspond to the two distinct choices of $L_R$ given by \eqref{LR S2S1 cases}. By a direct computation with the identifications \eqref{v to u S2S1}, we can check that:
\be\label{Zeps to Seifert id}
\CZ_{\epsilon={t\ov q}}(v, v_F)_{\m, \m_F} =(-2\pi i q)^\rk\,  e^{-2 \pi i \Omega(u, \nu)} \, \CG_{\epsilon}(u, \nu)_{\n_0, \n_1, \n_2, \n^F}~,
\ee
where the left-hand-side is the integrand \eqref{CZeps gen} evaluated at rational values of $\epsilon$.
  For instance, for a chiral multiplet, we have a ``factorization'' formula for \eqref{Zeps Phi gen r}, according to:
\bea\label{Zeps Phi factor}
&Z_{\epsilon={t\ov q}}^\Phi(v+ v_R r)_{\m-r} =\\
&\qquad \pif^\Phi(u +\nu_R r)^{\n_0+ 1- r+ \n_0^R r} \, \CG_{q,p}^\Phi(u+ \nu_R r)_{\n_1 + \n_1^R r} \, \CG_{q,-p}^\Phi(u+ \nu_R r)_{\n_2 + \n_2^R r}~.
\eea
Note that the left-hand-sides of \eqref{Zeps to Seifert id} and \eqref{Zeps Phi factor} only depend on $v$ and $\m$, and not on the fractional fluxes $\n_1$ and $\n_2$ individually.

We can now show that the Seifert result agrees with the localization result  \eqref{Zepsilon full}. Consider again the case $\GG= U(1)$, for simplicity. Using the identifications \eqref{v to u S2S1} and \eqref{Zeps to Seifert id}, we can  write  \eqref{Zepsilon full} as:
\be\nn
Z_{\epsilon}(v) = \sum_{\m \in \Z} \oint_{[0,q)} {du\ov 2\pi i q}\, \CZ_{\epsilon={t\ov q}}(v)_{\m}=
-\sum_{\n_0 \in \Z} \sum_{\n_1' \in \Z_q}\sum_{j \in \Z_q} \oint_{[0,1)}du\, e^{-2\pi i \Omega(u)} \CG_\epsilon(u-j)_{\n_0, \n_1', 0}~.
\ee
Here, the first integral is over the strip ${\rm Re}(u) \in [0,q)$, which is rewritten as a sum of integrals over the strip  ${\rm Re}(u) \in [0,1)$. Then, using \eqref{diff eq CG}, we may write:
\be \CG_\epsilon(u-j)_{\n_0, \n_1', 0} =  \CG_\epsilon(u)_{\n_0, \n_1'+p j, -p j} \ee
and we then find:
\be\label{Zeps intermed}
Z_{\epsilon}(v) = - \sum_{\n_0 \in \Z} \sum_{\n_1, \n_2 \in \Z_q} \oint_{[0,1)} du\; e^{-2\pi i \Omega(u)} \CG_\epsilon(u)_{\n_0, \n_1, \n_2}~.
\ee
This shows the equality of \eqref{Zepsilon full} with  \eqref{Zepsilon Seifert full}, whenever $\epsilon$ is rational. This proof is easily generalized to any gauge group $\GG$.~\footnote{Up to the unresolved difficulties in defining the correct JK-like contour.}

\section{Lens spaces and holomorphic blocks}
\label{sec:HB}

In this final section, we revisit the general lens spaces, $L(p,q)$.  We will recover some of the above results on $S^3_b$ and $S_\epsilon^2 \times S^1$ as special cases, and connect our results to general lens space partition functions.

As we already mentioned, three-dimensional lens spaces are rather special amongst Seifert manifolds, since they are the only three-manifolds that admit an infinite number of Seifert fibrations. 
The lens space $L(p,q)$ is defined as a $\Z_p$ quotients of the three-sphere:
\be \label{HBlpqdef} 
L(p,q) \cong S^3/\Z_p~, \qquad \qquad  \Z_p \, : \,\big(z_1~,\;  z_2\big) \sim \big(\omega_p^q \, z_1~,\;\omega_p\, z_2\big)~,\qquad \omega_p \equiv e^{2\pi i \ov p}~,
\ee
where we view $S^3$ as the set:
\be\label{label S3 in C2}
S^3 \cong \{ (z_1, z_2)\in \C^2 \; |\; |z_1|^2+|z_2|^2 =1\}~.
\ee
For $\gcd(p,q)=1$, this identification defines a free $\Z_p$ action and $L(p,q)$ is a smooth manifold, which depends only on $q$ modulo $p$.  For later convenience, we formally define:
\be 
L(0, \pm 1) = S^2 \times S^1~.
\ee
We described the $L(p,q)$ supersymmetric backgrounds in Section \ref{sec:lpqgeo}---see also Appendix~\ref{app: S3 and lens}.
The partition function of 3d $\cN=2$ gauge theories on spheres and lens spaces has been extensively studied in the literature---{\it e.g.} in \cite{Kapustin:2009kz,Jafferis:2010un,Hama:2011ea,Benini:2011nc,Alday:2012au,Beem:2012mb,Benini:2015noa, Nieri:2015yia};  see the reviews \cite{Willett:2016adv, Pasquetti:2016dyl}  and references therein.  
Lens spaces admit a family of supersymmetric backgrounds preserving two supercharges, which can be characterized by a complex ``squashing parameter,''  $b$ \cite{Hama:2011ea,  Closset:2013vra}.  To see how $b$ appears, let us introduce some angular coordinates on $L(p,q)$, with: 
\be 
z_1 = \sin{\theta\ov 2} e^{i \chi}, \;\;\; z_2 = \cos{\theta\ov 2} e^{i \varphi}~, \qquad
\big(\chi~,\; \varphi\big) \, \sim \,\big(\chi+ {2 \pi q\ov p}~,\; \varphi+ {2\pi \ov p}\big)~.
\ee
In this description, the covering space $S^3$ is viewed as a torus, with angles $(\chi, \varphi)$, fibered over the interval $\theta\in [0, \pi]$.  The lens space admits a metric where $\partial_\chi$ and $\partial_\varphi$ are generators of a $U(1) \times U(1)$ isometry.  Then, in the supersymmetric background $L(p,q)_b$ with squashing $b$, the anti-commutator of the two supercharges is an isometry along a Killing vector $K^{(b)}$:
\be \label{HBkdef}
\{ \CQ, \b \CQ\} =-2i \big(\CL_K^{(b)}+ Z\big)~,  \qquad K^{(b)} = b\; \partial_\varphi + b^{-1} \; \partial_\chi~,
\ee
where $\CL_{K^{(b)}}$ is the Lie derivative along $K^{(b)}$.  For generic $b$, the orbits of the $U(1)$ isometry generated by $K^{(b)}$ do not close, except at $\theta=0$ and $\theta= \pi$, where $\partial_\varphi$ and $\partial_\chi$, respectively, vanish.  However, in the special case where:
\be\label{rat squash}
b^2 = {q_1\ov q_2} \in \bQ~,
\ee
then all of the orbits close.  In this case, the orbits of the real Killing vector $K^{(b)}$ define a Seifert fibration with base $S^2(q_1, q_2)$, which generically has two exceptional fibers at the ``poles'' $\theta=0$ and $\theta=\pi$.

In the following, we start by reviewing earlier computations of supersymmetric partition functions on squashed spheres and lens spaces.  Interestingly, these partition functions are known to factorize into pairs of ``holomorphic blocks,'' or $D^2 \times S^1$ partition functions, reflecting the genus-one Heegaard splitting of the lens space into two solid tori~\cite{Pasquetti:2011fj, Beem:2012mb, Dimofte:2014zga, Nieri:2015yia}.   
We then explain how the holomorphic blocks can be directly related to the fibration operators, by considering a singular limit on the blocks. 
Using this correspondence, we demonstrate that,  in the case of rational squashing \eqref{rat squash}, the partition functions computed in the Seifert formalism reproduce the earlier computations on squashed spheres and lens spaces. Along the way, we will also clarify a number of points, and obtain some new explicit results for the $L(p,q)$ partition functions.

\subsection{Holomorphic blocks and lens space partition functions}
\label{sec:HBblockbackground}

Let us start by reviewing the definition of the holomorphic blocks \cite{Beem:2012mb}, and their role as building blocks of lens space partition functions.  This subsection is mostly a review, but we also clarify some details of the construction, in particular concerning the role of the R-symmetry background, which leads to some new observations.

\subsubsection{Definition and properties}
The holomorphic block of a $3d$ $\cN=2$ gauge theory is the partition function of the theory on a disk, $D^2$, fibered over a circle.  Specifically, one considers a space $D^2 \times_\tau S^1$ with a smooth metric of the form:
\be ds^2 = dr^2 + f(r)^2\left( \tau_2(r)^{-1}(d \varphi^2 + \tau_1(r) d\zeta)^2 + \tau_2(r) d\zeta^2\right)~.
\ee
Here, $(r, \varphi)$ are the disk coordinates and $\zeta\in [0, 2\pi)$ is the $S^1$ coordinate. The functions $\tau_1(r), \tau_2(r)$ become the constants $\tau_1, \tau_2$ at the boundary, $r=r_0$, while   $f(r_0)=1$. We then have a complex structure $\tau= \tau_1 + i \tau_2$ on the boundary torus, with complex coordinate $w= \varphi+ \tau \zeta$.
We must also perform a topological twist along $D^2$, so that the R-symmetry gauge field has flux $-\frac{1}{2}$ through the disk.  We will discuss the R-symmetry gauge field in more detail below.  As a consequence of the twist, the partition function is independent of the metric on $D^2$.  On the other hand, the space $S^1 \times_\tau D^2$ has a torus boundary, $T^2$, and we must specify some data at this boundary. The partition function on the solid torus $S^1 \times_\tau D^2$ will depend on this data, which consists of:
\begin{itemize}
	\item A choice of two-dimensional vacuum ({\it i.e.} a Bethe vacuum) on the disk, $\alpha \in \cS_{BE}$, which fixes the asymptotic behavior of the fields at the boundary.
	\item The complex structure parameter, $\tau$, of the boundary torus.
	\item The holonomy of background gauge fields, $A_F$, coupled to global symmetries.  Specifically, we define:
	\be \label{HBnudef} \nu = \oint_{S^1} A_F + \tau \oint_{\partial D^2} A_F~. \ee
	Here $\nu$ lives in the complexified Cartan subalgebra of the flavor symmetry group.  For simplicity of notation, we will focus on the case of a rank-one flavor symmetry group, $U(1)_F$, so that $\nu \in \C$, but the general case is a straightforward extension.
	\item The holonomy of the $U(1)_R$ background gauge field:
	\be\label{nuR hb} 
	\nu^R = \oint_{S^1} A_{(R)} + \tau \oint_{\partial D^2} A_{(R)}~,
	\ee
	and a corresponding choice of spin structure on $D^2 \times_\tau S^1$. We will discuss this point momentarily.
\end{itemize}
Given this data, the holomorphic block is a locally-holomorphic function:
\be\label{holo block Balpha}
B^\alpha(\nu,\tau)~,
\ee
The $\nu^R$ dependence  is kept implicit. 
Note that  \eqref{holo block Balpha} is neither a modular nor an elliptic function, since those symmetries of the boundary torus are broken by the fact that only one of its cycles is filled.  However, there is a residual symmetry under the independent shifts:
\be
\tau \rightarrow \tau +1~, \qquad \nu \rightarrow \nu+ 1~,
\ee
with the latter corresponding to large gauge transformations of $A_F$ along the $S^1$.  Thus, it is sometimes convenient to use the single-valued parameters:
\be
\q \equiv e^{2 \pi i \tau}~,\qquad\qquad  y \equiv e^{2 \pi i \nu}~. 
\ee

Note that shifting $\nu \rightarrow \nu + \n \tau$, or $y \rightarrow \q^\n y$, for any $\n \in \Z$, corresponds to a large gauge transformation on the boundary torus, but is not a symmetry of the block.  From \eqref{HBnudef} and Stoke's theorem, this is equivalent to a shift of the flux of $A^F$ through the disk by $\n$ units.  Thus we define:
\be \label{HBfluxdef} 
B^\alpha(\nu,\tau)_\n \equiv  B^\alpha(\nu+\n \tau,\tau) 
\ee
to be the block with $\n$ units of flux.  Here we have made an arbitrary choice of the zero of $\nu$, which can always be redefined by $\nu \rightarrow \nu + \alpha + \beta \tau$, with $\alpha, \beta \in \Z$. This freedom will be useful below. 

The $R$-charge dependence of the holomorphic blocks can be discussed similarly. In the following discussion, we assume that the $R$-charges are integer-quantized. (Later on, we will be able to relax this restriction on some compact three-manifolds.)
We have the background $U(1)_R$ gauge field $A_R$, and the corresponding parameter $\nu^R$ defined by \eqref{nuR hb}, similarly to the flavor parameter $\nu$. Naively, we have $\nu^R = -{\tau\ov 2}$ because the topological twist on $D^2$ introduces $-\half$ unit of $U(1)_R$ flux. More generally, we should consider:
\be
\nu^R = \t\nu^R- \half \tau~, \quad\qquad \t\nu^R \in \half \Z~.
\ee
Here, the parameter $\t\nu^R \in \half \Z$ mod $1$ gives a $\Z_2$ valued holonomy of $A_R$ through the $S^1$:
\be
e^{- i \int_{S^1} A_R}= (-1)^{2\t\nu^R}~.
\ee 
Similarly to the discussion in previous sections, in order to preserve supersymmetry, $\t\nu^R$ must be correlated with a choice of spin structure on $D^2 \times_\tau S^1$. If $\t\nu^R$ is an integer, we choose the periodic spin structure for fermions around the $S^1$; if $\t\nu^R$ is half-integer, we choose the anti-periodic spin structure.~\footnote{The parameters $\nu^R$ and $\t\nu^R$ here are closely related but distinct from the parameter $\nu_R$ on Seifert backgrounds. We will discuss the relation later in this section.}

If we mix the R-symmetry current with the flavor symmetry current according to:
\be
j_\mu^{(R)} \rightarrow j_\mu^{(R)}+  r j_\mu^F~,
\ee
this has the effect of shifting the $U(1)_F$ gauge field according to  $A_F \rightarrow A_F+   r A_R$.
The R-charge dependence of the blocks then appears through the shift:
\be\label{shift nu to nuR r}
\nu \rightarrow \nu +\nu^R r \qquad\quad
\leftrightarrow \qquad\quad y \rightarrow y \; (-1)^{2\tilde{\nu}^R r}\, \q^{-{r\ov 2}}~.
\ee
For $r \in 2 \Z$, this depends only on $\q$, or equivalently, on the choice of $\tau$ mod $1$, but for more general $r$ the symmetry under $\tau \rightarrow \tau+1$ is partially broken.  Namely, for general $r \in \Z$, we only have $\tau \sim \tau+2$. If we allow for general $r \in \R$, all shifts of $\tau$ are inequivalent.  We will discuss the R-charge dependence of the blocks more carefully below.

\subsubsection{Explicit construction of the blocks for a gauge theory}
To construct the blocks for a given $3d$ $\cN=2$ supersymmetric gauge theory, we first specify the contributions of the chiral multiplets and the Chern-Simons terms. Since we are not yet gauging any symmetry, we are considering a theory with a single vacuum and we can omit $\alpha\in \CS_{\rm BE}$ from the notation.

\paragraph{Chiral multiplets.} 
Let us introduce the function:
\be\label{Bphi def}
B^\Phi(\nu,\tau)_\n \equiv  \big(\q^{1 +\n} y; \q\big)_\infty~,
\ee
defined in terms of the (extended) $\q$-Pochhammer symbol:
\be\label{def qpochh}
(x;\q)_\infty \equiv  \begin{cases}
	\prod_{j=0}^\infty(1-x \q^j) \qquad&\text{if}\; \;\text{Im}(\tau)>0~, \\
	\prod_{j=0}^\infty(1-x \q^{-j-1})^{-1} \qquad&\text{if}\; \;\text{Im}(\tau)<0~. 
\end{cases}
\ee
The expression \eqref{Bphi def} is the contribution from a chiral multiplet in the $U(1)_{-\half}$ quantization, with unit charge under $U(1)_F$ and R-charge $r=0$, including also $\n$ units of flux through the disk.  A chiral multiplet of $R$-charge $r\in \Z$ contributes:~\footnote{For general $r$, we define $\q^{\frac{r}{2}} = e^{\pi i \tau r}$ and $(-1)^{2\t\nu^R r}= e^{2\pi i \t\nu^R r}$.}
\be \label{HBchiblock} 
B^\Phi(\nu+ \nu^R r, \tau)_{\n} = \big((-1)^{2\t\nu^R r}\q^{1 - {r\ov2} + \n} y;\q\big)_\infty~.
\ee
Note that the function  \eqref{def qpochh} is analytic for $|\q| <1$ or $|\q|>1$, with:
\be\label{qpoch invq}
(x;\q^{-1})_\infty = (\q x;\q)_\infty^{-1}~,
\ee
but it diverges at $|\q|=1$.
The holomorphic blocks are therefore divergent in the limit $\tau \rightarrow \R$, a limit we will discuss in detail below.

\paragraph{Chern-Simons contributions.} 
The classical Lagrangian contributes to the blocks through the supersymmetric Chern-Simons terms. These contributions are somewhat subtle, but can be inferred by consistency from the chiral multiplet contribution \cite{Beem:2012mb}.
The $U(1)$ Chern-Simons term at level $k=1$ contributes:
\be\label{HBCScont} 
B^{\rm GG}(\nu, \tau)_\n= {\theta\big((-1)^{2\t\nu^R} \q^{-\half}; \q\big)\ov \theta\big((-1)^{2\t\nu^R} \q^{\n-\half} y; \q\big)}~,
\ee
in terms of the Jacobi theta function:
\be
\theta(y,\q) \equiv (\q y;\q)_\infty (y^{-1};\q)_\infty~.
\ee
For a mixed CS term, we have:
\be
B^{\rm G_1G_2}(\nu_1, \nu_2, \tau)_{\n_1, \n_2} =  {\theta\big((-1)^{2\t\nu^R} \q^{\n_1-\half} y_1; \q\big) \theta\big((-1)^{2\t\nu^R} \q^{\n_2-\half} y_2; \q\big)\ov \theta\big((-1)^{2\t\nu^R} \q^{-\half}; \q\big) \theta\big((-1)^{2\t\nu^R} \q^{\n_1 + \n_2-\half} y_1 y_2; \q\big)}~.
\ee
This is useful, for instance, to insert the contribution of an FI term (which is a mixed $U(1)_T$-gauge CS term at level $1$) in an abelian theory.

We also find an interesting formula for the supersymmetric gravitational CS term, which contributes:~\footnote{As far as we know, this precise identification of $B^{\rm grav}(\tau)$ is new in the literature. We will provide some strong consistency checks of this claim---see equation \protect\eqref{Zgrav fact} below.}
\be\label{Bgrav def}
B^{\rm grav}(\tau)= \big((-1)^{2\t\nu^R} \q^{\half} ; \q\big)^{-1}_\infty~.
\ee
We have the obvious identity:
\be
\label{HBcommas} B^\Phi(\nu+ \nu_R, \tau)_{\n}\, B^\Phi(-\nu+ \nu_R, \tau)_{-\n} = B^{\rm GG}(\nu, \tau)_\n^{-1}\, B^{\rm grav}(\tau)^{-2}~.
\ee
This is simply the statement that a pair of chiral multiplets of $U(1)$ charge $\pm 1$ and $R$-charge $r=1$ can be given a superpotential mass, which generates the CS levels $k=-1$ and $k_g=-2$.

To fully specify the classical contribution to the blocks, we should also discuss the $R$-gauge and $RR$ CS contributions. To our knowledge, this has not been discussed precisely in the literature, and we leave it for  future work. We should also note that the holomorphic blocks are only defined up multiplications by certain elliptic functions $E(\nu, \tau)$, which cancel out from partition functions on closed three-manifolds upon gluing~\cite{Beem:2012mb}.

\paragraph{Vector multiplets.}
Next we have the contribution of the vector multiplets.  The contribution from the W-bosons, corresponding to the non-trivial roots of $\GG$, contribute in the same way as chiral multiplets of R-charge $2$ and $U(1)_a$ gauge charges $\alpha$. 
For the Cartan components $\GH= \prod_a U(1)_a$,  we propose the following contribution:
\be \label{HBcart} B^{\rm Cart}(\tau) = \left\{ \begin{array}{cc}  \tau^{-\rk}(\q;\q)_\infty^{-\rk} &\qquad\text{if} \; \; \text{Im}(\tau)>0~, \\
(\q^{-1};\q^{-1})_\infty^{\rk} &\qquad\text{if} \; \; \text{Im}(\tau)<0~. \end{array} \right.
\ee
We will argue below this has the appropriate limit as we take $\tau \rightarrow \Q$, reproducing the Seifert manifold formalism above.  We denote the total contribution from the Cartan and from the W-bosons by  $B^{\rm vec}(u,\tau)$.

\paragraph{Gauging the blocks.}
Given a $3d$ $\cN=2$ Lagrangian gauge theory, we may assemble the ``ungauged block:'' 
\be\label{ungauged B}
\tilde{B}(u,\nu,\tau) = B^{\rm GG}(u, \tau)^k  \,  B^{\rm vec}(u,\tau) \prod_{\rho, \omega}B^\Phi(\rho(u)+ \omega(\nu)+ \nu_R r_\omega, \tau) ~,
\ee
schematically,
by combining the above contributions for the chiral and vector multiplets and CS terms, as a function of both the gauge and the flavor symmetry parameters $u$ and $\nu$, respectively.

 Then the ``gauged block'' is determined by integrating this over certain middle-dimensional contours, $\Gamma^\alpha$, in the complex ``$u$-plane,'' which are naturally associated to the Bethe vacua, $\alpha \in \cS_{BE}$ \cite{Beem:2012mb}:
\be\label{HBgaugeblock} 
B^\alpha(\nu,\tau) = \int_{\Gamma^\alpha} du\, \tilde{B}(u,\nu,\tau)~,
\ee
The $\Gamma^\alpha$-contours are described in more detail in Appendix \ref{sec:HBA}.  We will see in a moment that we can also define the partition function of the gauge theory on a closed manifold directly in terms of the ungauged block, \eqref{ungauged B}.

\subsection{Lens space partition function from holomorphic blocks}
\label{sec:HBgluingblocks}
The holomorphic blocks can be used to construct supersymmetric partition functions on closed manifolds.  By gluing two solid tori along their boundary tori, we may obtain a general ``squashed'' lens space:
\be\label{HS Lpq}
L(p,q)_b \cong (-D^2\times_{\tau_1} S^1) \cup_g (D^2\times_{\tau_2} S^1)~.
\ee
Here, the gluing of the tori is through a large diffeomorphism:
\be\label{g def}
g = \mat{s & t \\ -p & q}\in SL(2, \Z)~, \qquad\qquad \tau_1 = - g \cdot \tau_2~.
\ee
Note that one should flip the orientation of one of the solid tori before gluing, to obtain a compact three-manifold.
Correspondingly, the lens space partition function of 3d $\CN=2$ gauge theories can be written as the ``fusion'' of two holomorphic blocks \cite{Beem:2012mb}.  
To see how this works, consider two holomorphic blocks with parameters $\nu_1,\tau_1, \n_1$ and $\nu_2, \tau_2, \n_2$, respectively.  To glue these along their boundaries, we must ensure that the boundary data is compatible.

\subsubsection{Identity gluing and $S^2\times S^1$}
Let us first consider the ``trivial gluing,'' with $g$ in \eqref{g def} the identity matrix.  Here, we simply identify the two boundary tori with a change of orientation of $D^2\times_{\tau_1} S^1$.  Thus we impose (ignoring the magnetic flux for the moment):
\be\label{identity fusion}
\nu_1 = \nu_2~, \qquad \quad  \tau_1 = - \tau_2~.
\ee
Topologically, this gives us the space $S^2 \times S^1$. Since each block includes a flux $-\half$ for the R-symmetry, there is a net $U(1)_R$ flux $-1$ through $S^2$. Thus we recover the background corresponding to the refined topological index, which we discussed in Section~\ref{sec:s2s1top}.  Explicitly, the fusion of blocks is achieved by taking an inner product in the basis of Bethe vacua. One finds \cite{Nieri:2015yia}:
\be \label{HBtrivglue} 
Z_{S^2 \times S^1}(\nu,\tau)_\m = \sum_{\alpha \in \cS^{BE}} B^\alpha(\nu,-\tau)_{\n_1} B^\alpha(\nu,\tau)_{\n_2}~. 
\ee
Here, we have set $\nu=\nu_1=\nu_2$ and  $\tau=-\tau_1=\tau_2$. By direct computation, one can check that the fusion \eqref{identity fusion} indeed reproduces the refined index \eqref{Zepsilon full}. The parameters used here and the ones in Section~\ref{sec:s2s1top} are related by:
\be
\epsilon= \tau~, \qquad v = u - {\tau\ov 2} (\n_1 - \n_2)~, \qquad  
v_F = \nu - {\tau\ov 2} (\n_1 - \n_2)~,
\qquad \m= \n_1 + \n_2~,
\ee
and similarly for the $R$-symmetry parameters. (In particular, $v_R= \t\nu_R$.)
For instance, for a chiral multiplet of $R$-charge $r=0$, we have:
\be
Z^\Phi_{S^2 \times S^1}(\nu,\tau)_\m = (\q^{-1-\n_1} y; \q^{-1})_\infty (\q^{1+\n_2} y; \q)_\infty = {(\q^{1+{\m\ov 2}} z; \q)_\infty\ov (\q^{-{\m\ov 2}}z; \q)_\infty}~,
\ee
where we defined $z= q^{-\half (\n_1-\n_2)} y= e^{2\pi i v}$. This is indeed equal to \eqref{Zeps v def}.

\paragraph{The supersymmetric index.}
Another ``trivial gluing,'' which was studied in \cite{Beem:2012mb}, corresponds to:
\be\label{identity fusion 2}
\nu_1 = -\nu_2~, \qquad \quad  \tau_1 = - \tau_2~.
\ee
This also gives $S^2\times S^1$ topologically, but it is distinct from \eqref{identity fusion}. When considering this gluing, the sign flip in $\nu_1$ must also be applied to the $U(1)_R$ flux through the disk, and the resulting background has a {\it vanishing} R-symmetry flux through the $S^2$. The corresponding supersymmetric background computes the ``ordinary'' supersymmetric index, also known as the 3d superconformal index \cite{Kim:2009wb, Imamura:2011su}. In terms of the gluing \eqref{HS Lpq}, it corresponds to:
\be
g= \mat{-1 & 0 \\ 0 & -1}~,
\ee
instead of the identity for the twisted index. We may denote this background by $L(0,-1)$, which is distinct from $L(0,1)$. The supersymmetric background $L(0,-1)$ is the exceptional case amongst the $L(p,q)$ backgrounds; it does not admit any Seifert limit, and therefore does not fit directly into the formalism of this paper. 

\subsubsection{Non-trivial gluing and the lens space $L(p,q)$}
Consider the more general gluing \eqref{HS Lpq}-\eqref{g def}. We should identify the parameters according to:
\be\label{id taus nus}
\tau_1= - {s \tau_2 + t\ov -p \tau_2 + q}~, \qquad \qquad \nu_1 = {\nu_2 \ov -p \tau_2 +q}~.
\ee
More precisely, the background gauge field parameters $\nu_i$ could be shifted by  elements of $\Z +\Z \tau_i$, for each block. As before, we may introduce in this way the fluxes $\n_i$ on each block:
\be
\nu_i \rightarrow  \nu_i + \alpha_i + \n_i \tau~, \qquad \qquad \alpha_i, \n_i \in \Z~.
\ee
Then, for the fluxes $\n_i$, we find the following consistency conditions upon gluing:~\footnote{Here we used the identities ${\tau_2 \ov -p \tau_2 + q}= - q\tau_1 -t$ and ${1 \ov -p \tau_2 + q}=-p \tau_1 +s$. }
\be
t\n_2 + \alpha_1 - s \alpha_2 \in \Z~, \qquad \quad \n_1+ q \n_2 + p \alpha_2  \in \Z~,
\ee
with $\alpha_1, \alpha_2 \in \Z$ and otherwise arbitrary. We then identify:
\be\label{id m ns blocks}
\m \equiv \n_1 + q\n_2 \in \Z_p~,
\ee 
as the torsion $U(1)$ flux on $L(p,q)$, since a shift of $\m$ by $p$ is equivalent to a shift of $\alpha_2$.   We may fix $\alpha_1= - t\n_2$ and $\alpha_2=0$, for definiteness.

\paragraph{Fusing the blocks.}
Given the identifications \eqref{id taus nus} and \eqref{id m ns blocks}, we directly find the partition function on a squashed lens space:
\be \label{HBlenspf} 
Z_{L(p,q)_b}(\sigma_\nu)_\m = \sum_\alpha B^\alpha(\nu_1,\tau_1)_{\n_1} \, B^\alpha(\nu_2,\tau_2)_{\n_2}~.
\ee
In order to directly compare to the more common notation on squashed spheres and lens spaces, we can solve the gluing constraints \eqref{id taus nus} by:
\bea\label{tau12 and nu12 to b}
& \tau_1 = -{b^{-2}-s\ov p}~, \qquad \qquad & \nu_1 = {b^{-1} \ov p} \left( i \h\sigma_\nu + b^{-1} \n_1 + b \n_2\right)~,\cr
& \tau_2 = -{b^2-q\ov p}~, \qquad \qquad & \nu_2 = {b \ov p} \left( i \h\sigma_\nu + b^{-1} \n_1 + b \n_2\right)~.
\eea
Here, $b^2 \in \C$ is the squashing parameter. Consider for instance the $L(p,q)_b$ partition function of a free chiral of $R$-charge $r=0$:
\be
Z^\Phi_{L(p,q)_b}(\h\sigma_\nu)_\m\equiv B^\Phi(\nu_1,\tau_1)_{\n_1} B^\Phi(\nu_2,\tau_2)_{\n_2} 
\ee
We directly find \cite{Dimofte:2014zga}:
\be
Z^\Phi_{L(p,q)_b}(\h\sigma_\nu)_\m ={\Big(e^{-{2\pi i {b^{-2}-s\ov p}}} \,e^{{2\pi i \ov p}(i b^{-1} \h\sigma_\nu + s \m)};\, e^{-{2\pi i {b^{-2}-s\ov p}}} \Big)_\infty \ov 
	\Big(e^{{2\pi i \ov p}(i b \h\sigma_\nu + \m)}; e^{{2\pi i {b^{2}-q\ov p}}} \Big)_\infty}~.
\ee
Here, we used \eqref{qpoch invq} to write the chiral-multiplet partition function in terms of the ``ordinary'' $\q$-Pochhammer symbol, for $|\q|<1$, which corresponds to ${\rm Im}(b^2)>0$. For $p=1$, this reduces to the $S^3_b$ one-loop determinant:
\be\label{ZS3b chiral2}
Z^\Phi_{L(1,1)_b}(\h\sigma)=Z^\Phi_{L(1,-1)_b}(\h\sigma)= Z_{S^3_b}^\Phi(\h\sigma) = \t\Phi_b\big(\h\sigma+ {i\ov 2}(b+b^{-1})\big)~. 
\ee
with $\t\Phi_b$ defined in \eqref{def Phib quantum dilog}.

\paragraph{$R$-symmetry dependence and spin structures.}
The $R$-symmetry background gauge fields:
\be
\nu_i^R = \t\nu_i^R - \half \tau_i~,
\ee
satisfy a consistency condition upon gluing, analogous to the one for flavor background gauge fields. Namely, we must have:
\be
\nu_1^R - {\nu_2^R\ov -p \tau_2 +q}   \in (\Z + \tau_1 \Z)~.
\ee
This is equivalent to:
\be\label{constraint on nuR12}
- {q+1\ov 2} + p \t\nu_2^R \in \Z~, \qquad \qquad -{t\ov 2} + \t\nu^R_1 - s\t\nu_2^R \in \Z~.
\ee
This fixes the parameters $\t\nu_i^R\in \half \Z$ (mod $1$). More precisely, there are three cases:
\bea
\begin{cases}
	(-1)^{2\t\nu^R_1}= (-1)^{s+1}~, \qquad (-1)^{2\t\nu^R_2}= -1~, \qquad & {\rm if} \; \; q \; \text{is odd,} \; p \; \text{is odd,}\cr
	(-1)^{2\t\nu^R_1}= (-1)^{t}~, \quad\qquad (-1)^{2\t\nu^R_2}= 1~, \qquad & {\rm if} \; \; q \; \text{is even,} \; p \; \text{is odd,}\cr
	(-1)^{2\t\nu^R_1}= (-1)^{2\t\nu^R_2 s+t}~, \quad (-1)^{2\t\nu^R_2}= \pm 1~, \qquad & {\rm if} \; \; q \; \text{is odd,} \; p \; \text{is even,}\cr
\end{cases}
\eea
We see that there is a unique consistent choice for the $\Z_2$ holonomies $\t\nu_i^R$ when $p$ is odd, while there are two distinct choices when $p$ is even. This is exactly as expected from our discussion of Seifert backgrounds. On $L(p,q)$ with $p$ even, there are two distinct choices of spin structures, which are probed by the two distinct supersymmetric backgrounds.

The $R$-symmetry $\Z_p$ torsion flux can be read off from \eqref{constraint on nuR12}, similarly to \eqref{id m ns blocks}:
\be
\m^R = -\half (q+1) + p \t\nu_2^R \in \Z_p~.
\ee
This is in agreement with the discussion in Section~\ref{sec:lpqgeo}.
For $p$ even (and therefore $q$ odd), there are two distinct choices in $\Z_p$, namely $\m^R= -\half(q+1)$ or $\m^R= -\half(q-p +1)$.  We also recover the fact that the canonical line bundle $\CK_{L(p,q)}\cong {\bf L}_R^2$ is trivial if and only if $q=-1$ (mod $p$). The $U(1)_R$ line bundle itself is trivial,  ${\bf L}_R\cong \CO$, if the two integers in \eqref{constraint on nuR12} vanish. This requires:
\be\label{t nuR conf}
\t\nu_1^R= {s+1\ov 2 p}~, \qquad\qquad \t\nu_2^R= {q+1\ov 2 p}~,
\ee
which is only possible for $q=-1$ mod $p$. On such a background, we can consider any real $R$-charges, $r\in \R$. In this case, combining \eqref{t nuR conf} with \eqref{tau12 and nu12 to b}, we see that:
\be
\nu_1^R = \t\nu_1^R -{\tau_1\ov 2}= {b^{-2} +1\ov 2 p}~, \qquad\qquad
\nu_2^R = \t\nu_2^R -{\tau_2\ov 2}= {b^{2} +1\ov 2 p}~, 
\ee
and therefore the shift \eqref{shift nu to nuR r} induced by a change of $R$-charge corresponds to:
\be
\h\sigma_\nu \rightarrow \h\sigma_\nu - i {b+b^{-1}\ov 2} r~.
\ee
This reproduces the well-known $R$-charge dependence on $S^3_b$, as in \eqref{def sigmaR}. In fact that result directly generalizes to $L(p,p-1)\cong S^3_b/\Z_p$ \cite{Closset:2014uda}, as we see here. Note that, for $p$ even, the ``superconformal'' background with $\m^R=0$ exists for a particular choice of spin structure. The other choice of spin structure gives rise to a distinct $L(p,p-1)_b$ background, with non-trivial $U(1)_R$ flux $\m^R= {p\ov 2}$ and Dirac-quantized $R$-charges.

\paragraph{Gravitational CS term.} As another consistency check of the above discussion, it is interesting to consider the gravitational Chern-Simons term. Consider the compact lens space $L(p,-1)_b$ with the the $R$-symmetry background such that ${\bf L}_R\cong \CO$, as described above. Then, we find~\cite{Dimofte:2014zga}:
\be
Z^{\rm grav}_{L(p,-1)_b} = e^{{\pi i \ov 24 p} \left(b^2+ b^{-2}\right)}\, e^{-{\pi i \ov 12} \left(p-{1\ov p} \right)}~.
\ee
For $p=1$, we recover the $S^3_b$ contribution as given on the last line of \eqref{CS terms S3b}. 
One can check that this term indeed factorizes as expected:
\be\label{Zgrav fact}
Z^{\rm grav}_{L(p,-1)_b} = B^{\rm grav}\Big(-{\tau\ov p\tau+1}\Big)\, B^{\rm grav}(\tau)~,\qquad \qquad \tau= -{b^2+1\ov p}~,
\ee
for $p\neq 0$. Here we used the gluing matrix $g$ with $(p,q)=(p,-1)$ and $(t,s)=(0,-1)$, which implies $\t\nu^R_{1, 2}=0$ due to \eqref{t nuR conf}. The identity \eqref{Zgrav fact} can be checked numerically. 
For the identity gluing, we obtain $B^{\rm grav}(-\tau) B^{\rm grav}(\tau)=1$ instead, as expected for the twisted index.

\paragraph{Integral formula.}
An alternative expression for the lens space partition function starts from the ungauged blocks, $\tilde{B}(u,\nu,\tau)$, introduced above.  We may fuse two ungauged blocks, thus obtaining the lens space partition function of the ungauged theory:~\footnote{The proportionality factor is a $\tau$-dependent factor, which can be thought of as contributing to the measure in the integral formula below. We will be somewhat imprecise about such measure factors in the following.}
\be\label{CZ ungauged}
\t\CZ_{L(p,q)_b}(\h\sigma_u,\h\sigma_\nu,\m,\m^F) \propto \tilde{B}(u_1,\nu_1,\tau_1)_{\n_1,\n_1^F} \tilde{B}(u_2,\nu_2,\tau_2)_{\n_2,\n_2^F}~.
\ee
Here, we denoted by $\n_i$ and $\n_i^F$ the gauge and flavor fluxes, respectively, and we identify the torsion fluxes:
\be
\m= \n_1 + q\n_2, \qquad \m^F= \n_1^F + q\n_2^F~,
\ee
as in \eqref{id m ns blocks}.  Then, for $p \neq 0$, the partition function of the gauge theory can be obtained by integrating \eqref{CZ ungauged} over the real $\h\sigma_u$ contour,~\footnote{More precisely, this is true for $b^2>0$.  For $b^2<0$ we should instead integrate over the imaginary $\sigma_\nu$ contour.
} and summing over fluxes $\m \in \Z_p$. Schematically, this gives:
\be \label{HBlenspfsc} 
Z_{L(p,q)_b}(\h\sigma_\nu)_{\m^F} =   \sum_{\m \in \Z_p} \int d\h\sigma_u\, \t\CZ_{L(p,q)_b}(\h\sigma_u, \h\sigma_\nu,\m,\m^F)~,
\ee
where we have included the contribution from the vector multiplets in the maximal torus, as above.  Indeed, one can show that this agrees with the expressions for the squashed sphere and lens space obtained directly by localization in  \cite{Kapustin:2009kz,Jafferis:2010un,Hama:2011ea,Benini:2011nc,Alday:2012au}.
When $p=0$, the sum in \eqref{HBlenspfsc} is over the integers, $\m\in \Z$---more generally, for a gauge group $\GG$, the sum is over the GNO-quantized magnetic fluxes on $S^2$---, and the integration is over the JK contour, as described in Section \ref{sec:localization}.

\paragraph{Geometric equivalences amongst lens spaces $\boldsymbol{L(p,q)}$.}
Before concluding this discussion of the $L(p,q)$ partition function, let us make a few more comments about the geometry. In the above, we constructed the lens space $L(p,q)$ using the arbitrary $SL(2,\Z)$ element $g$ in \eqref{g def}.
In particular, the integer $p$ could be positive or negative. In general, it is chosen positive. There is an obvious equivalence:
\be
L(-p,q) \cong - L(p,q)~,
\ee
where the minus on the right-hand-side stands for orientation reversal. This is clear, for instance, from the definition \eqref{HBlpqdef}-\eqref{label S3 in C2}, since sending $p$ to $-p$ is equivalent to sending $z_1, z_2$ to $\b z_1, \b z_2$. In terms of the Heegaard splitting \eqref{HS Lpq}, the lens space $L(-p,q)$ is realized by the gluing:
\be
\tau_1 = -\t g\cdot \tau_2~, \qquad  \t g =\mat{s & -t\\ p & q}~.
\ee 
Interestingly, we have:
\be\label{def tg}
\t g = \CC \cdot g \cdot \CC~, \qquad \CC= \mat{1 & 0 \\ 0 & -1}~,
\ee
and therefore:
\be
\tau_1 = -\t g\cdot \tau_2 \quad\qquad \leftrightarrow\quad \qquad \CC \cdot \tau_1 = - g \cdot (\CC\cdot \tau_2)~.
\ee
The last relation precisely realizes $-L(p,q)$. That is, $\CC$ acting on the boundary tori changes the sign of a single direction, and the gluing is with the same  $g$ that realizes $L(p,q)$, therefore we indeed obtain $L(p,q)$ with the opposite orientation.

Let us mention two other operations we can perform.  First, we may replace:

\be g =\left(\begin{array}{cc} s & t \\ -p & q \end{array} \right)\;\;  \rightarrow \;\;\cC g^{-1} \cC =\left(\begin{array}{cc} q & t \\ -p & s \end{array} \right) \ee
This is equivalent to exchanging the roles of the two blocks, \ie, $\tau_1 \leftrightarrow \tau_2$ and $\nu_1 \leftrightarrow \nu_2$, and so does not affect the partition function obtained by fusing the two blocks.  More precisely, we find the relation:

\be \label{ZLpsLpq} Z_{L(p,s)_b}(\h\sigma_\nu)_{\m^F} =Z_{L(p,q)_{b^{-1}}}(\h\sigma_\nu)_{s\m^F} \ee
This exhibits the equivalence $L(p,q) \cong L(p,q^{-1} \;(\text{mod} \; p))$, where we use $q^{-1} = s\; (\text{mod} \; p)$.

Next, we may replace $g \rightarrow -g$.  We see this has the effect of replacing $L(p,q) \rightarrow L(-p,-q)$.  From \eqref{id taus nus}, this has no effect on the identification of the $\tau_i$, but now the relation between $\nu_i$ incurs an additional sign.  Inspecting \eqref{tau12 and nu12 to b}, we see that, for $p \neq 0$, we may equivalently replace:

\be b \rightarrow i b, \;\;\; \sigma_\nu \rightarrow i \sigma_\nu , \;\;\;\; \m \rightarrow -\m \ee
In other words, we have the relation, for $p \neq 0$:

\be \label{Zlensgmg} Z_{L(-p,-q)_b}(\h\sigma_\nu)_{\m^F} =Z_{L(p,q)_{ib}}(i\h\sigma_\nu)_{-\m^F} \ee

For $p=0$, where $g=\text{Id}$, this operation takes us between the topological index, $L(0,1)$, and the ordinary index, $L(0,-1)$, but the partition functions on these two spaces do not obey a simple relation, as in \eqref{Zlensgmg}.  These two choices are the three dimensional analogue of the topological-topological ($tt$) and topological-anti-topological ($tt^*$) fusion of the holomorphic blocks, respectively \cite{Cecotti:1991me,Cecotti:2013mba}.  The statement of \eqref{Zlensgmg} is then that, once we non-trivially fiber the $S^1$ over $S^2$, the distinction between these two choices goes away.  This gives another perspective on why the partition function on the squashed sphere and lens spaces, which can be thought of as a $3d$ uplift of the $tt^*$ partition functions \cite{Cecotti:2013mba}, can also be computed in the $3d$ A-model, which is an uplift of the topological A-model.

\subsection{Rational squashing and the Seifert fibering operators}
\label{sec:HBrationallimit}

Given this discussion of the holomorphic block, we can now come back to the half-BPS Seifert manifold formalism.  Recall that,  although the orbits of the Killing vector $K^{(b)}$ in \eqref{HBkdef} do not close for generic $b \in \C$, they do close when
\be \label{b2 rat again}
b^2 = {q_1\ov q_2} \in \mathbb{Q}~. 
\ee
In this case, the orbits of $K^{(b)}$ are the fibers of one of the infinitely many inequivalent Seifert fibrations of the lens space, $L(p,q)$.  Thus we expect that, by taking a limit of the lens space partition function as $b^2$ approaches a rational number, we may express it in terms of the formalism introduced in earlier sections.

In fact, more directly, we expect there to be a simple relation between the holomorphic blocks and the Seifert fibering operators, which insert an exceptional $(q,p)$ fiber in a general Seifert manifold.  Namely, the holomorphic block is the partition function on a disk fibered over an circle, and in the limit where this fibration approaches a rotation by a rational angle, 
\be\label{tau to tq}
\tau \rightarrow \frac{t}{q}~,
\ee
this is precisely the solid fibered torus $T(q,t)$ introduced in Section \ref{sec: seifert backgs}---see equation \eqref{Tqt def}--, which is the local model of an exceptional fiber.  That is, schematically, we expect a relation of the form:
\be \label{HBbgschemrel} 
B^\alpha\Big(\nu,\tau= \frac{t}{q}\Big) \;\;\; \leadsto \;\;\; \cG^\alpha_{q,p}(\nu)~.
\ee
The precise relation is given in \eqref{HBntblim} below.
In the following, by establishing this correspondence more precisely, we demonstrate that the Seifert formalism of this paper indeed reproduces the known partition functions on squashed lens spaces in the rational-squashing limit, \eqref{b2 rat again}.

\subsubsection{Trivial gluing and the topological index}
Let us start with the ``trivial gluing'' of two blocks, as in \eqref{HBtrivglue} above, which gives us the topological index \cite{Benini:2015noa}.  First of all, in the limit $\tau \rightarrow 0$, we should recover the ordinary ({\it i.e.} non-refined) $S^2 \times S^1$ twisted index.  The precise claim is that, for each Bethe vacuum $\alpha \in \cS_{BE}$:
\be\label{HBhlim} 
\lim_{\tau \rightarrow 0} B^\alpha(\nu,-\tau)_{\n_1} B^\alpha(\nu,\tau)_{\n_2} = \cH^\alpha(\nu)^{-1} \, \pif^\alpha(\nu)^{\m^F}~.
\ee
Here $\m^F= \n_1+ \n_2$ is a flavor symmetry background flux, and $\pif$ the ordinary flavor flux operator. 
It then directly follows from \eqref{HBtrivglue} that:
\be 
Z_{S^2\times S^1}(\nu, \tau=0) = \lim_{\tau \rightarrow 0} \sum_{\alpha \in \cS_{BE}} B^\alpha(\nu,-\tau) B^\alpha(\nu,\tau)= \sum_{\alpha \in \cS_{BE}} {\cH^\alpha}(\nu)^{-1} \pif^\alpha(\nu)^\m~,
\ee
which is the expected result from the A-model computation of the twisted index.

We demonstrate \eqref{HBhlim} in Appendix \ref{sec:HBA}. Here, let us briefly illustrate the argument in the case of a free chiral multiplet with R-charge $r \in \Z$, and setting the fluxes to zero. The corresponding holomorphic block \eqref{HBchiblock} is given in terms of the $\q$-Pochhammer symbol:
\be 
B^\Phi(\nu,\tau) = ((-1)^{2\t\nu^R r} \q^{1-\frac{r}{2}} y;\q)_\infty~, \qquad
\q=e^{2 \pi i \tau}~, \; y=e^{2 \pi i \nu}~.
\ee
For $\tau \rightarrow 0$, we may use the following expansion of the $\q$-Pochhammer symbol \cite{Beem:2012mb}:
\be \label{HBQPlim} 
(\q y;\q)_\infty \underset{\tau \rightarrow 0}{\rightarrow} \exp \bigg(\frac{1}{2 \pi i \tau} \dilog(e^{2 \pi i \nu}) - \frac{1}{2} \log(1-e^{2 \pi i \nu}) + O(\tau) \bigg)~.
\ee
The leading behavior of the chiral multiplet block as $\tau \rightarrow 0$ is then:
\bea \label{HBchilim} 
B^\Phi(\nu,\tau) &= ((-1)^{2\t\nu^R r} \q^{1-\frac{r}{2}} x;\q)_\infty\cr
& \underset{\tau \rightarrow 0}{\rightarrow} \exp \bigg(\frac{1}{2\pi i\tau} \dilog(e^{2 \pi i (u+ \t\nu^R r - \frac{r}{2} \tau) }) - \frac{1}{2} \log(1-e^{2 \pi i (\nu+ \t\nu^R r)}) + O(\tau) \bigg) \cr
&=\exp \bigg(\frac{1}{2\pi i \tau} \dilog(e^{2 \pi i (\nu+ \t\nu^R r) }) + \frac{r-1}{2} \log(1-e^{2 \pi i (\nu+ \t\nu^R r)}) + O(\tau) \bigg)~.  \eea
When we multiply this with another block with $\tau \rightarrow -\tau$, as in the LHS of \eqref{HBhlim}, the leading divergence cancels and we obtain a finite limit:
\be \label{HBhlimchi} 
\lim_{\tau \rightarrow 0} B^\Phi(\nu,-\tau) B^\Phi(\nu,\tau)  = (1- e^{2 \pi i (\nu+ \t\nu^R r)})^{r-1} = \cH^\Phi(u)^{-1}~,
\ee
which agrees with the handle-gluing operator of a chiral multiplet of R-charge $r$, as claimed. Comparing \eqref{HBhlimchi} to \eqref{Omega Phi}, we also see that $\t\nu^R$ can be identified with $\nu_R$ in the Seifert formalism. This is expected in this case: given the identity gluing, we have $\t\nu^R_1= \t\nu^R_2$ and the choice of spin structure on the solid tori becomes a choice of spin structure on $S^2 \times S^1$.

\subsubsection{General lens spaces}\label{subsec: gen ls}
To consider similar limits for general lens spaces,  it will be useful to first introduce some notation.  For $g \in SL(2,\Z)$ given as in \eqref{g def}, we define:
\be
\t g = \CC \cdot g \cdot \CC= \mat{s& -t \cr p & q}~,\qquad \qquad \CC= \mat{1 & 0 \\ 0 & -1}~,
\ee
as in \eqref{def tg}. We also define:
\be \label{HBBgdef} 
B_{\t g}^\alpha(\nu,\tau) \equiv B^\alpha\Big(\frac{\nu}{p \tau+q} , \frac{s \tau -t}{p \tau+q}\Big)~.
\ee
This is simply a reparameterization of the variables $\nu$ and $\tau$ defining the blocks.  Specifically, these are the values measured after applying the $SL(2,\Z)$ transformation $\t g$ to the boundary torus.

\paragraph{Fibering operators from the holomorphic blocks.} Let us choose $q>0$ for definiteness. Then, the precise form of the relation \eqref{HBbgschemrel} between the holomorphic block and the Seifert fibering operator reads:
\be \label{HBntblim}
\lim_{\tau \rightarrow 0} \frac{B^\alpha_{\t g}(\nu,\tau)_\n}{B^\alpha(\nu,\tau)} 
= \lim_{\tau \rightarrow 0} \frac{B^\alpha\Big(\frac{\nu}{p \tau+q} , \frac{s \tau -t}{p \tau+q}\Big)_\n}{B^\alpha(\nu,\tau)} = \cG^\alpha_{q,p}(\nu)_\n~. 
\ee
We can understand this geometrically, as follows.  To glue an exceptional Seifert fiber of type $(q,p)$, we first cut out a tubular neighborhood of ``trivial fiber,'' $D^2 \times S^1$, and then glue the local model of an exceptional fiber, $T(q,t)$, to the boundary torus.  This is reflected in the equation above, where the denominator corresponds to the ``trivial block'' we are removing, and the numerator corresponds to the non-trivial block, with rational rotation angle $\t g\cdot \tau \rightarrow -\frac{t}{q}$.  More precisely, we must first regularize by taking non-zero $\tau \in \C\backslash\R$, as the blocks are divergent at real $\tau$, and we recover the above procedure in the limit $\tau \rightarrow 0$, reflected in \eqref{HBntblim}.

We prove the relation \eqref{HBntblim} in Appendix \ref{sec:HBA}. Let us again illustrate the argument in the case of a free chiral multiplet. For simplicity of notation, we take both the R-charge and flux to be zero.  Then, we are interested in the limit of:
\be \label{HBblocklimit}  {B}_{\t g}^\Phi(u;\tau) = (\tilde{\fq} \tilde{y};\tilde{\fq})~, \qquad
\tilde{y} = \exp \bigg( 2 \pi i\frac{\nu}{p\tau +q} \bigg)~, \quad
\tilde{\q}= \exp\bigg(2 \pi i   \frac{s \tau - t}{p\tau +q}\bigg)~,
\ee
as  $\tau \rightarrow 0$.
Using the identity:
\be \label{HBqpn} 
(\fq y;\fq) = \prod_{\ell=0}^{n-1} (\fq^n (\fq^{-\ell} y);\fq^n)~, \qquad \forall n \in \Z_{> 0}~,
\ee
and the limit \eqref{HBQPlim} for the $\q$-Pochhammer symbol, we find:
\bea
&(\tilde{\fq} \tilde{y};\tilde{\fq})\; &\underset{\tau \rightarrow 0}{\sim}& \;
\exp \Bigg( \sum_{\ell=0}^{q-1} \Bigg( \frac{q}{2 \pi i \tau} \dilog(e^{\frac{2 \pi i}{q} (\nu+t \ell)}) + \frac{p}{2 \pi i} \dilog(e^{\frac{2 \pi i}{q}(\nu + t \ell)})  \cr
&&&\qquad \qquad\qquad + \left({ p \nu +\ell\ov q}-\half\right)\log (1 -e^{\frac{2 \pi i}{q}(\nu+t \ell)}) \Bigg)+O(\tau)  \Bigg)~.
\eea
Using the identities \eqref{id sum dilog log app}, we may rewrite this as:
\bea \label{HBntbl}
& (\tilde{\fq}\tilde{y};\tilde{\fq})  \underset{\tau \rightarrow 0}{\sim} \exp \bigg(\frac{1}{2 \pi i \tau} \dilog(e^{2 \pi i \nu}) - \frac{1}{2} \log(1-e^{2 \pi i \nu})  \bigg) \\ & \times \exp \bigg({p\ov q}\left(\frac{1}{2 \pi i}  \dilog(e^{2 \pi i \nu})  +  \nu \log (1-e^{2 \pi i \nu})\right) +\sum_{\ell=0}^{q-1} \frac{\ell}{q}  \log (1 -e^{\frac{2 \pi i}{q}(\nu+t \ell)})  \bigg) \bigg)~.
\eea
We see this differs from the trivial block, \eqref{HBQPlim}, by a finite piece.  Thus we may divide by the trivial block to obtain a finite limit:
\bea \label{HBntblim2}
&\lim_{\tau \rightarrow 0} \frac{B^\Phi_{\t g}(\nu,\tau)}{B^\Phi(\nu,\tau)} =\cr
&\quad\exp \bigg({p\ov q}\left(\frac{1}{2 \pi i}  \dilog(e^{2 \pi i \nu})  +  \nu \log (1-e^{2 \pi i \nu})\right) +\sum_{\ell=0}^{q-1} \frac{\ell}{q}  \log (1 -e^{\frac{2 \pi i}{q}(\nu+t \ell)})  \bigg) \bigg)~.
\eea
Comparing to \eqref{CGqp0 Phi Bis}, we see the right-hand-side is precisely the $(q,p)$ fibering operator for the chiral multiplet:
\be\label{limit to GPhi of hb}
\lim_{\tau \rightarrow 0} \frac{B^\Phi_{\t g}(\nu,\tau)}{B^\Phi(\nu,\tau)} = \CG_{q,p}^\Phi(\nu)~.
\ee
This proves the relation \eqref{HBntblim} in this special case.

\paragraph{Fusing the blocks.}
Let us now consider the fusion of the blocks as in \eqref{HBlenspf}, with the gluing conditions:
\be\label{gluing cond again}
\tau_1=- g\cdot \tau_2 \equiv  - {s \tau_2 + t\ov -p \tau_2 + q}~, \qquad \qquad \nu_1 \equiv g\cdot \nu_2= {\nu_2 \ov -p \tau_2 +q}~,
\ee
where $g\in SL(2, \Z)$ acts on $\tau_2$ and $\nu_2$ in the obvious way.
It is then very convenient to introduce the two $SL(2, \Z)$ matrices:
\be
g_1=\mat{s_1 & t_1\cr -p_1 & q_1}~, \qquad g_2=\mat{s_2 & t_2\cr -p_2 & q_2}~, 
\ee
such that:
\be\label{g to g1g2}
g= g_1  \, \CC \,  g_2^{-1} \,  \CC \qquad \leftrightarrow \qquad \mat{s & t \cr-p &q}
= \mat{q_2 s_1-p_2 t_1 \; \; & s_1 t_1 + s_1 t_2\cr - q_1 p_2 - q_2 p_2\; \;& q_1 s_2 - p_1 t_2}~.
\ee
Then, it is clear that the conditions \eqref{gluing cond again} can be solved by:
\bea 
& \tau_1 = - g_1 \cdot \tau~, \qquad \qquad & \nu_1 =  g_1 \cdot \nu~,\cr
& \tau_2 = \t g_2 \cdot \tau~, \qquad \qquad & \nu_2 = \t g_2 \cdot \nu~,
\eea
where $\t g_2 = \CC g_2 \CC$, and $\tau$ and $\nu$ are free parameters. More explicitly, we have:
\bea\label{tau12 to tau nu}
& \tau_1 = {-s_1 \tau - t_1 \ov -p_1 \tau + q_1}~, \qquad \qquad & \nu_1 =  {\nu\ov -p_1 \tau + q_1}~,\cr
& \tau_2 = {s_2 \tau - t_2 \ov p_2 \tau + q_2}~, \qquad \qquad & \nu_2 = {\nu\ov p_2 \tau+ q_2}~.
\eea
In these new variables, the $L(p,q)$ partition function \eqref{HBlenspf} takes the suggestive form:
\bea\label{HBlenspf2}
&Z_{L(p,q)}(\nu, \tau)_\m &=&\; \sum_\alpha B^\alpha(\nu_1,\tau_1)_{\n_1} \, B^\alpha(\nu_2,\tau_2)_{\n_2}\cr
&&=&\; \sum_\alpha B_{\t g_1}^\alpha(\nu,-\tau)_{\n_1} \, B_{\t g_2}^\alpha(\nu,\tau)_{\n_2}~.
\eea
At this point this is simply a reparameterization.  However, if we now take the limit $\tau \rightarrow 0$ of \eqref{HBlenspf2}, we find:
\be 
\lim_{\tau \rightarrow 0} Z_{L(p,q)}(\nu, \tau)_\m=
\lim_{\tau \rightarrow 0} \sum_\alpha  { B_{\t g_1}^\alpha(\nu,-\tau)_{\n_1} \ov B^\alpha(\nu,-\tau)}  {B_{\t g_2}^\alpha(\nu,\tau)_{\n_2}\ov B^\alpha(\nu,\tau)}  \, B^\alpha(\nu,-\tau) B^\alpha(\nu,\tau)~.
\ee
Then using the limits \eqref{HBntblim} and \eqref{HBhlim},this becomes:
\be 
\lim_{\tau \rightarrow 0} Z_{L(p,q)}(\nu, \tau)_\m = \sum_\alpha \cG^\alpha_{q_1,p_1}(\nu)_{\n_1} \cG^\alpha_{q_2,p_2}(\nu)_{\n_2} \cH^\alpha(\nu)^{-1}~.
\ee
This is precisely the partition function of a Seifert manifold with two exceptional fibers, $(q_i,p_i)$, $i=1,2$, over a genus zero Riemann surface---namely, a lens space.
Recall from Section \ref{sec:lpqgeo} that the lens space $L(p,q)_b$ with $b^2={q_1\ov q_2}$ has the Seifert fibration:
\be\label{lens space M3 hbs}
L(p,q)_b  \cong [0~;\, 0~;\, (q_1, p_1)~,\; (q_2, p_2)]~,\qquad
p= p_1 q_2 + p_2 q_1~, \qquad q= q_1 s_2 - p_1 t_2~.
\ee
The identification of the lens space parameters $p, q$ with the Seifert invariants $(q_i, p_i)$ is precisely as in \eqref{g to g1g2}. Comparing \eqref{tau12 to tau nu} to \eqref{tau12 and nu12 to b}, we also find:
\be\label{HBbdef}
b^2 = {\nu_2 \ov \nu_1}= {-p_1 \tau+ q_1 \ov p_2 \tau + q_2} \quad \underset{\tau \rightarrow 0}{\longrightarrow}\quad {q_1\ov q_2}~,
\ee
as expected. Thus, we have proven that:
\be 
Z_{L(p,q)_b}(\sigma_\nu,\hat{\m})\Big|_{b^2= {q_1\ov q_2}}=  \sum_\alpha \cG^\alpha_{q_1,p_1}(\nu)_{\n_1} \cG^\alpha_{q_2,p_2}(\nu)_{\n_2} \CH^\alpha(\nu)^{-1}~,
\ee
which is our main result for Seifert manifolds, specialized to the lens space Seifert fibration \eqref{lens space M3 hbs}.

We should emphasize the condition $q>0$ stated before \eqref{HBntblim} is important for the argument above to go through.  Thus we must impose $q_i>0$ for both $SL(2,\Z)$ matrices $g_i$.  From \eqref{HBbdef}, we see $q_i=0$ is a singular limit of the squashing parameter, and this limit is not expressible in terms of the Seifert fibering operators.  When $q_i<0$, we may replace $g_i \rightarrow -g_i$, which (at most) takes $g \rightarrow -g$.  Then for $p \neq 0$, we have seen above, in \eqref{Zlensgmg}, that this is simply a reparameterization in terms of the equivalent space $L(-p,-q)$, and so there is no loss in restricting to $q_i>0$ in these cases. 

However, for $p=0$, taking $g \rightarrow -g$ is not a reparameterization, and takes us between the inequivalent backgrounds $L(0,1)$ and $L(0,-1)$, corresponding to the twisted index and the ordinary supersymmetric index, respectively.  Then, one can check that for $(p,q)=(0,-1)$, it is not possible to find matrices $g_i$ satisfying \eqref{g to g1g2} with both $q_i>0$.  Therefore, although the supersymmetric index can be constructed in terms of holomorphic blocks, it cannot be written in terms of the Seifert fibering operators.

\paragraph{Contour integral expression.}

Finally, let us return to the integral formula \eqref{HBlenspfsc}.  Applying the results above to the ungauged theory, we find:

\be \tilde{\CZ}_{L(p,q)_b}(\sigma_u,\sigma_\nu,\m,\m^F) \propto e^{-2 \pi i \Omega(u, \nu)}\;  \tilde{\cG}_{q_1,p_1}(u,\nu)_{\n_1,\n_1^F}\; \tilde{\cG}_{q_2,p_2}(u,\nu)_{\n_2,\n_2^F}~,
\ee
for $b^2$ rational. Here we used the fact that $\CH= e^{2 \pi i \Omega}$ for the ungauged theory. 
Then, plugging this into \eqref{HBlenspfsc}, we find:
\be \label{HBlenspfscrat} 
Z_{L(p,q)_b}(\sigma_\nu,\m^F)  \propto \sum_{\n \in \Z_p}\int d\sigma_u \;e^{-2 \pi i \Omega(u, \nu)}\;  \tilde{\cG}_{q_1,p_1}(u,\nu)_{\n_1,\n_1^F}\; \tilde{\cG}_{q_2,p_2}(u,\nu)_{\n_2,\n_2^F}~,
\ee
schematically,  in agreement with the ``$\sigma$-contour'' expression derived in \eqref{sigma contour}.

\section*{Acknowledgments} We would like to thank Clay Cordova, Dongmin Gang, Sergei Gukov, Diego Hofman and Hans Jockers for discussions, as well as Tudor Dimofte and Zohar Komargodski for discussions and comments on the draft. We especially thank Victor Mikhaylov for many comments and for collaboration on closely related projects.
  C.C.  gratefully acknowledges support from the Simons Center for Geometry and Physics, Stony Brook University, at which some of the research for this paper was performed, as well as the Galileo Galilei Institute for Theoretical Physics and the INFN for partial support during the completion of this work. The work of H.K. is supported by ERC Consolidator Grant 682608 ``Higgs bundles: Supersymmetric Gauge Theories and Geometry (HIGGSBNDL)."  B.W. was supported in part by the National Science Foundation under Grant No. NSF PHY11-25915.

\part*{Appendix}
\addcontentsline{toc}{part}{Appendix}
\appendix
\section{Parity anomaly and Chern-Simons contact terms}\label{app: parity anomaly}
In this appendix, we review some important properties of three-dimensional fermions and Chern-Simons terms,  following in particular \cite{Closset:2012vp,  Witten:2015aba, Witten:2016cio, Seiberg:2016gmd}.

\subsection{Three-dimensional fermions and the parity anomaly}
 It is well-known that three-dimensional fermions suffer from parity anomalies \cite{Redlich:1983dv, Niemi:1983rq, AlvarezGaume:1984nf}. Consider a 3d Dirac fermion $\psi$ coupled to a $U(1)$ background gauge field $A_\mu$ with charge $1$. On $\{x^\mu\} \cong \R^3$ with Euclidean signature, the parity operation is given by changing the sign of a single coordinate, say $x^3 \rightarrow -x^3$. (It is thus indistinguishable from time-reversal symmetry.) The classical Lagrangian of a massless fermion,
\be
\SL = - i \b\psi \gamma^\mu (\d_\mu- i A_\mu) \psi~,
\ee
preserves parity. The parity anomaly is the statement that there exists a mixed parity-$U(1)$ anomaly---in other words, one cannot preserve {\it both} parity and (background) gauge invariance. As usual in this type of situation, we have to give up some symmetry of the classical theory in the quantum theory. If we couple the theory to a metric, we have a similar parity anomaly with background diffeomorphism. In this paper, we always choose to preserve gauge invariance (and diffeomorphism invariance). Then, the ``parity anomaly'' is the statement that the quantum effective action:
\be
S_{\rm eff}[A_\mu] \equiv - \log \det\big(-i \gamma^\mu(\d_\mu-i  A_\mu)\big)~,
\ee
is a gauge-invariant (non-local) functional of $A_\mu$ which violates parity. The parity-violating term is an imaginary contribution to $S_{\rm eff}$, which arises because one needs to regulate carefully the infinite product over the eigenvalues of the Dirac operator.
This can be made rigorous on a closed three-manifold. A standard regularization of the phase of the Dirac determinant  gives \cite{AlvarezGaume:1984nf, Witten:2015aba}:
\be\label{detDA eta}
\det(-i\slashed D_A)= \left|\det(-i \slashed D_A)\right| e^{{\pi i\ov 2}{\bm\eta}(g, A)}~.
\ee
Here, the absolute value is unambiguous, while the phase is given by the APS ${\bm\eta}$-invariant.
Crucially, ${\bm \eta}(g, A)$ is a gauge-and diff-invariant functional of the gauge field $A$ and of the metric $g$,~\footnote{It is defined by a $\zeta$-regulated sum over the signs of the eigenvalues, $\lambda$, of the Dirac operator:
\be\nn
{\bm\eta}(g, A) =\lim_{s \rightarrow 0} \sum_\lambda  \sign(\lambda) |\lambda|^{-s}~.
\ee
}
which is generally non-local.
The ${\bm \eta}$ term is closely related to the $U(1)$ and gravitational CS terms:
\be\label{CS GG and grav def intro}
 S_{\rm GG}[A]=  {i \ov 4\pi}\int A \wedge d A~,\qquad\quad 
 S_{\rm grav}[g] = {i\ov 192 \pi}\int \Tr\Big(\omega\wedge d\omega-{2\ov 3} \omega\wedge \omega\wedge \omega  \Big)~,
\ee
with $\omega$ the spin connection. While these CS terms are not gauge (or diff) invariant unless their coefficient is a quantized level, $k\in \Z$ and $k_g\in \Z$, respectively, their infinitesimal variations are well-defined and coincide with the variations of the ${\bm\eta}$-invariant:
\be
\pi i \delta_{A} {\bm\eta}(\omega, A)= \delta_{A}  S_{\rm GG}[A]~, \qquad
\pi i \delta_{g} {\bm\eta}(\omega, A)= 2 \delta_{g}S_{\rm grav}[g]~.
\ee
By the Atiyah-Patodi-Singer (APS) index theorem, we also have:
\be\label{APS}
\exp{\left(-\pi i k {\bm\eta}(g, A)\right)}= \exp{\left(- k S_{\rm GG}[A]-2 k  S_{\rm grav}[g] \right)}~,
\ee
for any quantized integer $k\in \Z$.
Thus, for many purposes, the gauge-invariant phase $e^{{\pi i \ov 2} {\bm\eta}(g, A)}$ in \eqref{detDA eta} looks just like an improperly-quantized $U(1)$ CS term at level $-\half$ (plus a gravitational CS term at level $-1$).~\footnote{Nonetheless, we emphasize that it is not possible to ``cancel the parity anomaly'' (as it is sometimes stated in the supersymmetric localization literature)  by adding a CS term with level $k=\half$ to \protect\eqref{detDA eta}, since such a term violates gauge invariance. 
The point is that one cannot take the ``square-root'' of equation \protect\eqref{APS}.
}

In any 3d field theory coupled to background gauge fields and metric, the parity-violating terms are conveniently captured by the two-point functions of the conserved currents, whose parity-odd coefficients are denoted by $\kappa$.  As explained in \cite{Closset:2012vp}, the quantity:
\be\label{kappa mod 1}
\kappa \; {\rm mod}\; 1
\ee
is a  physical observable.
This is because $\kappa$ can always be shifted by some integer $k\in \Z$, by adding a background Chern-Simons term with integer level $k$ to the effective action. For instance, we may add the $U(1)$ CS term in \eqref{CS GG and grav def intro} to the UV action, which would shift the observable $\kappa$:
\be\label{A and omega CS app}
S_{\rm eff}[A] \rightarrow S_{\rm eff}[A]+ k\, S_{\rm GG}[A]\qquad \qquad \leftrightarrow\qquad\qquad
 \kappa \rightarrow \kappa +k~.
\ee
 Since the CS level $k$ must be quantized by gauge invariance, $\kappa$ mod $1$ is physical. By abuse of notation, we call $\kappa$ the ``CS contact term,'' but we should not loose sight of the fact that \eqref{kappa mod 1} is physical (unlike an ordinary contact term, which can be entirely canceled by a local term).
For the free fermion $\psi$ regularized as in \eqref{detDA eta}, we have:
\be\label{kappa m12}
\kappa= -\half~, \qquad\qquad \kappa_g=-1~.
\ee
Here, $\kappa$ is the parity-odd contact term in the two-point functions of the $U(1)$ current, and $\kappa_g$ is a similar contact term involving the stress-energy tensor \cite{Closset:2012vp}. The only ambiguity is in shifting $\kappa$ and $\kappa_g$ by integers, by adding the CS terms \eqref{CS GG and grav def intro} to the action. We must therefore make a choice in the UV. We call the choice \eqref{detDA eta}-\eqref{kappa m12} the ``$U(1)_{-\half}$ quantization,'' in agreement with standard notation.~\footnote{The term $\kappa=-\half$ is sometimes call the ``effective CS level.''} 
More generally, for a fermion with charges $Q^a\in \Z$ under some $U(1)_a$ symmetries, the ``$U(1)_{-\half}$ quantization'' \eqref{detDA eta} corresponds to the CS contact terms:
\be
\kappa_{ab}  = -\half Q^a Q^b~, \qquad\qquad \kappa_g= -1~.
\ee
The corresponding CS terms are the mixed $U(1)_a$-$U(1)_b$ CS interactions, with levels $k_{ab}\in \Z$. (The generalization to non-abelian symmetries is straightforward.)
Incidentally, let us note that, while the choice of background CS terms for global symmetries is unphysical, once we start gauging symmetries (making some $A_\mu$'s dynamical), their CS terms, of course, become an important part of the definition of the theory in the UV. For instance, a $U(1)$ gauge theory coupled to a single fermion $\psi$ of unit charge, with bare CS term $k\in \Z$ in the UV, together with our choice of quantization \eqref{detDA eta}, is generally denoted by $U(1)_{-\half +k} + \psi$, and its dynamics, of course, depends crucially on the level $k$ \cite{Seiberg:2016gmd}.

To conclude this overly detailed discussion of 3d fermions, we recall that the `real mass'' term:
\be\label{real mass intro}
\Delta\SL_m = -i m \b\psi \psi~,\qquad\qquad m\in \R~,
\ee
breaks parity explicitly. Integrating out the fermion $\psi$ with a large real mass, $m \rightarrow \pm \infty$, shifts the CS contact terms according to:
\be\label{shift kappa intro}
\delta \kappa_{ab}= \half \sign(m)\, Q^a Q^b~, \qquad\qquad \delta \kappa_g = \sign(m)~,
\ee
in our conventions. In particular, for a single fermion $\psi$ of $U(1)$ charge $1$ and with a mass term $m$, we have the CS contact terms \eqref{kappa m12} in the UV, while in the IR we can integrate out the fermion and obtain an empty (trivial, gapped) theory. If $m>0$,  that empty theory has vanishing CS contact terms, $\kappa=0$ and $\kappa_g=0$, while if $m<0$, we obtain the IR contact terms $\kappa=-1$ and $\kappa_g=-2$.

\subsection{Chern-Simons actions, contact terms and supersymmetry}
Now, let us specialize the above discussion to the case of $\CN=2$ supersymmetric theories with a $U(1)_R$ symmetry. We have to consider the various contact terms:
\be
\kappa_{ab}~,\qquad \quad \kappa_{aR}~, \qquad \quad  \kappa_{RR}~,\qquad  \quad \kappa_g~,
\ee
which correspond to the gauge, mixed gauge-$R$, $RR$, and gravitational contact terms, respectively.
The supersymmetrization of those terms was studied in \cite{Closset:2012vp}.

\paragraph{Supergravity CS terms.}
Let us briefly discuss the supersymmetric CS term $A^{(R)}dA^{(R)}$ for the $U(1)_R$ gauge field  $A_\mu^{(R)}$, and the supersymmetric gravitational CS term \cite{Closset:2012vp,Kuzenko:2013uya}.
Setting the fermions to zero, the $\CN=2$ supersymmetric version of the gravitational CS term reads:
\be\label{Sgrav App}
S_{\rm grav}={k_g \ov 192 \pi}\int d^3 x\sqrt{g} \Big( i\epsilon^{\mu\nu\rho} \Tr \big(\omega_\mu \d_\nu \omega_\rho - {2 \over 3} \omega_\mu \omega_\nu \omega_\rho\big)
+ 4 i \epsilon^{\mu\nu\rho} A^{(R)}_\mu  \d_\nu A^{(R)}_\rho\Big)~,
\ee
with $A_\mu^{(R)}$ the $U(1)_R$ gauge field. The level $k_g$ is integer-quantized. This action is conformally invariant. There also exists another, non-conformal CS-like term one can write down using the supergravity multiplet alone:
\be\label{kzz term}
S_{zz}={k_{zz}\ov 4\pi}\int d^3 x\sqrt{g} \Big( i \epsilon^{\mu\nu\rho} (A^{(R)}_\mu+V_\mu)  \d_\nu (A^{(R)}_\rho+V_\rho) -\half H R - H^3 - H V_\mu V^\mu \Big)~.
\ee
Here, $V_\mu$ and $H$ and auxiliary supergravity fields (as we briefly review in Section~\ref{subsec: M3 backgd}) and $R$ is the 3d Ricci scalar. The actual $U(1)_R$ Chern-Simons level is given by:
\be
k_{RR}= {k_g\ov 12}+k_{zz}~,
\ee
which is the net coefficient of the $A^{(R)}dA^{(R)}$ term. The $RR$ level $k_{RR}$ is integer-quantized if $U(1)_R$ is compact---that is, whenever all the $R$-charges are integer-quantized.

\paragraph{Quantizing  3d $\CN=2$ supersymmetric multiplets.}
Let $\Phi$ be a chiral multiplet of $U(1)_a$ charges $Q^a$ and $R$-charge $r$.
In this paper, we use the ``$U(1)_{-\half}$ quantization'' for the Dirac fermion $\psi$ in $\Phi$. Then, each chiral multiplet $\Phi$ contributes to the CS contact terms:
\bea\label{Phi U12 quant App}
&\kappa_{ab}=- \half Q^a Q^b~,\qquad\qquad &&  \kappa_{RR}=-\half (r-1)^2~,\cr
&\kappa_{aR}=- \half Q^a (r-1)~,\qquad &&\kappa_g=-1~.
\eea
If we give a large positive real mass, $m\rightarrow \infty$, to a free chiral multiplet in the $U(1)_{-\half}$ quantization, thus integrating it out, we get vanishing CS contact terms in the IR:
\be
m\gg 1\; : \; \quad\kappa_{ab}^{(\rm IR)}=0~, \qquad \kappa_{aR}^{(\rm IR)}=0~,\qquad
\kappa_{RR}^{(\rm IR)}=0~, \qquad \kappa_g^{(\rm IR)}=0~.
\ee
Operationally, and especially for the purpose of supersymmetric localization, this is as good a definition as any of what we mean by the ``$U(1)_{-\half}$ quantization'' of an $\CN=2$ chiral multiplet---it is the regularization of the one-loop determinant $Z^\Phi$ for $\Phi$ such that $Z^\Phi \rightarrow 1$ in the limit $m \rightarrow \infty$.

The vector multiplet $\CV$ also contains an adjoint fermion, the gaugino. To discuss its UV quantization, we decompose $\CV$ into abelian vector multiplets $\CV_a$ along a maximal torus $\GH\cong  \prod_a U(1)_a$, and into the components $\CV_\alpha$ along the non-trivial roots.
 Each $\CV_a$ and $\CV_\alpha$ contains a gaugino $\lambda$ or $R$-charge $1$ (and its charge conjugate $\b\lambda$ or $R$-charge $-1$), which we denote by $\lambda_a$ and $\lambda_\alpha$, respectively. The $\lambda_\alpha$'s come in pairs $\lambda_{\alpha}$, $\lambda_{-\alpha}$, which carry opposite gauge charges ($Q^a=\alpha^a$ and $Q^a=-\alpha^a$, respectively) under the Cartan subgroup $\GH$. We therefore choose the ``symmetric quantization'' with respect to the $U(1)_a$ gauge charges, which results in a vanishing net shift of the gauge contact terms:
\be\label{V quanti App 1}
\CV\; : \; \qquad \kappa_{ab}=0~, \qquad\qquad \kappa_{a R}=0~,
\ee 
for the full vector multiplet. (The elements $\CV_a$ are of course neutral.) We also have to specify the $U(1)_R$ and gravitational CS contact terms in the UV. We will quantize the gauginos $\lambda=(\lambda_a, \lambda_\alpha)$ such that they each induce the $RR$ and gravitational CS levels $\kappa_{RR}= \half$ and $\kappa_g=1$---that corresponds to a phase $e^{-{\pi i\ov 2} {\bm \eta}(A^{(R)}, g)}$ in their one-loop determinant, instead of \eqref{detDA eta}. Then, for the full vector multiplet, we obtain the UV CS contact terms:
\be\label{kappaRR App vec}
\CV\; : \; \qquad  \kappa_{RR}= \half {\rm dim}(\GG)~, \qquad \qquad
\kappa_{g}={\rm dim}(\GG)~.
\ee
Of course, we could always choose a different quantization in which $\kappa_g=0$, by adding the supersymmetric gravitational CS term \eqref{Sgrav App}, with level $k_g= - {\rm dim}(\GG)$, to the effective action. On the other hand, the UV contribution of the vector multiplet to $\kappa_{RR}$ is generally half-integer, and thus cannot be shifted to zero in any gauge-invariant scheme.
The quantization \eqref{kappaRR App vec} for the vector multiplet is particularly natural from the point of view of pure Chern-Simons theory, as we now explain.

\paragraph{Flowing to pure CS theory.}
Consider an $\CN=2$ supersymmetric Chern-Simons theory $\GG_k$,  consisting of a vector multiplet $\CV$ with a CS term at level $k\in \Z$:~\footnote{More generally, we have distinct levels $k_\gamma$ and $k_I$ for each factor in \protect\eqref{GG intro}. We could also consider distinct CS levels for the $SU(N)$ and $U(1)$ factors in $U(N)$.}
\be
{k\ov 4\pi}  \int d^3 x \sqrt{g} \left(i \epsilon^{\mu\nu\rho} \left(A_\mu \d_\nu A_\rho- {2 i \ov 3}A_\mu A_\nu A_\rho\right) - 2 D \sigma + 2 i  \b\lambda\lambda\right)~.
\ee
In such a theory, the gaugino has a real mass \eqref{real mass intro} given by the CS level:~\footnote{This is dimensionless, since the canonical dimension of the gaugino is ${3\ov 2}$.}
\be
m_\lambda = - {k\ov 2\pi}~.
\ee
It is useful to introduce a supersymmetric Yang-Mills (YM) term as a UV regulator, with gauge coupling $g^2$ (of mass dimension $1$). Then, the gauge field (and the gaugino) acquire a so-called topological mass, $m_T= g^2 m_\lambda$. In the infrared, at scales well below $m_T$ (and for $k$ large enough so that we do not break supersymmetry dynamically \cite{Witten:1999ds}), we can integrate out the gauginos and recover pure Chern-Simons theory for a gauge group $\GG$ and shifted levels:
\be
\h k^{ab}= k^{ab} - \half \sign(k^{ab}) \sum_{\alpha\in \Fg} \alpha^a \alpha^b~,
\ee
as follows from \eqref{shift kappa intro}. For a simple gauge group $\GG= \GG_\gamma$, this is:
\be
\h k_\gamma = k_\gamma - \sign(k_\gamma) h~,
\ee
with $h$ the dual Coxeter number. With the choice of quantization \eqref{kappaRR App vec} for the vector multiplet, we then obtain the infrared contact terms:
\be
\CV\; : \; \quad \delta\kappa_{RR}^{(\rm IR)}={1-\sign(k)\ov 2}  {\rm dim}(\GG)~, \qquad\quad
 \delta\kappa_{g}^{(\rm IR)}=\left(1-\sign(k)\right){\rm dim}(\GG)~.
\ee
In particular, if $k>0$ we have $\kappa_{RR}^{(\rm IR)}=0$ and $\kappa_{g}^{(\rm IR)}=0$. Since the $U(1)_R$ symmetry completely decouples in the pure CS theory, this is a very natural choice. It also simplifies the presentation of various supersymmetric dualities.

\section{Geometry conventions}\label{app:geom}
In this appendix, we briefly discuss some useful facts about the half-BPS Seifert geometries, and we set our conventions for spinors. Our geometric conventions closely follow \cite{Closset:2012ru, Closset:2017zgf}, to which we refer for further discussion.

\subsection{Seifert geometry and THF}
Consider the oriented Seifert three-manifold:
\be
S^1 \longrightarrow \CM_3 \longrightarrow \h\Sigma~,
\ee
with Riemannian metric and Killing vector:
\be\label{met M3 app}
ds^2(\CM_3)= \beta^2 \big(d\psi + \CC(z,\bz)\big)^2 + 2 g_{z\bz}(z, \bz) dz d\bz~,\qquad\quad
K= {1\ov \beta}\d_\psi~,
\ee
as discussed in the main text.  There is a natural metric-compatible transversely holomorphic foliation (THF) on $\CM_3$ generated by the one-form:
\be
\eta= \beta\big(d\psi + \CC(z,\bz)\big)~,
\ee
with $\eta_\mu= K_\mu$. The THF can be defined in terms of $\eta_\mu$ and:
\be\label{def phi app}
{\Phi_\mu}^\nu= - {\epsilon_\mu}^{\nu\rho}\eta_\mu~,
\ee
with $\epsilon_{\mu\nu\rho}$ the Levi-Civita tensor.
We have:
\be\label{eta phi prop app}
\eta^\mu \eta_\mu=1~, \qquad  {\Phi_\mu}^\nu {\Phi_\nu}^\rho =- {\delta^\mu}_\nu+ \eta^\mu \eta_\rho~,
\ee
together with an integrability condition which is automatically satisfied in this case~\cite{Closset:2012ru}. As one can see from \eqref{eta phi prop app}, the tensor $\Phi$ defines a complex structure $J$ on the space of leaves of the foliation---here, the space of leaves is the base $\h\Sigma$ of the Seifert fibration:
\be
\Phi\big|_{\h\Sigma} = J_{\h \Sigma}~.
\ee
The local coordinates $z, \bz$ that appear in \eqref{met M3 app} are the complex coordinates on $\h \Sigma$ adapted to $J_{\h \Sigma}$, while $\psi\in [0, 2\pi)$ is the local coordinate along the Seifert fiber.

\subsection{Conventions for spinors}
We define the canonical frame:
\be\label{canonical frame}
e^0= \beta\left(d\psi + p \CA\right)~, \qquad\quad  e^1=\sqrt{2 g_{z\bz}}dz~, \qquad \quad e^{\b1}=\sqrt{2 g_{z\bz}} d\bz~,
\ee
adapted to the Seifert fibration structure.  Here, $e^1$, $e^{\b 1}$ form a complex frame on $\h\Sigma$. 
The frame indices $a=0, 1, \b1$  are lowered using $\delta_{ab}$ with $\delta_{00}=1$ and $\delta_{1\b 1}= \half$. The orientation is such that $\epsilon^{0 1 \b 1}= -2 i$. We choose the $\gamma$-matrices:
\be
\left\{{(\gamma^a)_\alpha}^\beta\right\} = \left\{\gamma^0, \gamma^1, \gamma^{\b1}\right\}= \left\{ \mat{1 & 0 \cr 0 &-1}~,\; \mat{0& -2\cr 0&0}~,\; \mat{0& 0\cr -2&0}  \right\}~. 
\ee
When reducing to two dimensions along the fiber direction, $\gamma^{1}, \gamma^{\b 1}$ become the two-dimensional $\gamma$-matrices, with $\gamma^0 \equiv \gamma^3$ the chirality matrix. In particular, any three-dimensional Dirac fermion:
\be
\psi_\alpha = \mat{\psi_-\cr \psi_+}
\ee
naturally decomposes into the 2d Weyl fermions $\psi_\pm$. Dirac spinor indices are raised and lowered with $\epsilon^{\alpha\beta}$, $\epsilon_{\alpha\beta}$, with $\epsilon^{-+}=\epsilon_{+-}=1$. The covariant derivative is given by:
 \be
 \nabla_\mu \psi = (\d_\mu - {i\ov 4}\omega_{\mu ab}\epsilon^{abc}\gamma_c)\psi~.
\ee

\subsection{Decomposition and adapted connection}
Given the Seifert structure above, it is useful to introduce the projectors:
\bea\label{projectors}
& {{\rm P_0}^\mu}_\nu= \eta^\mu \eta_\nu~, \cr
& {\Pi^\mu}_\nu =\half \left({\delta^\mu}_\nu - i {\Phi^\mu}_\nu- \eta^\mu \eta_\nu\right)~, \cr
&{ \t\Pi^\mu}_{\phantom{\mu}\nu} = \half \left({\delta^\mu}_\nu +i {\Phi^\mu}_\nu- \eta^\mu \eta_\nu\right)~.
\eea
They allow a decomposition of any tensor into vertical, holomorphic and anti-holomorphic components, corresponding to the canonical frame \eqref{canonical frame}. For instance, for any one-form $\omega$, we have:
\be\label{dec omega}
\omega = \omega_0 \eta + \omega_z dz +\omega_\bz d\bz~.
\ee
In particular, a {\it holomorphic one-form} on $\CM_3$ is such that:
\be
  \omega_\mu {\Pi^\mu}_\nu= \omega_\nu~.
\ee
By definition, its single component $\omega_z$ is a section of the canonical line bundle of $\CM_3$:
\be
\omega_z \in \Gamma[\CK_{\CM_3}]~.
\ee
Importantly, the Levi-Civita connection $\nabla_\mu$ does not commute with $\eta_\mu$, and therefore does not preserve the decomposition \eqref{dec omega}. We define a Seifert-compatible {\it adapted connection} $\h \nabla$, such that
\be
\h \nabla_\mu g_{\nu\rho}=0 ~, \qquad \h \nabla_\mu \eta_\nu=0~.
\ee
It is given by:
\be
{\h \Gamma^\mu}_{\phantom{\nu}\mu\rho}= { \Gamma^\nu}_{\mu\rho}
+  {K^\nu}_{\mu\rho}~, \quad\qquad K_{\nu\mu\rho}\equiv  -\beta\, c_1(\CL_0) \left(\eta_\nu \Phi_{\mu\rho}- \eta_\rho \Phi_{\mu\nu}+ \eta_\mu \Phi_{\nu\rho}  \right)~,
\ee
with ${ \Gamma^\nu}_{\mu\rho}$ the Christoffel symbols. Here, $c_1(\CL_0)$ is the first Chern class of the defining line bundle on $\h\Sigma$, as defined in Section~\ref{sec: seifert backgs}.  The compatible spin connection is:
\be
\h\omega_{\mu\nu\rho}=  \omega_{\mu\nu\rho}- K_{\nu\mu\rho}~.
\ee
The adapted connection $\h\nabla$ commutes with the projectors \eqref{projectors}, thus it is compatible with the decomposition into vertical, holomorphic and anti-holomorphic components. 
The price to pay is that $\h\nabla$ has torsion \cite{Closset:2012ru}:
\be
{T^{\nu}}_{\mu\rho}={K^{\nu}}_{\mu\rho}-{K^{\nu}}_{\rho\nu}= -2 \beta \, c_1(\CL_0)\eta^\nu \Phi_{\mu\rho}~.
\ee
Using this geometric decomposition and the Killing spinors $\zeta, \b\zeta$, it is easy to rewrite all fields in terms of two-dimensional forms on $\h\Sigma$, thus providing a very explicit description of the topological A-twist pulled-back to $\CM_3$ \cite{Closset:2014uda,Closset:2017zgf}.


\section{Comments on $S^3_b$ and $L(p,q)_b$ as Seifert fibrations}\label{app: S3 and lens}
In this appendix, we discuss some properties of the three-sphere and lens space backgrounds seen as Seifert fibrations, complementing the discussion in the main text.

\subsection{The squashed three-sphere as a Seifert fibration}\label{app: S3b metric}
In Section~\ref{subsec: S3b background}, we saw that the squashed three-sphere $S^3_b$ with $b^2 \in \Q$ is given in terms of a Seifert fibration with two exceptional fibers:
\be\label{S3b Seifert app}
S^3_b \cong [0~;\, 0~;\, (q_1, p_1)~, (q_2, p_2)]~,\quad \qquad q_1 p_2+ q_2 p_1=1~, \quad\qquad b^2= {q_1\ov q_2}~.
\ee
Any ``squashed-sphere'' background preserving the supersymmetry algebra:
\be\label{QQbK algebra app}
\{\CQ, \b \CQ\}=-2 i(\CL_K^{(b)}+Z)~,
\ee
 with $K^{(b)}$ a real Killing vector,~\footnote{More general backgrounds with $b$ and $K^{(b)}$ complex exist. We can focus on $b$ real for our purposes.} can be written in the general Seifert form of Section \ref{subsec: M3 backgd}. For definiteness, let us consider the $U(1)\times U(1)$-isometric background of \cite{Hama:2011ea}, with a real squashing parameter $b$. Let us describe the three-sphere as a torus fibered over an interval, with $\theta\in [0, \pi]$ the interval and $\chi\in [0,2\pi)$ and $\varphi \in [0, 2\pi)$ the angular coordinates on the torus. The squashed-sphere metric reads:
\be\label{S3b hhl metric}
ds^2(S^3_b) = R_0^2 \left( {1\ov 4} h(\theta)^2  d\theta^2 + b^2  \sin^2{\theta\ov 2} d\chi^2 + b^{-2} \cos^2{\theta\ov 2} d\varphi^2\right)
\ee
The function $h(\theta)$ is some smooth positive function which behaves as:
\be
h(\theta) \sim b + \CO(\theta^2) \quad \text{as}\;\; \theta\sim 0~, \qquad 
h(\theta) \sim b^{-1} + \CO((\pi-\theta)^2) \quad \text{as}\;\; \theta\sim \pi~, \qquad 
\ee
near the ``poles'' $\theta=0, \pi$, and is otherwise arbitrary.
For $b=1$ and $h(\theta)=1$, \eqref{S3b hhl metric} is the round metric on $S^3$ with radius $R_0$.
The Killing vector $K^{(b)}$ appearing in \eqref{QQbK algebra app} is given by:
\be
K^{(b)}= {1\ov R_0}\left(b \d_\varphi + b^{-1} \d_\chi\right)~.
\ee
One can check that:
\be
\eta = K^{(b)}_\mu dx^\mu= R_0 \left(b \sin^2{\theta\ov 2} d\chi + b^{-1} \cos^2{\theta\ov 2} d\varphi\right)
\ee
defines a THF, and one finds the auxiliary fields:~\footnote{Here we choose $\kappa=0$ for the ``$\kappa$ parameter'' in \protect\cite{Closset:2012ru, Closset:2013vra}. In \protect\cite{Hama:2011ea}, the implicit choice was $\kappa=  -{2\ov R_0 h(\theta)}$, so that $H= -{i\ov R_0} h(\theta)$ and $V_\mu=0$. These are equivalent supersymmetric backgrounds~\protect\cite{Closset:2013vra}.}
\be\label{sugra HHL}
H= {i \ov R_0 h(\theta)}~, \qquad V_\mu= -{2 \ov R_0 h(\theta)} \eta_\mu~, \qquad A_\mu^{(R)}dx^\mu= {1\ov 2 h(\theta)} \left(b d\chi + b^{-1} d\varphi\right)~. 
\ee
Now, consider the case:
\be\label{b as q1q2 app}
b= \sqrt{q_1\ov q_2}~,\qquad q_1, q_2\in \Z_{>0}~.
\ee 
To bring the supersymmetric background \eqref{S3b hhl metric}-\eqref{sugra HHL} to the general form of Section~\ref{subsec: M3 backgd}, we need to perform a change of coordinates from the angles $\chi, \varphi$ to some new angles $\phi, \psi$ such that:
\be
K^{(b)}= {1\ov \beta}\d_\psi~,
\ee
with $\beta$ the radius of the generic Seifert fiber. One can check that,  in the new variables:
\be
\mat{\phi\\ \psi}= M \mat{\chi\\ \varphi}~, \qquad M= \mat{q_1 & -q_2\\ p_1 & p_2}\in SL(2, \Z)~, \qquad R_0= {\beta\ov \sqrt{q_1 q_2}}~,
\ee
the metric  \eqref{S3b hhl metric} takes the standard form:~\footnote{Note that $H$ and $V_\mu$ in \protect\eqref{sugra HHL} differ from \eqref{AVH sugra} by a factor ${1\ov h(\theta)}$. Relatedly, we have $d\eta= {2\ov R_0 h(\theta)} d\vol(\h\Sigma)$ instead of \protect\eqref{deta explicit}. To land exactly on the  background \protect\eqref{AVH sugra}, we need to do a Weyl rescaling of the metric on the spindle $S^2(q_1, q_2)$. This does not affect supersymmetric observables~\protect\cite{Closset:2013vra}.}
\be
ds^2(S^3_b)= \beta^2 \left(d\psi+ \CC\right)^2 + ds^2(\h\Sigma)~, 
\ee
with the Seifert connection:
\be\label{CC S3b}
\CC= \half\left(-{p_1\ov q_1}+ {p_2\ov q_2}- {\cos\theta\ov q_1 q_2}\right) d\phi~,  
\ee
and the orbifold metric:
\be
ds^2(\h\Sigma) = R_0^2\, h(\theta)^2 \left(d\theta^2 + {\sin^2\theta \ov f(\theta)^2} \right)~, \qquad {\rm with} \quad f(\theta)= \sqrt{q_1 q_2} h(\theta)~.
\ee
Comparing to \eqref{met spindle}, this is clearly a metric on the spindle $S^2(q_1, q_2)$. The connection \eqref{CC S3b} satisfies:
\be
{1\ov 2\pi}\int_{\omega_1} \CC= {p_1\ov q_1}~, \qquad 
{1\ov 2\pi}\int_{\omega_2} \CC= {p_2\ov q_2}~, \qquad
{1\ov 2\pi}\int_{\h\Sigma} d\CC= {1\ov q_1q_2}= {p_1\ov q_1}+ {p_2\ov q_2}~,
\ee
with $\omega_1,\omega_2$ the generators of the orbifold fundamental group on $S^2(q_1, q_2)$.
This shows that the $S^3_b$ background with $b^2= {q_1\ov q_2}$ corresponds to the Seifert fibration \eqref{S3b Seifert app}.

\subsection{All the Seifert fibrations of $L(p,q)$}\label{app:lens space algo}
As we mentioned in Section \ref{sec:lpqgeo}, any genus-zero Seifert fibration with $n \leq 2$ exceptional fibers is a lens space. Conversely,  for any lens space $L(p,q)$, we would like to find all of its possible Seifert fibrations.
Here we review the algorithm of \cite{2016arXiv160806844G}---Theorem 4.10 therein---, which constructs all the Seifert fibrations of a given $L(p,q)$.  (We assume that $p\neq 0$. The case $L(0,1)\cong S^2\times S^1$ was treated separately in the main text.)
Consider $p$ and $q$ two mutually-prime integers. Given the mutually-prime non-zero integers $q_1^0$ and $q_2^0$, we can construct the Seifert fibration:
\be\label{M3 lens app}
\CM_3 \cong  [0~;\, 0~;\, (q_1, p_1)~,\; (q_2, p_2)]
\ee
on $L(p,q)$, as in \eqref{M3lens}-\eqref{p and q for lens},  in the following way:
\begin{itemize}
\item Choose some integers $s, t$ such that $q s+ p t=1$.
\item Define:
\be\label{q1q2 algo}
q_1 = \alpha q_1^0~, \quad q_2= \alpha q_2^0~, \qquad \quad \alpha \equiv {p\ov \gcd(p, s q_1^0-q_2^0)}~.
\ee
This defines $q_1, q_2$ in \eqref{M3 lens app}.
\item Define: $$t_1= -{ s q_1^0 - q_2^0\ov \gcd(p, s q_1^0-q_2^0)}~.$$
 Then there exists some integers $s_1, p_1$ such that $q_1 s_1 + p_1 t_1=1$. This defines $p_1$ in  \eqref{M3 lens app}. 
\item Finally, we define $p_2\equiv -s p_1 + s_1 p$.
\end{itemize}
One can check that this reproduces \eqref{p and q for lens}, for $q$ mod $p$. Due to \eqref{q1q2 algo}, we also see that:
\be
b^2 ={q_1 \ov q_2}= {q_1^0\ov q_2^0}~,
\ee
so that a choice of $(q_1^0, q_2^0)$ is equivalent to a choice of $b^2\in \Q$.

\section{Supersymmetric one-loop determinants on Seifert manifolds}\label{app:oneloop det}
In this appendix, we further explain the computation of one-loop determinant of a chiral multiplet on $\CM_3$, generalizing the discussion on $\CM_3 = \Mgp$ in \cite{Closset:2017zgf}. We then study some of its properties, providing additional details about computations that we alluded to in the main text.

\subsection{Derivation and general properties of $Z^\Phi_{\CM_3}$}
Consider a 3d $\CN=2$ chiral multiplet $\Phi$, of $U(1)$ gauge charge $1$ and $R$-charge $r$, on the half-BPS Seifert background $(\CM_3, L_R)$ discussed in Section~\ref{sec: seifert backgs}. Let $D_\mu$ be the  covariant derivative:
\be\label{Dmu def app}
D_\mu = \h\nabla_\mu - i A_\mu - i r_\varphi A_\mu^{(R)}~,
\ee
with the adapted connection $\h\nabla$ introduced in Appendix~\ref{app:geom}, acting on some field $\varphi$ of $R$-charge $r_\varphi$. On the supersymmetric locus for the vector multiplet, with holomorphic parameters:
\be
u = i \beta \sigma - a_0~,\qquad a_0 = {1\ov 2 \pi }\int_\gamma A~,
\ee
 and flux $\n$, most modes cancel out between the bosons and fermions. By a standard argument---see in particular \cite{Pestun:2007rz, Hama:2011ea, Closset:2015rna}---we then find:
\be
Z_{\CM_3}^\Phi = {{\rm det}_{{\rm coker}D_{\b1}}(-\sigma+ D_0)  \ov  {\rm det}_{{\rm ker}D_{\b1}}(-\sigma+ D_0) }~.
\ee
Here, $(D_0, D_{1}, D_{\b 1})$ denote the covariant derivative \eqref{Dmu def app} in the canonical frame basis.  We then expand any field along the Seifert fiber:
\be\label{varphi exp app}
\varphi = \sum_{\kk\in \Z} \varphi_\kk\, e^{i \kk \psi}~,
\ee
with the modes $\varphi_\kk$ the two-dimensional fields on $\h\Sigma$. In particular, the modes $\CA_\kk$ in the kernel of $D_{\b 1}$ are the holomorphic sections of the orbifold line bundle:
\be
L_{(r,k)}\equiv L \otimes L_R^{r} \otimes \CL_0^\kk \cong L_0^{\n_0 + \t \n_0^R r+ \kk {\bf d}}\, \bigotimes_{i=1}^n L_i^{\n_i + \n_i^R r+ \kk p_i}~,
\ee
Note that:
\be
\dim {\rm ker}D_{\b1} = h^0(L_{(r,k)})~, \qquad\quad \dim {\rm coker}D_{\b1} = h^0(L_{(r,k)}^{-1} \otimes \CK)~.
\ee
Using the Riemann-Roch-Kawasaki formula \eqref{RRK thm}, we directly find the formal product:
\be
Z_{\CM_3}^\Phi = \prod_{\kk \in \Z} \left({1\ov u + \nu_R r + \kk}\right)^{{\rm deg}(L_{(r,k)}) +1 -g}~,
\ee
which is the result \eqref{ZM3 Phi product} quoted in the main text.

\subsection{Regularizing $\CG_{q,p}^\Phi$}
As a formal infinite product, the $(q,p)$ fibering operator for $\Phi$ with $R$-charge $r=0$ reads:
\be\label{CGqp app}
\CG^\Phi_{q,p}(u)_\n = \prod_\kk \left({1\ov u + \kk}\right)^{\floor{p \kk + \n \ov q}}~.
\ee
By rewriting the product over $\kk\in \Z$ as:
\be
\kk = q n + t l~, \qquad n \in \Z~, \qquad l=0, \cdots, q-1~, \qquad qs+pt =1~,
\ee
 we obtain:
\be\label{Gqp reg inter app}
\CG^\Phi_{q,p}(u)_\n =\prod_{l=0}^{q-1} \CF^\Phi\left({u + t l\ov q}\right)^p \, \pif^\Phi\left({u + t l\ov q}\right)^{\floor{t p l + \n\ov q}}~,
\ee
with the ordinary fibering and flux operators $\CF^\Phi$ and  $\pif^\Phi$, respectively, as in  \eqref{flux fiber Phi}:
\be\label{phi F app}
\pif^\Phi(u) \equiv  \prod_{\kk\in \Z} {1\ov u+ \kk}~, \qquad\qquad 
\CF^\Phi(u)  \equiv  \prod_{\kk \in \Z}\left(1\ov u+ \kk\right)^\kk~.
\ee
 which are regularized to:
\be\label{phi F reg app}
\pif^\Phi(u) \equiv {1\ov 1-e^{2\pi i u}}~, \qquad
\CF^\Phi(u)  \equiv  \exp\left({1\ov 2 \pi i} \dilog(e^{2\pi i u}) + u \log\left(1- e^{2\pi i u}\right)\right)~,
\ee
in the $U(1)_{-\half}$ quantization for the Dirac fermions, as discussed in \cite{Closset:2017zgf}.
The regularized expression \eqref{Gqp reg inter app} can then be written as:
\be
\CG^\Phi_{q,p}(u)_\n = \pif_{q,p}^\Phi(u)_\n\, \CG^\Phi_{q,p}(u)
\ee
with:
\bea\label{CGqp0 Phi app}
&  \pif_{q,p}^\Phi(u)_\n  &\equiv&\;  \left(e^{2\pi i u \ov q}; e^{2\pi i t\ov q}\right)_{-\n}~, \cr
&\CG_{q,p}^\Phi(u) &\equiv&\;  \exp \sum_{l=0}^{q-1}\left\{ {p\ov 2 \pi i} \dilog(e^{2\pi i {u+ t l\ov q}}) + {pu + l \ov q} \log\left(1-e^{2\pi i {u+ t l\ov q}}\right)\right\}~,
\eea
which is the expression discussed in the main text. Using the identities:
\be\label{id sum dilog log app}
\sum_{l=0}^{q-1} \dilog(e^{2\pi i {u+ t l\ov q}})= {1\ov q}  \dilog(e^{2\pi i u})~,\qquad
\sum_{l=0}^{q-1} \log(1-e^{2\pi i {u+ t l\ov q}})=  \log(1-e^{2\pi i u})~,
\ee
 one can also write $\CG_{q,p}(u)$ as:
\be\label{CGqp0 Phi app 2}
\CG_{q,p}^\Phi(u) = e^{{p\ov q}\left({1\ov 2 \pi i} \dilog(e^{2\pi i u})+ u \log\left(1-e^{2\pi i u}\right)\right) } \, \prod_{l=1}^{q-1} \left(1-e^{2\pi i {u+ t l\ov q}}\right)^{l \ov q}~.
\ee

\subsection{The Chern-Simons limit of $\CG^\Phi_{q,p}$}\label{app:sec ZPhi and limit}
Let us consider the limit $u \rightarrow \pm i \infty$ on the chiral-multiplet fibering operator. We obviously have:
\be
\lim_{u\rightarrow i \infty} \CG^\Phi_{q,p}(u)_\n = 1~,
\ee
consistently with our choice of quantization of $\Phi$. In the opposite limit, $u \rightarrow- i \infty$, we should generate the CS terms:
\be\label{k shifts sigma app}
k_{GG} = -1~, \qquad  k_{GR}= 1~, \qquad k_{RR}= -1~, \qquad k_{g}=-2~,
\ee
corresponding to plugging $r=0$ into \eqref{k shifts sigma}. Let us first consider the case $\n=0$. Using the expression \eqref{Gqp reg inter app} and the limits \cite{Closset:2017zgf}:
\be
\lim_{u\rightarrow -i \infty} \CF^\Phi(u) \sim e^{\pi i \left(u^2 - {1\ov 6}\right)}~, \qquad
\lim_{u\rightarrow -i \infty} \pif^\Phi(u) \sim -e^{-2\pi i u}~,
\ee
we  find:
\be
\lim_{u\rightarrow -i \infty} \CG^\Phi_{q,p}(u)  \sim e^{\pi i {p\ov q} u^2} \, e^{\pi i {q-1\ov q} u }\, e^{i \varphi^{(0)}_{q,p}}~,
\ee 
with the phase:
\be\label{varphi0 def}
e^{i \varphi^{(0)}_{q,p}} \equiv (-1)^{s {q(q-1)\ov 2}} \,  e^{{\pi i\ov 6} \left(t \left({1\ov q}+s\right)(q-1)(2q-1) - p q \right)}~.
\ee
This expression only depends on the coprime integers $(q,p)$, not on the choice of $s,t$ such that $qs +pt=1$. We  claim that, for $q>0$ and $p$ coprime to $q$, \eqref{varphi0 def} can  be written as:
\be\label{varphi0 eq}
e^{i \varphi^{(0)}_{q,p}}=\left(\CG^{(0)}_{q,p}\right)^{-2} = \exp\left(-2\pi i \left({p\ov 12 q}- s(p,q)\right)\right)~,
\ee
with $s(p,q)$ the Dedekind sum and $\CG^{(0)}_{q,p}$ defined as in \eqref{def CG0}.
This is equivalent to:
\be\label{eval sqp}
\exp\Big({2\pi i\, s(p,q)}\Big) =
\exp\Big({{\pi i (q-1) \ov 6 q} \Big(3 s q^2-p(q+1)+ t (q s+1)(2q-1) \Big)}\Big)~,
\ee
which is an evaluation formula for the Dedekind sum $s(p,q)$ modulo integers, when $q>0$ and $\gcd(q,p)=1$. It would be interesting to prove \eqref{eval sqp} directly.~\footnote{We discovered this relation  by comparing our results with known results in pure Chern-Simons theory. It can be checked ``experimentally.'' We leave the proof as an exercise for the interested number theorist, as it were.}
As a sanity check, we note that the following known evaluation formulas for the Dedekind sum at $p=1$ and $p=2$:
\be
 s(1,q) = {(q-1)(q-2)\ov 12 q}~,\qquad  s(2,q) = {(q-1)(q-5)\ov 24 q}~,
\ee
are consistent with \eqref{eval sqp}.
Now, re-introducing the fractional flux $\n\in \Z$, we also have the limit:
\be
\lim_{u\rightarrow -i \infty}\pif_{q,p}^\Phi(u)_\n \sim (-1)^\n\, e^{-{2\pi i\ov q} \n u} \, e^{\pi i {t\ov q}\n(\n+1)}~.
\ee
Therefore, we find:
\be\label{lim Gphiqpn app}
\lim_{u\rightarrow-i \infty} \CG^\Phi_{q,p}(u)_\n\, \sim \,   (-1)^\n\, e^{-{2\pi i\ov q} \n u} \, e^{\pi i {t\ov q}\n(\n+1)} \, 
e^{\pi i {p\ov q} u^2} \, e^{\pi i {q-1\ov q} u }\, \left(\CG^{(0)}_{q,p}\right)^{-2}~.
\ee
For a general $R$-charge $r\in \Z$, we simply replace:
\be
u \rightarrow u+\nu_R r~, \qquad \n\rightarrow \n + \n^R r~,
\ee
in \eqref{lim Gphiqpn app}. The resulting expression must be equal to the correct CS terms, as indicated in \eqref{CS limit of GPhi}. This is indeed the case, as one can check by direct computation. In fact, that is the method we first used to derive the expression of Section~\ref{subsec: CSterms} for the Chern-Simons contributions to the fibering operator. In Appendix~\ref{app: CS}, we give an independent consistency check of those results.

\subsection{Comments on $Z^\Phi_{S^3_b}$ and its Seifert factorization}\label{app:subsec:factorization of S3b}
In this subsection, we provide some additional details about the one-loop determinant $Z^\Phi_{S^3_b}$ for a chiral multiplet on the squashed three-sphere, $S^3_b$,  discussed in Section~\ref{sec: S3b}, and we demonstrate the Seifert factorization of $Z^\Phi_{S^3_b}$ at rational values of $b^2$.

The supersymmetric one-loop determinant on $S^3_b$ is given by the formal product \cite{Hama:2011ea}:
\be\label{ZS3b prod app}
Z_{S^3_b}^\Phi(\h\sigma+ \h\sigma_R r) =  \prod_{n_1=0}^\infty  \prod_{n_2=0}^\infty {n_2  b + n_1b ^{-1}- i\h\sigma + {b+b^{-1}\ov 2}(2-r) \ov n_2 b + n_1 b^{-1} +i \h\sigma +{b+b^{-1}\ov 2} r }~.
\ee
This is naturally regularized in terms of the quantum dilogarithm (to be discussed below):
\be\label{ZS3b dilog app}
Z_{S^3_b}^\Phi(\h\sigma+ \h\sigma_R r) = \t \Phi_b\big(\h\sigma + \h\sigma_R(r-1)\big)~, \qquad
\h\sigma_R = -i {b+ b^{-1}\ov 2}~,
\ee 
or, equivalently:
\be\label{ZS3b dilog app 2}
Z_{S^3_b}^\Phi(\h\sigma+ \h\sigma_R r) = \Phi_b\big(\h\sigma + \h\sigma_R(r-1)\big) e^{-\pi i \h\sigma^2}e^{-2\pi i (r-1)\h\sigma \h\sigma_R} e^{-\pi i (r-1)^2 \h\sigma_R^2} e^{-{\pi i \ov 12}\left(b^2+ b^{-2}\right)}~.
\ee
The expression \eqref{ZS3b dilog app} corresponds to the chiral multiplet in the $U(1)_{-1}$ quantization, which includes the contact terms:
\be
\kappa_{GG}= -\half~, \qquad \kappa_{GR}= -\half(r-1)~,\qquad \kappa_{RR}=-\half(r-1)^2~,
\qquad \kappa_g=-1~. 
\ee
Then, since the CS terms on $S^3_b$ are given by \eqref{CS terms S3b}, it is clear from \eqref{ZS3b dilog app 2} that $\Phi_b\big(\h\sigma + \h\sigma_R(r-1)\big)$ corresponds to the chiral multiplet in the $U(1)_\half$ quantization.

\subsubsection{Some properties of the quantum dilog $\Phi_b(\h\sigma)$.}
The quantum dilogarithm is generally defined as:
\be
\Phi_b(\h\sigma)\equiv    \left(e^{2\pi b\h\sigma}\, e^{\pi i \left(b^2+1\right)}; e^{2\pi i b^{2}}\right)_\infty\;
\left(e^{{2\pi\ov b} \h\sigma}\, e^{-\pi i \left({1\ov b^2}+1\right)}; e^{-2\pi i b^{-2}}\right)^{-1}_\infty~,
\ee
for ${\rm Im}(b^2)>0$, which is related to the function $\t\Phi_b(\h\sigma)$ introduced in \eqref{def Phib quantum dilog} by:
\be\label{Phi Phit id}
\Phi_b(\h\sigma)=  \t\Phi_b(-\h\sigma)^{-1} = \t\Phi_b(\h\sigma) e^{\pi i \h \sigma^2}e^{{\pi i \ov 12}\left(b^2+ b^{-2}\right)}~.
\ee
In fact, $\Phi_b(\h\sigma)$ and $\t\Phi_b(\h\sigma)$ correspond to a chiral multiplet of $R$-charge $r=1$ in the $U(1)_\half$ or $U(1)_{-\half}$ quantization, respectively, on $S^3_b$. Then, the identity:
\be
\Phi_b(\h\sigma)  \t\Phi_b(-\h\sigma)=1~,
\ee
 is simply the statement that, for two chiral multiplets of $R$-charge $r=1$ and gauge charges $\pm 1$ in the ``symmetric'' quantization, we have no leftover CS contact terms in the IR.

The quantum dilogarithm satisfies many interesting identities, which have  nice interpretations in the field theory \cite{Dimofte:2011ju}. 
The simplest relation is:
\be
\Phi_b(\h\sigma)\Phi_b(-\h\sigma)= e^{\pi i \h \sigma^2}e^{{\pi i \ov 12}\left(b^2+ b^{-2}\right)}~,
\ee
which is equivalent to \eqref{Phi Phit id}. (Here the statement is that, integrating out two chiral multiplets in the $U(1)_\half$ quantization, we are left with the CS levels $k=1$ and $k_g=2$.)
Interestingly, $\Phi_b(\h\sigma)$ has a simple Fourier-transform (see {\it e.g.} \cite{Faddeev:2000if}):
\be\label{Fourrier Phib}
e^{-{\pi i \ov 12}(b^2 + b^{-2} +3)} \int d\h\sigma\, \Phi_b(\h\sigma) \, e^{2\pi i \h\xi \h \sigma} = e^{-\pi i \h \xi^2}\, \Phi_b(\h\xi -\h \sigma_R)~,
\ee
with $\h\sigma_R$ as defined in \eqref{ZS3b dilog app}.
This corresponds to the elementary $\CN=2$ mirror symmetry \eqref{EMS duality} between the $U(1)_\half$ gauge theory with one chiral multiplet (with R-charge $r=1$) and the free chiral $T^+$ (with $R$-charge $0$). Indeed, since $\Phi_b(\h\sigma)$ corresponds to the ``$U(1)_\half$ quantization,'' we have $\kappa=\half$ (for the gauge symmetry) and $\kappa_g=1$ on the left-hand-side of \eqref{Fourrier Phib} (with the phase in front of the integral corresponding to the gaugino), while in the right-hand-side we have the contributions $\delta\kappa_{TT}=\half$, $\delta\kappa_{TR}=-1$, $\delta\kappa_{RR}=1$ and $\kappa_g=1$ from $\Phi_b(\h\xi -\h \sigma_R)$, and a contribution $\delta\kappa_{TT}=-1$ from the bare CS term $e^{-\pi i \h \xi^2}$. This is therefore equivalent to the duality \protect\eqref{EMS duality}.

Another interesting relation for $\Phi_b(\h\sigma)$ is the pentagon identity \cite{Faddeev:1993rs}. Let us use the function $\t\Phi_b(\h\sigma)$, for convenience. Then, the pentagon identity can be written as: 
\bea\label{pentag id app}
&e^{-{\pi i \ov 12}(b^2 + b^{-2} +3)} \int d\h\sigma\, \t\Phi_b(\h\sigma+ \h m_A)\t\Phi_b(-\h\sigma+ \h m_A)e^{\pi i \h\sigma^2}e^{2\pi i \h\xi\h\sigma}\,=\, e^{{\pi i \ov 12}(b^2 + b^{-2})}\cr
& \times e^{\pi i \h\xi^2}e^{2\pi i (\h m_A + \h\sigma_R)^2} \, \t\Phi_b(\h\xi+ \h m_A-\h\sigma_R) \t\Phi_b(-\h\xi+ \h m_A-\h\sigma_R) \t\Phi_b(2 \h m_A+\h\sigma_R)
\eea
This identity corresponds to the well-known mirror symmetry between SQED, a $U(1)$ theory with two chiral multiplets of charge $\pm1$,  and the $XYZ$ model, consisting of three chiral multiplets $(X,Y,Z)\equiv(M, T^+, T^-)$ coupled by cubic superpotential $W=M T^+ T^-$. This is also a special case of Aharony duality discussed in Section~\ref{subsec: aha duality}, when $N_f=N_c=1$ (and $r=1$, here). The parameters $\h m_A$ and $\h\xi$ in \eqref{pentag id app} are the complexified chemical potentials for the global symmetry $U(1)_A \times U(1)_T$ of SQED.

\subsubsection{Seifert factorization of $Z^\Phi_{S^3_b}$}
Finally, let us further comment on the factorization of the one-loop determinant \eqref{ZS3b dilog app} into fibering operators, when $b^2$ is rational. It is convenient to start with the unregularized product \eqref{ZS3b prod app} at $r=0$. (The general $r$ case can be obtained by shifting $\h\sigma$.) Given:
\be
b^2 ={q_1\ov q_2}~, \qquad q_1, q_2\in \Z_{>0}~, \qquad u = i \sqrt{q_1 q_2} \h\sigma~,
\ee
the expression \eqref{ZS3b prod app} becomes:
\be
Z_{S^3_b}^\Phi(u)\big|_{r=0} =  \prod_{n_1=0}^\infty  \prod_{n_2=0}^\infty {(n_2+1)q_1 + (n_1+1)q_2-u \ov n_2 q_1 +q_2 n_1 + u}~.
\ee
By reordering the infinite product, this can brought to the form:
\be
Z_{S^3_b}^\Phi(u)\big|_{r=0} =\prod_{\kk \in \Z}\prod_{l_1=0}^{q_1-1}\prod_{l_2=0}^{q_2-1} \left({1\ov u + \kk q_1 q_2 + l_1 q_2 + l_2 q_1}\right)^{\kk+1}~.
\ee
The product over $\kk \in \Z$ can be interpreted as a product over the momentum modes $\varphi_\kk$ along the Seifert fiber. Using \eqref{phi F app}-\eqref{phi F reg app}, that product can then be immediately regularized to:
\be
Z_{S^3_b}^\Phi(u)\big|_{r=0} = \prod_{l_1=0}^{q_1-1}\prod_{l_2=0}^{q_2-1}  \pif^\Phi\left({{u\ov q_1 q_2}+ {l_1\ov q_1}+ {l_2\ov q_2}}\right)\,  \CF^\Phi\left({{u\ov q_1 q_2}+ {l_1\ov q_1}+ {l_2\ov q_2}}\right)
\ee
By using the identities \eqref{id sum dilog log app} repeatedly, it is easy to show that:
\be
 \prod_{l_1=0}^{q_1-1}\prod_{l_2=0}^{q_2-1}  \pif^\Phi\left({{u\ov q_1 q_2}+ {l_1\ov q_1}+ {l_2\ov q_2}}\right)= \pif^\Phi(u)~,
\ee
and:
\be
 \prod_{l_1=0}^{q_1-1}\prod_{l_2=0}^{q_2-1}  \CF^\Phi\left({{u\ov q_1 q_2}+ {l_1\ov q_1}+ {l_2\ov q_2}}\right) = \CG^\Phi_{q_1,p_1}(u)\, \CG^\Phi_{q_2,p_2}(u)~, \qquad \text{if}\; \;q_1 p_2+ q_2 p_1=1~.
\ee
This gives a physicist's proof of the factorization formula \eqref{ZPhiS3b factorization},  which can be written as a property of the quantum dilogarithm:
\be
\Phi_b\left({i u \ov \sqrt{q_1 q_2}} -{i\ov 2} {q_1+q_2\ov \sqrt{q_1 q_2}}\right) =\pif^\Phi(u)^{-1} \,  \CG^\Phi_{q_1, p_1}(u)^{-1}\,  \CG^\Phi_{q_2, p_2}(u)^{-1}~, \qquad b^2= {q_1\ov q_2}~,
\ee
with $q_1 p_2+ q_2 p_1=1$.
To the best of our knowledge, this property was first discussed in \cite{Garoufalidis:2014ifa} from a mathematical perspective. Here, we give it a new physical interpretation, by viewing $S^3_b$, with $b^2$ rational, as a Seifert fibration.

\section{Chern-Simons actions on Seifert manifolds}\label{app: CS}
 In this appendix, we collect some comments about the classical Chern-Simons functional on Seifert manifolds. In Appendix~\ref{app:sec ZPhi and limit}, we explained how we derived the classical Chern-Simons contribution $Z^{\rm CS}_{\CM_3}$ to the supersymmetric partition function $Z_{\CM_3}$, for any supersymmetric Seifert background $\CM_3$, by taking appropriate limits  of the one-loop determinant for free chiral multiplets coupled to background vector multiplets. 
 
 This is a convenient but oddly roundabout way to compute  $Z^{\rm CS}_{\CM_3}$. Indeed, the straightforward computation would be to evaluate the known supersymmetric actions on the Seifert background:
 \be\label{ZCS app}
 Z^{\rm CS}_{\CM_3}= \exp\left(- \int_{\CM_3} d^3 x \sqrt{g} \, \SL_{\rm CS}\right)~,
 \ee
with $\SL_{\rm CS}$ the sum of the various supersymmetric CS Lagrangians. 
However, in the presence of non-trivial flat connections on $\CM_3$, the direct evaluation of \eqref{ZCS app} is not entirely straightforward. In the following, we make some further comments on the evaluation of \eqref{ZCS app}, and we compare our results from Section~\ref{subsec: CSterms}  to previously-known results. 

\subsection{The $U(1)_k$ Chern-Simons functional}
The $\CN=2$ supersymmetric Chern-Simons functional for a gauge group $U(1)$ at CS level $k\in \Z$ takes the form:
\be
S_{U(1)_k} ={k \over 4 \pi}   \int d^3 x \sqrt{g} \left(i \epsilon^{\mu\nu\rho} A_\mu \d_\nu A_\rho - 2 D \sigma + 2 i  \b\lambda\lambda\right)~.
\ee
On the supersymmetric locus, $\lambda=\b\lambda=0$, $\sigma$ is constant and:
\be
D= 2 i f_{1\b 1}+ \sigma H~, \qquad H= i \beta c_1(\CL_0)~.
\ee
Let us assume that $c_1(\CL_0)\neq 0$, so that we can expand the gauge field $A_\mu$ as:
\be
A_\mu = \h a_0 \eta_\mu + a^{(\rm flat)}_\mu~,
\ee
with $\h a_0$ constant.
Then, defining the quantity:
\be
u_0 \equiv i \beta\left(\sigma + i  \h a_0\right)~,
\ee
it is easy to check that:
\be\label{SU1k app}
S_{U(1)_k} =  S_{U(1)_k}^{\rm flat} +  \pi i  k  c_1(\CL_0)\, u_0^2~,
\ee
including the contribution from the flat connection:
\be\label{SU1k flat def}
S_{U(1)_k}^{\rm flat}\equiv {i k\ov 4 \pi} \int_{\CM_3} a^{(\rm flat)} d a^{(\rm flat)}~.
\ee
In principle, this latter quantity (modulo $2\pi i$) can be computed for any $a^{(\rm flat)}$ valued in the torsion group $H^1(\CM_3, \Z)$, for instance by extending to a four-manifold with compatible spin structure \cite{Dijkgraaf:1989pz}:
\be
 {i k\ov 4 \pi} \int_{\CM_3} a^{(\rm flat)} d a^{(\rm flat)}= {i k\ov 4 \pi}  \int_{\CM_4}  F^{(\rm 4d)}\wedge  F^{(\rm 4d)}~.
\ee
In practice, this is rather non-trivial to compute.

\paragraph{Extracting the CS functional from the fibering operator.} 
Consider then a non-trivial torsion line bundle:
\be
{\bf L} \in \t\Pic(\CM_3)~,
\ee
with flat connection $A^{\rm flat}$. It can be represented as:
\be\label{bf L as L}
{\bf L}= \pi^{\ast}(L)~,\qquad L=  \bigotimes_{i=0}^n L_i^{\n_i}\;  \in \; \Pic(\h\Sigma)~,
\ee
with $L$ an orbifold line bundle on $\h \Sigma$, and $\n_i$ the ``fractional fluxes''.~\footnote{Here $\n_0$ is an ``ordinary flux''. We find it convenient to let the index $i$ run from $0$ to $n$.}
Our supersymmetric result \eqref{CS GG} then gives:
\be\label{Zcs from G}
 Z^{U(1)_k}_{\CM_3} = e^{-S_{U(1)_k}} = \left(\prod_{i=0}^n \CG^{\rm GG}_{q_i,p_i}(u)_{\n_i}\right)^k~,
\ee
for an arbitrary Seifert manifold. Here we have:
\be
u= u_0 + {c_1(L)\ov c_1(\CL_0)}~, \qquad\qquad c_1(L)= \sum_{i=0}^n {\n_i\ov q_i}~,
\ee
as discussed in Section \ref{sec:localization}. One can check that \eqref{Zcs from G} agrees with \eqref{SU1k app} if and only if:
\be\label{SU1k explicit}
\exp\left(-S_{U(1)_k}^{\rm flat}\right) = (-1)^{{\bf s}(\n, l^R, \nu_R) k} \, \exp\left(\pi i k \left( {c_1(L)^2\ov c_1(\CL_0)} - \sum_{i=0}^n {t_i \n_i^2\ov q_i} \right)\right)~.
\ee
Here we defined the sign:
\be\label{def sign sL}
(-1)^{{\bf s}(\n, l^R, \nu_R)}=\prod_{i=0}^n (-1)^{\n_i (1+ t_i+ l^R_i t_i +2 \nu_R s_i)}~.
\ee
Note that \eqref{SU1k explicit} is determined by both the choice of $A^{\rm flat}$ and of the spin structure on $\CM_3$. As a small consistency check, one can verify that  \eqref{SU1k explicit} is invariant upon $L \rightarrow L \otimes \CL_0$---this follows directly from \eqref{GG lgt}.~\footnote{Under a shift $\n_i \rightarrow \n_i +p_i$, for every $i$ at once, the sign \protect\eqref{def sign sL} transforms as $(-1)^{\bf s}\rightarrow (-1)^{\sum_i p_i s_i} (-1)^{\bf s}$, while the full answer \protect\eqref{SU1k explicit} is invariant.}

The formula \eqref{SU1k explicit} therefore gives us a completely explicit formula for the $U(1)_k$ CS action \eqref{SU1k flat def}, for any non-trivial torsion line bundle ${\bf L}$ over $\CM_3$.

\subsection{Non-abelian Chern-Simons functional}
The above considerations can be generalized to the case of a non-abelian, simply connected (or unitary) gauge group $\GG$ at level $k\in \Z$, by ``abelianization'' as in equation  \eqref{GG nonabelian}.  Let us define:
\be\label{SGGk def}
S_{\GG_k}^{\rm flat}\equiv {i k\ov 4 \pi} \int_{\CM_3} \Tr\left(A^{\rm flat} d A^{\rm flat}- {2 i\ov 3} A^{\rm flat}\wedge A^{\rm flat}\wedge A^{\rm flat} \right)~,
\ee
the non-abelian generalization of \eqref{SU1k flat def}. Here $A^{\rm flat}$ is the connection of a given flat principal $\GG$-bundle ${\bf E}$. On the ``2d Coulomb branch'', $\GG$ is broken to its maximal torus $\GH = \prod_a U(1)_a$, and the bundle ${\bf E}$ can be described in terms of $U(1)_a$ line bundles ${\bf L}_{(a)}$, which can be treated as in \eqref{bf L as L}. Let $L_{(a)}$ denote some line bundles in $\Pic(\h\Sigma)$ such that $\pi^\ast(L_{(a)})= {\bf L}_{(a)}$, and let  $\n_{i,(a)}$ denote the corresponding fractional fluxes. Here, the non-trivial step is to identify the correct fractional fluxes for a given $\GG$-bundle ${\bf E}$. We will not discuss that step here.

Let us focus on the case $\GG$ semi-simple and simply-connected.  Upon abelianization, the signs $(-1)^{\bf s}$ in front of \eqref{SU1k explicit} cancel out in that case, consistent with the fact that \eqref{SGGk def} is independent of the spin structure. We then find the explicit formula:
\be\label{SGk explicit nonab}
\exp\left(-S_{\GG_k}^{\rm flat}\right)=    \exp\left(\pi i k h^{ab} \left( {c_1(L_{(a)}) c_1(L_{(b)})\ov c_1(\CL_0)} - \sum_{i=0}^n {t_i \n_{i,(a)} \n_{i,(b)}\ov q_i} \right)\right)~,
\ee
with $h^{ab}$ the Killing form. 
In the next subsection, we compare \eqref{SGk explicit nonab} to known results for lens spaces. This already provides a non-trivial consistency check.

\subsection{The CS action on $L(p,q)$}
Consider the lens space $L(p,q)$, with $p\neq 0$. For simplicity, let us first compare \eqref{SGk explicit nonab} to previous results in the case of a gauge group $SU(2)$. In that case, there is a single variable $u_a=u$ and the Killing form gives $h^{aa}=2$. Considering the Seifert fibration:
\be
L(p,q)  \cong [0~;\, 0~;\, (q_1, p_1)~,\; (q_2, p_2)]~, \qquad
p= p_1 q_2 + p_2 q_1~, \quad q= q_1 s_2 - p_1 t_2~.
\ee
We also define:
\be
q' = q_2 s_1 - p_2 t_1~,
\ee
which satisfies $q q'=1$ mod $p$.
The formula \eqref{SGk explicit nonab} then gives:
\be\label{SU2 CS flat}
S_{SU(2)_k}^{\rm flat}\big(L(p,q)\big)=- 2\pi i k {q' \n_1^2 +2 \n_1 \n_2+ q \n_2^2  \ov p} \mod 2\pi i~,
\ee
with $\n_1, \n_2$ the fractional fluxes. As we explained in Section~\ref{sec:lpqgeo}, any 3d line bundle ${\bf L}$ with first Chern class $\m\in \Z_p$ can be represented by $L_1^\m$ in $\Pic\big(S^2(q_1, q_2)\big)$. We also have the relation $L_2 \cong L_1^{q}$ in $\t\Pic\big(L(p,q)\big)$. Therefore:
\be\label{SU2 from L}
\pi^{\ast}\big(L_1^{\n_1} L_2^{\n_2}\big) = {\bf L}  \qquad\text{with} \qquad c_1({\bf L})= \m= \n_1 + q \n_2 \in \Z_p~.
\ee
Since $\pi_1(L(p,q))$ is abelian, any $SU(2)$ flat connection is labeled by its torsion flux $\m \in \Z_p$. From \eqref{SU2 CS flat} and \eqref{SU2 from L}, we find:
\be
S_{SU(2)_k}^{\rm flat}\big(L(p,q)\big)= - 2\pi i k  {q'\ov p} \m^2 \mod 2\pi i~.
\ee
This is in perfect agreement with known results \cite{jeffrey1992}.
The generalization to any $\GG_k$ is straightforward: 
\be
S_{\GG_k}^{\rm flat}\big(L(p,q)\big)=- \pi i k  {q'\ov p} \Tr(\m^2) \mod 2\pi i~, \qquad \m\in \Gamma_{\GG^\vee}(p)~,
\ee
with $\Tr(\m^2)= h^{ab} \m_a \m_b$ and the flux lattice $\Gamma_{\GG^\vee}(p)\cong \Gamma_{\rm cochar}\otimes \Z_p$. This agrees with the Conjecture 5.6 of \cite{2002math......9403H}.


\section{Seifert operators from holomorphic blocks}
\label{sec:HBA}

In this appendix, we complete the proof of \eqref{HBhlim} and \eqref{HBntblim}. This gives the relation between the holomorphic blocks and the Seifert fibering operators, as discussed in detail in Section \ref{sec:HB}.

\subsection{Proof for the trivial gluing}

We start by considering \eqref{HBhlim}, which reads:
\be\label{HBAhlim}
\lim_{\tau \rightarrow 0} B^\alpha(\nu,-\tau)_{\n_1} B^\alpha(\nu,\tau)_{\n_2} = \cH^\alpha(\nu)^{-1} \, \pif(\nu)^{\m^F}~.
\ee
This corresponds to the case of trivial gluing.  

\paragraph{Trivial gluing for the ungauged blocks.} 
We start by proving the relation \eqref{HBAhlim} on the building blocks of the ungauged theory.  The proof for the chiral multiplet was given in the main text in the case of zero flux.  To incorporate flux, we simply note that:
\be 
B^\Phi(\nu,\tau)_{\n} = B^\Phi(\nu,\tau)_{0} \,(\q \t y;\q)_\n^{-1}~,
\ee
with $\q=e^{2 \pi i \tau}$, $\t y=e^{2 \pi i (\nu+\t\nu^R r)}$, and $(y;\q)_\n$ defined as in \eqref{qpochn}, namely:
\be\label{def xqn}
(x;\q)_\n\equiv   \begin{cases} \prod_{l=0}^{\n-1}\left(1- x q^l\right) &  {\rm if}\;\; \; \n>0~,\\ \\
	\prod_{l=1}^{|\n|} \left(1- x q^{-l}\right)^{-1}	\qquad& {\rm if} \;\; \; \n<0~.								\end{cases}
\ee
  Then we have the limit:
\be \lim_{\tau \rightarrow 0} \;(\q \t y;\q)_\n = (1-\t y)^{-\n} = \pif^\Phi(\nu)^{\n}~, 
\ee
reproducing the ordinary flux operator.  Thus, the extra contribution from the fractional fluxes $\n_1$ and $\n_2$ of the two blocks is simply $\pif^{\Phi}(\nu)^{\n_1+\n_2}=\pif^{\Phi}(\nu)^{\m^F}$, as in \eqref{HBAhlim}.
Consider next the Chern-Simons contributions \eqref{HBCScont} and \eqref{Bgrav def}, which read:
\be \label{HBAcs} B^{\rm GG}(\nu, \tau)_\n= {\theta\big((-1)^{2\t\nu^R} \q^{-\half}; \q\big)\ov \theta\big((-1)^{2\t\nu^R} \q^{\n-\half} y; \q\big)}~,\;\;\;\;\quad\quad
B^{\rm grav}(\tau)= \big((-1)^{2\t\nu^R}  \q^{\half} ; \q\big)^{-1}_\infty~.
\ee
 The gravitational CS term does not contribute directly to the handle-gluing operator, therefore it should drop out in the limit above.  
As we already mentioned, we have:
\be B^{\rm grav}(-\tau) B^{\rm grav}(\tau)=1 \ee
for any $\tau$, and in particular in the limit \eqref{HBAhlim}.  
For the $U(1)$ CS term, we similarly have:
\be
\lim_{\tau \rightarrow 0}  B^{\rm GG}(\nu, -\tau)_{\n_1} B^{\rm GG}(\nu, \tau)_{\n_2} = (-\t y)^{\m^F},
\ee
as expected. This simply follows from the relation  \eqref{HBcommas} for a pair of massive chiral multiplets.
We should also note that fusing the Cartan contribution \eqref{HBcart} gives:
\be
B^{\rm Cart}(-\tau) B^{\rm Cart}(\tau) = \tau^{-\rk}~.
\ee
This factor should be compensated by the measure factor in the integral formula \eqref{HBlenspfsc}.

 Thus, up to this last subtlelty, we find that \eqref{HBAhlim} indeed holds true for the ungauged blocks.
More generaly, for any ``ungauged'' theory, we have the following expansion at small $\tau$:
\be \label{HBAldw} 
B(\nu,\tau)_{\n}  \underset{\tau \rightarrow 0}{\sim} \exp \bigg( \frac{2\pi i}{\tau} \cW(\nu) -  \pi i \Omega(\nu) +2 \pi i \n \d_\nu \CW(\nu) + O(\tau)   \bigg)~.
 \ee
Here, $\cW(\nu)$ is the twisted superpotential and $\Omega(\nu)$ is the effective dilaton.
When we glue two blocks with opposite orientations, as in \eqref{HBAhlim}, the divergences cancel out, and we indeed recover \eqref{HBAhlim}, with $\pif= e^{2\pi i \d_\nu \CW}$ the flux operator and $\cH=e^{2 \pi i \Omega}$  the handle-gluing operator for the ungauged theory.

\paragraph{Trivial gluing for the gauged blocks.} Let us now consider the blocks of the gauged theory. 
Let us first note the following limit, for real $x$ and $\text{Im} (\tau) >0$ \cite{2016arXiv160201085B}:
\be \label{HBAqxlim} 
(\q^x;\q)_\infty \underset{\tau \rightarrow 0}{\rightarrow} \exp \bigg( -\frac{\pi i}{12 \tau} + (\frac{1}{2}-x) \log (-i \tau) + (1-x) \log (2 \pi) - \log \Gamma(x) + O(\tau) \bigg)~.
\ee
In particular, we have:
\be \label{HBABc} 
(\q;\q)_\infty \rightarrow \frac{1}{\sqrt{-i\tau}} \exp \bigg(-\frac{\pi i}{12 \tau} + O(\tau) \bigg)~,
\ee
which gives the limit for $B^{\rm Cart}(\tau)$ as
\be  \label{HBABd} 
B^{\rm Cart}(\tau) \rightarrow \frac{1}{\sqrt{i\tau}} \exp \bigg(\frac{\pi i}{12 \tau} + O(\tau) \bigg)~,
\ee 
The divergent $\tau^{-\half}$ scaling will be important below.

As shown in \cite{Beem:2012mb}, the holomorphic blocks of the gauge theories are obtained by integrating the holomorphic block $\t B$ of the ungauged theory over certain contours, $\Gamma^\alpha$, in the $u$-plane, which are in one-to-one correspondence with the Bethe vacua, $\hat{u}^\alpha$ of the theory:
\be \label{HBAgaugeblock} B^\alpha(\nu,\tau) =\int_{\Gamma^\alpha} du\, \tilde{B}(u,\nu,\tau)~.
 \ee
Here and below we work in the rank-one case, for simplicity.
The contour $\Gamma_\alpha$ is determined by the following conditions:
\begin{itemize}
	\item It passes through the point $\hat{u}^\alpha$, which is a critical point of the twisted superpotential, $\cW(u)$, and near this point the contour follows the path of steepest descent of $\cW(u)$.
	\item It is invariant under the shifts $u \rightarrow u+\tau$. 
	\item It does not cross lines of poles.
\end{itemize}

\begin{figure}[t]
	\begin{center}
		\subfigure[\;]{
			\includegraphics[height=7cm]{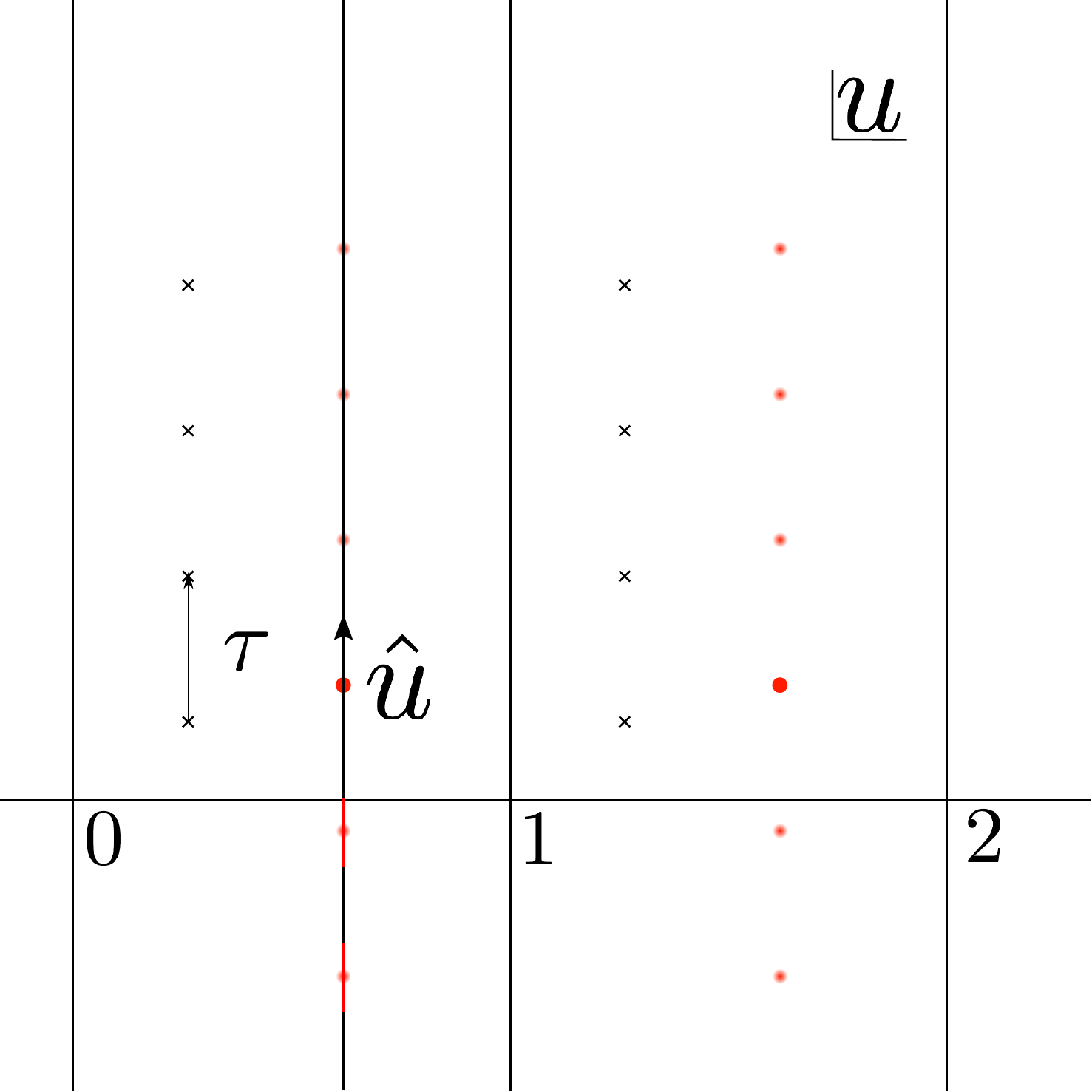}\label{fig:block5}}\qquad
		\subfigure[\;]{
			\includegraphics[height=7cm]{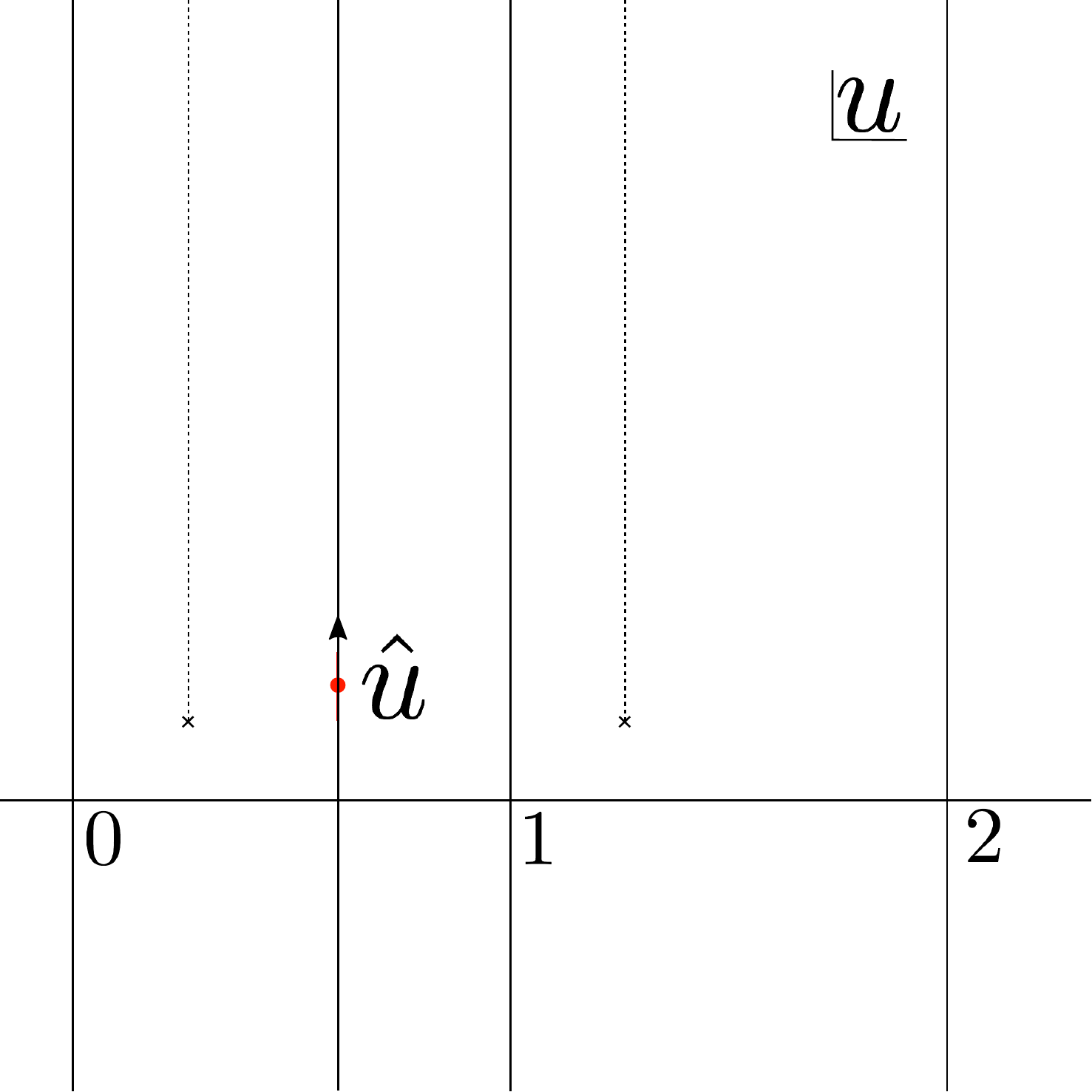}\label{fig:block6}}\quad
		\caption{{On the left, the contour $\Gamma_\alpha$ is shown for finite $\tau$, with towers of poles separated by $\tau$.  As $\tau$ becomes small, the contributions from $\hat{u}$ and its images are dominant, shown in red. On the right, we take the $\tau \rightarrow 0$ limit, where we may approximate the answer by the contribution from $u=\hat{u}$.  Here the towers of poles have collapsed to form the branch cuts of $\cW(u)$.}}
	\end{center}
	
\end{figure} 

\noindent A cartoon of the contour $\Gamma_\alpha$ is shown in Figure \ref{fig:block5}, and we consider the contribution near a single Bethe vacua, $\hat{u}$.  In the limit $\tau \rightarrow 0$, the dominant contribution comes from the neighborhood of the critical point, $\hat{u}_\alpha$, as in Figure \ref{fig:block6}, and we may write:~\footnote{Here and in the following, we omit the background fluxes for flavor symmetries, to avoid clutter. It is straightforward to include them, using \protect\eqref{HBAldw}.}
\be\label{Blimit intermed F}
B^\alpha(\nu,\tau) \;\;\;  \underset{\tau \rightarrow 0}{\longrightarrow}  \;\;\; \frac{1}{\sqrt{i\tau}} \int_{u \approx \hat{u}_\alpha} du\exp \bigg( \frac{2\pi i}{\tau} \cW(u,\nu) - \pi i \Omega(u,\nu) \bigg) \ee
where $\hat{u}_\alpha$ is the $\alpha$th solution to the Bethe equation in the region $0<\text{Re}(u)<1$.   Here we have used the observation \eqref{HBAldw}, and included the extra divergence due to $B^{\rm Cart}(\tau)$, as in \eqref{HBABc}.  The saddle-point approximation at the saddle $u= \h u_\alpha$ becomes exact in this limit. The $\tau^{-1/2}$ prefactor in \eqref{Blimit intermed F} cancels, leaving us with:
\be \label{HBAtza} 
B^\alpha({\nu}, {\tau}) \approx  \bigg(-\frac{\partial^2 \cW}{\partial u^2}\bigg)^{-\half}\bigg|_{u=\hat{u}_\alpha}\exp \bigg( \frac{2\pi i}{\tau} \cW(u_\alpha,\nu) - \pi i \Omega(u_\alpha,\nu)\bigg)~. \ee
A similar argument holds for the block with $\tau \rightarrow -\tau$.  Then, when we fuse these blocks as in \eqref{HBAhlim}, we see that the divergences in $\tau$ cancel out, and we find the finite result:
\be\label{HBAhlimproof}
 \lim_{\tau \rightarrow 0} B^\alpha(\nu,-\tau) B^\alpha(\nu,\tau)  = \bigg(\frac{\partial^2 \cW}{\partial u^2}\bigg)^{-1}\bigg|_{u=\hat{u}_\alpha} e^{-2 \pi i \Omega(\hat{u}_\alpha,\nu)}= \cH^\alpha(\nu)^{-1}~,
\ee
as claimed.
The case of a higher-rank gauge group is a staightforward generalization, by the same saddle-point argument, with the handle-gluing operator given by:
\be
 \cH = e^{2 \pi i \Omega(u,\nu)} \det_{a,b} \frac{\partial^2 \cW}{\partial u_a \partial u_b}~,
  \ee
in the general case. This concludes the proof of \eqref{HBAhlim}.

\subsection{Fibering-operator limit of the holomorphic block}
We now turn to the proof of the limit \eqref{HBntblim}, namely:
\be \label{HBAntblim} \lim_{\tau \rightarrow 0} \frac{B^\alpha_{\t g}(\nu,\tau)_\m}{B^\alpha(\nu,\tau)}= \cG^\alpha_{q,p}(\nu)_\n~. \ee
Recall that $B_{\t g}(\nu,\tau)_\n$ is defined in \eqref{HBBgdef}.  

\paragraph{Ungauged block.}
The proof of \eqref{HBAntblim}  for a chiral multiplet with zero flux was given in Section~\ref{subsec: gen ls}.  For non-zero flux, we again note that:
\be 
B^\Phi_{\t g}(\nu,\tau)_\m =(\t\q^{1+\m} \t y;\t \q) = (\t \q \t y,\t \q)_\m^{-1}\, B^\Phi_{\t g}(\nu,\tau)_{0}~,
 \ee
where $\t\q=e^{2 \pi i \frac{s \tau-t}{p \tau+q}}$ and $\t y=e^{\frac{2 \pi i \nu}{p \tau+q}}$, and $(x;\q)_\m$ is defined as in \eqref{def xqn}.~\footnote{Here, we are also setting $\t\nu^R=0$ to avoid clutter.} Then, the flux contribution gives an extra finite piece in the limit \eqref{HBAntblim}, in addition to \eqref{limit to GPhi of hb}. We find:
\be
\lim_{\tau \rightarrow 0}    (\t \q \t y,\t \q)_\m^{-1} = (e^{2\pi i (\nu-t)\ov q}; e^{-{2\pi i t\ov q}})_\n^{-1}\, =\,
(e^{2\pi i \nu\ov q} ; e^{2\pi i t\ov q})_{-\n}= \pif_{q,p}^\Phi(u)_\n~,  
\ee
reproducing the fractional-flux contribution  \eqref{qpochn}. Therefore, we indeed find:
\be	\lim_{\tau \rightarrow 0} \frac{B^\Phi_{\t g}(\nu,\tau)_\n}{B^\Phi(\nu,\tau)} = \cG^\Phi_{q,p}(\nu)_{\n}~. \ee
The proof for the Chern-Simons contribution can be shown using the behavior of the Jacobi theta function under modular transformations, or more simply from their relation to two chiral multiplets of opposite charges, as noted above.  

The contribution from the vector multiplets and gravitational CS term can also be worked out straightforwardly.  
It is interesting to consider the Cartan component contribution in detail. Consider the modular transformation properties of the Dedekind eta function:~\footnote{This is valid for $p>0$. The transformation for $p<0$ can be obtained by sending $(p,q,s,t)$ to $(-p,-q,-s,-t)$ on the right-hand-side of \protect\eqref{eta modular}.}
\be\label{eta modular}
\eta\Big({s\tau - t\ov p \tau +q}\Big) = e^{-\pi i \left({p\ov q} - s(p,q)\right) } e^{{\pi i \ov 12} \left({s\ov p} - {1\ov p q}\right)} \, \sqrt{p \tau + q} \; \eta(\tau)~.
\ee
Using the fact that $\eta(\tau) = \q^{-1/24} (\q;\q)_\infty$, we find:
\be \label{qqmod limit} 
(\t q; \t q)_\infty \;\underset{\tau \rightarrow 0}{\rightarrow}  \;
\sqrt{q}\left(\cG_{q,p}^{(0)}\right)^{-1} \; (-i \tau)^{-1/2} \exp \bigg( - \frac{\pi i}{12 \tau} + O(\tau) \bigg)~, 
\ee
with $\cG_{q,p}^{(0)}$ defined as in \eqref{def CG0}.

\paragraph{Gauged blocks.} Next, let us consider the holomorphic blocks of a gauge theory. We again consider $\GG$ of rank one.
 We have:
\be B_{\t g}^\alpha(\nu;\tau)_{\n^F} = \int_{\Gamma_\alpha} d{u} B_g(u,\nu;{\tau})_{0,\n^F}=  \int_{\Gamma_\alpha} d\tilde{u} B(\tilde{u},\tilde{\nu};\tilde{\tau})_{0,\n^F} \ee
where we identify $\tilde{u}=\frac{u}{p\tau+q}$ and $\tilde{\tau}=\frac{s\tau-t}{p\tau+q}$.  Here, by convention, we work at zero gauge flux, which is allowed since the contour is invariant under shifts $\tilde{u} \rightarrow \tilde{u}+\tilde{\tau}$ by assumption.  The contour at finite $\tau$ is shown in Figure \ref{fig:block7}.

\begin{figure}[t]
	\begin{center}
		\subfigure[\;]{
			\includegraphics[height=7cm]{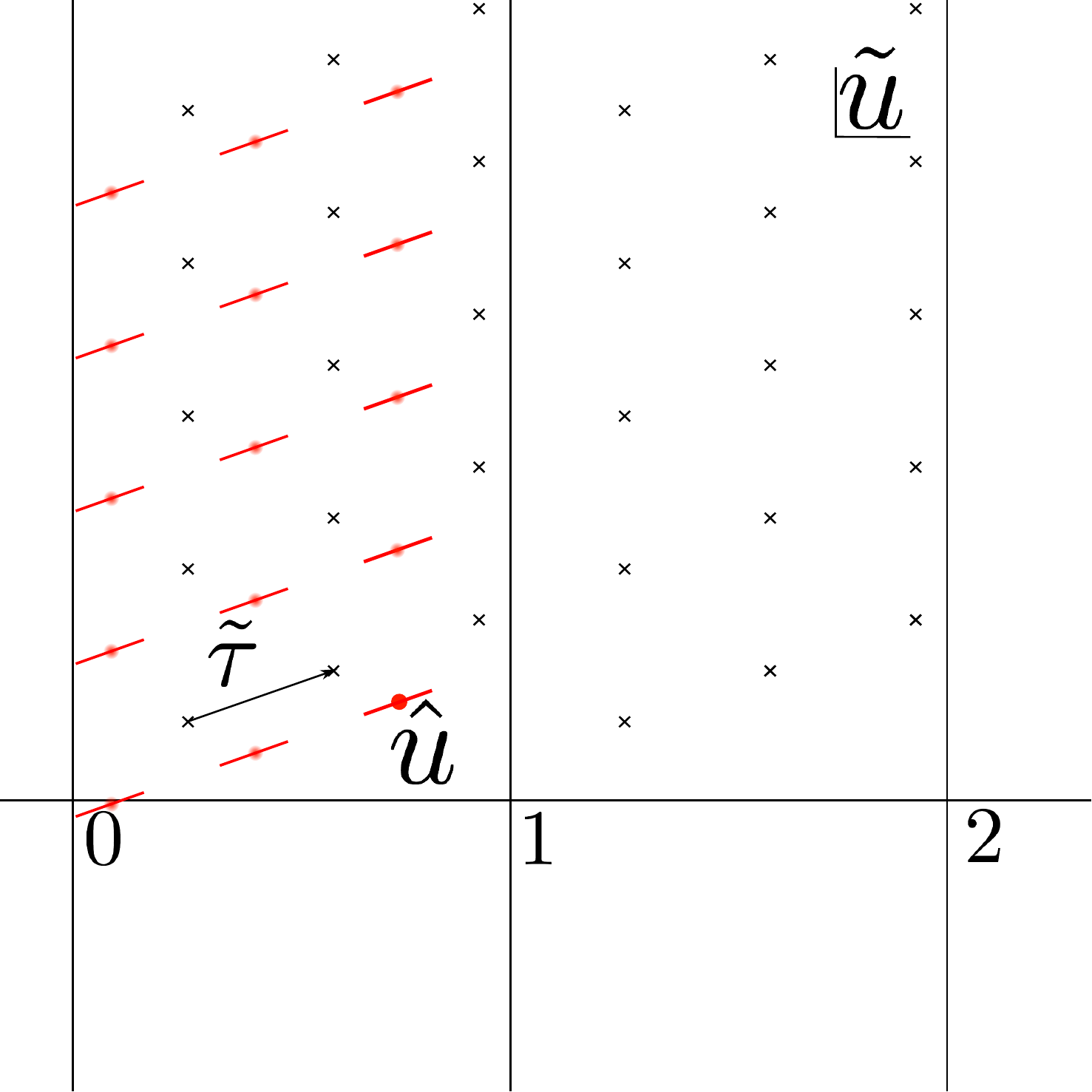}\label{fig:block8}}\qquad
		\subfigure[\;]{
			\includegraphics[height=7cm]{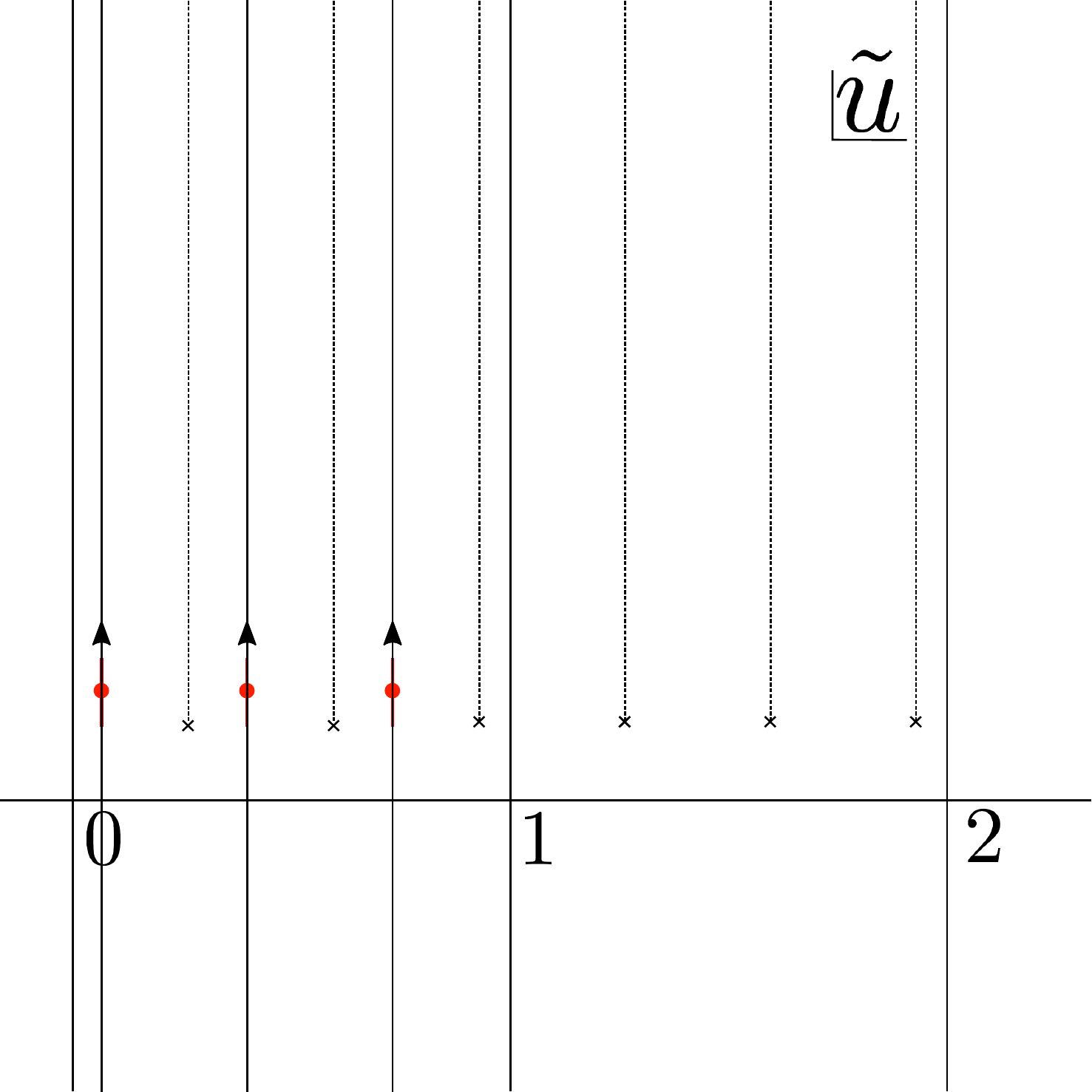}\label{fig:block9}}
		\caption{{In the limit $\tau \rightarrow 0$, the dominant contributions to the integral over $\Gamma_\alpha$ come from the regions around $\hat{u}$ and their images.  These can be reassembled into a series of $q$ shifted copies of contours passing through the critical points $u=q\tilde{u} = \hat{u}, \hat{u}+1,...,\hat{u}+q-1$.}}
	\end{center}
	
\end{figure}

As shown in Figure \ref{fig:block8}, as $\tau \rightarrow 0$, this contour can be deformed to one that passes through $q$ images of the Bethe vacua at $\hat{u}_\alpha$:
\begin{figure}[t]
	\begin{center}
		\includegraphics[height=7cm]{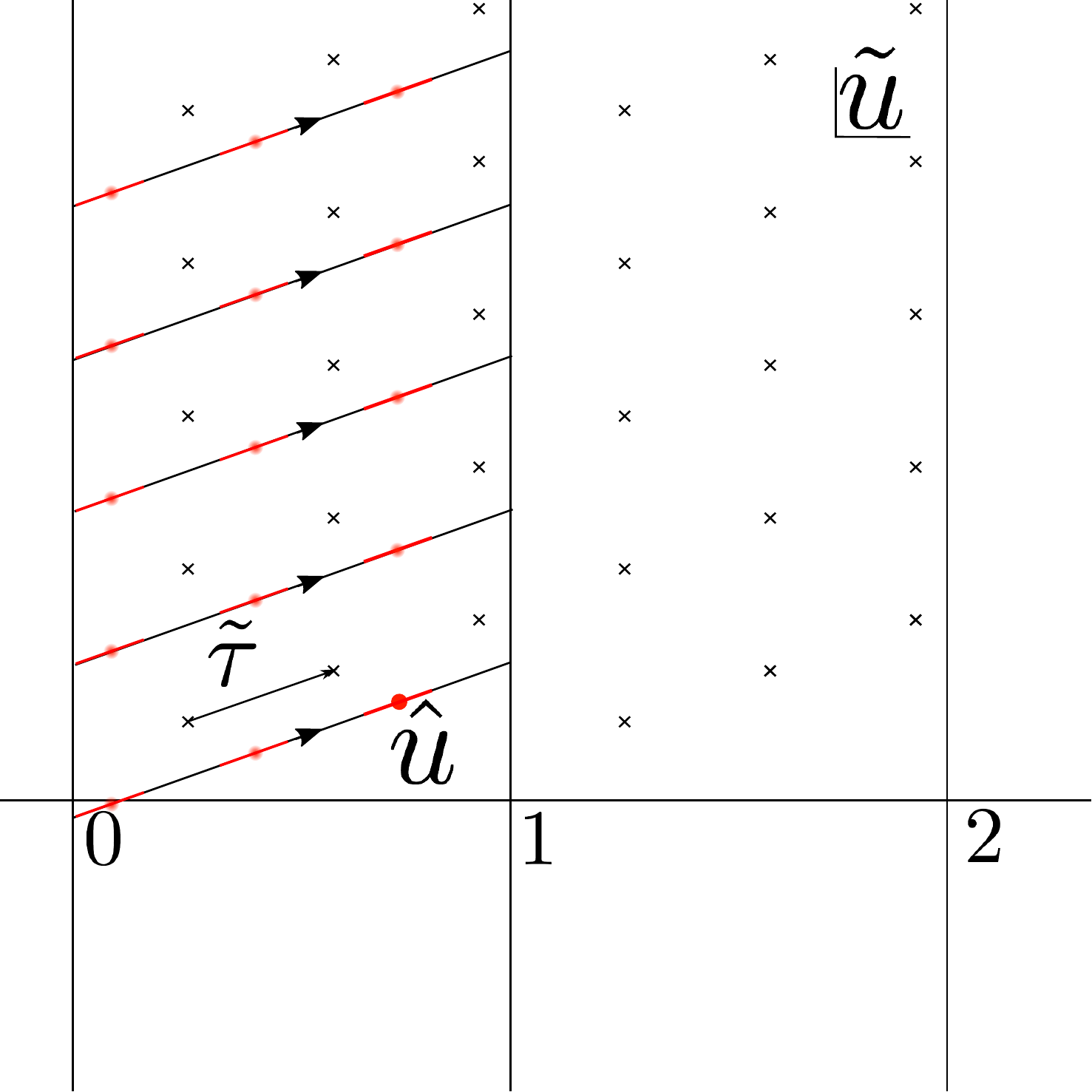}
		\caption{{Contour $\Gamma_\alpha$ corresponding to a block $B_{\t g}$ with non trivial $(q,p)$, at finite $\tau$.\label{fig:block7}}}
	\end{center}
	
\end{figure} 

\be \label{HBAtusp} \tilde{u} = \frac{\hat{u}_\alpha+j}{q}, \;\;\; j=0,...,q-1 \ee
For the ungauged block, we can use the result above to write:
\be \label{B as BG interm}
B(\tilde{u},\tilde{\nu};\tilde{\tau})_{0,\n^F} \sim q\, B(u,\nu;\tau)\, \cG_{q,p}(u,\nu)_{0,\n^F}~, \quad \;\;\; \text{as}\;\tau \rightarrow 0~.
\ee
Here, the prefactor of $q$ on the right-hand-side correspond to the fact that we have a factor of $1/\sqrt{q}$ in \eqref{qqmod limit}.~\footnote{This is also modulo a factor of $(\CG^{(0)}_{q,p})^2$, which is a matter of convention---unlike $\CG^{(0)}_{q,p}$ itself, the factor $(\CG^{(0)}_{q,p})^2$  corresponds to properly quantized background CS terms $k_{RR}= 1$ and $k_g=2$, and thus depends on our choice of quantization for the gaugino.}
By a similar argument as above, we find:
\be B_{\t g}^\alpha({\nu};{\tau})_\n =
\int_{\Gamma_\alpha} d\tilde{u}\, B(\tilde{u},\tilde{\nu};\tilde{\tau})_{0,\n} \, \underset{\tau \rightarrow 0}{\approx}   \int_{\hat{\Gamma}_\alpha} d{u} \,B(u,\nu;\tau)\, \cG_{q,p}(u,\nu)_{0,\n}~,
 \ee
where, in the second equation, we changed variables from $\tilde{u}$ to $u$, incurring a factor of $q^{-1}$, which precisely cancels the prefactor in \eqref{B as BG interm}.  Performing the saddle point approximation about the points \eqref{HBAtusp}, the factor of $\tau^{-1/2}$ cancels as in \eqref{HBAtza}, and we obtain:
\bea
&B_{\t g}^\alpha({\nu};{\tau})_{\n^F} & \sim&\; \sum_{j=0}^{q-1} B({u_\alpha}-j,{\nu};{\tau}) \, \cG_{q,p}(u_\alpha-j,\nu)_{0,\n^F}  \cr
& &=&\; B^\alpha({\nu};{\tau}) \sum_{\n=0}^{q-1} \cG_{q,p}(u_\alpha,\nu)_{\n,\n^F}~, 
\eea
where we used the fact that $B(u,\nu;\tau)$ is invariant under $u_\alpha \rightarrow u_\alpha-j$, together with the relation $\cG_{q,p}(u_\alpha-j,\nu)_{0,\n^F} = \cG_{p,q}(u_\alpha,\nu)_{p j,\n^F}$, 
to replace the sum over $j$ by a sum over fractional fluxes.  Finally, we divide by the trivial gauged block, obtaining:
\be
 \lim_{\tau \rightarrow 0} \frac{B^\alpha_{\t g}(\nu;\tau)_{\n^F}}{B^\alpha(\nu;\tau)} = \sum_{\n=0}^{q-1}  \cG_{q,p}(\h u_\alpha,\nu)_{\n,\n^F}= \cG^{\alpha}_{q,p}(\nu)_\n~.  \ee
This expression for the ``on-shell'' gauged $(q,p)$ fibering operator, including the sum over fractional fluxes, agrees nicely with the discussion in Section~\ref{sec: Bethe vac sum}. This completes the proof of \eqref{HBAntblim}.

\bibliographystyle{utphys}
\bibliography{bib3d}{}

\end{document}